\documentclass[a4paper,twoside,12pt]{book}

\usepackage{psfig}
\usepackage{graphicx}
\usepackage{afterpage}
\usepackage{psfrag}
\usepackage{epsf}
\usepackage{picinpar}

\RequirePackage{multicol}
\def\lsi{\raise0.3ex\hbox{$<$\kern-0.75em\raise-1.1ex\hbox{$\sim$}}}
\def\gsi{\raise0.3ex\hbox{$>$\kern-0.75em\raise-1.1ex\hbox{$\sim$}}}
\newcommand{\lsim}{\mathop{\lsi}}
\newcommand{\gsim}{\mathop{\gsi}}
\newcommand{\aob}{\mbox{\footnotesize{a,obs}}}
\newcommand{\asc}{\mbox{\footnotesize{a,src}}}
\newcommand{\sob}{\mbox{\footnotesize{s,obs}}}
\newcommand{\ssc}{\mbox{\footnotesize{s,src}}}
\newcommand{\src}{\mbox{\footnotesize{src}}}
\newcommand{\obs}{\mbox{\footnotesize{obs}}}
\newcommand{\fa}{\mbox{\footnotesize{a}}}
\newcommand{\fs}{\mbox{\footnotesize{s}}}
\newcommand{\fm}{\mbox{\footnotesize{max}}}
\newcommand{\off}{\mbox{\footnotesize{off}}}
\newcommand{\sdp}{\subset\hspace{-0.42cm}\times}

\renewcommand{\thefootnote}{\fnsymbol{footnote}}

\begin{document}

\begin{titlepage}
\centering
       {\LARGE {\bf Astronomical Tests of the Einstein}}
        
       \vspace{0.5cm}
       {\LARGE {\bf Equivalence Principle}}\\
	    \vspace{3cm}
       {\Large Dissertation}\\[0.3cm]
       for the degree of \\
       Doctor of Natural Science (Dr.rer.nat.)
   
            \vspace{1cm}
       Presented by \\[0.5cm]	    
       Oliver Preu\ss \footnote[2]{Present address: Max-Planck-Institut f\"ur Aeronomie,
       Max-Planck-Strasse 2, 37191 Katlenburg-Lindau, Germany. Email: opreuss@linmpi.mpg.de}
       \\Universit\"at Bielefeld\\Fakult\"at f\"ur Physik \\[2cm]
       November 2002
   	
\end{titlepage}
\renewcommand{\thefootnote}{\arabic{footnote}}
\thispagestyle{empty}

\vspace*{20cm}
Gedruckt auf alterungsbest\"andigem Papier $^{\circ\circ}$ ISO 9706
\newpage

\thispagestyle{empty}
\begin{window}[0,l,{\hspace{0mm}\psfig{file=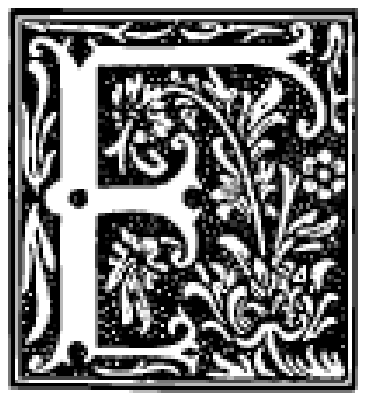}},{ }] 
 \noindent {\sc rom this fountain (the free will of God) it is those laws, which we call 
                the laws of nature, have flowed, in which there appear many traces of the 
		most wise contrivance, but not the least shadow of necessity. These therefore 
		we must not seek from uncertain conjectures, but learn them from observations 
		and experimental. He who is presumptuous enough to think that he can find the 
		true principles of physics and the laws of natural things by the force alone 
		of his own mind, and the internal light of his reason, must either suppose the 
		world exists by necessity, and by the same necessity follows the law proposed; 
		or if the order of Nature was established by the will of God, the [man] himself, 
		a miserable reptile, can tell what was fittest to be done.}
\end{window}		
  \vspace{1cm}
  \hspace*{8cm} {\sc Isaac Newton}

\newpage  
\thispagestyle{empty}
\vspace*{10cm}

\hspace*{7cm}{\large {\sc Dedicated to my parents}}

\newpage
\pagenumbering{roman}
\tableofcontents
\chapter*{Summary}
\pagenumbering{arabic}  
\addcontentsline{toc}{chapter}{Summary}
  
  \vspace*{2cm}
  \noindent Based on the assumption of the Einstein 
  equivalence principle and the principle of general
  covariance general relativity describes the gravitational field 
  successfully as a purely geometrical property of  
  four dimensional spacetime on Riemannian manifolds.   
  However, despite the so far remarkable accuracy in 
  its experimental verification, general relativity 
  remains a classical theory. So the necessity of finding 
  a quantum mechanical description of the gravitational 
  field implies the need to embed or to modify the above 
  principles in a more general framework, which is one of 
  the major challenges in modern theoretical physics.
  
  In this thesis we investigate specific predictions of such a
  class of more general theories, the so called nonmetric 
  theories of gravity. Within this framework theories based on
  e.g., a metric-affine geometry of spacetime predict that 
  in a gravitational field a pair of orthogonal linear 
  polarisation states of light propagates with different 
  phase velocities. This gravity-induced birefringence could 
  in principle be measured in local test experiments and, 
  hence, violates the Einstein equivalence principle.   
  Therefore we have used polarization measurements in solar spectral 
  lines, as well as in continua and lines of various isolated 
  magnetic white dwarfs and of cataclysmic variables
  (interacting binary systems) to constrain the essential coupling 
  constants for this effect predicted by metric-affine gravity and 
  other prototypes of nonmetric theories. These measurements provide
  an empirical formula which predicts the upper limit on the metric-affine coupling
  constant, measured for a particular celestial body as a function of its
  Schwarzschild radius and its physical, stellar radius. By modelling the lightcurves
  of a certain cataclysmic variable system, the results could, in principle,
  be interpreted as a direct detection of gravitational birefringence,
  although alternatives exist.
   
  This thesis provides the first systematic search for signals of gravitational 
  birefringence in astronomical polarimetric data. As an outlook I propose 
  further promising tests which also have the potential for setting strong upper 
  limits on gravity-induced birefringence.

\newpage
\thispagestyle{empty}

\chapter{Introduction}
    The Einstein equivalence principle plays the role of a key element in the development
    of new improved theories of gravity. Although being an important building block in Einstein's 
    general relativity, theoretically predicted violations of its validity are an important feature in 
    alternative, nonmetric gravitation theories if they are to incorporate quantum mechanical 
    principles. Hence, the intention of this chapter is to motivate the conviction, grown 
    within the last few years, that violations of the equivalence principle must be an essential 
    part of every theory of gravity which pays attention to the quantum mechanical character of matter.
    
    After a brief historical outline of the weak and the Einstein equivalence principle and its
    implications, this chapter presents a theoretical framework which admits the analysis as well as the 
    development of experimental tests for a broad class of gravitation theories. This purpose requires a
    critical examination of the underlying, mostly classical, concepts and notions. So to say as a side effect
    one is led to the possibility of looking at EEP violations as violations of spacetime symmetries in the 
    spirit of modern quantum field theory. Indeed the principle of gauge symmetries, taken from 
    the Standard model of elementary particle physics is used as a cornerstone in the mathematical 
    formalism of the metric-affine gauge theory of gravity (MAG), which is the second theory considered
    here besides Moffat's nonsymmetric gravitation theory (NGT). Metric-affine theories can be regarded as
    extensions of Einstein-Cartan type theories. Becoming nonmetric when the additional gravitational
    potentials couple directly to matter, MAG as well as NGT predict that a gravitational field singles 
    out an orthogonal pair of polarization states of light that propagate with different phase velocities.
    This gravity-induced birefringence implies that propagation through a gravitational field can alter
    the polarization of light and, so, violate the Einstein equivalence principle. Quantitative predictions
    for this phase shift are given which are used in the following chapters for setting strong limits
    on this effect by utilizing astrophysical spectropolarimetry of compact stellar objects. 
    
    The polarization of an electromagnetic wave is completely and consistently described by a system of
    four real valued quantities, called Stokes parameters. Since the possible influence of gravity-induced
    birefringence on polarized light shows up in an alteration of these parameters, a brief introduction 
    to this topic is given at the end of this chapter.

  \newpage
  \section{Equivalence Principles}
    The significance of the principles of equivalence for the development of modern physics 
    can hardly be overestimated. For example, Galilei's famous free fall experiments performed from
    the leaning tower of Pisa marked the beginning of the development from the medivial, aristotelic 
    way of science, up to modern physics \cite{galilei}. Also, Newton realized very soon that
    his new ideas about the principles of motion and a universal gravitational force are basically
    founded on the equivalence between gravitational and inertial mass, so that he performed numerous
    pendulum experiments to have an experimental justification for his new laws. The importance 
    this equivalence had for Newton can easily be estimated from the fact that he later devoted 
    the opening paragraph of his {\em Principia} \cite{newton} to it.     
    What Newton and also Galilei had introduced into modern physics is today known as the    
    
    \vspace{0.5cm}
    \noindent\fbox{\begin{minipage}{\textwidth}
    \begin{center}
    
    \vspace{0.5cm}
    {\bf Weak Equivalence Principle (WEP):} \\
    In a gravitational field all bodies fall with the same acceleration regardless of their
    mass or internal structure.
    
    \vspace{0.5cm}
    \end{center}
    \end{minipage}}
    \vspace{0.5cm}
   
    The weak equivalence principle currently belongs to the physical predictions with the most accurate 
    empirical underpinning. Beginning with the torsion balance experiments, performed by Baron von
    E\"otv\"os and collaborators in the 19th. century an accuracy of approximately $10^{-9}$ is currently 
    reached (in comparison to $10^{-3}$ of Newton's pendulum tests), while an accuracy of $10^{-12}-10^{-15}$ 
    is theoretically expected for free fall experiments in orbit, e.g. for STEP \cite{step}. Thus the 
    refinement of WEP tests still continues. For a detailed summary of the current experimental status and 
    historical overviews see \cite{will1}.
 
    Historically, the next important theoretical development after Newton with respect to the 
    equivalence principle was given by Einstein in the context of his theory of general relativity in
    1915 \cite{einstein}. While the WEP formulated in the language of special relativity demands that in a
    sufficiently small, free falling laboratory all {\em mechanical laws} of physics are the same as if 
    gravity is absent, Einstein generalized this statement from mechanical to {\em all laws} of
    physics which is formulated as the

    \vspace{0.5cm}
    \noindent\fbox{\begin{minipage}{\textwidth}
    
    \begin{center}\vspace{0.5cm}
      {\bf Einstein Equivalence Principle (EEP):} \\
      All physical laws of special relativity are valid in the presence of a gravitational field
      in an infinitesimally small, free falling laboratory.
    
    \vspace{0.5cm}    
    \end{center}
    \end{minipage}}
    \vspace{0.5cm}
   
    Together with the principle of general covariance, the EEP provides the foundation of general
    relativity and hence of the idea that gravity is a phenomenon of curved spacetime. For 
    this reason the necessity of having a solid experimental verification as well as a critical
    analysis of the underlying concepts and possible connections between the WEP and the EEP becomes 
    very clear. These topics will be developed within the next sections. 
   
    \subsection{Schiff's Conjecture}
      The clear distinction that was made in the early days between the basic concepts of the 
      WEP and the EEP has become increasingly blurry today. Test masses are composed of atoms where the 
      constituents, the protons, neutrons and electrons, interact via the mass-energy of the 
      electromagnetic, strong and weak interaction. Validity of the WEP in this context implies 
      that all nongravitational fields couple in the same universal way to gravity so that, finally, 
      measurements of the freefall accelerations turn out to be profound tests of the 
      EEP as well as gravitational redshift measurements. This plausibility argument also further 
      supports Schiff's Conjecture, originally invented by Leonard I. Schiff in 1960 
      \cite{schiff}: {\em Any complete, self-consistent theory of gravity that embodies WEP 
      necessarily embodies EEP.}
      
      Here a theory is defined as being ''complete'' when it is capable of making definite 
      predictions about the result of any experiment, within the scope of a theory of gravity, 
      that the current technology is able to perform. In this sense Milne's kinematic relativity 
      \cite{milne} must be considered as an uncomplete theory since it makes no gravitational 
      redshift prediction. A theory is called ''self-consistent'' if the prediction for the 
      outcome of any experiment within its scope is unique and does not depend on the way  
      it was derived. According to this definition Kustaanheimo's various vector theories 
      \cite{kust} must be seen as being inconsistent since the results for light propagation 
      are different for light viewed as waves and for light viewed as particles. Following 
      the argumentation above general relativity provides an example of where Schiff's 
      conjecture is validated since it describes gravity by a second-rank symmetric tensor 
      $g_{\mu\nu}$ to which all matter fields couple universally.

      It is obvious that Schiff's conjecture, if valid, would have a strong impact on gravitational 
      research since, e.g., E\"otv\"os experiments could then be seen as direct tests of the EEP and
      the idea of gravity as a phenomenon of curved spacetime. It is generally recognized, that a 
      rigorous proof in a mathematical sense of such a conjecture is impossible, since such a proof
      would require an at least moderately deep understanding of all gravitation theories that satisfy 
      WEP, including those not yet invented, and never destined to be invented \cite{ll73}. Nonetheless, 
      a number of plausibility arguments have been formulated within the past decades. The most 
      general and elegant of these consists of a simple cyclic gedanken experiment under the 
      assumption of energy conservation and was first formulated by Dicke in 1964 \cite{dk64} 
      and subsequently developed by Nordvedt (1975)\cite{nd75} and Haugan (1979)\cite{hg79}. 
      A more qualitative argumentation given by Thorne, Lee and Lightman \cite{tll73} is founded 
      on Lagrangian-based theories of gravity and is similar to the one mentioned above.   
  
  \section{Theoretical context for analyses of EEP tests}
    The relativity revolution and the quantum revolution are certainly among the greatest
    successes of 20th century physics. Both have changed our view of space, time and matter in 
    a radical way but the underlying concepts of the theories they produced are unfortunately 
    fundamentally incompatible. For example general relativity is a purely classical theory where
    each particle has simultaneously a definite position and momentum in a given spacetime point,
    whereas quantum mechanics tells us that this is only approximately true for macroscopic
    objects within a region where the spacetime curvature is much greater than the position
    uncertainty.       
    This conceptual tension becomes even more obvious by including the weak equivalence principle.
    The WEP, statet in an alternative formulation, says that {\em if an 
    uncharged test body is placed at an initial event in spacetime and given an initial velocity 
    there, then its subsequent trajectory will be independent of its internal structure and 
    composition} \cite{dk64}. Here, an uncharged test body is meant to describe an electrically neutral 
    test mass with negligible self-gravitational energy that is small enough that inhomogenities 
    and therefore tidal effects of the external gravitational field can be ignored. So, given 
    two test masses which should be used for testing the WEP, the locality of this principle 
    requires that the volume of space between both trajectories must go to zero before the 
    statement becomes exact \cite{chiao}. That is exactly the point where the WEP comes into 
    conflict with the uncertainty principle since this limiting process causes an infinite 
    uncertainty in their momenta and, hence, makes any prediction about the trajectory 
    impossible. A simple gedanken experiment which reveals a violation of the WEP in this 
    sense was given by R. Chiao in \cite{davis}: Two perfectly elastic balls with different 
    chemical compositions were dropped from the same height above a perfectly elastic table. 
    WEP predicts that the vertical trajectories as well as the subsequent oscillations are 
    identical and indistinguishable since the total amount of kinetic and potential energy 
    remains constant. However the time-energy uncertainty in quantum mechanics indeed allows 
    the balls to propagate into the classicaly forbidden regions above their turning points. 
    Since tunneling depends on the mass and therefore on the chemical composition of the object 
    this effect would represent a quantum violation of the weak equivalence principle.          
    This example of a possible quantum violation of the WEP underlines again the discomfort many
    physicists feel with having two fundamental theories, both so far experimentally verified with
    an outstanding precision but without a satisfying common interface. It would certainly go beyond 
    the scope of this thesis to summarize, even in a brief way, all the approaches physicists and
    philosophers have taken within the last 70 years to resolve this conflict. The interested reader 
    may therefore have a look at the exellent reviews by, e.g., Rovelli \cite{ro00} and Carlip 
    \cite{ca01} which also provide a huge list of references for further reading.   
    
    However, from the viewpoint of a relativist a key role in understanding the unification problem is 
    certainly played by the EEP and its incorporated WEP (see \cite{will1,will2}). Since theories 
    which predict violations of the EEP are numerous and the experimental guidance is so far negligible, 
    it is important to establish a systematic theoretical framework for analysing various experiments 
    and theoretical concepts of different gravitational theories. For this reason the first step must
    consist of providing careful definitions of general concepts and notions that every valid theory 
    of gravity has to obey. This procedure can easily be regarded as pedantic and even as superfluous 
    since most of the notions from everyday gravitational physics and experience seems to be more than 
    obvious without further need of explanations. Indeed the next section will show that even the 
    distinction between what is a gravitational and what is a nongravitational phenomenon is highly
    nontrivial. Taking into consideration that every theory in physics for historical reasons is build 
    up on notions of everyday experience, one has to be very careful by using these concepts in more
    sophisticated theories. Nevertheless, from this starting point general schemes for analysing gravitation 
    theories will be developed. This schemes encompass that class of theories which predict violations 
    of the EEP, which are relevant to this work.
    \subsection{Basic concepts and notions} 
      {\sc Spacetime and Gravitational Theories:} Following the notions and definitions given by 
      Thorne, Lee and Lightman \cite{tll73}, all
      gravitation theories can be regarded as a subclass of the more general spacetime theories.
      A spacetime theory basically possesses a mathematical formalism which is constructed from
      a 4-dimensional manifold and from geometric objects defined on that manifold \cite{trm65}.
      Two different mathematical formalisms will be called {\em different representations of the 
      same theory} if the predictions they produce are identical for every experiment. The geometric 
      objects of a particular representation are called its {\em variables}. The equations which 
      these variables have to satisfy will be called the {\em physical laws} of the representations,
      e.g. in the case of general relativity the physical laws are the Einstein field equations.
      
      \noindent {\sc Gravitational phenomenon:} Certainly, the above general scheme is able to 
      encompass a rich variety of different theories
      for various physical phenomena. To restrict ourself to gravitation theories one simply has
      to demand that the physical laws the spacetime theory provides, must correctly match with
      generally recognized laws based entirely on gravitational phenomena like Keppler's law.
      This definition immediately requires a clear distinction between what is a gravitational 
      and what is a nongravitational phenomenon. Already Thorne, Lee and Lightman \cite{tll73} 
      mentioned that there seems to be a variety of ways in which such a distinction could be made.
      They suggested to define gravitational phenomena as {\em ''those which are either absolute or 
      'go away' as the amount of mass-energy in the laboratory (isolated from external influences)
      decreases''}. In other words, they suggested that gravitational phenomena are either prior
      geometric effects or generated by mass-energy. Concerning the first issue one has to reply
      that the interpretation of gravity as a geometric phenomenon is entirely based on the validity
      of the EEP and, so, is inappropriate for a general theory of gravitational theories. Concerning 
      the second point it is important to note that this definition also includes electromagnetic 
      phenomenon since the electromagnetic charge is so far restricted to massive elementary particles.
      If the amount of mass-energy could be totally removed from a shielded laboratory, then also all
      electromagnetic phenomena would vanish. It is therefore more appropriate to define the 
      classical gravitational phenomena as those which are generated purely by mass-energy, regardless 
      of charges. Taking also quantum mechanical properties of matter like charges and spins into 
      consideration could therefore certainly lead to a modification of the above definition, important for
      quantum gravity approaches.

      \noindent {\sc Local nongravitational test experiment:} An experiment performed in an arbitrary
      spacetime point is called local and nongravitational if the following conditions are satisfied      
      \begin{itemize}
        \item  Performed in the center of a freely-falling laboratory.
        \item  Inhomogenities of the external field can be ignored.
        \item  Self-gravitational effects are negligible.
      \end{itemize}  
      In addition the laboratory must be impermeably shielded against external electromagnetic and other
      (real or virtual) particle fields. To make sure that external inhomogenities of the gravitational
      fields are unimportant, one has to perform a sequence of experiments with decreasing size, until
      the experimental results reaches asymtotically a constant value. An example of a local 
      nongravitational test experiment is a measurement of the fine structure constant $\alpha$, while a
      Cavendish experiment is not.
    \subsection{The EEP revised}  
      Using the last definition it is possible to give an alternative formulation of the EEP which is
      capable of providing a large variety of new sophisticated tests and also reveals the important and
      far reaching symmetries which are inherent to the EEP.  
 
      \vspace{0.5cm}
      \noindent\fbox{\begin{minipage}{\textwidth}
      \begin{center}
    
      \vspace{0.5cm}
      
        {\bf Einstein Equivalence Principle:}
        \begin{enumerate}
          \item WEP is valid.
          \item The outcome of {\em any} local nongravitational experiment is independent
                of the velocity of the freely-falling reference frame in which it is performed.
          \item The outcome of {\em any} local nongravitational experiment is independent of 
                where and when in the universe it is performed.	    
        \end{enumerate}
      
      \vspace{0.5cm}
      \end{center}
      \end{minipage}}
      \vspace{0.5cm}
    
    The second aspect in this definition demands that in two frames moving relative to each other,
    all the nongravitational laws of physics must make the same predictions for identical experiments.
    This is therefore called {\sc Local Lorentz Invariance} (LLI) \cite{will1,gh90}. The third point in the 
    above formulation of the EEP requires a homogenity of spacetime since the outcome of any local
    nongravitational test experiment must be independent of the spacetime location of the laboratory and
    is therefore called {\sc Local Position Invariance} (LPI). It is important and interesting to make clear 
    that the local position invariance not only refers to the position in space but also to the position 
    in time. Validity of LPI forces the fundamental constants of nongravitational physics like the
    fine structure constant $\alpha$ or the weak and strong interaction constants not to change
    throughout the lifetime of the universe. For a detailed review and references on this topic see
    \cite{dy72,dy78}.
    
    The time variation of fundamental nongravitational contants or the gravity-induced 
    birefringence of light in gravitational fields, which is the main topic in this thesis, 
    are only two examples of tests of certain aspects of the EEP. Indeed many of the 
    experiments in gravitational and nongravitational physics are direct tests of the 
    symmetries defined by the principles of equivalence. Nongravitational test experiments 
    respond to their external gravitational environment during their free-fall and, so, the 
    presence or absence of local Lorentz or local position invariance is entirely determined
    by the form of the coupling of the gravitational field to matter \cite{gh90}. Therefore,
    after defining the group of gravitational theories which incorporate the EEP, a formalism
    capable of representing the coupling between gravitational and matter fields for a whole class
    of gravitational theories will be presented. For this purpose, Lagrangian field theory provides
    a natural setting for general considerations. 
        
  \subsection{Metric theories of gravity}
     The validity of the EEP is a crucial distinctive feature regarding the classification of
     various gravitational theories. If EEP is valid then, according to the second and the third
     point of the definition given in Sect.1.2.2, the laws of physics which govern a certain 
     experiment must be independent of the velocity of the free falling laboratory (local Lorentz 
     invariance) and also of its position in spacetime (local position invariance) which demands 
     time-independent physical constants. The only laws which are known to satisfy these requirements 
     are those of special relativity. According to the first point, validity of WEP, it follows 
     that test bodies within the laboratory are moving unaccelerated on locally straight lines 
     which can be regarded as geodesics of a metric ${\bf g}$, i.e. geodesics in a curved spacetime 
     \cite{will1,will2}. These inferences from the EEP are commonly summarized in the
       
     \vspace{1cm}
      \centerline{{\bf Metric Postulates}}
    
    \begin{itemize}
      \item Spacetime is endowed with a symmetric metric {\bf g}.
      \item The trajectories of freely falling bodies are geodesics of that metric.
      \item In local freely falling reference frames, the nongravitational laws of physics 
            are those of special relativity.
    \end{itemize}
            
    \vspace{1cm}
    Every theory which embodies the EEP necessarily includes the metric postulates and is 
    therefore called a {\em metric theory} of gravity. As a consequence, in every metric theory
    all nongravitational fields couple in the same way to a single second rank symmetric tensor 
    field which is called {\em universal coupling}. This means that the metric itself can be 
    viewed as a property of spacetime itself rather than as a field over spacetime. For this reason
    one can say, that EEP serves not only as the foundation of general relativity but of the more 
    general idea of gravity as a curved spacetime phenomenon. However, it is important to note
    that it tells us nothing about {\em how} spacetime is curved, i.e. how the metric is generated.
    Actually for this reason it is possible that besides the metric other gravitational fields
    such as scalar, vector or tensor fields could exist which only modifies the way in which matter
    and nongravitational fields generate the metric. Nevertheless, in order to preserve universal
    coupling only the metric acts back on matter and nongravitational fields in the way, prescribed
    by EEP. For example in general relativity the metric is generated directly by the stress-energy
    tensor of matter and nongravitational fields, whereas in the Brans-Dicke theory \cite{bd60}, 
    besides general relativity the most famous representative from the class of metric theories, 
    matter and nongravitational fields first generate a scalar field $\phi$. Then $\phi$ acts together
    with matter and other fields to generate the metric but it only couples indirectly to matter
    so that the theory remains metric.
      
    From these two examples one can see that the main feature which distinguishes different metric 
    theories is the number and the kind of the additional gravitational fields they contain and, in turn,
    the equations which govern the evolution and the structure of these fields. Whether or not a theory
    of gravity exhibits the symmetries defined by EEP depends therefore entirely on the manner in which
    the theory couples the metric to matter and nongravitational fields. This aspect later becomes 
    very important with respect to nonmetric couplings which lead to gravity-induced birefringence of
    polarized light. 
      
    Nonmetric theories of gravity like Moffat's nonsymmetric gravitation theory (NGT) \cite{mof79} or 
    the metric-affine gauge theory of gravity (MAG) \cite{he94} violate, by definition, one or more 
    of the metric postulates and hence it is more necessary than surprising that they predict novel 
    couplings between gravitational and nongravitational fields. An appropiate framework for general 
    considerations of theses aspects is provided by Lagrangian field theory which will be discussed 
    in the next section.
                   
    \subsection{Lagrangian based theories of gravity}
      According to Thorne, Lee and Lightman \cite{tll73}, a generally covariant representation of a 
      spacetime theory is called Lagrangian-based if
      \begin{enumerate}
        \item There exists an action principle that is extremized only with respect to variations of
	      all dynamical variables (For a detailed definition of dynamical variables see p.8 of
	      Thorne et al.).
	\item The dynamical laws of the representation follow from the action principle.
      \end{enumerate}
      A theory is called Lagrangian-based if it possesses a generally covariant, Lagrangian-based 
      representation. Examples are general relativity as well as the Brans-Dicke theory and also
      NGT and MAG. An action principle of the form    
      \begin{equation}\label{action1}
           \delta\int{\cal L}(\psi_m,\psi_g)\,d^4x = 0
      \end{equation}
      encompasses metric as well as nonmetric theories of gravity. Here, $\psi_m$ and $\psi_g$ denotes
      the corresponding (quantummechanical) matter and gravitational fields respectively of a given 
      theory. Then the final objective is, in the end, to break down the explicit structure of a 
      particular action. Like in conventional Langrangian field theory this strategy is supported by 
      the existence of certain symmetries which enforce definite restrictions on the form of the 
      action. In the case of gravitational theories the symmetries of LLI and LPI are consequences 
      of EEP so that many experiments in gravitational physics are direct tests of the structure of 
      a given lagrangian density. The investigation of this aspect can be deepend by splitting a 
      given lagrangian density into a purely gravitational part ${\cal L}_g$ and a nongravitational 
      part ${\cal L}_{ng}$.      
      \begin{equation}
        {\cal L} = {\cal L}_g + {\cal L}_{ng} \quad .
      \end{equation}
      The gravitational density depends entirely on the gravitational potentials and their derivatives, 
      so that its structure determines the dynamics of the free gravitational fields in the theory. 
      The nongravitational part ${\cal L}_{ng}$ also depends on the gravitational potentials
      and their derivatives but additionally on the matter fields and their derivatives so that the form of
      ${\cal L}_{ng}$ specifies the coupling between matter and gravity, i.e. how matter responds to gravity
      and how matter acts as a source of gravity \cite{hla01}. 
      
      \begin{figure}[t]
        \centerline{\psfig{figure=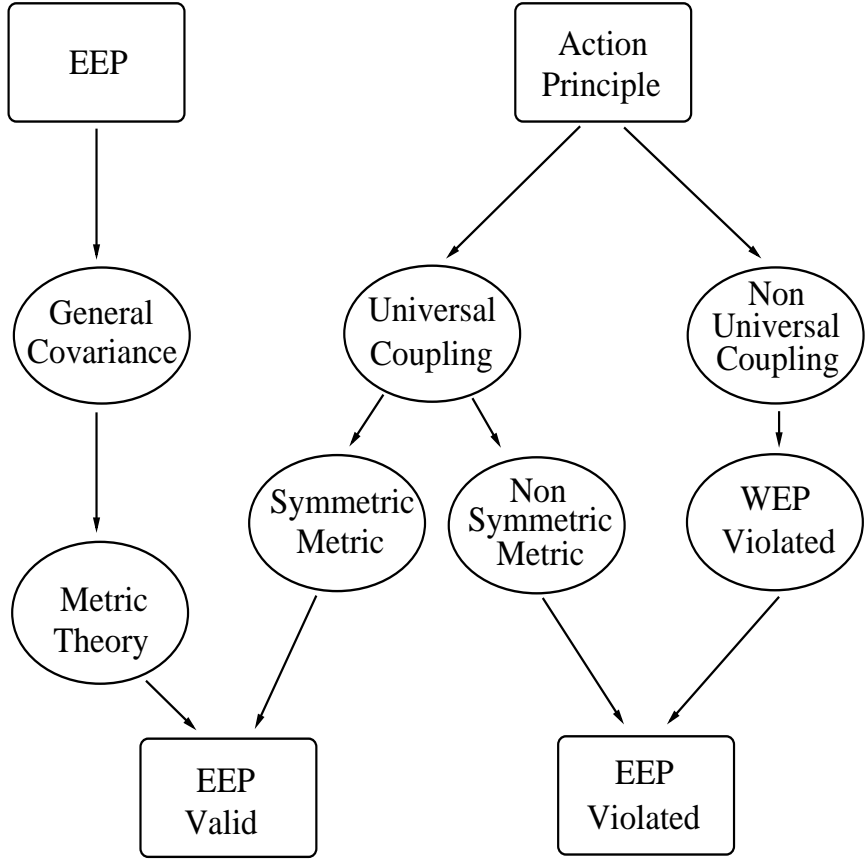,width=4.5in,height=4.5in}}
	\caption{Starting from the two known approaches which can lead to a relativistic theory of gravity,
	         EEP and action principle, different ways could result in a theory which violates EEP. 
		 Details are given in the text.}
	\label{eep}
      \end{figure}       
      Matter equations of motion which predict the outcome of a local, nongravitational test experiment
      are determined only by the form of ${\cal L}_{ng}$ and, therefore are derived from the action 
      principle       
      \begin{equation}\label{action2}
        \delta\int{\cal L}_{ng}(\psi_m,\psi_g^{(e)})\,d^4x = 0 \quad .
      \end{equation}
      In this equation, only the matter fields $\psi_m$ are variable in a true sense, whereas the external
      gravitational fields in which the experiment is performed are viewed as static. 
      In the case of metric theories, the symmetries defined by EEP forces ${\cal L}_{ng}$ to a metric
      form, i.e. ${\cal L}_{ng}$ have to couple a single symmetric tensor gravitational field universally
      to all nongravitational fields. For nonmetric theories ${\cal L}_{ng}$ can admit several EEP
      violating couplings so that detailed investigations of the structure of this lagrangian density
      with respect to experimental tests requires a broader framework which is able to encompass classes 
      of conceivable nonmetric couplings. This will be discussed within the next sections.
      
      A broad class of experiments in gravitational physics are those where the self-gravi\-ta\-tio\-nal 
      effects are not negligible. The underlying symmetries are defined by the {\sc Strong equivalence
      principle} \cite{dk64} which, as a generalization of EEP, also refers to local {\em gravitational}
      test experiments and demands that the symmetries are analogous to those of EEP. Then, the equations 
      of motion follow from the action principle  
      \begin{equation}\label{action3}
        \delta\int{\cal L}(\psi_m,\psi_g^{(l)}+\psi_g^{(e)})\,d^4x = 0 
      \end{equation}  
      where the matter field $\psi_m$, and the local gravitational fields $\psi_g^{(l)}$ are varied
      but the external field $\psi_g^{(e)}$ is assumed to be static. General relativity provides an 
      example of a theory which exhibits the symmetries defined by the strong equivalence principle.

      Having laid down the basic classes of relativistic theories of gravity, the remaining part of this
      section is devoted to the question in which class one could expect violations of WEP and EEP
      respectively. Basically only two ways are known in which a set of gravitational laws could be 
      combined with the special relativistic, nongravitatational laws of physics. Classical starting 
      point of the first approach is the EEP where gravity is described by one or more fields of tensorial 
      character including the metric. By requiring that in local Lorentz frames the nongravitational 
      laws of physics take their special relativistic form one arrives with the principle of general
      covariance at a metric theory of gravity which, by construction, includes EEP. Starting point of the
      second approach is a general action principle with a relativistic lagrangian where one arrives
      at a relativistic theory of gravity in the way described aboved. Assuming a universal coupling,
      there are again two possibilities: Taking the usual second rank symmetric tensor one arrives at 
      a metric theory which obeys EEP. Whereas taking a nonsymmetric metric, Will has shown \cite{will3}
      that even if coupled to matter fields in a universal way, those theories violate WEP and thus EEP. 
      Rejecting the idea of a universal coupling leads to a nonmetric theory which, by definition, 
      violates at least one of the metric postulates. Now, since Schiff's conjecture states that any 
      complete, self-consistent theory of gravity that obeys WEP also includes EEP, it therefore suggests 
      that nonmetric, relativistic, Lagrangian based theories should always violate WEP.
      
    \subsection{General theoretical frameworks}
      Although every experiment performed so far is in nearly perfect agreement with general relativity,
      its conceptual problems (i.e. incompability with quantum mechanics, see Sect.1.2) have led to the 
      development of numerous alternative theories of gravity within the last decades. Since the outcome 
      of a certain experiment is, in most cases, not only relevant for a special but for a whole class of 
      theories with similar characteristics it has become essential to have theoretical frameworks for 
      the classification and comparison of various approaches. Within such a framework one can seek for 
      conceptual differences and similarities and also compare predictions for a variety of experiments. 
      Among these schemes the Dicke framework could be seen as a starting point for enhanced models like 
      the $TH\epsilon\mu$-formalism and the $\chi g$-formalism which have become crucially important for 
      designing and interpreting experiments that have the ability to reveal possible violations of the EEP.  
      \subsubsection{Dicke Framework} 
        The Dicke framework, given 1964 in Appendix 4 of Dicke's Les Houches lectures 
	\cite{dk64} imposes several very fundamental constraints that every acceptable 
	theory of gravity has to obey. Assuming that nature likes things as simple as 
	possible he suggests that the only geometrical concepts introduced {\em a priori} 
	in a spacetime theory are those of a differentiable four-dimensional manifold, with
	each point in the manifold corresponding to a physical event. The manifold need not 
	have either a metric or an affine connection, whereas one would hope that experiments 
	will lead to the conclusion that it has both. Furthermore, the dynamical equations 
	should be constructed in a generally covariant form to avoid arbitrary subjective 
	elements in the equations of motion which merely reflects properties of a particular 
	coordinate system.   
	
	After formulating these mathematical constraints, Dicke requires two aspects to be 
	fulfilled by every viable theory of gravity:
	\begin{enumerate}
	  \item Gravity must be associated with one or more fields of tensorial character. Since 
	        nature abhors complicated situations, interactions involving one field will occur 
		before involving two or more, which favours the viewpoint that gravity is associated
		with only one field.
	  \item Having in mind the close connection between variational principles and conservation 
	        laws, all dynamical equations that govern gravity must be derivable from an invariant
		variational principle.
	\end{enumerate} 
	
	The Dicke framework can be seen as the prototype of all successive classification schemes. 
	The assumptions and constraints placed on all acceptable gravitational theories are often
	used as a basic building block in the development of enhanced schemes like the $TH\epsilon\mu$
	and the $\chi$g-formalism as will be shown below.
	The main achievement of the Dicke framework is that it supports the design and interpretation 
	of experiments which address fundamental questions on the nature of gravity, like what 
	types of fields (scalars, vectors, tensors of various rank) are associated with gravity. 
	For specific questions especially about the motion of electromagnetically charged particles 
	in gravitational fields the $TH\epsilon\mu$-formalism was developed.
      \subsubsection{$TH\epsilon\mu$-formalism}
        The $TH\epsilon\mu$-formalism, developed in 1973 by Lightman and Lee \cite{ll73} encompasses
	all metric and many nonmetric theories of gravity. It describes motions and electromagnetic 
	interactions of charged particles in an external static spherically symmetric (SSS) gravitational 
	field which is described by a potential $U$. The equations of motion of charged particles in 
	the external potential are characterized by two arbitrary functions $T(U)$ and $H(U)$ while 
	the response of the electromagnetic field to the external potential, the gravitationally 
	modified Maxwell equations, are characterized by the functions $\epsilon(U)$ and $\mu(U)$.
	Clearly, the explicit forms of the phenomenological gravitational potentials $T$, $H$, 
	$\epsilon$ and $\mu$ varies from theory to theory.
	One of the most important results of this formalism is that certain combinations of these 
	functions reflect different aspects of EEP \cite{will2} which could be shown by means of the 
	corresponding action.
	
	Within the $TH\epsilon\mu$-formalism the nongravitational laws of physics can be derived
	from an action $I_{\mbox{\footnotesize{NG}}}$ for a structureless test particle $a$ with 
	restmass $m_{0a}$, charge $e_a$ and electromagnetic fields coupled to gravity, given by the sum
	\begin{equation}
	  I_{\mbox{\footnotesize{NG}}} = I_{\mbox{\footnotesize{0}}} + I_{\mbox{\footnotesize{int}}} 
	                               + I_{\mbox{\footnotesize{em}}}
	\end{equation} 
	where the motion of a free (neutral particle) with coordinate velocity $v^{\mu}_a=dx^{\mu}_a/dt$
	on the particle world line $x^{\mu}_a(t)$ is described by
	\begin{equation}
	  I_0 = -\sum_a m_{0a}\int (T-H v_a^2)^{1/2}\, dt \quad .
	\end{equation}
	The interaction of the particle $a$ with the electromagnetic field follows from
	\begin{equation}
	  I_{\mbox{\footnotesize{int}}} = \sum_a e_a \int A_{\mu}v_a^{\mu}\,dt
	\end{equation}
        where $A_{\mu}$ represents the electromagnetic vector potential $F_{\mu\nu}\equiv A_{\nu,\mu}-
	A_{\mu,\nu}$. Finally, the coupling of the electromagnetic to the gravitational field is given by
	\begin{equation}
	  I_{\mbox{\footnotesize{em}}} = (8\pi)^{-1}\int (\epsilon E^2 - \mu^{-1}B^2)\, d^4x \quad .
	\end{equation} 
	with ${\bf E}=\nabla A_0-\partial{\bf A}/\partial t$ and ${\bf B}=\nabla\times{\bf A}$.
	
	Basically, $I_{\mbox{\footnotesize{NG}}}$ violates Local Lorentz Invariance \cite{will1}.
	A metric theory is obtained if and only if $\epsilon$ and $\mu$ satisfies
	\begin{equation}
	  \epsilon = \mu = (H/T)^{1/2} \quad \mbox{or} \quad T^{-1}H\epsilon^{-1}\mu^{-1} = 1
	\end{equation}
	for all $U$. The quantity $(T^{-1}H\epsilon^{-1}\mu^{-1})^{1/2}$ plays the role of the ratio of
	the speed of light $c_{\mbox{\small{light}}}$ to the limiting speed of neutral test particles $c_0$.
	Analogously, one gets that $I_{\mbox{\footnotesize{NG}}}$ is Local Position Invariant if and only
	if
	\begin{eqnarray}
	  \epsilon T^{1/2} H^{-1/2} &=& \mbox{constant} \\
	  \mu T^{1/2} H^{-1/2} &=& \mbox{constant}
	\end{eqnarray}
	independent of the position ${\cal P}(x^{\mu})$ \cite{will1}. Nonconstant combinations like 
	above therefore directly lead to preferred-positions effects like gravitational birefringence.
      \subsubsection{$\chi g$-formalism}
        Invented by W.-T. Ni in 1977 \cite{ni77}, the $\chi g$-formalism provides similarly to the 
	$TH\epsilon\mu$-formalism a framework for the analysis of electrodynamics in a background 
	gravitational field. However, one of the main differences is that the $\chi g$-formalism is
	not restricted to static, spherically symmetric gravitational fields. The $\chi$ of its name
	refers to a tensor density which provides a phenomenological representation of the gravitational
	fields. The structure of the $\chi g$-formalism is in agreement with the basic assumptions
	and constraints of the Dicke framework. Furthermore it is assumed that in the absence of gravity,
	the nongravitational Lagrangian density ${\cal L}_{\mbox{\footnotesize{NG}}}$ reduces to the
	special relativistic form. Respecting these assumptions together with demanding electromagnetic 
	gauge invariance and linearity of the electromagnetic field equations the most general Lagrangian
	density becomes
	\begin{equation}\label{lchi-g}
	 {\cal L}_{\mbox{\footnotesize{NG}}} =  -\frac{1}{16\pi}\chi^{\alpha\beta\gamma\delta}
	                                        F_{\alpha\beta}F_{\gamma\delta} \quad .
	\end{equation}   
	Within this general formalism, the independent components of the tensor density 
	$\chi^{\alpha\beta\gamma\delta}$ comprise 21 phenomenological gravitational potentials which 
	allows one to represent gravitational fields in a very broad class of nonmetric theories. 
	In the case of metric theories, the tensor density is constructed alone from the metric tensor
	\begin{equation}
	  \chi^{\alpha\beta\gamma\delta} = \frac{1}{2}\sqrt{-g}\left(g^{\alpha\gamma}g^{\beta\delta} - 
	                                    g^{\alpha\delta}g^{\beta\gamma}\right) \quad .
	\end{equation} 
	The Lagrangian density has then the usual metric form like the one from general relativity and,
	therefore, incorporates the coupling between the electromagnetic field and a metric gravitational
	field. As already mentioned, nonmetric theories involve various other gravitational potentials
	in addition or instead of the metric one. In this case the Lagrangian density of the 
	$\chi g$-formalism describes the coupling between the electromagnetic field and the various, mostly
	nonmetric, gravitational potentials which could lead to EEP violating effects.
  \section{Nonsymmetric Gravitation Theory}
    The idea of describing the gravitational field by means of a nonsymmetric second rank tensor field
    traces back to the work of Einstein and Strauss in 1946 who tried to formulate a unified theory of gravitation
    and electromagnetism \cite{es46,ae56}. By using the decomposition
    \begin{equation}\label{ngt-gmn}
      g_{\mu\nu}=g_{(\mu\nu)} + g_{[\mu\nu]}
    \end{equation}
    where
    \begin{equation}
      g_{(\mu\nu)}=\frac{1}{2}(g_{\mu\nu}+g_{\nu\mu}),\quad g_{[\mu\nu]}=\frac{1}{2}(g_{\mu\nu}-g_{\nu\mu})
    \end{equation}
    their intention was to interpret the symmetric $g_{(\mu\nu)}$ as an expression of the gravitational field
    and the antisymmetric part $g_{[\mu\nu]}$ as an expression of the electromagnetic field.
    Unfortunately, it was soon realized that $g_{[\mu\nu]}$ could not describe physically the electromagnetic
    field without serious contradictions, so that the nonsymmetric ansatz vanished for more than 30 years.
    
    In 1979 Moffat picked up this issue again and published the first version of his nonsymmetric gravitation 
    theory (NGT) \cite{mof79} which is based on a non-Riemannian geometry according to (\ref{ngt-gmn}) and the 
    analog expression for the affine connection
    \begin{equation}
     \Gamma^{\lambda}_{\mu\nu}=\Gamma^{\lambda}_{(\mu\nu)}+\Gamma^{\lambda}_{[\mu\nu]} \quad .
    \end{equation}
    The motivation of NGT is to construct the most general classical description of space-time that contains
    general relativity as a special, low-energy case. The main conceptual difference to Einstein-Strauss theory
    is that the nonsymmetric field structure now rather describes a generalization of Einstein gravity than 
    a unified field of gravitation and electromagnetism. We consider here the physical implications
    which follows from the NGT version, described by Moffat in 1990 \cite{mof91}, and set sharp limits on the
    essential coupling constant $\ell^2$, responsible for violations of EEP. Although it became clear 
    in the meantime that this theory as well as a later published modification \cite{mof94} suffers from 
    serious problems like ghost poles, tachyons and higher order poles (see \cite{mof00} for a list of 
    references) NGT nevertheless serves as a prototype for a whole class of nonmetric gravitational theories 
    which predict spatial anisotropy and birefringence. Setting sharp and reliable limits on $\ell^2$ 
    is therefore not only a further test of NGT but rather adresses the question on the physical relevance 
    of gravity-induced birefringence in principle. 
    
    Defining the contravariant tensor $g^{\mu\nu}$ in terms of the equation
    \begin{equation}
      g^{\mu\nu}g_{\sigma\nu} = g^{\nu\mu}g_{\nu\sigma} = \delta^{\mu}_{\sigma} 
    \end{equation} 
    the Lagrangian density with matter sources is given by
    \begin{equation}\label{ngt-lagr}
      {\cal L}_{\mbox{\footnotesize{NGT}}} = {\cal L}_R + {\cal L}_M
    \end{equation}
    with
    \begin{equation}
      {\cal L}_R = {\bf g}^{\mu\nu}R_{\mu\nu}(W)-2\Lambda\sqrt{-g}
    \end{equation}
    and
    \begin{equation}
      {\cal L}_M = -\frac{8\pi G}{c^4}g^{\mu\nu}{\bf T}_{\mu\nu}+\frac{8\pi}{3}W_{\mu}{\bf S}^{\mu} \quad .
    \end{equation}
    Here, $\Lambda$ is the cosmological contant and ${\bf g}^{\mu\nu}=\sqrt{-g}g^{\mu\nu}$, while the other 
    constants have their usual meaning. $R_{\mu\nu}(W)$ denotes the NGT contracted curvature tensor
    \begin{equation}
      R_{\mu\nu}(W) = W^{\beta}_{\mu\nu,\beta}-\frac{1}{2}(W^{\beta}_{\mu\beta,\nu}+W^{\beta}_{\nu\beta,\mu})
      - W^{\beta}_{\alpha\nu}W^{\alpha}_{\mu\beta}+W^{\beta}_{\alpha\beta}W^{\alpha}_{\mu\nu} \quad ,
    \end{equation}
    defined in terms of the unconstrained nonsymmetric connection
    \begin{equation}
      W^{\lambda}_{\mu\nu}=\Gamma^{\lambda}_{\mu\nu}-\frac{2}{3}\delta^{\lambda}_{\mu}W_{\nu} \quad , 
    \end{equation}
    where
    \begin{equation}
      W_{\mu}=\frac{1}{2}(W^{\lambda}_{\mu\lambda}-W^{\lambda}_{\lambda\mu}) \quad .
    \end{equation}
    The full NGT field equations with matter source therefore become
    \begin{equation}
      G_{\mu\nu}(W)=\frac{8\pi G}{c^4}T_{\mu\nu}+\Lambda g_{\mu\nu} \quad ,
    \end{equation}
    \begin{equation}
      {\bf g}^{[\mu\nu]}_{\quad \, ,\nu} = 4\pi{\bf S}^{\mu} \quad ,
    \end{equation}
    where
    \begin{equation}
      G_{\mu\nu}(W)=R_{\mu\nu}-\frac{1}{2}g_{\mu\nu}R  \quad .
    \end{equation}
    So, in addition to a conserved nonsymmetric energy-momentum tensor $T^{\mu\nu}$ (and in contrast to 
    general relativity), NGT also contains a conserved-vector-current density $S^{\mu}$, where the current 
    conservation arises by Noether's theorem from the invariance of the Lagrangian density (\ref{ngt-lagr}) 
    under the transformations of an Abelian $U(1)$ group, i.e.         
    \begin{equation}
      {\bf g}^{[\mu\nu]}{}{}_{,\mu,\nu}\equiv  4\pi{\bf S}^{\mu}{}_{,\mu} = 0 \quad .
    \end{equation}
    It suggests itself to interpret $S^{\mu}$ as the conserved particle number of the fluid
    \begin{equation}
      S^{\mu} = \sum_i f^2_i n_i u^{\mu} \quad .
    \end{equation}
    Here, $f_i^2$ is a coupling constant for each species $i$ of fermions, $n_i$ is the constant fermion
    particel number and $u^{\mu}=dx^{\mu}/d\tau$ denotes the proper-time velocity of the particle. The
    so-called NGT charge $\ell^2$ is defined as
    \begin{equation}   
      \ell^2=\int{\bf S}^0\, d^3x \quad ,
    \end{equation}
    having the dimension of [length$]^2$. Since $\ell^2$ arises from a conserved current, it has been 
    postulated that it is proportional to conserved particle number \cite{mof91}. In order to understand
    the exceptional position of fermions in NGT one has to recall that in the NGT scheme, the nonsymmetric 
    tensor $g_{\mu\nu}$ leads to a nontrivial extension of the homogeneous Lorentz group $SO(3,1)$ of 
    general relativity to the local gauge group $GL(4,R)$. This extension describes the most general 
    transformations of the linear frames that contains the 
    homogeneous Lorentz group as a subgroup \cite{mof88}. Furthermore, the group $GL(4,R)$ contains only 
    infinite-dimensional spinor representations which leads to the idea that particles are described as 
    extended objects in contrast to general relativity where point particle fermions are conventionally 
    described by finite, nonunitary representations of the homogeneous Lorentz group $SO(3,1)$ 
    \cite{mof88,mof88a}. Therefore, fermions play an important role in NGT and are given special 
    consideration as a source of the (antisymmetric part of the) gravitational field \cite{mof89}.
    
    Despite various arguments, like beauty and simplicity of the theory, one could have in favour of
    NGT it is merely a, perhaps elegant, hypothesis because of the complete lack of any experimental 
    verification. One therefore has to look for the physical implications of this idea which implies
    constraining the essential coupling constant $\ell^2$ since history is full of elegant hypothesis, later
    contradicted by nature.     
  \section{Metric-Affine Gravity} 
    Questioning critically the foundations of a certain physical theory 
    is often the first, promising step for getting a deeper insight into the basic 
    mechanisms ruling the corresponding phenomena. In this sense it is certainly
    important to realize that the weak equivalence principle as it is formulated by Newton  
    only gives a prediction for the behaviour of macroscopic objects and, in the 
    language of general relativity, their interaction and influence on the structure
    of spacetime. Microscopic properties of matter like the spin angular momentum 
    of particles are totally neglegted as they average out on the macroscopic scale
    which justifies the accusation that general relativity is blind to the
    microscopic structure of matter. So, the hypothesis is near at hand that spin
    angular momentum might be the source of a gravitational field too, since it certainly
    characterizes matter dynamically in the microphysical realm. 
    
    Although this issue seems to be evident, the question immediately arises what kind 
    of ''interface'' or ansatz one has to choose in order to extend the concepts of 
    general relativity to emcompass quantum mechanically relevant observables. An answer
    to this is very likely given by looking at gravity in a gauge theoretical way which
    is favoured by the successfull description of the other three fundamental interactions
    by means of gauge theories of underlying local symmetry groups. In this sense Utiyama 
    \cite{uti56} has shown in 1956 that general relativity could be recovered by gauging 
    the Lorentz $SO(1,3)$ group. Nevertheless this procedure is unsatisfactory since 
    the conserved current associated to the Lorentz group via the Noether theorem is 
    merely the angular momentum current which alone cannot represent the source of 
    gravity. This problem was solved by Sciama and Kibble in 1961 \cite{sc61,sc64,kib61} who 
    prooved that it is really the Poincar\'{e} group as the semi-direct product 
    of the translation and the Lorentz groups, which underlies gravity. This scheme now 
    allows spin angular momentum to be included. In analogy to the coupling of energy momentum 
    to the metric, spin is coupled to a geometrical quantity which is related to 
    rotational degrees of freedom in spacetime. This concept leads to a generalization 
    of the Riemannian spacetime of general relativity to the Riemann-Cartan spacetime 
    $U_4$. In a $U_4$ space the affine connection $\Gamma^{\lambda}_{\alpha\beta}$ is not
    symmetric in the lower indices $\alpha$ and $\beta$ which leads to a nonzero torsion tensor
    \begin{equation}
      T^{\lambda}_{\alpha\beta} := \frac{1}{2}(\Gamma^{\lambda}_{\alpha\beta}-
      \Gamma^{\lambda}_{\beta\alpha}) \quad ,
    \end{equation}
    initially introduced by E. Cartan \cite{car20}, as the antisymmetric part of the affine 
    connection. Analogously to Riemannian spacetime it is required that the local Minkowski 
    structure is preserved, i.e. that the line element is invariant under parallel transfer. 
    The deformation of length and angle standards during parallel transport is measured by 
    the so-called {\em nonmetricity} one-form, defined by
    \begin{equation}
      Q_{\alpha\beta} := Dg_{\alpha\beta} \quad .
    \end{equation}     
    In Riemann-Cartan spacetime $U_4$ as well as in Riemannian spacetime $V_4$ of general relativity
    and in the Minkowski spacetime $R_4$ of special relativity the nonmetricity vanishes
    \begin{equation}     
	Q_{\alpha\beta} = 0 \quad .
    \end{equation} 
    The geometrical structure of $U_4$ is that of an $n$-dimensional differentiable manifold $M_n$
    where at each point of $M_n$, there is an $n$-dimensional tangent vector space $T_P(M_n)$. The 
    local vector basis $e_{\alpha}$ can be expanded in terms of the local coordinate basis 
    $\partial_i :=\partial/\partial x^i$ 
    \begin{equation}
      e_{\alpha} = e^i{}_{\alpha}\partial_i
    \end{equation}
    where $\alpha,\,\beta = 0,1,2,\ldots,(n-1)$ are anholonomic or frame indices and $i,j,k,\ldots,(n-1)$
    are holonomic or coordinate indices. In the cotangent space $T^{\star}_P(M_n)$ there exists a one-form
    or a coframe
    \begin{equation}
      \vartheta^{\beta}=e_j{}^{\beta}dx^j \quad .
    \end{equation}
    Defining further the one-form $\eta_{\alpha\beta\gamma}= \,^*(\vartheta_{\alpha}\wedge\vartheta_{\beta}\wedge
    \vartheta_{\gamma})$ with the Hodge star * and the three-form $\eta_{\alpha}=\,^*\vartheta^{\alpha}$ the field 
    equations of Einstein-Cartan theory read
    \begin{eqnarray}
      \frac{1}{2}\eta_{\alpha\beta\gamma}\wedge R^{\beta\gamma}+\Lambda\eta_{\alpha} &=& \ell^2 \sum\nolimits
                                                                                        _{\alpha} \\
      \frac{1}{2}\eta_{\alpha\beta\gamma}\wedge T^{\gamma} &=& \ell^2\tau_{\alpha\beta}
    \end{eqnarray}
    where $\Lambda$ denotes the cosmological constant, $\ell^2$ Einstein's gravitational constant and 
    $R^{\beta\gamma}$ the curvature-2-form. $\sum_{\alpha}$ denotes the canonical energy-momentum current of matter. 
    While the first equation relates curvature to energy momentum, the second equation provides a link between 
    the spin angular momentum tensor $\tau_{\alpha\beta}$ to Cartan's Torsion. It is now obvious how 
    Einstein-Cartan gravity, general relativity and special relativity are connected. Starting from
    a Riemann-Cartan spacetime $U_4$ the usual Riemannian spacetime of general relativity is recovered
    by neglegting torsion. If additionally curvature vanishes one gets the Minkowski spacetime $R_4$ of 
    special relativity.
    
    \begin{equation}
      U_4\stackrel{T=0}{\longrightarrow}V_4\stackrel{R=0}{\longrightarrow}R_4
    \end{equation}
        
    The Einstein-Cartan theory of gravity is a viable theory since it is in agreement with all experiments
    performed so far \cite{he76,mil87,trt99}. Under usual conditions the spin $\tau_{\alpha\beta}$ averages 
    out and can be neglegted which in turn, according to the second field equation, implies vanishing 
    torsion and, so, general relativity is recovered. The additional spin-spin contact interaction shows up 
    only at extremely high matter densities $(\sim 10^{54}$ g/cm$^3)$ and therefore has not been seen so 
    far since even typical neutron stars have only densities of the order of $(10^{15}$ g/cm$^3)$.  
    
    The reason why the Einstein-Cartan theory has been considered here is, that it provides the simplest
    model of the so-called metric-affine gauge theory of gravity (MAG)\cite{gh96,gh97,he94}, representing  
    the most general canonical gauge theory. Metric-affine theories are build upon the more general affine 
    group $A(n,R)$ which is the semidirect product of the translation group and the group of linear 
    transformations, i.e. $A(n,R)=T^n \sdp GL(n,R)$. This transformation group acts on an affine n-vector 
    $\{\xi^{\alpha}\}$ according to
    
    \begin{equation}\label{affin}  
       \xi \longrightarrow \xi'=\Lambda\,\xi + \tau 
    \end{equation}
    where $\Lambda = \{\Lambda^{\alpha}{}_{\beta}\}\in GL(n,R)$ and $\tau = \{\tau^{\alpha}\}\in R^{\alpha}$.
    The transformations (\ref{affin}) of the affine group represent a generalization of the Poincar\'{e}
    group where the pseudo-orthogonal group $SO(1,n-1)$ is replaced by the general linear group $GL(n,R)$.
    The reason for introducing $A(n,R)$ is the assumption that physical systems are indeed invariant under 
    the action of the entire affine group and not only invariant under its Poincar\'{e} subgroup.
    The physical symmetries which are added by the general affine invariance are the dilation and shear 
    invariance. Both are of physical importance since dilation invariance is a crucial component of particle
    physics in the high energy regime and shear invariance was shown to yield representations of hadronic matter.
    Further, the corresponding shear current can be related to hadronic quadrupole excitations \cite{he94,gh97}.
    Therefore, the additional appearance of symmetries in the high energy regime could be taken as an indication
    that metric-affine gravity has played an important role at the early stages of the universe and reduces to 
    general relativity and translational invariance in the low-energy limit after some symmetry breaking mechanism
    \cite{he94,tre95a}. 
    
    Metric affine gravity uses the metric $g_{\alpha\beta}$, the coframe $\vartheta^{\alpha}$, and the linear
    connection $\Gamma_{\alpha}{}^{\beta}$ to represent independent gravitational potentials. This is summarized
    in Table 1.
    
    \begin{center}
      \begin{tabular}{||l|l||}\hline\hline
                  &                \\
        Potential & Field strength \\ 
	          &                \\ \hline
		  &                \\
        metric $g_{\alpha\beta}$               & nonmetricity $Q_{\alpha\beta}=-D g_{\alpha\beta}$ \\
	          &                \\
        coframe $\vartheta^{\alpha}$           & torsion $T^{\alpha}=D\vartheta^{\alpha}$          \\
                  &                \\
	connection $\Gamma_{\alpha}{}^{\beta}$ & curvature $R_{\alpha}{}^{\beta}=d\Gamma_{\alpha}{}^{\beta}
        -\Gamma_{\alpha}{}^{\mu}\wedge\Gamma_{\mu}{}^{\beta}$ \\ 
	          &                \\ \hline\hline
    \end{tabular}
    \end{center}
    \centerline{{\small Tab. 1: Gauge fields in metric-affine gravity.}}
    \vspace{0.5cm}
    
    \noindent In contrast to NGT, metric-affine theories do not allow for an antisymmetric part 
    in the metric tensor, since it does not lend itself to a direct geometrical interpretation \cite{he94}.
    Although the symmetric tensor is referred to as the metric, metric-affine gravity becomes 
    nonmetric when the new gravitational potentials or their derivatives (the nonmetricity, torsion 
    and curvature gravitational fields) couple directly to matter, as they generally do. Nonmetric 
    couplings to the electromagnetic field are what can lead to gravity-induced birefringence which
    will be discussed in the next section.

  \section{Electrodynamics in a background gravitational field}
    \subsection{Birefringence in nonsymmetric theories}    
      Moffat´s nonsymmetric gravitation theory (NGT) can be regarded as the prototype of a diverse
      class of Lagrangian-based nonmetric theories where a nonsymmetric tensor field does not couple 
      universally to matter stress energy. Especially the coupling of the antisymmetric part
      of the nonsymmetric gravitational field to the electromagnetic field leads to a polarization 
      dependent delay and, so, to an alteration of a polarization signal which is a consequence of 
      the breakdown of EEP in these theories \cite{gab91,gea91}. The following analysis was first
      published by Gabriel et al. \cite{gab91} from whom most of the notations are adopted.
      
      The electromagnetic field equations which govern the propagation of light through a nonsymmetric
      gravitational field can be derived from an action principle. A general form for this action, quadratic
      in both the electromagnetic field strength $F_{\mu\nu}\equiv A_{[\mu,\nu]}$ and the inverse 
      metric, was given by Mann et al. \cite{mpm89}

      \begin{equation}\label{act1}
        I_{\mbox{\footnotesize{em}}}=-\frac{1}{16\pi}\int d^4x \sqrt{-g}{\cal F}g^{\mu\alpha}g^{\nu\beta}
        \left(ZF_{\mu\nu}F_{\alpha\beta}+(1-Z)F_{\alpha\nu}F_{\mu\beta}+YF_{\mu\alpha}F_{\nu\beta}\right)
      \end{equation}
      The matrix $g^{\mu\nu}$ denotes the inverse of the nonsymmetric gravitational field $g_{\mu\nu}$ 
      defined by $g^{\mu\alpha}g_{\nu\alpha}=g^{\alpha\mu}g_{\alpha\nu}=\delta^{\mu}_{\nu}$. $Y$ and $Z$
      are constants while ${\cal F}$ is a scalar function which cannot depend on the electromagnetic 
      field and which must be unity in the Einstein-Maxwell limit $g_{[\mu\nu]}\to 0$. This implies that
      ${\cal F}={\cal F}(\sqrt{-g}/\sqrt{-\gamma})$, where $g\equiv \det g_{\mu\nu}$ and $\gamma \equiv
      \det g_{(\mu\nu)}$.
    
      Within this scheme a static, spherically symmetric gravitational field like that of the Sun is
      described by an isotropic coordinate system, centered on the sun. The symmetric part of the field
      takes the form $g_{00}=-T(r),\, g_{(0i)}=0$, and $g_{(ij)}=H(r)\delta_{ij}$ where $T$ and $H$ are 
      functions of $r\equiv |x|$. The theories which are encompassed by this formalism like NGT provide
      a representation of the antisymmetric part of the gravitational field that can be expressed in 
      an isotropic coordinate system as $g_{[0i]}=L(r)n_i$ and $g_{[ij]}\equiv 0$, where $n_i \equiv
      x_i/r$. At this point it should be mentioned that the polarization dependent delay that this analysis
      reveals reflects this special form of the nonsymmetric gravitational field. Gabriel et al. point out
      in their paper \cite{gab91} that a similar analysis, based on a more general representation
      also reveals polarization dependence.
      
      Employing this special representation together with definitions for the electric and magnetic fields
      via $E_{j0}\equiv E_j$ and $F_{jk}\equiv\epsilon_{jkl}B_l$ this yields 
      \begin{equation}\label{act2}
        I_{\mbox{\footnotesize{em}}}=-\frac{1}{8\pi}\int d^4x\left[\epsilon E^2+X\epsilon\alpha
        (\hat{{\bf n}}\cdot E)^2-\frac{1}{\mu}B^2+\frac{\Omega}{\mu}(\hat{{\bf n}}\cdot B)^2\right]
      \end{equation}
      with
      \begin{eqnarray}
        \epsilon &\equiv& {\cal F}\left[\frac{H}{T}\right]^{1/2}\left[1-\frac{L^2}{TH}\right]^{-1/2} \\
        \mu &\equiv& {\cal F}^{-1}\left[\frac{H}{T}\right]^{1/2}\left[1-\frac{L^2}{TH}\right]^{1/2} \\
        \alpha &\equiv& \frac{2L^2}{TH}\left[1-\frac{L^2}{TH}\right]^{-1} \\
        \Omega &\equiv& \frac{L^2}{TH}
      \end{eqnarray}
      and
      \begin{equation}
         X \equiv 1 - Y - Z \quad .
      \end{equation}
      In order to derive the electromagnetic field equations from the general action (\ref{act2}) the
      special form of the coupling between the nonsymmetric gravitational field and the electromagnetic
      field in terms of $X,\, Y, \, Z$ and ${\cal F}$ must be given. Gabriel et al. used the form $Y=1-Z$
      and ${\cal F}=(1-L^2/TH)^{1/2}=\sqrt{-g}/\sqrt{-\gamma}$ by demanding $\epsilon = \mu$ in accordance
      with NGT. However, it is important to note that similar analyses of theories having other values of
      $Y$ and $Z$ also reveals a polarization dependence in delay measurements.
      
      Using this special coupling, the action (\ref{act2}) reduces to
      \begin{equation}\label{act3}
        I_{\mbox{\footnotesize{em}}} = \frac{1}{8\pi}\int d^4x \left(\epsilon E^2 - \frac{1}{\mu}
        (B^2- \Omega({\bf n \cdot B})^2)\right)
      \end{equation}
      This action differs from the usual Einstein-Maxwell action mainly because of the presence of the
      $\Omega({\bf n \cdot B})^2$ term. Will \cite{will3} has pointed out, that such a term may produce
      pertubations in the energy levels of an atomic system, depending on the relative orientation
      of the system's wave function to the direction ${\bf n}$. Such pertubations could be constrained by
      ultraprecise energy-isotropy experiments of the Hughes-Drever type \cite{hug60,drv61}, using 
      trapped atoms and magnetic resonance techniques \cite{stb87} which is presently being investigated. 
      Given this action $I$, the field equations that govern the propagation of light through the
      nonsymmetric gravitational field are
      \begin{eqnarray}
        & &\nabla \times {\bf E} + \frac{\partial {\bf B}}{\partial t} = 0 \\
        & &\nabla \cdot {\bf B} = 0 \\
        & &\nabla \cdot (\epsilon {\bf E}) = 0 \\
        & & \nabla \times \left[\frac{{\bf B}}{\mu}\right]-\frac{\partial(\epsilon {\bf E})}{\partial t}
        - \nabla \times \left[\frac{\Omega {\bf n}({\bf n \cdot B})}{\mu} \right] = 0
      \end{eqnarray}
      The first two pairs follow from the usual definitions of ${\bf E}$ and ${\bf B}$ in terms of electromagnetic
      vector potentials while only the last two pairs follow from the action (\ref{act3}).
      
      To investigate the propagation of electromagnetic waves with respect to polarization dependend delay, the 
      appropiate representations of a locally plane wave are
      \begin{equation}\label{eab}
        {\bf E} = {\bf A}_E e^{i\Phi}, \quad {\bf B} = {\bf A}_B e^{i\Phi}
      \end{equation}
      Denoting $k_{\mu}$ as the gradient of the phase function 
      \begin{equation} 
        \partial_{\mu}\Phi \equiv (\partial \Phi/\partial t,\nabla\Phi)\equiv (-\omega,{\bf k}) 
      \end{equation}
      the eikonal equation which governs the propagation of a locally plane wave in the limits of geometric 
      optics can be derived by inserting the representations (\ref{eab}) into the above set of Maxwell equations,
      ignoring all derivatives other than those of the phase function. Therefore, one gets
      \begin{equation}\label{eiko}
        k^2{\bf A}_B-\epsilon\mu\omega^2{\bf A}_B-\Omega{\bf \hat{n} \cdot A_B}[k^2{\bf \hat{n}}-{\bf k}
       ({\bf \hat{n} \cdot {\bf k}})] = 0
      \end{equation}
      under the assumptions that the wavelength $\lambda$ is much smaller than the typical scale on which any of the
      fields $\epsilon,\, \mu,\, \Omega,\, {\bf k},\, {\bf A_E}$ and ${\bf A_B}$ vary significantly.        
      This equation is similar to the usual dispersion relation, exept the term which is proportional to
      $\Omega$. Since the speed of an electromagnetic wave (\ref{eab}) is given by $\omega/k$, it is apparent
      from the structure of (\ref{eiko}) that this speed depends on the orientations of ${\bf k}$ and ${\bf A}_B$
      relative to ${\bf \hat{n}}$. Therefore, the velocity of an electromagnetic wave propagating through a 
      nonsymmetric gravitational field depends on its orientation which, in turn, implies a violation of EEP!
      
      \begin{figure}[t]
        \centerline{\psfig{figure=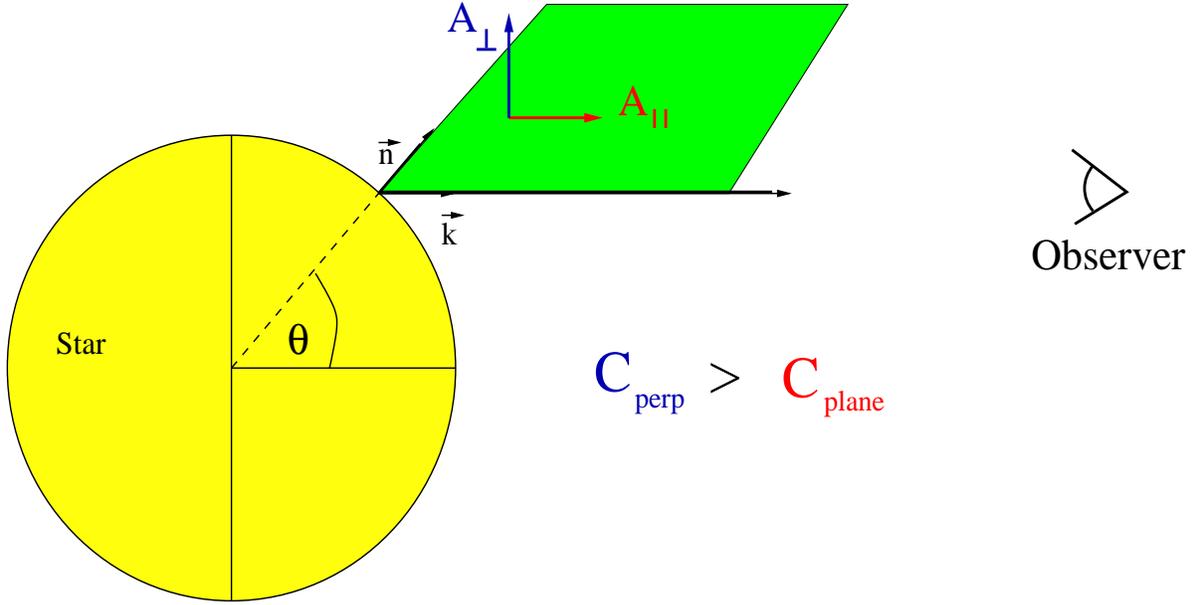,height=3.15in,width=6.2in}}
	\caption{Visualization of gravitational birefringence in the case of Moffat's nonsymmetric gravitation
	         theory (NGT).}
      \end{figure}
      The speed of a linearly polarized wave with its magnetic field lying perpendicular to ${\bf \hat{n}}$
      follows directly from the eikonal equation with ${\bf \hat{n}}\cdot {\bf A_B}$ so that one gets 
      \begin{equation}
        c_{\perp}\equiv \omega/k=(\epsilon\mu)^{-1/2} \quad .
      \end{equation}
      Light propagating in any other direction travels at $c_{\perp}$ only if ${\bf A_B}$ is perpendicular
      to the plane spanned by ${\bf \hat{n}}$ and ${\bf k}$.
      The same wave, this time with the magnetic field vector lying in the plane spanned by ${\bf \hat{n}}$
      and ${\bf k}$, propagates according to $k^2(1-\Omega\sin^2\theta)=\epsilon\mu\omega$, where $\theta$ 
      denotes the angle between ${\bf \hat{n}}$ and ${\bf k}$. The coordinate speed of a wave, having this 
      polarization is
      \begin{equation}
        c_{\theta}=\left[\frac{1-\Omega\sin^2\theta}{\epsilon\mu}\right]^{1/2} \quad .
      \end{equation}
      Since both waves, parallel and perpendicular to the plane, are independent solutions of the modified 
      Maxwell equations, they can propagate independently through spacetime without changing their polarization.
      However, because they travel at different velocities the ''medium'' has now two refractive indices, one
      for each such eigenmode so that spacetime has become (linearly) birefringent. 
      
      Now, since light having any other polarization can be viewed as a coherent superposition of the two eigenmodes
      where the difference in the propagation velocities between the components changes their initial relative phase
      and, thus, the light's polarization. The accumulated phase shift is calculated by integrating the eikonal
      equation along a straight line, extended far past the Sun (or other astronomical body). The result for light 
      with frequency $\omega$, given by Gabriel et al. \cite{gab91}, is
      \begin{equation}\label{dphi}
        \Delta\Phi=\frac{1}{2}\omega\int_{t_0}^{t_1}\Omega\sin^2\theta(t)\,dt \quad .
      \end{equation}
      The integration of (\ref{dphi}) requires a ray parametrization ${\bf x}(t)={\bf b}+\hat{{\bf k}_0}t$
      where the unit vector $\hat{{\bf k}_0}$ denote the ray direction and ${\bf b}$ is the impact vector
      that connects the center of the sun with the closest point on the ray. When ${\bf b}$ is smaller than
      the radius $R$ of the Sun (or the star), the portion of the ray inside the object is, of course, of no
      interest. The integration of (\ref{dphi}) is performed from the Sun's surface with $t_0=(R^2-b^2)^{1/2}$ 
      along a straight line up to an observer in an infinite distance $t_1=\infty$.
      
      Restricting the investigations of birefringence to Moffat's NGT with $\Omega = \ell^4_{\odot}/r^4$ the 
      integration yields  
      \begin{equation}\label{ngt-form}
        \Delta\Phi(\mu)=\frac{\pi \ell^4_{\odot}}{\lambda R^3}\left\{\frac{3\pi}{16(1-\mu^2)^{3/2}}
        -\frac{\mu}{4}-\frac{3\mu}{8(1-\mu^2)}-\frac{3}{8(1-\mu^2)^{3/2}}\arcsin(\mu)\right\}\quad ,
      \end{equation}
      where $\mu$ denotes the cosine of the heliocentric angle $\theta$ between the ray's source and the 
      Sun's center. Although in a different context, the basic calculations which lead to (\ref{ngt-form}) 
      are given in Appendix B.
      
      The accumulated phase shift is inversely proportional to the observed wavelength $\lambda$ 
      and the radius $R$. Therefore with respect to experimental tests of gravitational birefringence, only those
      objects can be utilized which emit polarized radiation at short wavelengths from a sufficiently small 
      radius for a given mass. For a source located at the center of the solar disc $(\mu=1)$, the phase shift
      vanishes, whereas for light emitted from the solar limb $(\mu=0)$ we have $\Delta\Phi=3\pi^2 \ell^4_{\odot}
      /16\lambda R^3$.
      
    \subsection{Birefringence in metric-affine gravity}
      Soon after Gabriel et al. \cite{gab91} has shown that nonsymmetric gravitation theories like Moffat's NGT 
      predict gravitational birefringence and, so, a violation of the Einstein equivalence principle, Haugan
      and Kauffmann \cite{hkf95} remarked that this phenomenon could be extended to the far more diverse class of 
      nonmetric theories encompassed by the $\chi g$-formalism. Although already discovered in 1984 by Ni 
      \cite{ni84} this result was overlooked up to that time because no gravitation theories predicting such
      birefringence were known and the available techniques for testing these predictions were not sufficient.
      
      Starting with the nongravitational Lagrangian density (\ref{lchi-g}) of the $\chi g$-formalism in the limit
      of weak gravitational fields, Haugan and Kauffmann \cite{hkf95} gave a general prediction for the 
      accumulated phase shift $\Delta\phi$ by using the methodology of geometric optics. The detailed 
      calculations are given in Appendix A.
      Basically, the relative phase shift is simply a function of the difference in velocity between the two
      eigenmodes denoted by $\delta c/c$ plus a small second order correction from the tensor density 
      $\chi^{\alpha\beta\gamma\delta}$.
      \begin{equation}\label{chi-form}
        \Delta\Phi = \omega \int\frac{\delta c}{c}\, dt + {\cal O}(\delta\chi^2)
      \end{equation}  
      The explicit form of (\ref{chi-form}) depends of course on the phenomenological representation of the 
      gravitational potential and their couplings which is provided by $\chi^{\alpha\beta\gamma\delta}$.
      In the case of metric-affine gravity one first has to answer the question which of the gravitational
      fields one has to couple to electromagnetism. Nonmetricity is a rather exotic possibility since it 
      is assumed that it only plays a relevant role in the very high energy regime like the early stages 
      of the Big Bang. It is therefore more suggestive to think of torsion which couples to electromagnetic 
      fields. However, the form that such coupling could have is not exactly clear and only very little
      has been done is this direction so far. Indeed there are numerous nontrivial ways in which torsion 
      could couple to the electromagnetic field. However, we decided to use
      \begin{equation}\label{mag-lem}
        \delta{\cal L}_{EM} = k^2 {}^{\star}(T_{\alpha}\wedge F){}^{\star}(T^{\alpha}\wedge F)
      \end{equation}
      since this form is equivalent to a particular fourth-rank $\delta\chi$ with tensorial character and such a 
      term could, as we have shown in \cite{sph01}, induce birefringence. In analogy to the nonsymmetric 
      charge $\ell^2$, the strength of the coupling is described by a possibly material dependent constant $k^2$ 
      (see Appendix A) having the dimension of length. Our intention is to set strong limits on $k^2$ and, so, to 
      decide about the physical relevance of gravity-induced birefringence. Since different astrophysical 
      objects (Sun, white dwarfs, active galatic nuclei) may have different chemical compositions, it is 
      important to set and to compare limits on $k^2$ for a variety of different objects which is one of the objective
      targets of this thesis. 
      
      Since we are going to look for birefringence in the spherically symmetric gravitational fields of
      stars, we are interested in static and spherically symmetric solutions of the metric-affine field
      equations with respect to torsion. Such a solution was given by Tresguerres in 1995 \cite{tre95a,tre95b}
      (see (\ref{torsion}) in Appendix A) which can be split into a nonmetricity dependent and a nonmetricity
      independent part which is assumed to couple to the electromagnetic field via (\ref{mag-lem}).
      Using the method of Haugan and Kauffmann \cite{hkf95} the phase shift (\ref{chi-form}) takes after tantalizing 
      computations, given in detail in Appendix A, the form
      \begin{equation}
        \Delta\Phi = -\sqrt{6}\,\omega\, k^2 \,M_{\star} \int\frac{\sin^2\theta(t)}{r^3(t)}\,dt
      \end{equation} 
      where $\omega$ denotes the light's circular frequency and $M_{\star}$ the mass of the star in geometrized 
      units.              
      Using the same integration technique as outlined in the case of the nonsymmetric theories this leads
      to the explicit form
      \begin{equation}\label{mag-form}
         \Delta\Phi = \sqrt{\frac{2}{3}}\cdot \frac{2\pi\cdot k^2 M_{\star} }{(\lambda R_{\star}^2)} \left(
	 \frac{(\mu + 2)(\mu - 1)}{\mu + 1}\right) \quad . 
      \end{equation} 
      The concept of torsion has become an important tool in many present time gravitation theories. Quite 
      recently it has been suggested to identify torsion with the field strength of a second rank symmetric 
      Kalb-Ramond tensor field which also appears in the low-energy, effective field theory limit of string 
      theory \cite{gup01,string}. The rank and symmetry of this field are similar to that of the torsion 
      field in metric-affine theories so that it is not unreasonable to expect analogous couplings to the 
      electromagnetic field and, consequently, birefringence. Explicit calculations on this subject has not 
      been done so far.      
      
  \section{Description of Polarized Radiation}
    If the concept of gravity-induced birefringence is in fact realized by nature, the required ingredients
    for a chance to observe it are certainly given by a source of strong gravitational fields, emitting
    a reasonable amount of polarized electromagnetic radiation. Since every polarized wave can be decomposed
    into two orthogonal modes, a speed difference between them due to birefringence would lead to a phase shift
    and, hence, to an alteration of the initial polarization state. A measurement of this effect, or at least 
    establishing upper limits on it, therefore requires some basic knowledge of astrophysical spectropolarimetry.
    A brief introduction to this subject is given here so that the results given in the following chapters are 
    understandable. In this section I will not go into details concerning the diverse generation processes of 
    polarized radiation since it is from a pedagogical point of view by far more useful to do this later in the 
    individual chapters when the corresponding knowledge is needed. Instead I will briefly discuss the three 
    main types of polarization which are, in practice, best described by means of Stokes parameters.
      
    For this purpose, we consider the time harmonic representation of a plane, electromagnetic wave, i.e. when each 
    Cartesian component of ${\bf E}$ is of the form
    \begin{equation} 
      a \, \cos(\tau + \delta) = \Re(a\,e^{-i(\tau+\delta)})   
    \end{equation}
    with $\tau = \omega t - {\bf k}{\bf x}$. Since the electric and magnetic components are related via
    \begin{equation} 
     {\bf B_i}=\sqrt{\mu\epsilon}\frac{{\bf k}\times {\bf E}_i}{k} \quad ,
    \end{equation}
    it is sufficient to consider in the following only the components of ${\bf E}$. Hence, a transversal wave
    propagating in the $z$-direction can be written as
    \begin{eqnarray}\label{wave}
      E_x &=& a_1\cos(\tau+\delta_1) \nonumber \\ 
      E_y &=& a_2\cos(\tau+\delta_2) \\
      E_z &=& 0 \nonumber \quad .
    \end{eqnarray} 
    On the basis of this equations one can distinguish between three types of polarization states by considering
    the nature of the curve which is described by the end points of (\ref{wave}).  
	
    \vspace{0.5cm}
    \noindent {\bf Elliptic Polarization:}
    The most general form of elliptic polarization is recovered by squaring and adding the components of 
    (\ref{wave}) which yields
    \begin{equation}\label{ellipse}
      \left(\frac{E_x}{a_1}\right)^2+\left(\frac{E_y}{a_2}\right)^2-2\frac{E_x}{a_1}\frac{E_y}{a_2}\cos\delta
      =\sin^2\delta
    \end{equation} 
    with $\delta = \delta_2 - \delta_1$.
    Since the associated determinant is not negative, this equation describes an ellipse where, in general, 
    the axes of the ellipse do not coincide with the $x$ and $y$ direction. Instead, the ellipse is characterized
    by (a) the tilt angle $\varphi$ between the $x$-axis and the major axis and (b) by the ratio between major 
    and minor axis described by the quantity $\tan\beta=\xi_1/\xi_2$.

    \begin{figure}[t]
      \centerline{\psfig{figure=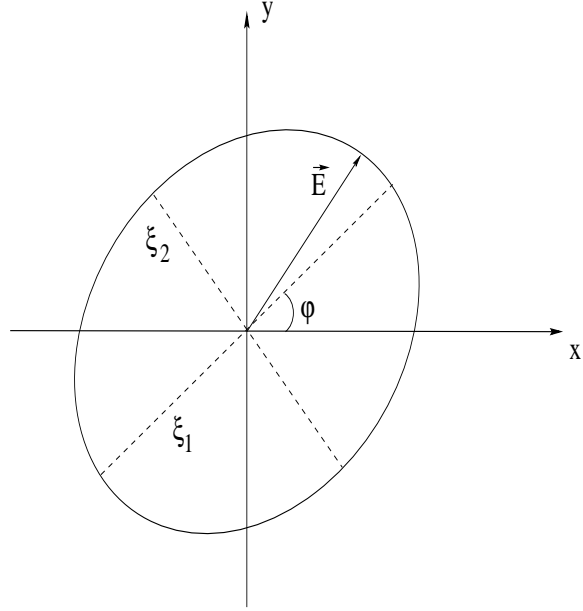,width=3.in,height=3.2in}}
      \caption{Polarization ellipse. The geometry and, hence, the polarization state of a light beam is completely
               characterized by means of $\xi_1$ and $\xi_2$ and the tilt angle $\varphi$. Positive helicity is defined 
	       for an electric vector rotating counterclockwise.}
    \end{figure}
    \vspace{0.5cm}	  
    \noindent {\bf Circular Polarization:}
    The case of circular polarization is recovered if the polarization ellipse degenerates into a circle.
    A necessary condition for this is that $a_1=a_2=a$. In addition always one component of ${\bf E}$ has to be 
    zero while the other has its maximum which is fulfilled by
    \begin{equation}
       \delta = \delta_2 - \delta_1 = m \pi/2 \quad (m=\pm 1,\, \pm 3,\, \pm 5 \ldots)\quad ,  
    \end{equation}	
    so that (\ref{ellipse}) reduces to the equation of a circle $E_x^2+E_y^2=a^2$.
    The wave is said to have positive helicity if $\sin\delta > 0$ and negative helicity for  $\sin\delta < 0$.
    
    \vspace{0.5cm}
    \noindent {\bf Linear Polarization:} 
    A linearly polarized wave is obtained if the ellipse (\ref{ellipse}) reduces to a straight line. This is the 
    case for
    \begin{equation}
       \delta = \delta_2 - \delta_1 = m \pi \quad (m=0,\, \pm 1,\, \pm 2,\, \pm 3 \ldots)\quad ,  
    \end{equation}	
    such that $E_x/E_y=(-1)^m a_2/a_1$.	
  \section{Stokes Parameters}      
    The geometry of the polarization ellipse can be completely characterized by means of three independent parameters,
    e.g. the amplitudes $a_1$, $a_2$ and the phase difference $\delta$, or the major and minor axes $\xi_1$ and $\xi_2$
    and the orientation angle $\varphi$. For the practical use in astronomy it is convenient to use a set of four real
    valued parameters, the so-called Stokes parameters $I$, $Q$, $U$ and $V$, first invented by Sir G.G. Stokes 
    in 1852. In terms of the amplitudes and phases in (\ref{wave}) they can be defined as  
    \begin{eqnarray}\label{stokesdef}
      I &=& a_1^2 + a_2^2 = I_0 + I_{90} = I_{45} + I_{135} = I_{circ}(+) + I_{circ}(-) \\
      Q &=& a_1^2 - a_2^2 = I_0 - I_{90}\\ 
      U &=& 2a_1a_2\cos(\delta_2-\delta_1) = I_{45} - I_{135}\\
      V &=& 2a_1a_2\sin(\delta_2-\delta_1) = I_{circ}(+) - I_{circ}(-)\label{stokesdefv} \quad .              
    \end{eqnarray}
    For completely polarized radiation only three of them are independent so that we have
    \begin{equation}
      I^2 = Q^2 + U^2 + V^2 \quad .  
    \end{equation}        	
    \begin{figure}[t]
      \centerline{\psfig{figure=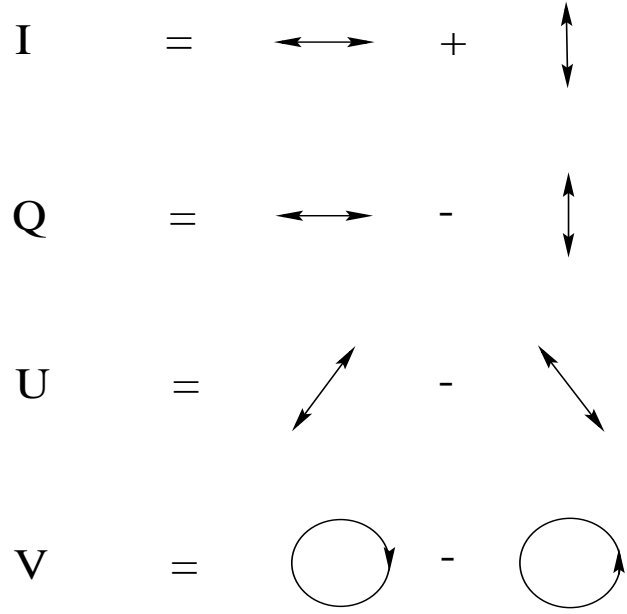,width=3.2in,height=3.2in}}
      \caption{Pictorial representation of the Stokes parameters. The observer is taken to face the radiation
               source (adapted from Landi Degl'Innocenti \cite{xii02}).}
      \label{stoke2}	       
    \end{figure}
    The Stokes parameters have the great advantage that they are quadratic in the amplitudes and, hence, easily
    obtained from a telescope which is equipped with a polarizer. A very evident, operational definition can be given 
    by defining a reference direction in a plane perpendicular to the light beam of interest. Setting the transmission
    axis of an ideal polarizer along this reference direction, a measurement at the exit of this polarizer yields the 
    value $I_0$. This procedure is repeated three times after rotating the polarizer clockwise by the angles $45^{\circ}$,
    $90^{\circ}$ and $135^{\circ}$, respectively, obtaining the values $I_{45}$, $I_{90}$ and $I_{135}$. The linear
    polarizer is then replaced by an ideal filter for positive circular polarization which gives $I_{circ}(+)$ at exit
    and, afterwards, by an ideal filter for negative circular polarization, measuring $I_{circ}(-)$. Then, the 
    operational definition of the Stokes parameters is given by (\ref{stokesdef}) - (\ref{stokesdefv}) which is 
    pictorally summarized in 
    Fig.\ref{stoke2}. Following this definition the fractional degree (or percentage) of linear polarization is 
    given by $(Q^2+U^2)^{1/2}/I$ while the fractional degree of circular polarization is simply $|V|/I$.

\chapter{Solar Observations}

  Since Eddingtons famous measurement of the bending of light at the solar limb during
  a total eclipse in 1919 \cite{ded20}, the sun is counted among the most important 
  objects concerning experimental tests of theories of gravity. Due to it's relative 
  spatial proximity it is possible to determine the relevant initial parameters for a 
  particular experiment like the mass and the distance with high accuracy. 
  This, together with the high gravitational potential of the sun, opens the possibility 
  to look for often tiny effects which makes the difference in predictions between 
  competing theories. 
  
  In this sense we utilize solar polarimetric data to constrain birefringence induced by the 
  gravitational field of the Sun and set limits on the coupling constants $\ell^2$ and $k^2$ 
  required by NGT and metric-affine gravity. The initial parameter mentioned above which is 
  relevant for the major part of this present work represents a prediction about the fractional 
  percentage of Stokes profiles with anomalous symmetry properties in solar polarimetric 
  data. Based on intensive numerical simulations of the creation of Stokes profiles in the 
  solar atmosphere as well as on observations, it is found that the fraction of such 
  profiles is always less than 10\% of all observed profiles (if this number is sufficiently 
  large), independent of the spatial resolution of the instrument. Since a sufficiently 
  strong gravity-induced birefringence could produce any desired amount of anomalous profiles 
  up to 100\%, this value serves as an upper limit for our subsequent analysis. To test this 
  approach, we have developed a second technique which is independent of the previous 
  assumptions and gives comparable constrains.

  The following chapter starts with a description of the two independent statistical 
  approaches, the so-called 'Stokes asymmetry technique' and the 'profile difference
  technique'. The employed data sets, recorded 1995 at the Iza\~na Observatory in Tenerife and 
  in 2000 in Locarno, are described in section 2.2 together with the relevant technical
  details of the instruments. We have measured the line profiles of the full Stokes vector
  in four spectral lines which gave us a total of 1120 spectra. The Stokes asymmetry 
  technique yields $\ell^2_{\odot} < (57.1\, \mbox{km})^2$ in line Cr I 5247.56{\AA}
  in the case of NGT and $k^2_{\odot} < (1.1\, \mbox{km})^2$ in the same line for 
  metric-affine gravity, respectively. These results are consistent with the profile 
  difference technique. In the case of NGT, the result is seven orders of magnitude smaller
  than the $\ell^2_{\odot}$ value favored by Moffat (Moffat 1979 \cite{mof79}).

\newpage
\section{Technique}
  We follow two strategies to test for gravitational birefringence. One of these
  was outlined by Solanki \& Haugan (1996)\cite{sah96}. It is summarized and its implementation
  is described in Sec. 2.1.1. The other technique is new and is described in Sec. 
  2.1.2.
    
  \subsection{Stokes asymmetry technique}   
    The strategy proposed by Solanki \& Haugan (1996)\cite{sah96} makes use of the symmetry 
    properties of the Stokes profiles produced by the Zeeman splitting of atomic 
    spectral lines. In the absence of radiative transfer effects in a dynamic 
    atmosphere net circular polarization, Stokes $V$, is antisymmetric in wavelength 
    and net linear polarization aligned at $45^{\circ}$ to the solar limb, Stokes $U$, 
    is symmetric. Gravitational birefringence changes the phase between orthogonal 
    linear polarizations and thus partly converts Stokes $V$ into Stokes $U$ and 
    vice versa. However, $U$ produced from $V$ by gravitational birefringence 
    still has the symmetry of $V$ and can thus be distinguished from the Zeeman signal.
  
    Let $\Delta\Phi$ be the phase shift which accumulates at the central wavelength
    of a spectral line between Stokes $V$ and $U$ as light propagates from a point on 
    the solar surface to the observer. Formulae for $\Delta\Phi$ as predicted by
    metric affine theories were already given in chapter 1. For Moffat's NGT 
    (Moffat 1979 \cite{mof79}) a corresponding expression has been published by 
    Gabriel et al. (1991)\cite{gab91}. Further, let subscripts '$a$' and '$s$' 
    signify the antisymmetric and symmetric parts of the Stokes profiles, respectively, 
    and the subscripts 'src' and 'obs' the Stokes profiles as created at the source 
    and as observed, respectively. Then,
    \begin{eqnarray}
      \frac{U_{\asc}}{U_{\ssc}} &=& 
      \frac{V_{\aob}\;\, \sin\Delta\Phi + U_{\aob}\;\, \cos\Delta\Phi}
           {V_{\sob}\;\, \sin\Delta\Phi + U_{\sob}\;\, \cos\Delta\Phi} \label{U_{src}}\\ 
      \frac{V_{\ssc}}{V_{\asc}} &=& 
      \frac{V_{\sob}\;\,\cos\Delta\Phi + U_{\sob}\;\,\sin\Delta\Phi}
           {V_{\aob}\;\,\cos\Delta\Phi + U_{\aob}\;\,\sin\Delta\Phi} \label{V_{src}} \quad .
    \end{eqnarray}
    Thus for observed symmetric and antisymmetric fractions of $U$ and $V$ Eqs.(\ref{U_{src}})
    and (\ref{V_{src}}) predict the ratios $U_{\fa}/U_{\fs}$ and 
    $V_{\fs}/V_{\fa}$ at the solar source.
  
    If the solar atmosphere were static these ratios would vanish $(U_{\asc}=V_{\asc}=0)$, 
    so that any observed $U_{\fa}$ or $V_{\fs}$ would be due to either $\Delta\Phi$ 
    or noise: $U_{\aob} = V_{\asc} \sin\Delta\Phi$, $V_{\sob} = U_{\ssc}\sin\Delta\Phi$. 
    The solar atmosphere is not static, however, and consequently the Stokes profiles do 
    not fulfill the symmetry properties expected from the Zeeman effect even for rays 
    coming from solar disc centre, which are unaffected by gravitational birefringence. 
    This asymmetry has been extensively studied, in particularly for Stokes $V$, which 
    most prominently exhibits it (e.g. Solanki \& Stenflo 1984 \cite{sos84}, Grossmann-Doerth et al. 
    1989 \cite{gd89}, Steiner et al. 1999 \cite{sgs99}, Mart\'{\i}nez Pillet et al. 1997 
    \cite{mp97}). Although most profiles 
    have $V_{\fs}/V_{\fa} \lsim 0.2$, a few percent of $V$ profiles exhibit $V_{\fs}/V_{\fa}$ 
    values close to unity, even at solar disc centre. Such profiles occur in different 
    types of solar regions, e.g. the quiet Sun (Steiner et al. 1999 \cite{sgs99}), active region 
    neutral lines (Solanki et al. 1993 \cite{sea93}) and sunspots (S\'{a}nchez Almeida 
    \& Lites 1992 \cite{sal92}). 
    The magnitude of $V_{\fs}/V_{\fa}$ decreases rapidly with increasing $V_{\fa}$ and 
    profiles with $V_{\fs}/V_{\fa} \gsim 1$ are all very weak. They are often associated 
    with the presence of opposite magnetic polarities within the spatial resolution element 
    and a magnetic vector that is almost perpendicular to the line of sight, situations
    which naturally give rise to small $V$ (e.g. S\'{a}nchez Almeida \& Lites 1992 \cite{sal92}, 
    Ploner et al. 2001 \cite{plo01}).     
  
    The observed Stokes $U$ asymmetry is on average smaller than the $V$ asymmetry. This is 
    true in particular for extreme asymmertic values, i.e. $(U_{\aob}/U_{\sob})_{\mbox{\footnotesize{max}}} 
    \ll (V_{\sob} /V_{\aob})_{\mbox{\footnotesize{max}}}$. Since this relation also holds at 
    $\cos\theta = \mu = 1$ where $\theta$ denotes the heliozentric angle between the source
    on the solar surface and the line-of-sight, it is valid for source profiles as well. 
    Thus, S\'{a}nchez Almeida \& Lites (1992) \cite{sal92} point out that Stokes $U$ retains 
    $U_{\aob}/U_{\sob} \ll 1$ throughout a sunspot, although $V_{\fs}/V_{\fa} > 1$ is 
    invariably achieved at the neutral line. 
    The reason for the smaller maximum asymmetry lies in the fact that Stokes $U$ senses the 
    transverse magnetic field. Since velocities in the solar atmosphere are directed mainly 
    along the field lines they generally have a small line-of-sight component when $U$ has a
    significant amplitude. Sizable line-of-sight velocities are needed, however, to produce a
    significant asymmetry (Grossmann-Doerth et al. 1989 \cite{gd89}). Another reason for the smaller 
    maximum $U$ asymmetry is that, unlike Stokes $V$, it does not distinguish between oppositely 
    directed magnetic fields.
  
    Thus it is not surprising that in the following analysis Stokes $U$ provides tighter 
    limits than Stokes $V$. Another reason is that due to the on average stronger observed $V$
    profiles asymmetries introduced in $U$ (through gravitationally introduced cross-talk from $V$)
    are larger than the other way round. However, we also analyse Stokes $V$ as a consistency
    check. 

    In order to seperate the asymmetry produced by solar effects from that introduced 
    by gravitational birefringence, one strategy to follow is to consider large amplitude
    Stokes profiles only.
    Another is to analyse data spanning a large range of $\mu$ values, since gravitational 
    birefringence follows a definite centre-to-limb variation, as predicted by particular 
    gravitation theories. Finally, the larger the number of analysed line profiles, the 
    more precise the limit that can be set on $\Delta\Phi$. Better statistics not only 
    reduce the influence of noise, they are also needed because for a single profile 
    gravitational birefringence can both increase or decrease $V_{\fs}/V_{\fa}$ and 
    $U_{\fa}/U_{\fs}$. The latter may become important if the source profiles are strongly 
    asymmetric. Thus a small observed $V_{\fs}/V_{\fa}$ or $U_{\fa}/U_{\fs}$ is in itself
    no guarantee for a small gravitational birefringence. However, since almost all source
    profiles are expected to have $V_{\fs}/V_{\fa} \ll 1$ and $U_{\fa}/U_{\fs} \ll 1$, on average 
    we expect gravitational birefringence to increase these ratios.

  \vspace*{0.5cm}
  \subsection{Profile difference technique}
    The profile difference technique relies on the fact that $\Delta\Phi$ is expected 
    to be a strong function of $\mu$ which is confirmed in the two concrete cases of 
    NGT and metric affine theories (see chapter 1). This means that for any sufficiently large NGT 
    charge $\ell_{\odot}$ or equivalent metric affine parameter $k$ a mixture of Stokes 
    $V$ and $U$ profiles will be observed across the solar disc irrespective of 
    the {\it relative} numbers and strengths of $V$ and $U$ profiles leaving the solar
    photosphere.              
    Thus, irrespective of the value of $<|V_{\src}|>-<|U_{\src}|>$
    for sufficiently large $\ell_{\odot}$ or $k$ $<|V_{\obs}|>-<|U_{\obs}|>$ will tend to zero.
    The averaging is over all $\mu$ values and the total number of profiles is assumed to 
    be very large.
  
\begin{figure}[t]     
      \centerline{\psfig{figure=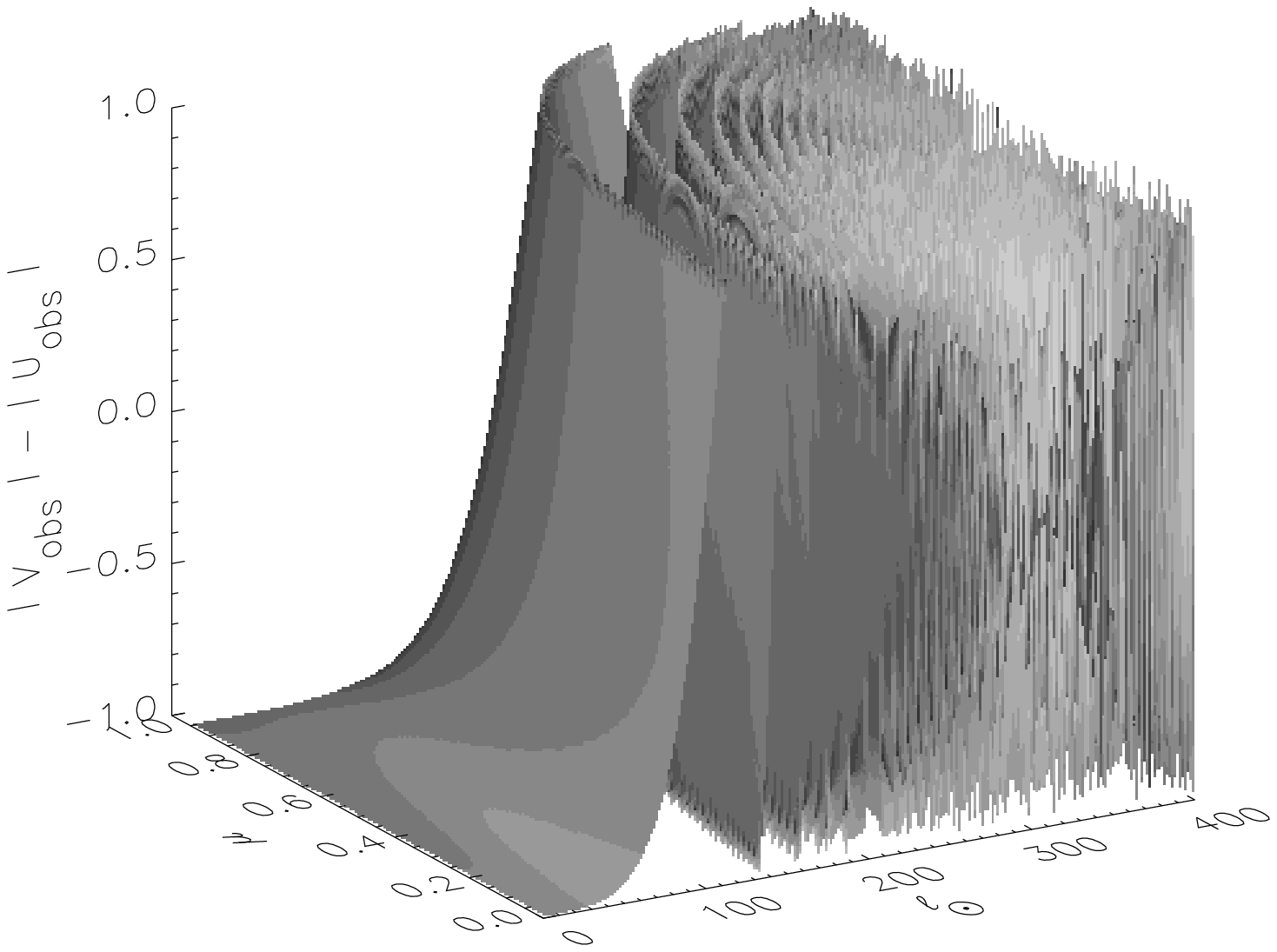,height=2.7in,width=2.7in}
                  \psfig{figure=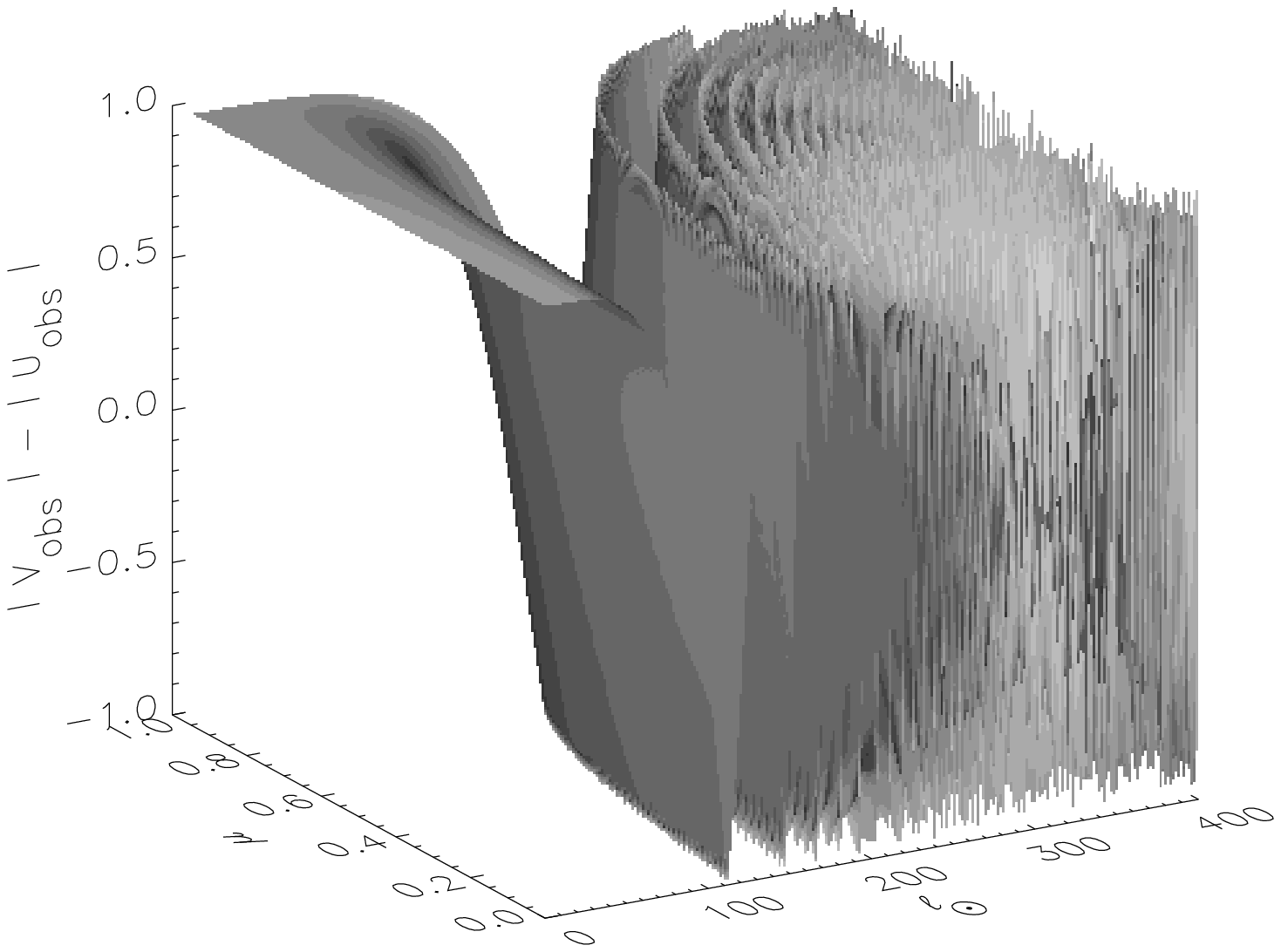,height=2.7in,width=2.7in}}
      \centerline{\psfig{figure=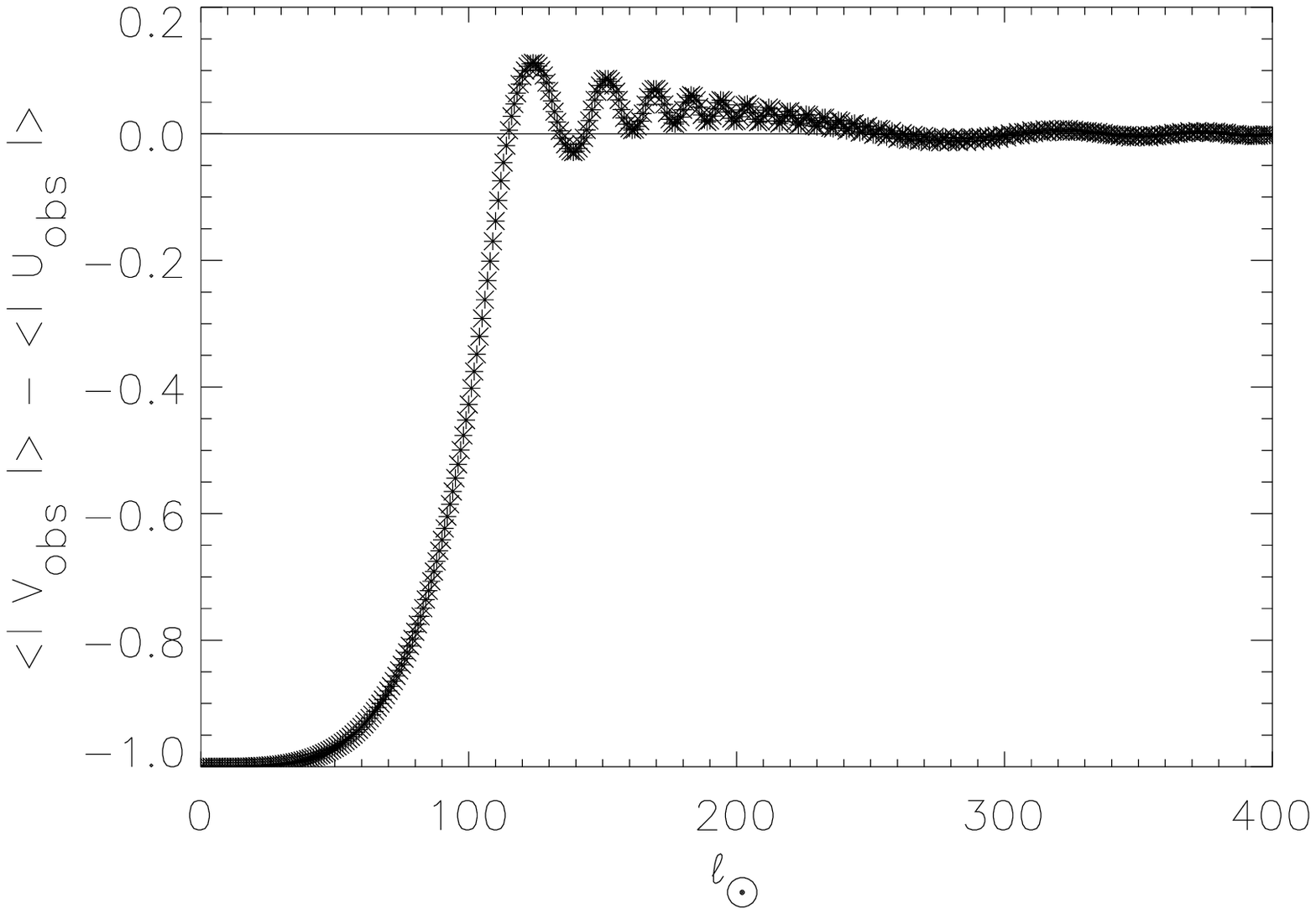,height=2.7in,width=2.7in}
                  \psfig{figure=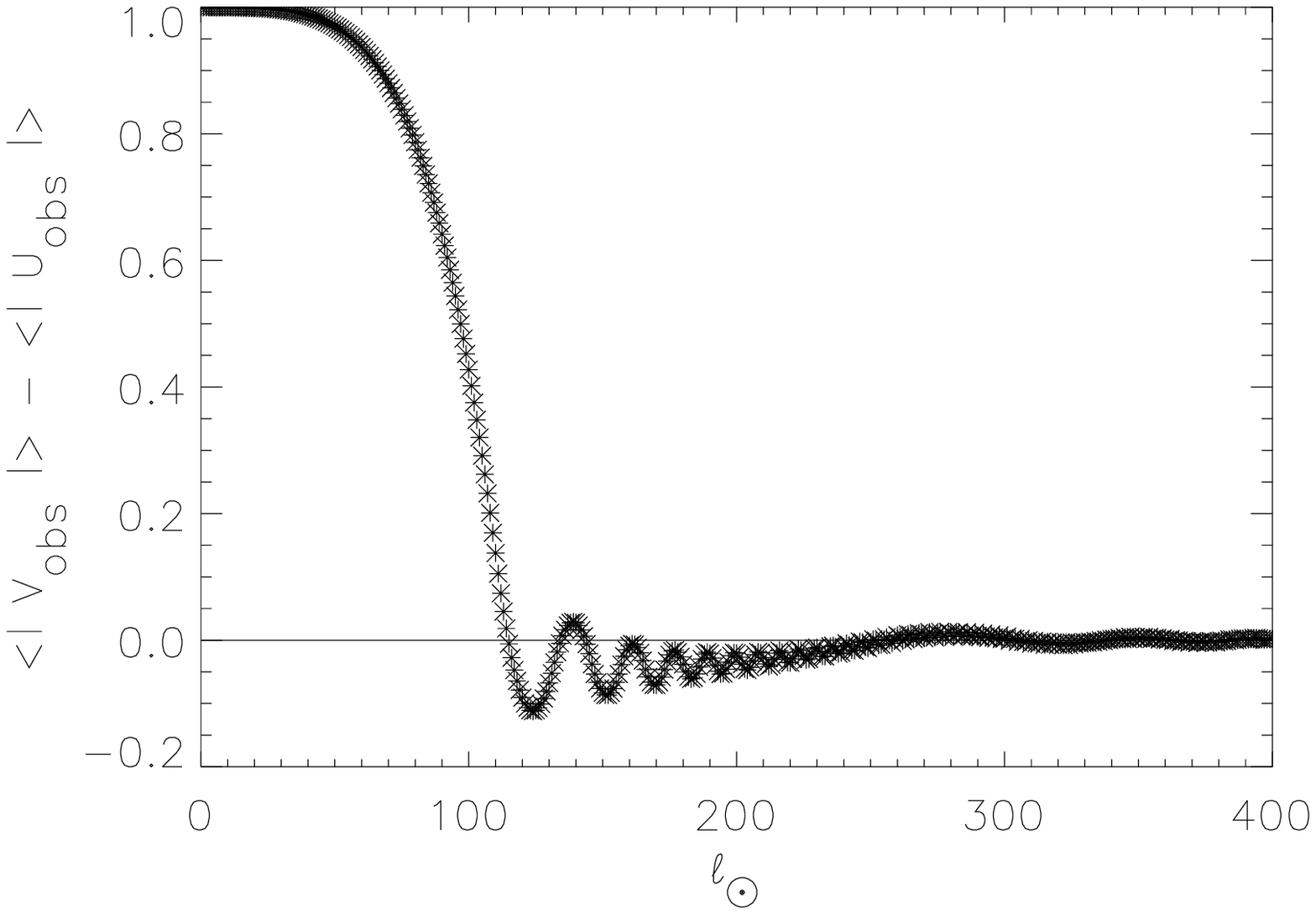,height=2.7in,width=2.7in}}		  
      \caption{Top: $|V_{\obs}| - |U_{\obs}|$ vs. $\ell_{\odot}$ and $\mu$ for 
               $|V_{\src}| = 0$ (left) and $|U_{\src}| = 0$ (right).
	       Bottom:$|V_{\obs}| - |U_{\obs}|$ averaged over all $\mu$ for 
	       the cases above. }
      \label{pdt1}
    \end{figure}

    This effect is illustrated Fig. \ref{pdt1}. In in the top of Fig.\ref{pdt1} 
    $|V_{\obs}| - |U_{\obs}|$ is plotted vs. $\ell_{\odot}$ and $\mu$ for the extreme cases 
    $|V_{\src}| = 0$ (Fig.\ref{pdt1} top left) and $|U_{\src}| = 0$ (Fig.\ref{pdt1} top right). 
    Other combinations of $|V_{\src}(\mu)|$ and $|U_{\src}(\mu)|$ give qualitatively similar results.
    
    The $|V_{\obs}| - |U_{\obs}|$ surface oscillates even more rapidly with increasing
    $\ell_{\odot}$ and with decreasing $\mu$. Lines of equal $|V_{\obs}| - |U_{\obs}|$
    are strongly curved in the $\ell_{\odot} - \mu$ plane. These two points combine to lead
    to decreasing $<|V_{\obs}|>-<|U_{\obs}|>$ with increasing $\ell_{\odot}$.    
    This is shown in the bottom of Figs.\ref{pdt1} left and right for the cases illustrated 
    in the figures above. 
    As expected the $<|V_{\obs}|>-<|U_{\obs}|>$ vs. $\ell_{\odot}$ curves exhibit a rapidly 
    damped oscillation around zero.
    This effect can be used to set upper limits on gravitational birefringence if the
    observations exhibit a $ <|V_{\obs}|>-<|U_{\obs}|>$ that differs significantly from
    zero. As is later shown, this is indeed the case.
    

\section{Observations and data}
  Two sets of data have been analysed in the present work. They are described
  below.
  \subsection{Data obtained in 1995}
    Observations were carried out from 7 - 13th Nov. 1995 with the Gregory 
    Coud\a'e Telescope (GCT) at the Iza\~na Observatory on the
    Island of Teneriffe. For the polarimetry we employed the original version 
    of the Z\"urich Imaging Polarimeter (ZIMPOL I), which employs 3 CCD cameras,
    one each to record Stokes $I\pm Q,$ $I\pm U$ and $I\pm V$ simultaneously
    (e.g. Keller et al. 1992 \cite{kea92}). 
    
    The recorded wavelength range contains four prominent spectral lines, 
    Fe I 5247.06{\AA}, Cr I 5247.56{\AA}, Fe I 5250.22{\AA} \& Fe I 5250.65{\AA}.  
    Three of these spectral lines are among those with the largest Stokes amplitudes 
    in the whole solar spectrum and are also unblended by other spectral lines (Solanki 
    et al. 1986 \cite{sea86}). Blending poses a serious problem since it can affect the blue-red 
    asymmetry of the Stokes profiles. By analysing more than one such line it 
    is possible to reduce the influence of hidden blends and noise.
    Nowhere else in the visible spectrum are similar lines located sufficiently
    close in wavelength that they can be recorded simultaneously
    on a single detector. Also, compared to other lines with large Stokes
    amplitudes the chosen set lies at a short wavelength. This is important since
    the influence of gravitational birefringence on line polarization is proportional 
    to $1/\lambda$. The sum of the above properties make the chosen range almost 
    uniquely suited for our purpose.

    In order to image all 4 spectral lines of interest onto a single CCD we introduced
    reduction optics between the image plane of the spectrograph and the detectors.
    They produce an image-scale reduction by a factor of 3.2. The final spectral 
    resolving power $\lambda/\Delta\lambda$ corresponded to 210'000. The spatial 
    scale corresponding to a pixel was $1.13''$ (or 860 km on the sun). However, 
    the spatial resolution of the data is limited by turbulence in the Earth's 
    atmosphere, the so-called seeing. This varied somewhat in the course of the 
    observing run, so that the estimated angular resolution of the observations 
    lies between $1.1''$ and $3''$.
    
    The modulator package, composed of 2 piezoelectric modulators oscillating
    at frequencies of around 50'000 \& 100'000 and a linear polarizer (glass prism),
    was placed ahead of the entrance slit to the spectrograph, but was nevertheless
    (unavoidably) located after 2 oblique reflections in the telescope. Oblique 
    reflections produce cross-talk between Stokes parameters, i.e. they partially
    convert one form of polarization into another. Since we are trying to observe, 
    or at least set limits on ''cross-talk'' between Stokes $U$ and $V$ due to 
    gravitational birefringence we took some trouble to reduce the instrumental 
    cross-talk to the extent possible. A first step was the choice of the telescope. 
    With only two oblique reflections, whose relative angles change only slowly in
    the course of a year, the GCT is relatively benign compared to most other large 
    solar telescopes. Secondly, a half-wave plate was introduced between the two 
    oblique reflections. S\'{a}nchez Almeida et al. \cite{smw91,smk95} have 
    pointed out that a half-wave plate at that location should, under ideal circumstances, 
    completely eliminate all instrumental cross-talk. To test the efficiency of the
    half-wave plate in suppressing instrumental cross-talk between Stokes $Q$, $U$ and $V$ 
    we first carried out a series of observations of a sunspot near the centre of the 
    solar disc both with and without a half-wave plate introduced in the light path. 
    Such tests were necessary since the half-wave plate available at the GCT is not
    optimized for the observed wavelength. Note that at solar disc centre $(\mu = 1)$ 
    gravitational birefringence disappears, so that we test for instrumental cross-talk 
    only. The half-wave plate was indeed found to significantly reduce instrumental 
    cross-talk. Remaining cross-talk was removed during data reduction using a numerical 
    model of the telescope that includes an imperfect half-wave plate (adapted 
    from a model kindly provided by V. Mart\'{\i}nez Pillet). The parameters of the 
    model were adjusted slightly using the observations of the sunspot umbra close to 
    disc centre. We estimated that the residual cross-talk after this procedure is 
    at the level of a few percent. Since Stokes $V$, $Q$, $U$ profiles generally have 
    amplitudes of $0.1I_c$ or less, the influence of the cross-talk is of the same order as
    the noise, which is roughly $1-2\times10^{-3} I_c$, where $I_c$ is the continuum 
    intensity. Photon noise is by far the largest contributor to this noise level. 
    At this level instrumental cross-talk caeses to be of concern for our analysis.
    
    The Zimpol polarimeters are unique in that they combine CCD detectors with a very
    high modulation frequency $(\geq 50'000 $ Hz) and hence preclude distortion of the
    Stokes profiles and cross-talk between them due to seeing fluctuations. This again 
    improves the accuracy of the profile shapes and hence the accuracy of our results.
    
    A total of 106 recordings were made at different locations on the solar disc in 
    an attempt to cover a large range of $\mu$ homogenously. Particular emphasis was 
    placed on observations close to the limb since gravitational birefringence is expected 
    to be largest for such rays.
     
    Since only a single sunspot was present on the solar disc during the observing run 
    most recordings refer to faculae and network features, i.e. magnetic features with
    lower Stokes $Q$, $U$, $V$ signals. 
    \begin{figure} 
      \vspace{0.5cm}
      \centerline{\psfig{figure=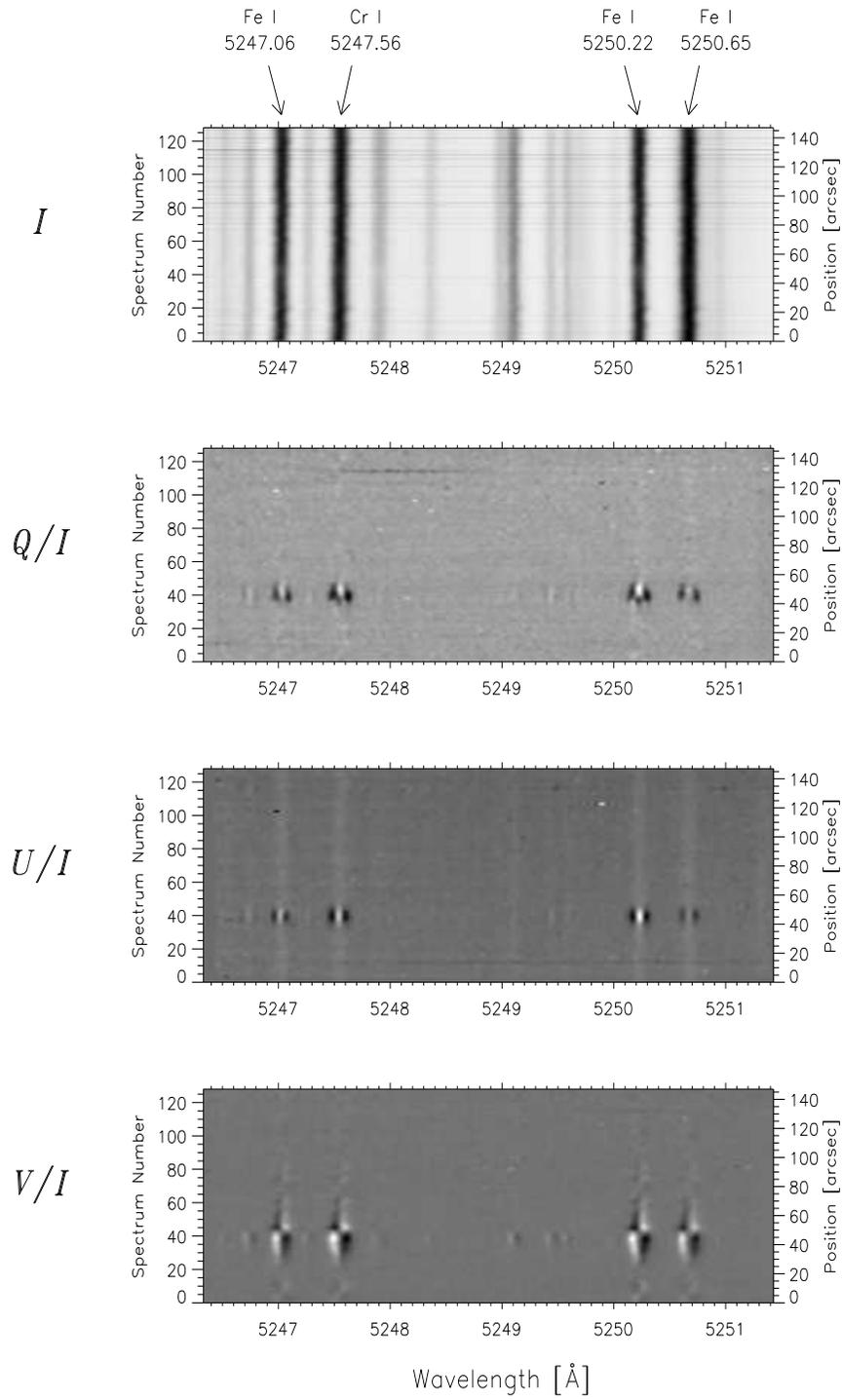,width=4.37in,height=7.02in}}
      \vspace{0.5cm}
      \caption{Sample spectrum of Stokes $I, Q, U$ and $V$.}
      \label{n4fac}
    \end{figure}       
    In Fig.\ref{n4fac} a sample Stokes $I,Q,U,V$ spectrum of a facular region near the solar 
    limb is plotted. The 4 analyzed spectral lines are identified. These data were fully 
    reduced following the tedious, but well-tested procedures described by Bernasconi (1997)
    \cite{ber97}.
    
  \subsection{Data set of March 2000}

    In order to improve the statistics and the $\mu$ coverage a second observing run was 
    carried out in March 2000 with the Coud\a'e - Gregory Telescope in Locarno, Switzerland. 
    This telescope is almost identical to the GCT on Tenerife and the parameters such as 
    spectral resolution, noise level etc. are very similar to those of the 1995 observations.
    
    The second generation, ZIMPOL II polarimeter (Gandorfer \& Povel 1997, Povel 1998 
    \cite{gap97,pov98}) was employed
    for the polarization analysis and data recording. It simultaneously records three
    of the four Stokes parameters, either Stokes $I,Q,V$ or $I,U,V$ on a single CCD
    detector chip. Observations in these two modes were interlaced, such that alternate
    exposures record Stokes $I,Q,V$ and $I,U,V$, respectively.
    Exposures of the same Stokes parameters were then added together to reduce noise.
    Thus the final data set consists of all four Stokes parameters. The only differences 
    with respect to the recordings made in 1995 are that the number of 
    spatial pixels is reduced and that the noise level of Stokes $V$ in the newer 
    data is a factor of $\sqrt{2}$ lower than of Stokes $Q$ and $U$, whereas Stokes 
    $Q$, $U$ and $V$ had the same noise level in the earlier recordings. Due to the 
    superior modulation scheme implemented in ZIMPOL II Stokes $Q$ and
    $U$ achieve a noise level of $10^{-3}I_c$ after roughly the same exposure time 
    during the observations made in 2000 as during the earlier campaign.
    
    These observations were carried out on the day of the equinox, at which time
    the two mirrors producing oblique reflections of the beam ahead of the modulator package
    are oriented such that their polarization cross-talk cancels out. Hence for
    these observations the instrumental cross-talk is essentially zero and no further
    treatment of the data for this effect is required.

    The Sun was very active at the time of these observations with many active regions
    harboring sunspots and faculae present on the solar disc. Since active regions
    generally give larger amplitude Stokes signals we concentrated on observing them. 
    A total of 7 exposures were made. The typical seeing during these observations was estimated
    to be $2 - 3''$, while the spatial pixel size was $1.13''$. 
   
    \vspace{0.5cm}  
      \begin{figure}[h]                 
       
\centerline{\psfig{figure=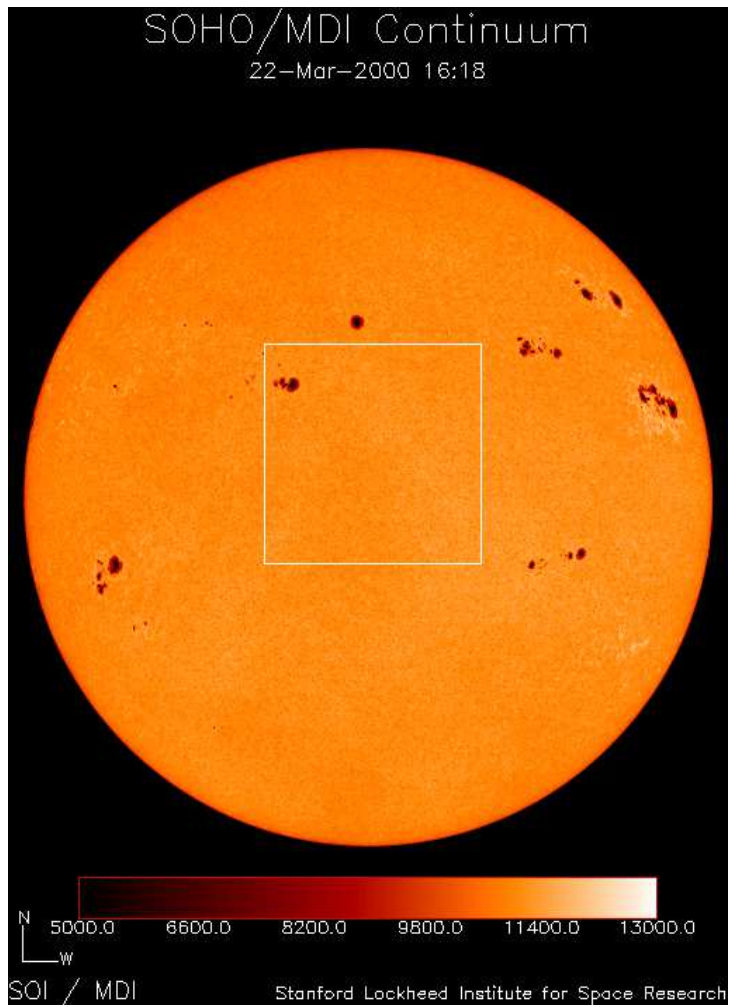,height=5.22in,width=3.76in}}
        \caption{SOHO MDI intensity image from 22nd March 2000.}
      \end{figure}       
 
    \newpage

\section{Data analysis}
  Each exposure gives us the line profiles of the 4 spectral lines in Stokes $I, Q, U$
  and $V$ at a set of 94 (128 in the 1995 data) positions on the solar disc. Once the
  reduction and calibration procedure is completed we select from a given frame those 
  spectra for further analysis for which the $S/N$ ratio for either Stokes $U$ or $V$ 
  in at least one of the four spectral lines is above 12 in the 1995 data and above 15 
  in the 2000 data.

  This criterion gave us a total of 4480 profiles from the four lines and were analysed further.
  The non-orthogonal wavelet-packets smoothing scheme of Fligge \& Solanki (1998) \cite{fas98}
  was employed to enhance the $S/N$ ratio by a factor of 1.5 - 2 without significantly 
  affecting the profile shapes. 
  
  Then a set of parameters was determined of all Stokes profiles of all 4 spectral lines. 
  Of these parameters only the signed amplitudes of the blue and red wings, $a_{\mbox{\footnotesize{b}}}$ and 
  $a_{\mbox{\footnotesize{r}}}$ (i.e. of the blue and red Zeeman $\sigma$-component) are 
  of relevance for the present study. In Fig. 2 we plot a Stokes $U$ and $V$ profile 
  of the Fe I line at 5250.22{\AA}. $a_{\mbox{\footnotesize{b,r}}}(V)$ and 
  $a_{\mbox{\footnotesize{b,r}}}(U)$ are indicated in the figure. Using these we can 
  form the symmetric and antisymmetric parts of the Stokes $V$ and $U$ 
  profile amplitudes, $V_{\fs}=(a_{\mbox{\footnotesize{b}}} + a_{\mbox{\footnotesize{r}}})/2$,
  $V_{\fa}=(a_{\mbox{\footnotesize{b}}} - 
  a_{\mbox{\footnotesize{r}}})/2$, $U_{\fs}=(a_{\mbox{\footnotesize{b}}} + 
  a_{\mbox{\footnotesize{r}}})/2$ and $U_{\fa}=(a_{\mbox{\footnotesize{b}}} - 
  a_{\mbox{\footnotesize{r}}})/2$, respectively, which enter Eqs. (\ref{U_{src}}) 
  and (\ref{V_{src}}). In the following all $V$ and $U$ values and parameters are normalized to the 
  continuum intensity, although this is often not explicitely mentioned.
  \begin{figure}[h]     
    \centerline{\psfig{figure=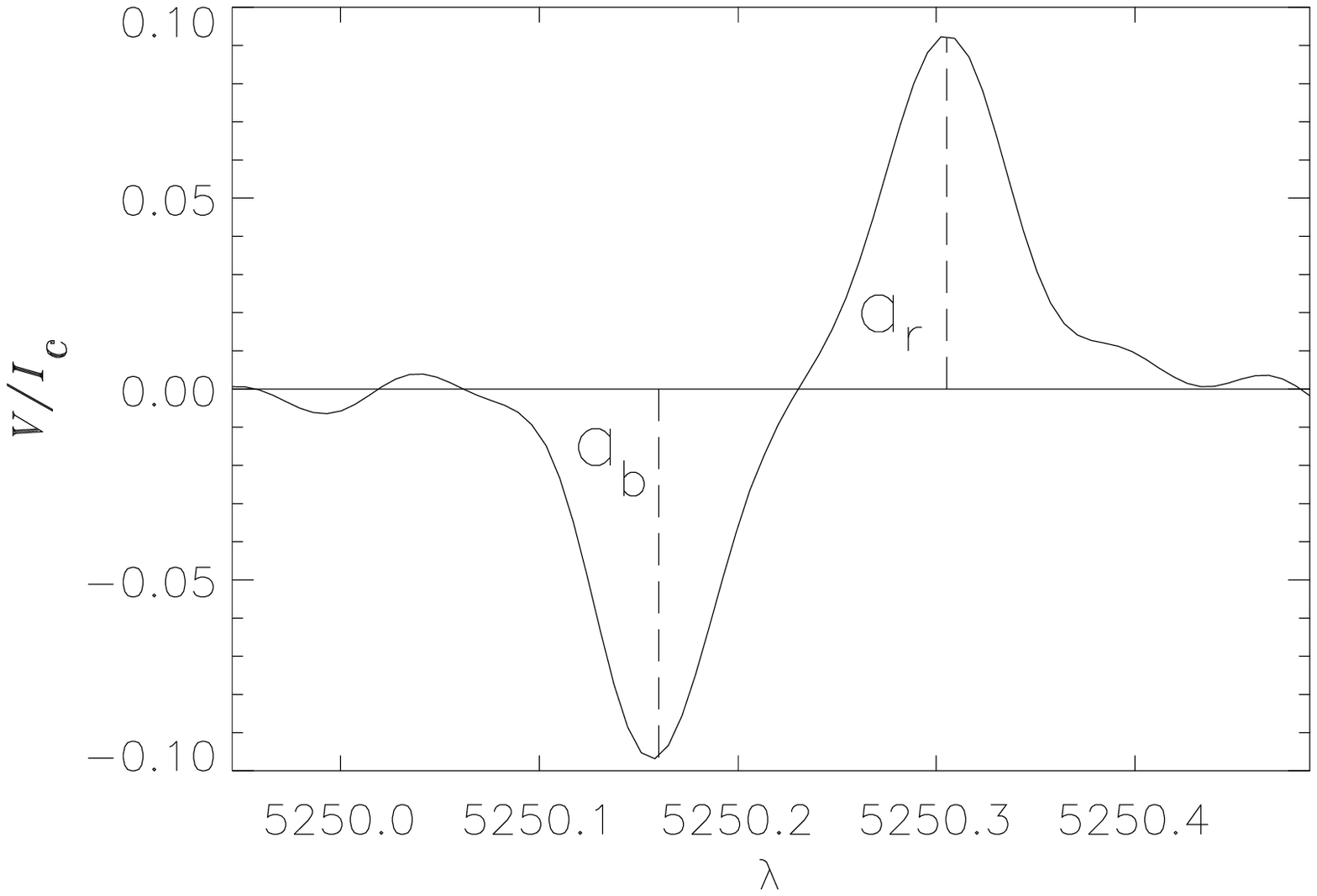,height=3.2in,width=3.in}
                \psfig{figure=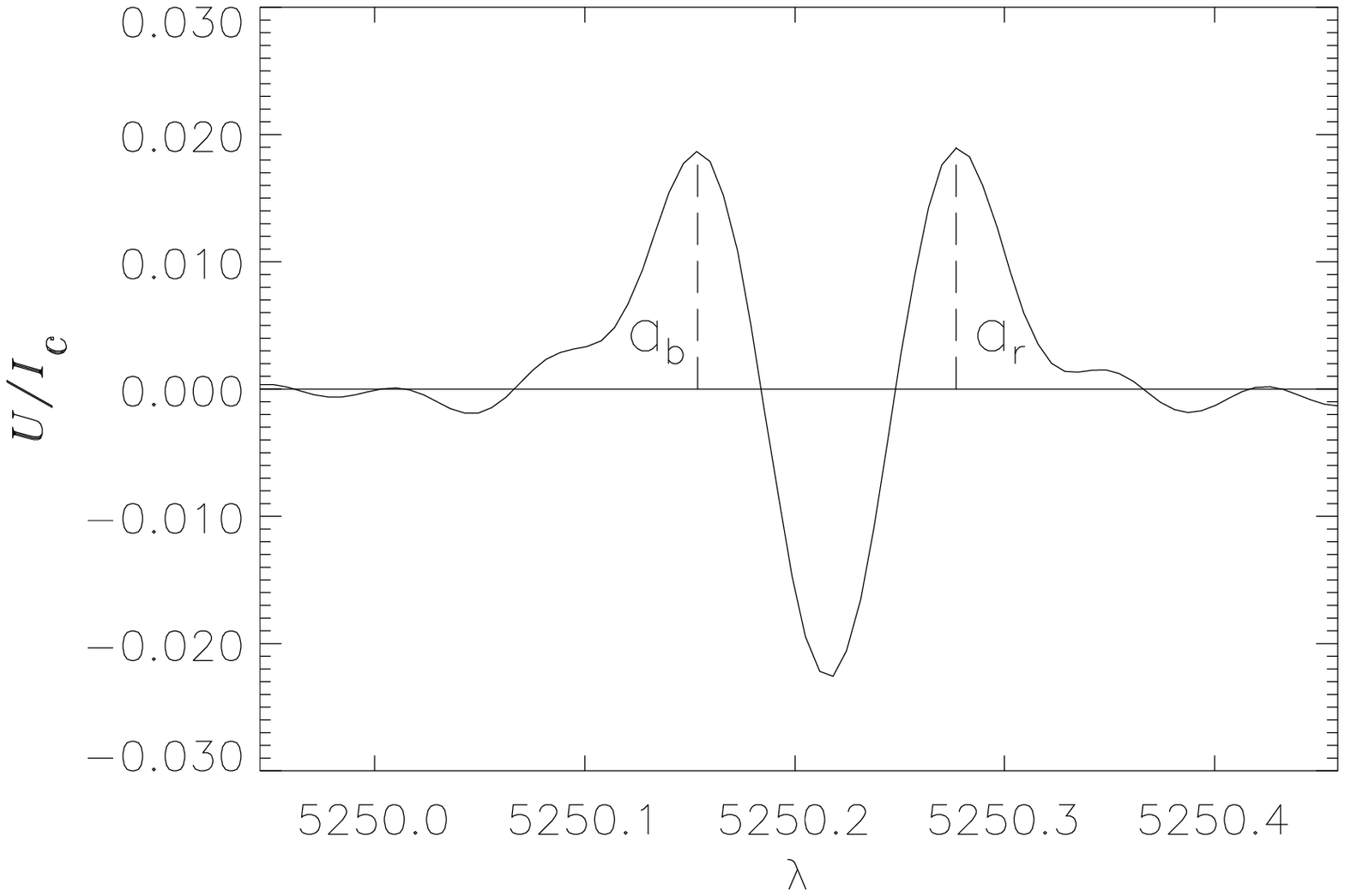,height=3.2in,width=3.in}}
    \caption{Simultaneously measured profiles of Stokes $V$ and Stokes $U$
             of the Fe I line at 5250.22{\AA}. }
    \label{fig2}
  \end{figure}       

  \newpage
  \subsection{Stokes asymmetry technique}
\begin{figure}[t]     
      \centerline{\hspace*{-1.cm}\psfig{figure=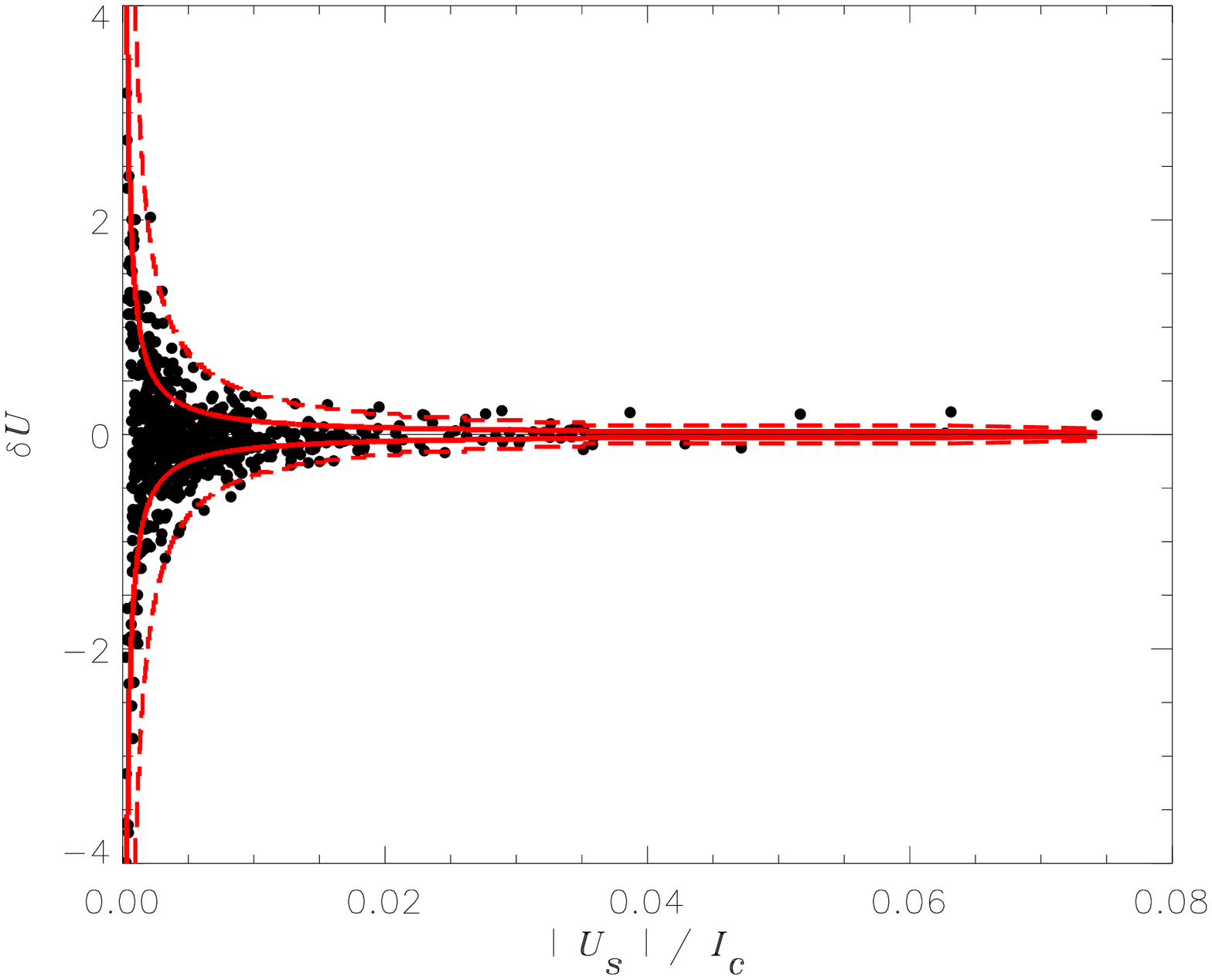,height=3.3in,width=3.3in}\hspace{-0.3cm}
                  \psfig{figure=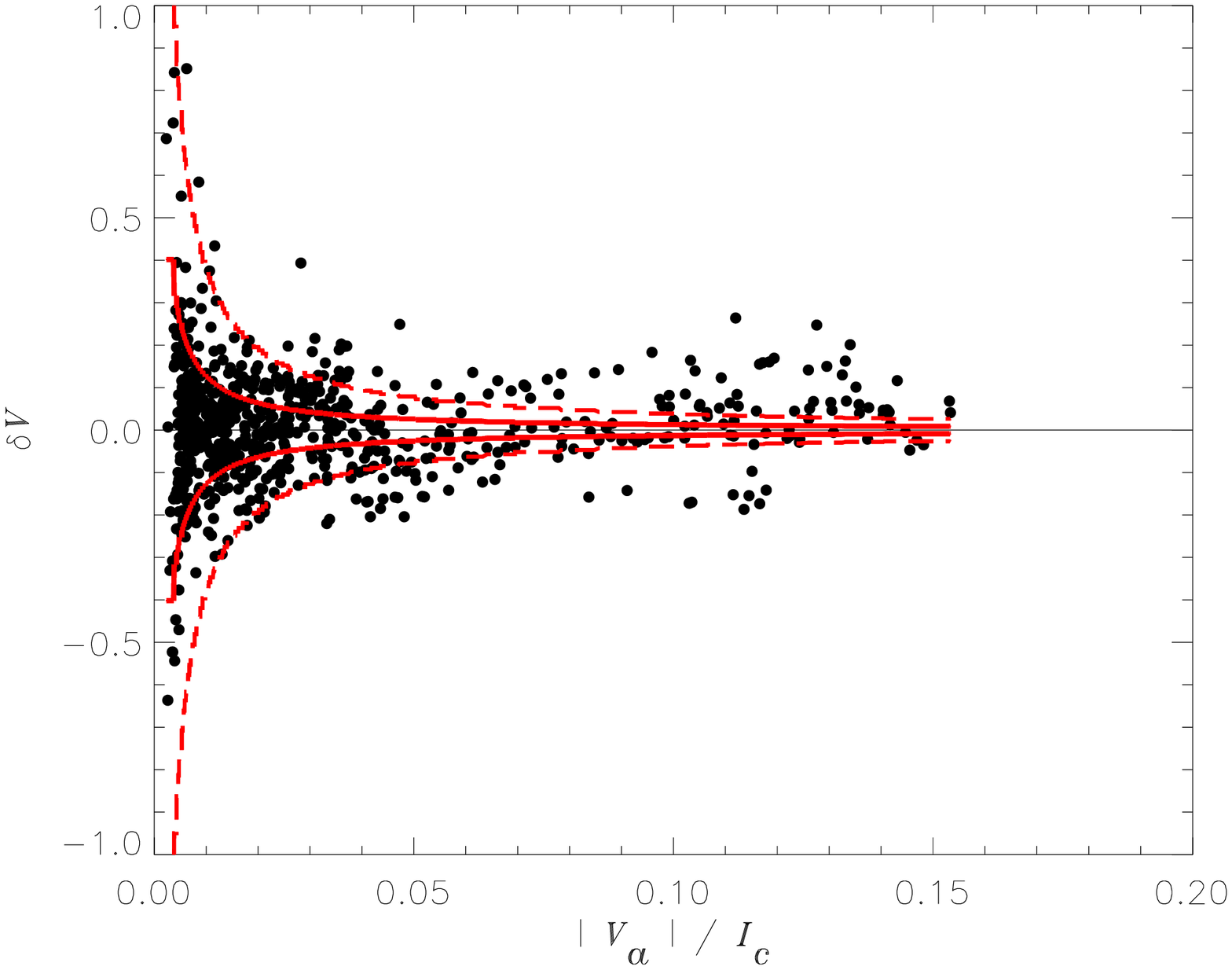,height=3.3in,width=3.3in}}
      \caption{Measured amplitude asymmetries for Stokes $U$ and Stokes $V$. }
      \label{sat1}
    \end{figure}

\begin{figure}[h]     
      \centerline{\psfig{figure=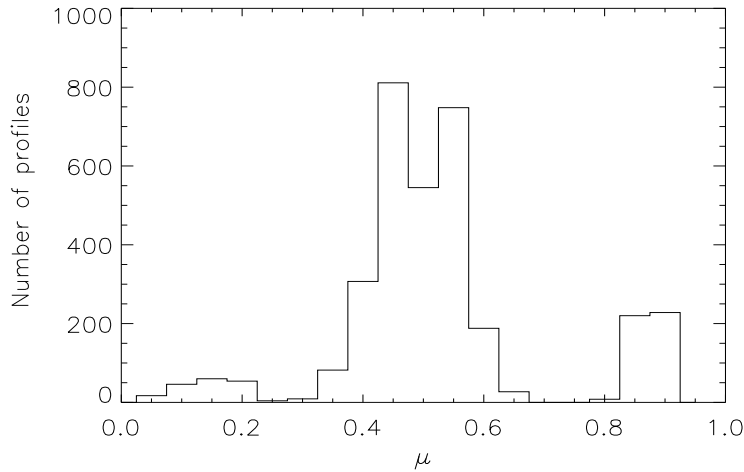,height=2.75in,width=4.18in}}
      \caption{Histogram of the number of profiles as a function of $\mu$.}
      \label{histo}
    \end{figure}

    A measure of the asymmetry of a Stokes profile is given by the ratio 
    $\delta V = V_{\fs}/V_{\fa}$, respectively $\delta U = U_{\fa}/U_{\fs}$ 
    (e.g. Solanki \& Stenflo 1984 \cite{sos84}, Mart\'{\i}nez Pillet et al. 1997 \cite{mp97})
    In Fig. \ref{sat1} we plot these quantities vs. $|V_{\fa}|$ and $|U_{\fs}|$, respectively. 
    Each point in these plots refers to a Stokes profile of the Fe I 5250.65{\AA} line. 
    Although our observations cover a range of $\mu$ values, Fig. \ref{sat1} is very 
    similar to corresponding figures based on data obtained near solar disc centre (Grossmann-Doert 
    et al. 1996 \cite{gd96}, Mart\'{\i}nez Pillet et al. 1997 \cite{mp97}), where the influence of
    gravitational birefringence is expected to be negligible. For large amplitudes ($V_{\fa}$, 
    $U_{\fs}$) the relative asymmetry ($V_{\fs}/V_{\fa}$,$U_{\fa}/U_{\fs}$) is small, 
    while for weaker profiles it shows an increasingly large spread. For the weaker profiles
    this spread is mainly due to noise as can be judged from the solid and dashed curves in Fig. \ref{sat1},
    which outline the $1\sigma$ and $3\sigma$ spread expected due to photon noise, respectively. 
    The curves reveal that Stokes profiles with amplitudes ($V_{\fa}$, $U_{\fs}$) smaller than one percent 
    of the continuums intensity are so strongly affected by noise that they are of little use 
    for the present purpose. This leaves us with 1966 individual profiles for further analysis. 
    In Fig.\ref{histo} we plot a histogram of the number of these profiles as a function of $\mu$. 
    The distribution is uneven, being determined by the position on the solar disc of magnetic 
    features at the time of the observations.
  
    The further analysis is made more complicated by the fact that a $V_{\fs}$ and a $U_{\fa}$
    signal can be produced not just by gravitational birefringence, but also by radiative
    transfer processes acting in the dynamic solar atmosphere, as described in Sect. 3.
    To circumvent this problem we consider all profiles satisfying the criterion that
    $|V_{\fa}|$ or $|U_{\fs}| \geq 0.01$. For all these profiles $|\delta V_{\obs}| < 0.7$
    and $|\delta U_{\obs}| < 0.6$. A similar picture is also obtained at the center of the solar 
    disc. Hence one way to limit $\ell_{\odot}^2$ is to require $|\delta V_{\src}| < 1$ and
    $|\delta U_{\src}| < 1$ for all positions on the solar disc. This condition is strengthened
    by the fact that $|\delta V_{\obs}|$ and $|\delta U_{\obs}|$ decrease with decreasing $\mu$,
    (Stenflo et al. 1987 \cite{sta87}, Mart\'{\i}nez Pillet et al. 1997 \cite{mp97}), whereas 
    gravitational birefringence increases towards the limb, so that one would expect exactly 
    the opposite behaviour if gravitational birefringence had a significant effect on the Stokes 
    $V$ or $U$ profiles.

\begin{figure}[h]     
    \centerline{\psfig{figure=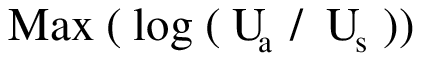,angle=90,height=1.98in,width=.24in}\hspace{-0.4cm}
                \psfig{figure=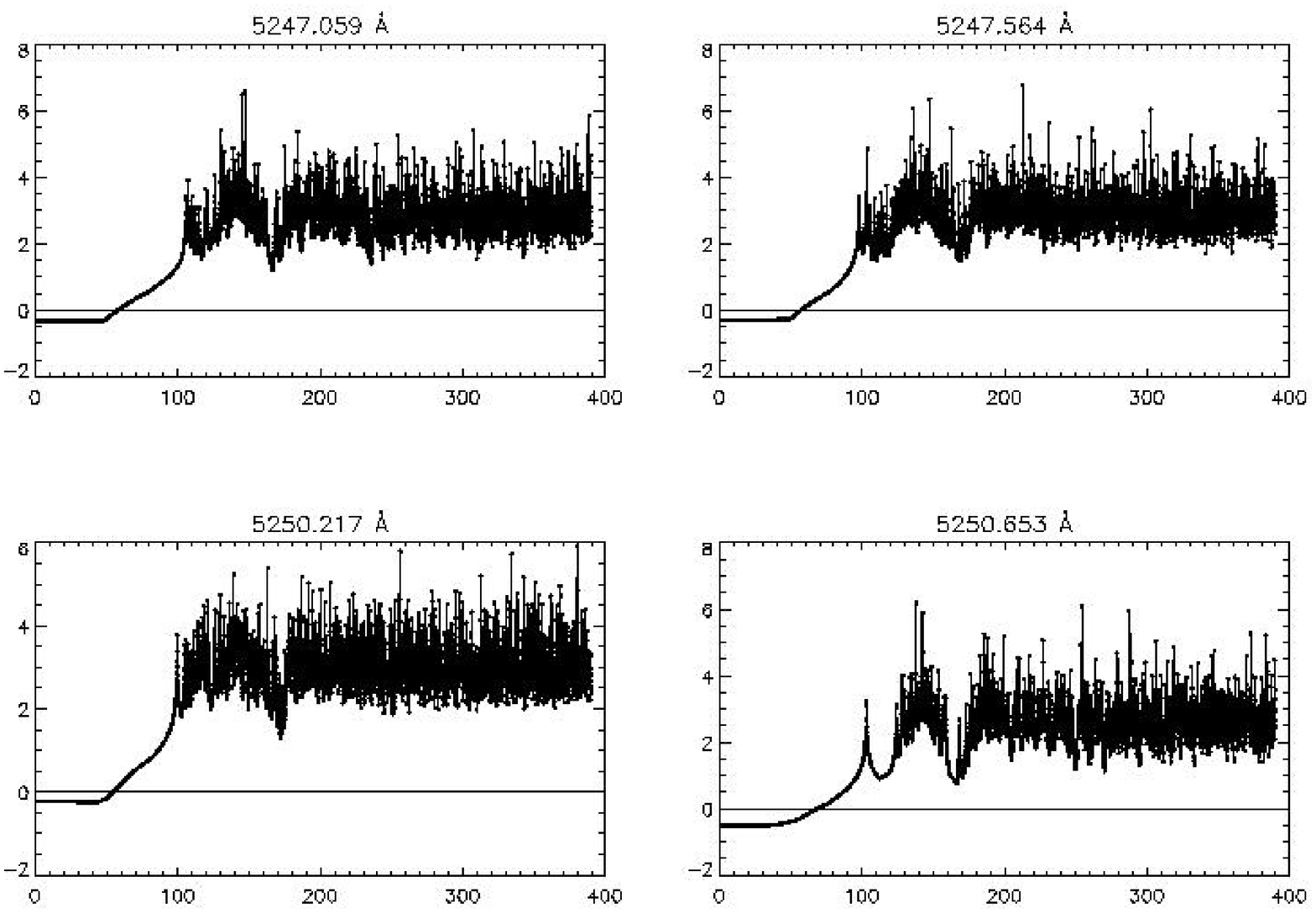,angle=90,height=3.3in,width=4.62in}}
    \caption{Maximum of $(U_{\asc}/U_{\ssc})$, on a logarithmic scale, values for all 4 spectral 
             lines vs. $\ell^2_{\odot}$.}
    \label{max}
    \centerline{\psfig{figure=max.eps,angle=90,height=1.98in,width=.24in}\hspace{-0.4cm}
                \psfig{figure=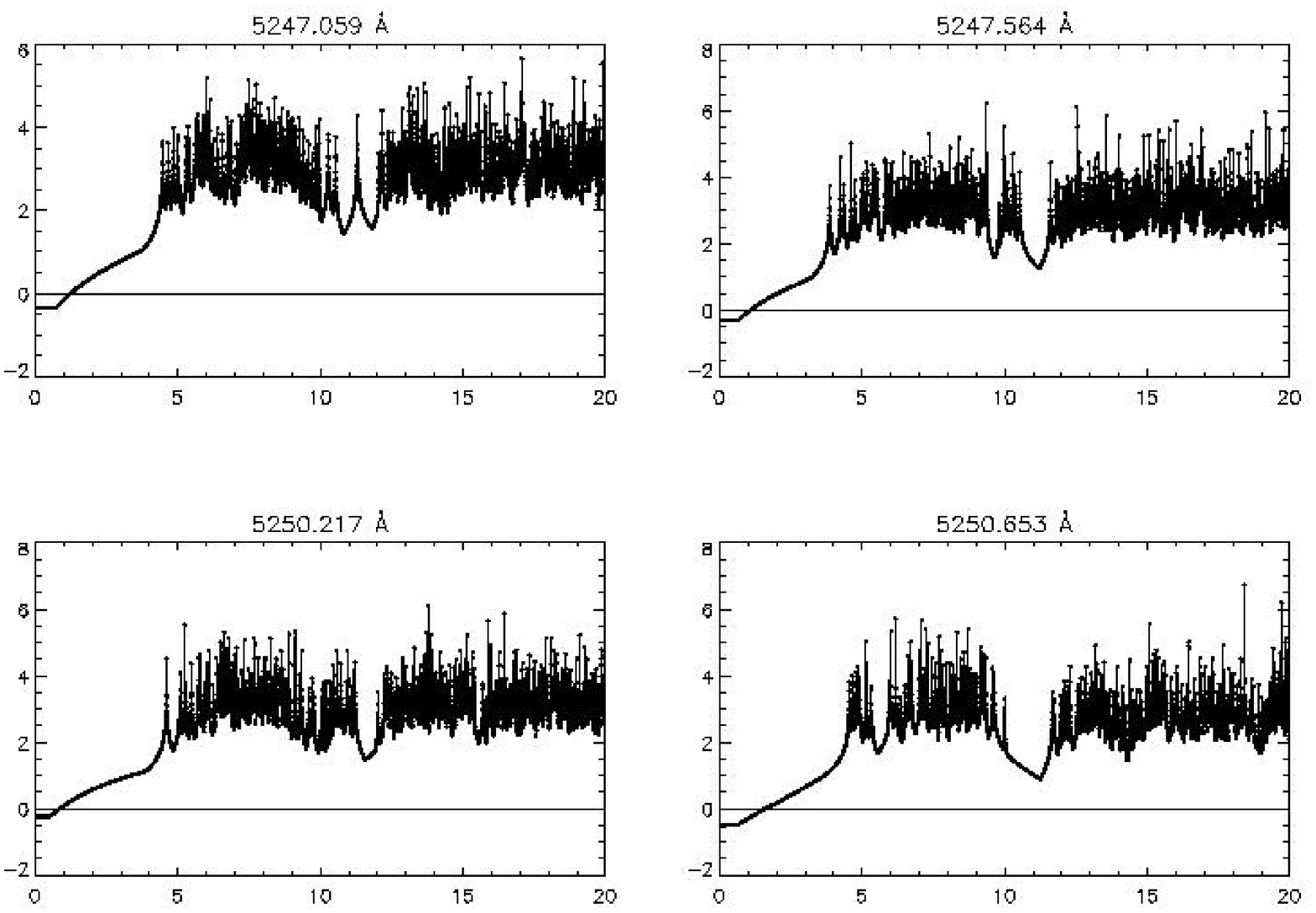,angle=90,height=3.3in,width=4.62in}}
    \caption{Maximum of $(U_{\asc}/U_{\ssc})$, on a logarithmic scale, values for all 4 spectral 
             lines vs. the metric affine parameter $k$.}
    \label{mag1}
  \end{figure}

  In Fig.\ref{max} we plot the maximum $(U_{\asc}/U_{\ssc})$ value predicted for 
  each of the 4 spectral lines, based on all analysed data, vs. $\ell^2_{\odot}$. 
  The horizontal line represents the value $\log(U_{\asc}/U_{\ssc} = 1)$, a limit 
  above which this ratio is not observed at $\mu \approx 1$. Clearly, as $\ell^2_{\odot}$ 
  increases $(U_{\asc}/U_{\ssc})_{\fm}$ initially remains almost 
  equal to $(U_{\aob}/U_{\sob})_{\fm}$, but begins to increase 
  for $\ell^2_{\odot} \gsim (50\,\mbox{km})^2$, becoming $\gsim 10$ at $\ell^2_{\odot} < (100\, 
  \mbox{km})^2$ and finally oscillating randomly around $(U_{\asc}/U_{\ssc})
  _{\fm}$ of 100 - 1000. All 4 spectral lines exhibit a similar behaviour, implying 
  that the influence of noise is very small.The largest effect of gravitational birefringence 
  is exhibited by the two most strongly Zeeman split lines, Cr I 5247.56{\AA}, Fe I 5250.22 
  {\AA}, which also produce the largest Stokes $V$ and $U$ signals, while the line with 
  the smallest splitting, Fe I 5250.65 {\AA} provides the weakest limit.  
    
  It is in principle sufficient to limit $\ell_{\odot}$ by requiring that none of the observed
  spectral lines has $(U_{\asc}/U_{\ssc}) > 1$ within the range of allowed 
  $\ell_{\odot}$ values. This gives $\ell^2_{\odot} < (57.1\, \mbox{km})^2$. A limit obtained 
  similarly from Stokes $V$ is both larger $\ell^2_{\odot} < (80.3\, \mbox{km})^2$ and less 
  reliable, since we cannot completely rule out $V_{\ssc}/V_{\asc} > 1$ to be present, 
  although we expect such profiles to be very rare, among large amplitude profiles.
  
  Fig.\ref{mag1} shows the analogous plot for metric affine theories. As in the above case
  for Moffat's NGT it can be seen that the strongest limit for gravity-induced birefringence
  is exhibited in the Cr I 5247.56{\AA} and Fe I 5250.22{\AA} line.
  We get $k^2 < (1.1\,\mbox{km})^2$ from Cr I 5247.56{\AA} and $k^2 < (0.9\,\mbox{km})^2$ from
  Fe I 5250.22{\AA}.

  \newpage
  
  \clearpage
 
\begin{figure}[t]   
  \vspace{-0.3cm}  
    \centerline{\psfig{figure=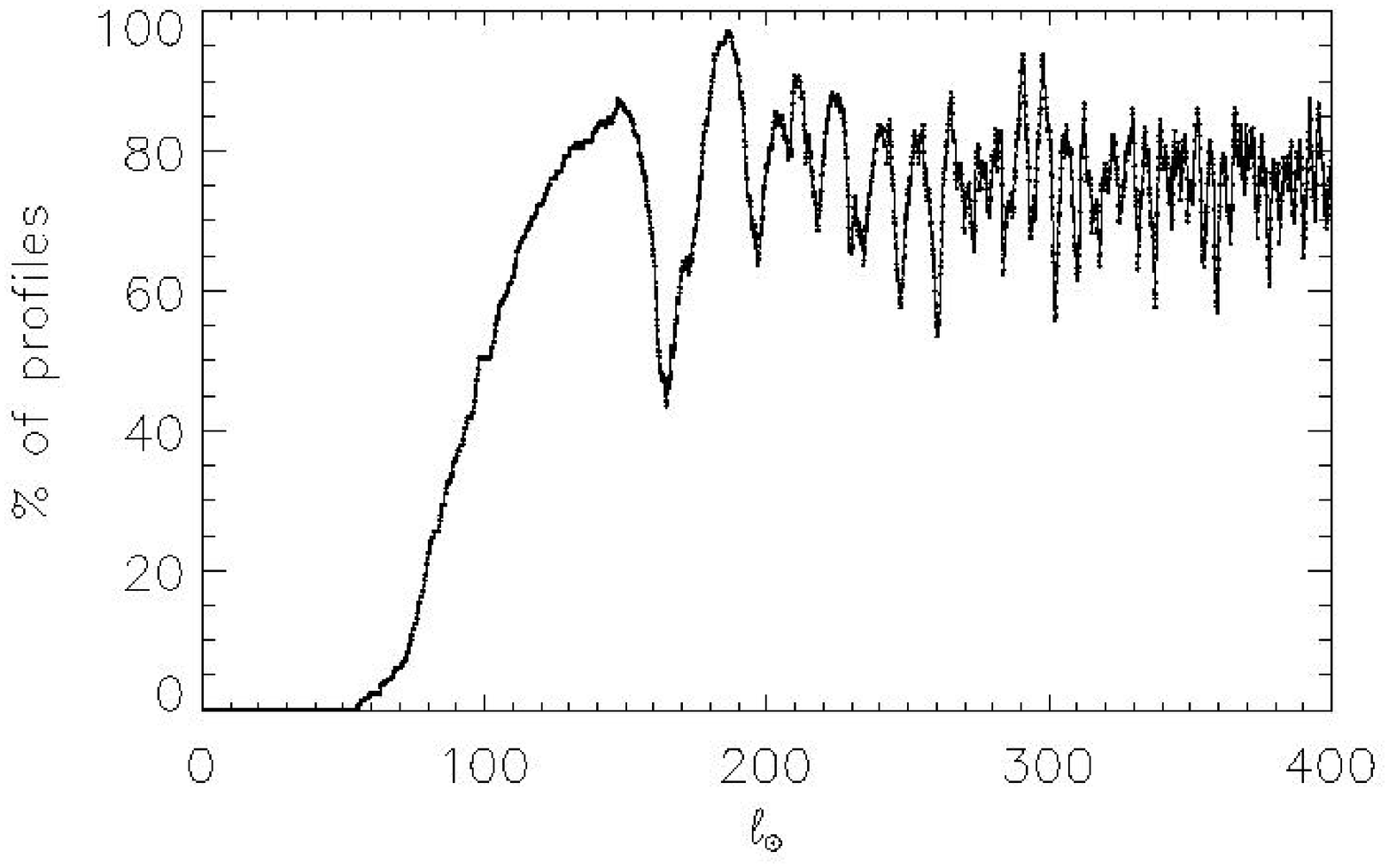,angle=90,height=2.75in,width=4.18in}}
    \centerline{\psfig{figure=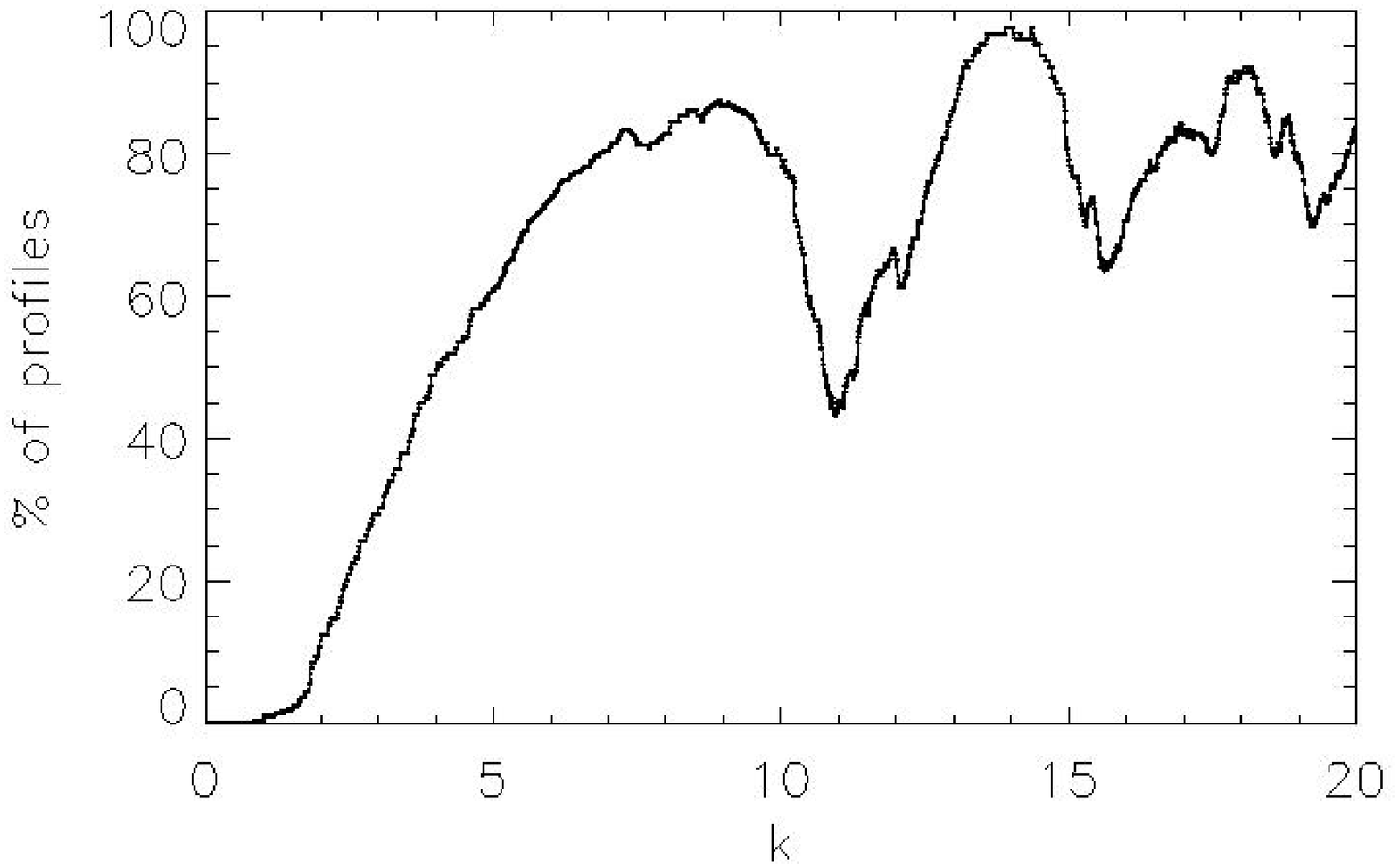,angle=90,height=2.75in,width=4.18in}}
    \caption{Fraction of profiles with $(U_{\asc}/U_{\ssc}) > 1$ for NGT (top) and MAG (bottom).}
    \label{ungt}
  \end{figure}

  \noindent An alternative test is to determine the fraction of profiles with $(U_{\asc}/
  U_{\ssc}) > 1$ (Fig.\ref{ungt})
  Fig.\ref{ungt} reveals that initially no $U$ profile satisfies
  the criterion, above $\ell^2_{\odot} = (57.5\, \mbox{km})^2$ 1.5 \% of the profiles does. 
  This number keeps increasing with $\ell_{\odot}$, before finally oscillating around 70\% 
  at large $\ell_{\odot}$. Thus 10\% of all data points have $(U_{\asc}/U_{\ssc}) > 1$ for 
  $\ell^2_{\odot} = (74.0\, \mbox{km})^2$, 20\% for $\ell^2_{\odot} = (79.4\, \mbox{km})^2$.
  We are not aware of any solar observations of Stokes $U$ with $\delta U > 1$ for which
  instrumental cross-talk is negligible (see S\'{a}nchez Almeida \& Lites 1992 \cite{sal92}, 
  Skumanich et al. 1990 \cite{sku90}). $\ell_{\odot}^2 < (80 \mbox{km})^2$ is thus a very conservative 
  upper limit.
  The lower picture shows the analogous plot for metric affine theories. Also, initially no profile
  satisfies the criterion, above $k^2 = (1.33\, \mbox{km})^2$ 1.7 \% of the profiles does.
  This number keeps increasing with $\ell_{\odot}$, before finally oscillating around 70\% 
  at large $k$. Thus 10\% of all data points have $(U_{\asc}/U_{\ssc}) > 1$ for 
  $k^2 = (1.95\, \mbox{km})^2$, 20\% for $k^2 = (2.43\, \mbox{km})^2$.
  $k^2 < (2.5\mbox{km})^2$ is thus a very conservative upper limit.
  \subsection{Profile difference analysis}
    We now apply the technique outlined in Sect. 2.1.2 to our data. Due to the limited 
    number of profiles and their irregular distribution over $\mu$ (see Sect. 2.3.1)
    any limit on gravitational birefringence will be less tight than what is achievable
    with ideal data presented in Sect. 2.1.2. In order to be able to compare the present 
    results with literature values we first set limits on the $\ell_{\odot}$-parameter in 
    Moffat´s NGT, since earlier work has concentrated on constraining this theory.
    
\begin{figure}[t]
      \centerline{\psfig{figure=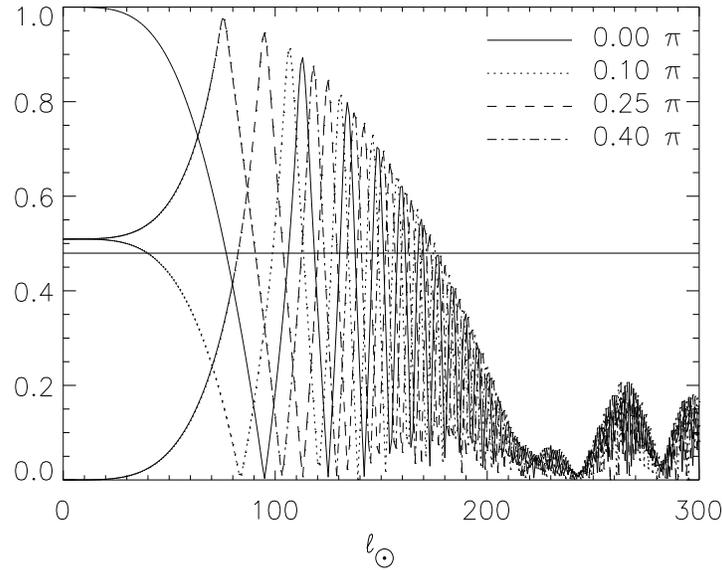,width=4.4in,height=3.3in}}
      \caption{Plot of the observable (Stokes $V$, Stokes $U$) mixture for initial phase
               differences of $0\pi$, $0.1\pi$, $0.25\pi$ and $0.4\pi$. Note that $\Delta\Phi$ values 
	       bigger than $0.5\pi$ give cyclic results.}
      \label{pdtl3}
    \end{figure}

    In Fig.\ref{pdtl3} we plot 
    \begin{equation}
      \frac{|<|V_{\obs}|>-<|U_{\obs}|>|}{<|V_{\obs}|>+<|U_{\obs}|>} 
      \quad \mbox{vs.} \quad \ell_{\odot},
    \end{equation} 
    for different initial phase differences $\Delta\Phi$ between the orthogonal modes of
    line Fe I 5250.65{\AA}. The averaging has been done over the $\mu$ values at which 
    observations are available. 
    This line is chosen, since it gives the tightest limits on $\ell_{\odot}^2$. 
    The thick horizontal line represents the value obtained from observations. Obviously 
    above $\ell_{\odot}^2 = (178\, \mbox{km})^2$ the curve obtained from theory always lies below
    the observed value, hence ruling out such $\ell_{\odot}^2$ values. Note that it is 
    sufficient to test for $\ell^2_{\odot} < (305\, \mbox{km})^2$, since this upper limit has 
    been set using independent data and another technique by Solanki \& Haugan (1996) \cite{sah96}.
    
    Similarly we can also set a limit on the coupling constant $k$ of the metric affine
    theories. Using this technique we obtain $k^2 < (0.69\, \mbox{km})^2$ measured in line
    Fe I 5250.65{\AA}.
\newpage    
    \subsection{Brief history of constraints on $\ell_{\odot}^2$}
      Up to this point we have presented our results with respect to sharp constraints on the coupling
      constants $\ell_{\odot}^2$ and $k^2$. In order to get an estimate of the quality of these results
      compared to previous upper limits we will present here a brief overview on the history of
      constraints on $\ell_{\odot}^2$ achieved with a variety of different techniques.
      
      The first upper bound on the value of the NGT charge $\ell^2_{\odot}$ was given by Moffat himself
      in 1982 \cite{mof82}. Based on a new determination of the quadrupole moment of the sun he concluded
      that this result would lead to a deviation of 1.6\% from Einstein's prediction for the precession 
      of the perihelion of Mercury. Moffat claimed that NGT could fit the measured precession with the new
      quadrupole moment by using a value of $\ell_{\odot}=3.1\cdot 10^3\, \mbox{km}$ close to the upper bound
      of $\ell_{\odot} \leq 2.9\cdot 10^3\, \mbox{km}$ obtained in 1982 by using an average value on the available
      Mercury data.
      
      The next significant improvement was made by Gabriel et al. in 1991 \cite{gea91}. By using for the 
      first time the NGT prediction of gravity-induced birefringence they presented a quantitative prediction
      for the phase difference between the orthogonal modes. Consequently, birefringence leads to a depolarization
      of the Zeeman components of spectral lines emitted from extended, magnetically active regions on the
      sun. Since the solar-physics data which were available at that time limited the extend of such 
      depolarization, they could only conclude that the Sun's antisymmetric charge must be less than 
      $(535 \, \mbox{km})^2$. 
      
      This result was based on extremely conservative assumptions regarding the polarization of light
      emitted by the observed solar feature and an inapplicable requirement regarding the determination of the
      magnetic filling factor, which is defined as the fraction of the aperture covered by magnetic field.
      In 1995 Solanki and Haugan \cite{sah96} improved the limit on $\ell^2_{\odot}$ by using
      more realistic values of the parameters that characterize the solar polarization source and a refined 
      analysis procedure which yielded an upper bound of $\ell^2_{\odot}< (305\, \mbox{km})^2$.
      
      The current work represents the state of the art with respect to upper limits on gravitational birefringence
      induced by the nonsymmetric field of the sun. Since the physical consequences of the nonmetric NGT description
      relevant for our purpose are proportional to $\ell^4_{\odot}$ this new constraint on the Sun's NGT charge
      is three orders of magnitude smaller than the previous limit by Solanki and Haugan and 7 orders of magnitude
      smaller than the value favoured initially by Moffat. This gives a further significant restriction to the
      viability of NGT. 
\newpage
\section{Possible tests using the {\sl Solar Probe} spacecraft}
  \begin{figure}[t]
    \centerline{\hspace*{-0.5cm}\psfig{figure=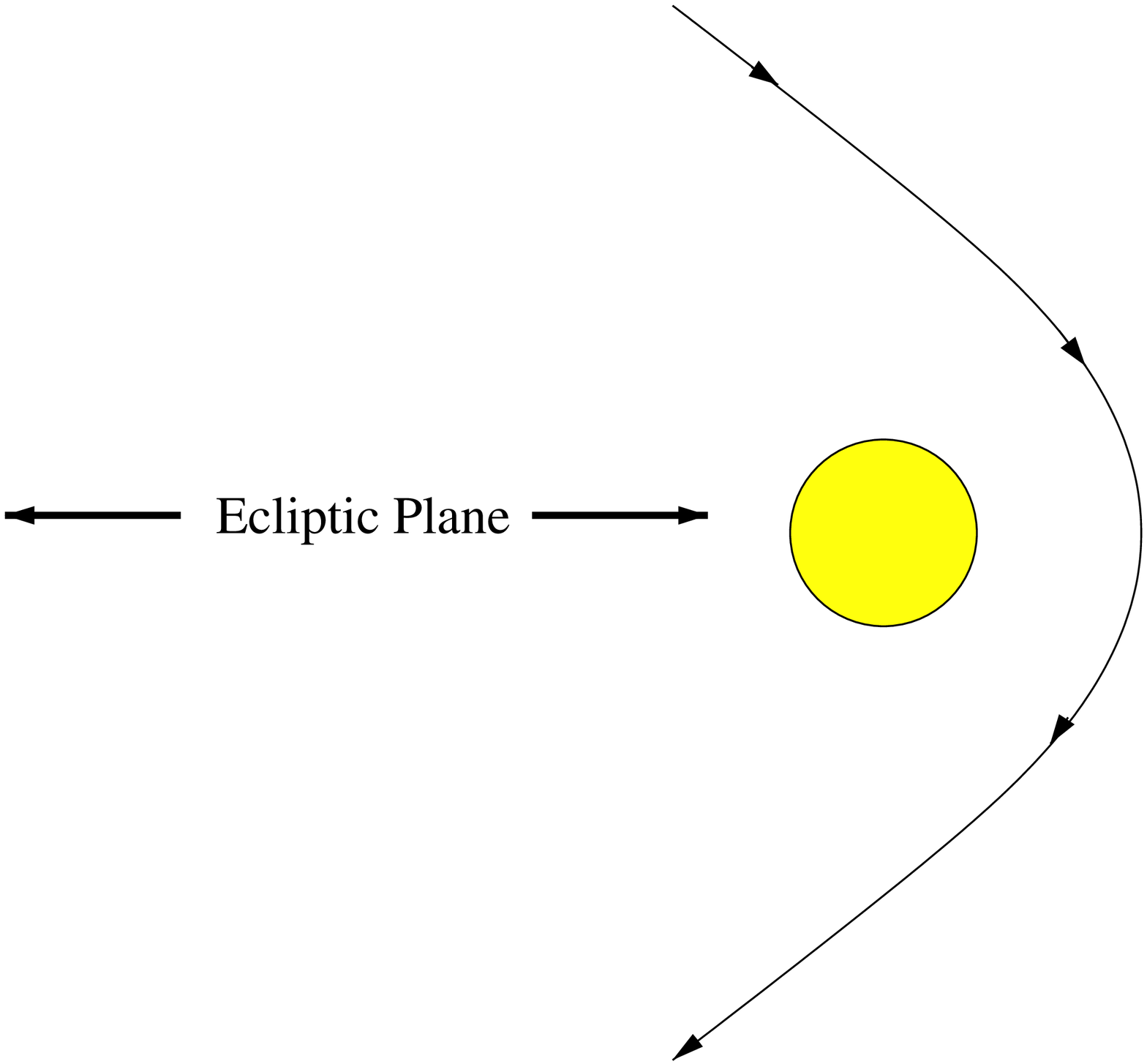,width=3.in,height=3.in}\hspace{0.8cm}
                \psfig{figure=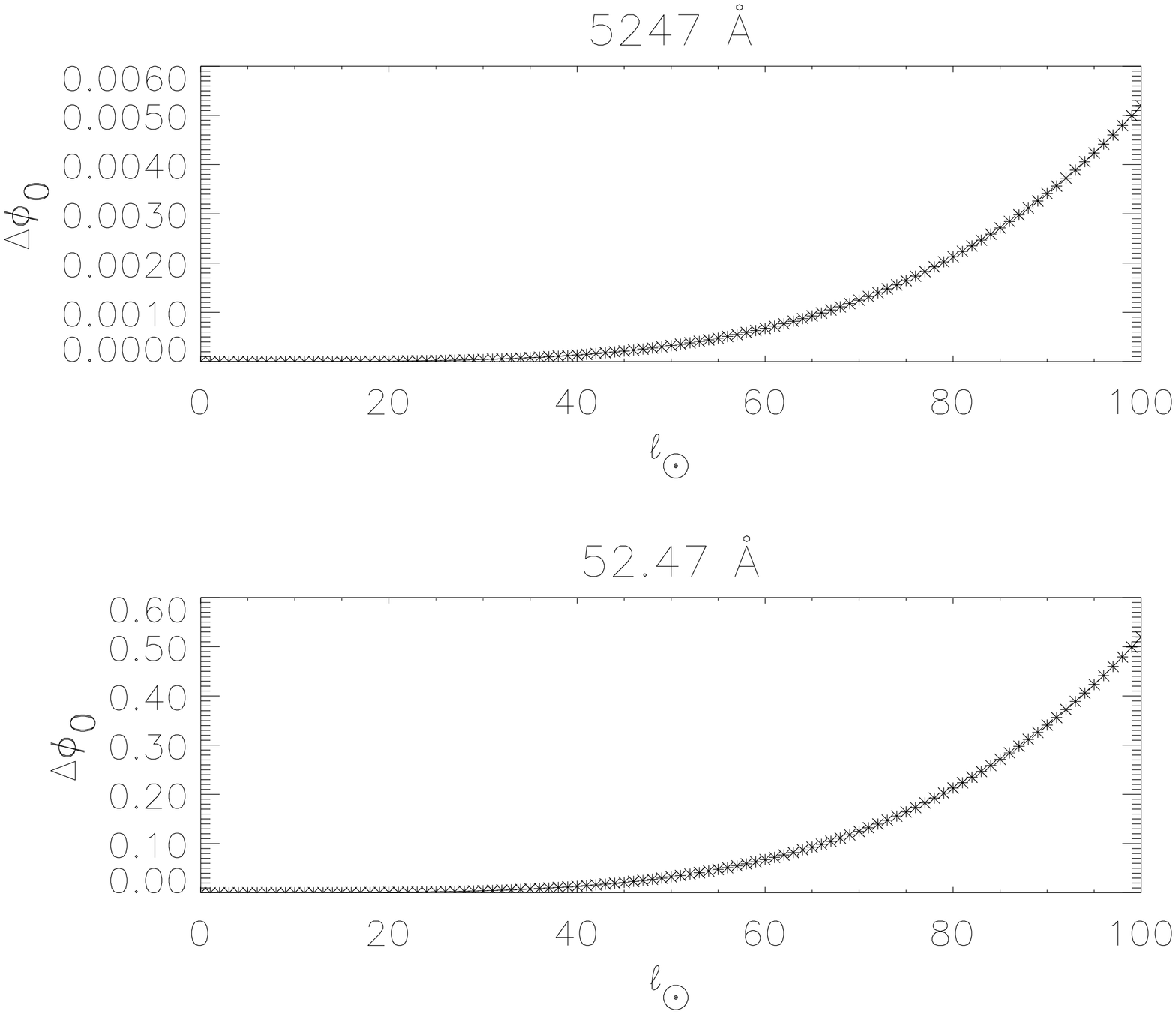,width=3.in,height=3.in,angle=90}}
    \caption{Left figure: Sketch of the {\sl Solar Probe} trajectory near the sun as seen from
             Earth. Right figures: Expected phase shifts for a signal emitted at $3R_{\odot}$ 
	     from the sun. Top: Optical wavelength. Bottom: Soft X-ray.}
    \label{shifts}     		
  \end{figure}
  The possibility of setting strong limits on gravitational birefringence with the tests which 
  we have outlined so far suffered from the major drawback that they all depend strongly on our
  knowledge of solar magnetic features. For example we have assumed that none of the observed
  spectral lines has $(U_{\asc}/U_{\ssc}) > 1$ which is also based on current observations with
  a certain spatial resolution. At the moment nobody can say if this statement will hold if the
  instrumental resolution could be highly increased in the future, although simulations of solar
  magnetoconvection suggest that $U_{\asc} < U_{\ssc}$ remains valid even if the resolution is
  improved by an order of magnitude. 
  
  Taking this into account, it is obvious that drastic improvements of our limits for $\ell^2_{\odot}$ 
  could be achieved by using an artificial source for the polarized radiation with well defined 
  properties, placed in close vicinity to the sun. Such a source could possibly be provided by
  the {\sl Solar Probe} spacecraft \cite{probe}, which is part of NASA's mid-term plans.
  Designed mainly as a mission to explore the Sun's Corona, its trajectory (sketched in Fig.a) 
  lies in a plane perpendicular to the ecliptic with a target perihelion distance at 3-4 solar 
  radii. This close
  perihelion could provide a unique opportunity to look for gravitational birefringence if
  it is possible to receive a well defined polarized signal from an instrument onboard the
  spacecraft when it passes near the solar limb.
    
  To see, if we could expect a reasonable gravitational effect in the signal, we simply have to 
  take the case $\mu=0$ in the phase shift formula (\ref{ngt-form}) and replace the solar radius $R_{\odot}$
  with $(R_p+1)\cdot R_{\odot}$ where $R_p$ is the perihelion distance of the spacecraft in units of the
  solar radius. This leads to
  \begin{equation}
    \big.\Delta \Phi\;\big|_{\mu=0} = \frac{3\,\ell^4_{\odot}\,\pi^2}{16\,\lambda\,((R_p+1)\cdot R_{\odot})^3}
  \end{equation}
  Of course this result depends strongly on the wavelength of the signal. The shorter the wavelength the
  bigger is the possible phase shift that we could expect. This is illustrated in Fig.(\ref{shifts}b), 
  which shows the phase difference in units of $\pi$ for a range of $\ell_{\odot}$ values with 
  $R_p = 3$. Optical wavelength is used in the upper diagram and X-ray in the lower. 
  One can see that for optical wavelength a gravitational effect on the polarization of the signal 
  would be extremely difficult to observe under realistic conditions with a certain noise level.
  Whereas for wavelengths of the order $10^{-10}$m, we could expect a reasonable transformation
  of linear to circular polarization and vice versa for $\ell^2_{\odot}$ values which are clearly
  below our current upper limit. However, at the moment it is out of reach from the technological 
  point of view, to place a source of polarized X-rays in the near vicinity of the sun. Nevertheless
  it is worth to keep this promising possibility for very precise tests of the Einstein equivalence 
  principle in mind.  
  \subsection{Alternative Spacecraft Trajectories}
    \begin{figure}[t]
      \centerline{\psfig{figure=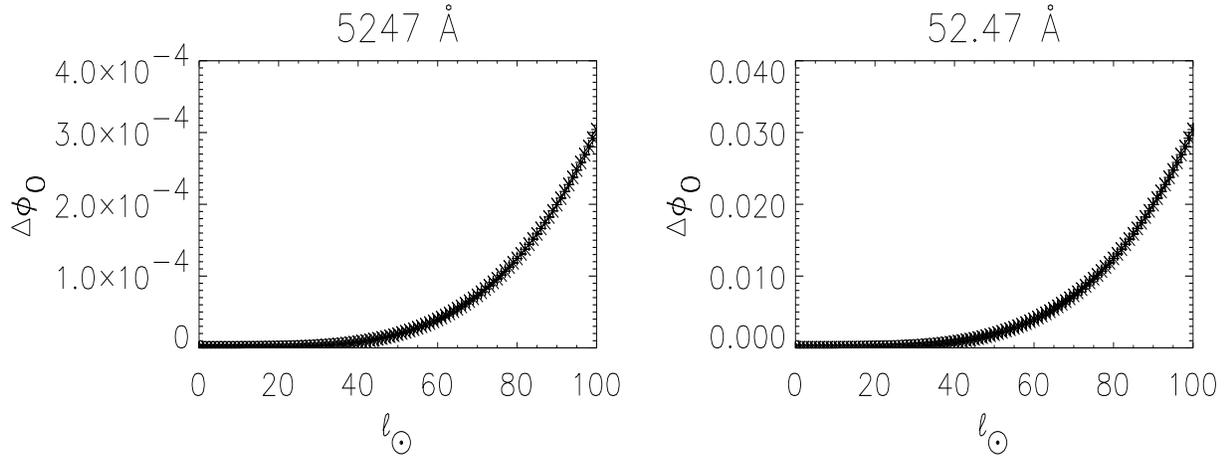,width=6.5in,height=3.in,angle=90}}\vspace{-0.6cm}
      \caption{Expected phase shifts for a signal emitted at $3R_{\odot}$ from the sun when the spacecraft
               passes in front of the solar disc on an alternative trajectory. The wavelengths are the same as 
	       in Fig.\ref{shifts}}		
      \label{shifts-alt}		       
    \end{figure}
    Interestingly the planned spacecraft trajectory is also a very suitable one with respect to possible 
    tests of gravity-induced birefringence. This is due to the fact that when the spacecraft reaches 
    its perihelion position the phase shift is maximized by the mimimal distance to the sun, i.e. highest 
    gravitational potential, and also by the maximum distance that the signal could travel through the 
    gravitational field of the sun.
    
    If the craft reaches on a different trajectory its closest position to the sun which is also closer
    towards earth this would mean a shorter distance for the signal to travel and, hence, a smaller total 
    phase shift. This is shown for a spacecraft passing in front of the visible solar disc at distance $R$ 
    to the sun. The phase shift of a signal  with wavelength $\lambda$ emitted at this position is given by
    \begin{equation}
      \big.\Delta \Phi\;\big|_{ \atop {\mu=0 \atop R \rightarrow \infty}} = 
      \frac{\pi\l^4_{\odot}}{\lambda}\left(\frac{3\pi}{16 R_0^3}-\frac{R}{2}\left(
      \frac{1}{2(R_0^2+R^2)^2}+\frac{3}{4R_0^2(R_0^2+R^2)}\right)-\frac{3}{8R_0^3}\arctan
      \frac{R}{R_0}\right)\quad .
    \end{equation}
    The detailed calculation is given in Appendix B.
    Fig.\ref{shifts-alt} shows the result for the same wavelength as in Fig.\ref{shifts} and $\mu=0$.
    The cumulative phase shift for the alternative trajectory is smaller by a factor of 15 although the
    distance to the sun is the same in both cases. The crucial point is that in the first case the signal
    covers a bigger distance within that region of the gravitational field where it is strong enough to
    induce a reasonable phase shift. This relevant distance has an extend of approximately 5-6 solar radii.  
    A different trajectory where the spacecraft passes the visible limb behind the sun as seen from earth 
    would give no improvement since the phase shift which accumulates from behind the sun up to a position
    above the poles has an opposite sign compared to the phase shift of a signal which is emitted above the 
    pole, propagating towards earth. This can be seen from equation (B.8) in Appendix B which gives the 
    phase shift for a signal running from the surface up to a distant point $R$. If the signal runs in the
    opposite direction we have to switch the integration limits which gives a negative sign for the phase shift.

\section{Discussion and Conclusions}
  Using two techniques (the Stokes asymmetry technique and the new profile difference technique)
  we have improved previous limits on $\ell^2_{\odot}$ given by Solanki \& Haugan (1996) \cite{sah96} by 
  nearly one order of magnitude. It will be difficult to set much tighter limits on gravitational birefringence 
  than those found here using solar data in the visible spectral range. To obtain a significant
  improvement on the basis of solar data one would need to observe at shorter wavelengths. The line 
  at the shortest wavelength that is strong enough to provide a hope of detecting Stokes $U$ and $V$ at
  sufficient $S/N$ is Ly$\alpha$ at 1216{\AA}. The maximum gain that one could expect relative 
  to the current analysis is a factor of
  \begin{equation}
   \frac{\lambda_{\mbox{\footnotesize{visible}}}}{\lambda_{\mbox{\footnotesize{Ly}}\alpha}} = 
   \frac{5250}{1216} = 4.32 \quad .
  \end{equation}
  Although the idea of a possible utilization of future space missions for enhanced new tests 
  is very appealing because of its high potential for further improvements of the current limits 
  on  $\ell^2_{\odot}$ and $k^2$, the technological realization of such an experiment is currently
  out of reach. For setting stronger limits on gravitational birefringence or, if birefringence really has
  a physical relevance, for having the chance of a direct detection of this effect it is therefore more
  promising to proceed the investigations with more compact astrophysical objects. This will be the
  content of the following chapters.
  

\chapter{Magnetic White Dwarfs} 

  \vspace*{1cm}
  The conjecture that gravity-induced birefringence is most distinctive for high 
  gravitational potentials suggests itself from the structure of the phase shift 
  formulas in NGT and MAG. For this reason the next logical step in this work
  is to continue the analysis with suitable, compact objects like magnetic white 
  dwarfs. These stars provide a versatile tool for testing predictions of nonmetric 
  theories of gravity because of their Megagauss magnetic fields and high surface 
  gravity. Currently 65 or $\sim 5$\% of all known white dwarfs are classified 
  as isolated magnetic stars with field strengths in the range $ 3 \times 10^4 - 
  10^9$G with the field strength distribution peaking at $1.6 \times 10^7$G \cite{wf00}. 
 
  However, within this small class only a fraction of stars is of interest for
  our purpose. In addition to wavelength resolved polarimetric data we also need
  information about the mass and the radius of an object as well as 
  measurements of the magnetic field structure to set strong limits on gravitational 
  birefringence. With these restrictions we are left with just four magnetic 
  white dwarfs which display high levels of circular polarization between 8 and 
  22 percent: Grw $+70^{\circ}8247$, REJ0317-853, PG2329+267 and 40 Eridani B.  
  The degree of circularly polarized light of wavelength $\lambda$, 
  $(V/F)_{\mbox{\footnotesize{obs}}}$, reaching the observer from any of these 
  stars is to first order simply a function of the strength of the local longitudinal 
  magnetic field \cite{aw89} and the coupling constant $k^2$.
  Since gravitation theories based on a metric-affine geometry of space-time predict
  a depolarization of light emitted from extended astrophysical sources we have looked 
  for the largest $k^2$ that predicts values of $(V/F)_{\mbox{\footnotesize{obs}}}$ 
  larger than or equal to that observed. This sets an upper limit on $k$.
  In the case of Grw $+70^{\circ}8247$ such a limit was already set by Solanki et al.\cite{shm99} 
  based on NGT predictions. We redo the analysis using both an improved model of the
  magnetic field distribution of the star, as well as the predictions of metric affine
  theory, which had not been predicted earlier.

  However, we encounter a self-consistency problem since the relevant 
  source properties like masses, inclination angles, etc. we use have been determined with 
  models which, of course, neglect the possible influence of birefringence and, so, need 
  not be valid. We circumvent this problem by assuming worst-case properties of the source 
  that minimize depolarization caused by gravity-induced birefringence.

\newpage  

  \section{Polarization Modelling Technique} 
    The search for a possible influence of gravity-induced birefringence on 
    astronomical signals, i.e. the modelling technique which is used to set 
    strong upper limits, depends of course heavily on the amount of appropiate 
    information that we can gain on the source of polarized radiation.
    This becomes decisively clear in the transition from the techniques which 
    we have used to evaluate the solar observations of the last chapter to 
    the magnetic white dwarfs. Here, a crucial point is basically played by 
    the spatial resolution of the observations, since the individual calculation 
    of the phase shift $\Delta\Phi$ as a function of $\mu$ for a given light 
    source requires, at least theoretically, an infinite resolution of the source.
          
    So, due to the finite spatial resolution of observations, the effect the 
    phase shift has on a light ray depends on whether one observes a pointlike 
    source, e.g. small sunspots, or an extended source like white dwarf 
    magnetospheres. In the case of a pointlike source, all light received 
    from it suffers the same phase shift $\Delta \Phi(\mu_p)$. Introducing 
    Stokes parameters to describe polarized light, with Stokes $Q$ defined 
    to represent the difference between linear polarization parallel and 
    perpendicular to the part of the stellar limb closest to the source, one 
    therefore finds a crosstalk between Stokes $U$ and Stokes $V$. This crosstalk 
    is such that although the observed values $U_{\obs}$ and $V_{\obs}$ differ 
    from the values emitted by the source, $U_{\src}$ and $V_{\src}$, the 
    composite degrees of polarization remains equal: $(U_{\obs}^2 + V_{\obs}^2)
    ^{1/2} = (U_{\src}^2 + V_{\src}^2)^{1/2}$. If an extended source covering 
    a range of $\mu$ values is observed then light emitted from different 
    points suffers different phase shifts and, so, adds up to an incoherent 
    superposition. Summing over the different contributions, using the 
    additive properties of Stokes parameters yields a reduction of the 
    observed polarization relative to the light emitted from the source:
    $(U_{\obs}^2 + V_{\obs}^2)^{1/2} < (U_{\src}^2 + V_{\src}^2)^{1/2}$. 
    Since the stellar sources of polarization we consider in this chapter 
    are spatially extended, any observed (nonzero) degree of polarization 
    provides a limit on the strength of gravity-induced birefringence that 
    the star's gravitational field could induce.
  
    It is generally agreed that the polarized radiation from white dwarfs 
    is produced at the stellar surface as a result of the presence of Megagauss 
    dipolar magnetic fields \cite{cha92,lan92}. To first approximation the flux 
    of net circularly polarized light at wavelength $\lambda$ emitted toward 
    the observer from the surface is directly proportional to the strength of 
    the line of sight component of the magnetic field at the stellar surface.
    Therefore, this net flux can be written as         
    
    \begin{equation}\label{vl}
      V_{\lambda,\src} = 2\pi\int_{0}^1I_{\lambda}(\mu)\cdot w(\mu) \mu\,d\mu \quad,
    \end{equation}         
   
    \vspace{0.4cm}
    \noindent where $I(\mu)$ is the intensity at wavelength $\lambda$. The weightfunction $w(\mu)$
    describes the strength of the vertical magnetic field component at position $\mu$. 
    Since our analysis encompasses different white dwarfs with different types of dipole 
    field geometries, $w(\mu)$ is calculated using an oblique dipolar rotator model which 
    is introduced in the next section.
        
    To define a degree of circular polarization, we have to divide (\ref{vl}) by the total 
    stellar flux emitted to the observer at wavelength $\lambda$    
   
    \begin{equation}\label{fl}
      F_{\lambda} = 2\pi\int_0^1I_{\lambda}(\mu)\mu\,d\mu \quad .
    \end{equation}   
    
    The function $I_{\lambda}(\mu)$ also describes the limb darkening of the star. Since
    the surface of a white dwarf cannot be resolved the limb darkening cannot be
    measured directly. Because the broadband spectrum of Grw +70$^{\circ}$8247 is well 
    represented by blackbody radiation \cite{gok82} and radiative equilibrium models 
    \cite{wf88,shp71} it is reasonable to assume a simple law like the one describing 
    the directly observed solar limb darkening. For this, we have chosen to use \cite{aln73}
    
    \begin{equation}\label{limb} 
      \frac{I_{\lambda}(\mu)}{I_{\lambda}(\mu=1)}=1+(\mu-1)g+(\mu^2-1)h \quad ,
    \end{equation}
    with
    \begin{displaymath}
      0 \leq g+h \leq 1 
    \end{displaymath}

    for Grw +70$^{\circ}$8247. Since this kind of a linear interpolation is of a very 
    general form we have taken the limb darkening model of Grw +70$^{\circ}$8247 as a 
    prototype for the white dwarfs within our sample.

    In practise, we have evaluated the degree of circular polarization $(V_{\lambda,\src}/
    F_{\lambda})$ numerically for increasing values of $k_{\star}$ and, in the case of 
    Grw +70$^{\circ}$8247, of $l_{\star}$. In contrast to the earlier work
    of Solanki, Haugan and Mann \cite{shm99} (1999) we assume that the light emitted 
    by the source is polarized proportional to ${\bf B}$, i.e. to $w(\mu)$. 
  \section{Oblique dipolar rotator model}
    The model we use was already described by Stift in 1975 \cite{stift} from 
    whom we adopted most of our notations. 
    
    Starting with the observer's system where the z-axis is defined 
    by the line-of-sight, a surface point on the visible hemisphere
    of the star is given by ${\bf z}$.
    This vector is transformed into the rotation system of the star by 
    the matrix 
    \vspace{0.3cm}
    \begin{equation}
      S_i = \left(
            \begin{array}{rrr}
	      1&0&0\\
	      0&\cos i&\sin i\\
	      0&-\sin i&\cos i
	    \end{array}
            \right) \quad ,
    \end{equation}
    
    \vspace{0.3cm}
    \noindent which is followed by a rotation around the rotational z-axis\vspace{0.3cm}
    \begin{equation}
      S_{\phi} = \left(
                 \begin{array}{rrr}
	           \cos\phi&\sin\phi&0 \\
	           -\sin\phi&\cos\phi&0\\
	           0&0&1
	         \end{array}
                 \right) \quad .
    \end{equation}
    
    \vspace{0.3cm}
    \noindent The position of the dipole system relative to the rotational system
    is uniquely described by the three Eulerian angles, e.g. by
    \begin{equation}
      S = S_{\xi}S_{\zeta}S_{\chi}
    \end{equation}
    \newpage
    \noindent where $S_{\xi}$ and $S_{\chi}$ are generating rotations around 
    the x-axis and $S_{\zeta}$ around the z-axis. Since the only measurable
    quantity in observations is the resulting tilt angle $\beta$ between 
    the dipole and rotation axis, the effect of the three Eulerian rotations 
    can be summarized without loss of generality by a rotation with $\beta$ 
    around the x-axis. 
       
    Given the offset coordinates of the dipole in the rotation system
    by ${\bf x}$, we obtain the coordinates of the surface point ${\bf z}$
    relative to the dipole system by
    \begin{equation} 
      {\bf r} = S_{\beta}(S_{\phi}S_i{\bf z}-{\bf x}) \quad ,
    \end{equation}
    so that the field strength in the dipole system is given by
    \begin{equation}
      {\bf B}_{dp} = - \nabla({\bf m}\cdot {\bf r}/r^3) \quad .
    \end{equation}
    From this we get the field strength in the observer's system through
    \begin{equation}\label{bobs}
       {\bf B}_{obs} = S_i^T S_{\phi}^T S_{\beta}^T {\bf B}_{dp}
    \end{equation}
    where $S_i^T,\, S_{\phi}^T,\, S_{\beta}^T$ are the transpose matrices of
    $S_i,\,S_{\phi},\,S_{\beta}$. The longitudinal field component in the 
    observer's system is then of course the ${\bf B}_{obs,z}$ component of 
    (\ref{bobs}).
    
    The effective field ${\bf B}_e$ for the visible hemisphere is then given by
    \begin{equation}
      {\bf B}_e = \int\int {\bf B}_zI\, dA \bigg/ \int\int I\,dA \quad .
    \end{equation}  
    ${\bf B}_z$ denotes the line-of-sight component of ${\bf B}_{obs}$, while the
    limb darkening is described, as above, by the empirical function $I$. 
    \begin{figure}[t]
      \centerline{\psfig{figure=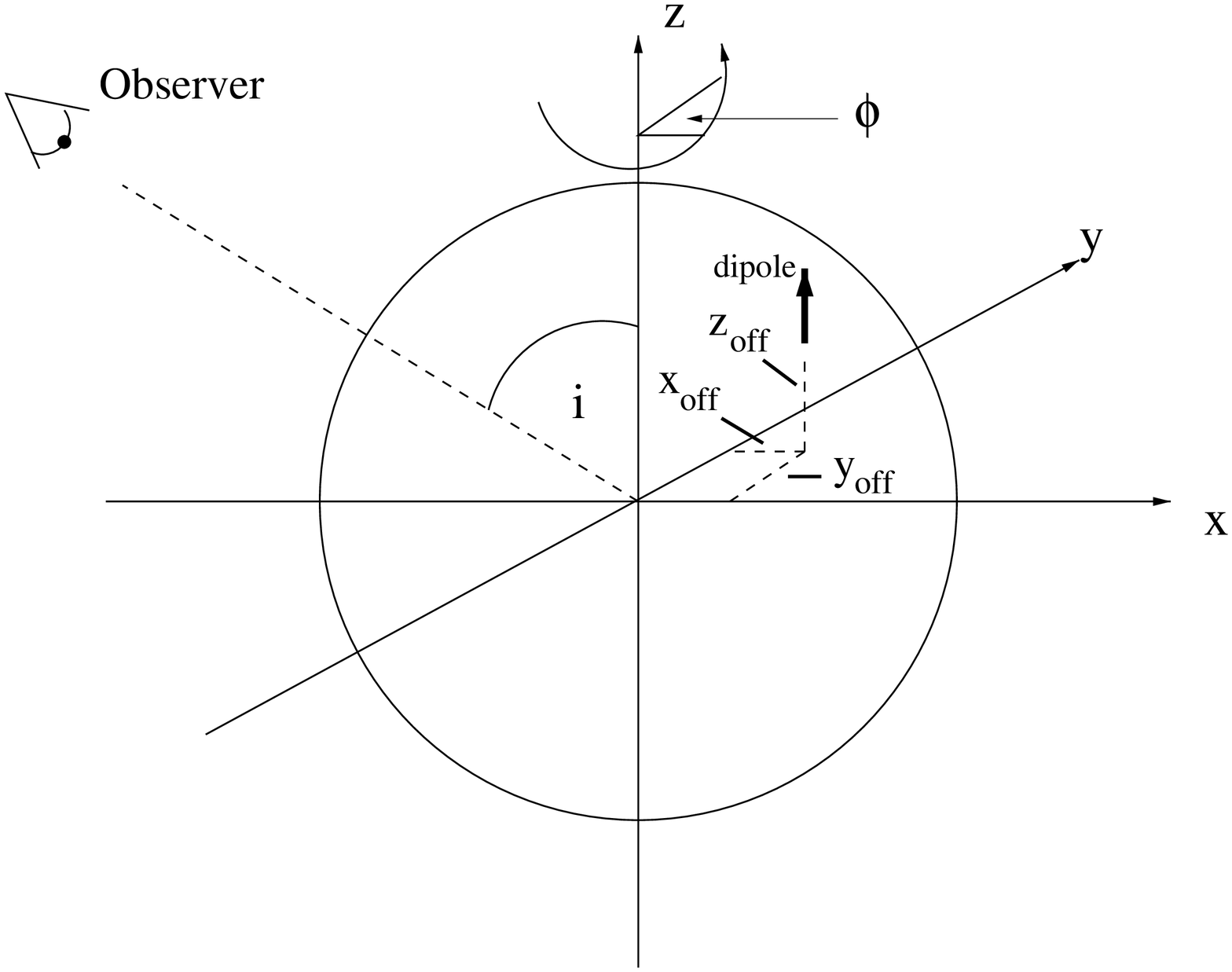,width=3.7in,height=3in}}
      \caption{Geometrical representation of the dipole- relative to the 
               rotation system of the star with offsets. The rotation axis is inclined 
	       to the observer by an angle $i$. In general, the offset dipole 
	       is tilted against the rotation axis as described in Fig.3.2.}
    \end{figure}
             
 \section{Grw +70$^{\circ}$8247}
  
    Within the class of magnetic white dwarfs, Grw +70$^{\circ}$8247 has played the role of a 
    ''Rosetta stone'' in the analysis and interpretation of their spectra. Although the first 
    measurement of circular and linear polarization in the spectrum of a white dwarf \cite{kem70} 
    gave a strong indication for a magnetic field, the presence of broad absorption features in 
    its blue spectrum remained unexplained until the 1980s. A solution of this puzzle was only 
    given after detailed calculations of hydrogen energy levels at field strengths $B > 100$ MG 
    \cite{hoc84,hoc85,fea84,rea84} so that, as a result, all spectral features could consistently 
    be explained in terms of stationary lines, i.e. lines whose wavelengths vary only slowly with
    the field strength, of atomic hydrogen in a centered dipolar magnetic field 
    with a polar field strength of $\sim 320$ MG and a viewing angle (angle between the line 
    of sight and the magnetic axis) $i = 0^{\circ} - 30^{\circ}$ \cite{wf88,wun90,jor92}.
    
    In this way it was possible to determine the physical parameters of the stellar atmosphere
    and of the magnetic field from a best fit of synthetic to observed spectra.  
    The observations can be best explained by an effective temperature of 14,000 K 
    \cite{wf88,gok82} which implies a radius of 0.0076 $R_{\odot}$ (determined from photometry and 
    parallax distance measurements), where $R_{\odot}$ is the solar radius, and a mass of about 
    1.0 $M_{\odot}$, where $M_{\odot}$ is the solar mass \cite{shp79}. Since the characteristics 
    of the polarization curve have remained invariant for almost $\sim 25$ yr now \cite{wf00} 
    we can infer a very long rotation period of much more than 25 years. This slow rotation 
    is fitted into the obique rotator model by setting $\beta=0$, implying a dipole aligned
    with the rotation axis. Thus at a given point in the observer system with polar coordinates 
    $(r,\theta,\phi)$, the Cartesian field components in the case $i = 0^{\circ}$ are 
    simply given by
   
    \begin{eqnarray}
      B_x &=& 3 B_d \sin\theta\cos\theta\cos\phi/2r^3 \quad ,\\
      B_y &=& 3 B_d \sin\theta\cos\theta\sin\phi/2r^3 \quad ,\\
      B_z &=& B_d(3\cos^2\theta -1)/2r^3 \quad ,\label{bz}    
    \end{eqnarray}   
    where $B_d$ denotes the surface polar field strength for the case of a centered dipole.
    In our case, the z-axis is also the line of sight. Since we focus on measurements of circular 
    polarization, only the longitudinal Zeeman-Effect is relevant for our purpose.
    
    Polarization measurements of Grw +70$^{\circ}$8247 yielded a level of $-6 \pm 0.25$\% 
    in the visible spectrum at 449 nm \cite{aea85,la75,aj94}. Since the constraints 
    on $k_{\star}^2$ and $l_{\star}^2$ are stronger for larger observed degrees of 
    polarization and shorter wavelengths we make use of Hubble Space Telescope 
    spectropolarimetry in the ultraviolet, which revealed high levels of circular 
    (12\%) and linear (20\%) polarization. This was measured between 130 and 140 nm 
    and is related to the absorption feature at 134.7 nm \cite{aj94}.
    To be conservative, we assume a large absolute error of 1\% on these measurements, and
    use a degree of circular polarization of 11\% at 134.7 nm.    
    
    In 1999 Solanki, Haugan and Mann \cite{shm99} have used the NGT prediction to set 
    strong upper limits on $l^2_{\star}$ of the magnetic white dwarf Grw +70$^{\circ}$8247 
    with a very simple model of the magnetic field distribution. Although it became clear 
    in the meantime that NGT suffers from technical problems that render it mathematically 
    inconsistent, we also carry out the analysis based on Eq.(\ref{ngt-form}) for 
    Grw +70$^{\circ}$8247, in order to have a quantitative comparison between the old, 
    unrealistically simple and the improved model incorporating an oblique rotator of the 
    magnetic field. 
    
    The results of our analysis can be seen in Fig.(\ref{grw-ngt}) and Fig.(\ref{grw-mag})
    Our search for the largest values of $k_{\star}$ and $l_{\star}$ compatible with the
    observed degree of cicular polarization at 134.7 nm yields the constraints 
    $k^2_{\star} \leq (0.074\, \mbox{km})^2$ and $l^2_{\star} \leq (4.1\, \mbox{km})^2$ 
    for a viewing angle $i=0^{\circ}$. This limit on $l^2_{\star}$ is to be compared with 
    the former limit $l^2_{\star} \leq (4.9\, \mbox{km})^2$. 
    These results were obtained for limb darkening coefficients $(g,h)=(0,1)$, which provide the most
    conservative constraints. Neglecting 
    limb darkening yields $k^2_{\star} \leq (0.06\, \mbox{km})^2$ and $l^2_{\star} 
    \leq (3.7 \,\mbox{km})^2$ compared with the old constraint $l^2_{\star} \leq (4.6\, \mbox{km})^2$. 
    For other values of the $(g,h)$ pair the constraint on $k_{\star}$ and $l_{\star}$ falls 
    between these extremes.
    
    For increasing viewing angle $i$ to the dipole axis the maximum degree of circular 
    polarization decreases, due to the dipole field geometry. From this it follows that constraints
    on $k_{\star}$ become stronger for larger $i$, although the dependence turns out to be weak. 
    E.g. for $i=30^{\circ}$, we obtain the tiny improvement $k^2_{\star} \leq 
    (0.072\, \mbox{km})^2$ with limb darkening and $k^2_{\star} \leq (0.058\, 
    \mbox{km})^2$ without limb darkening. Hence the most conservative estimate is for $i=0$, 
    which we adopt.
    
    \begin{figure}[t]
      \centerline{\psfig{figure=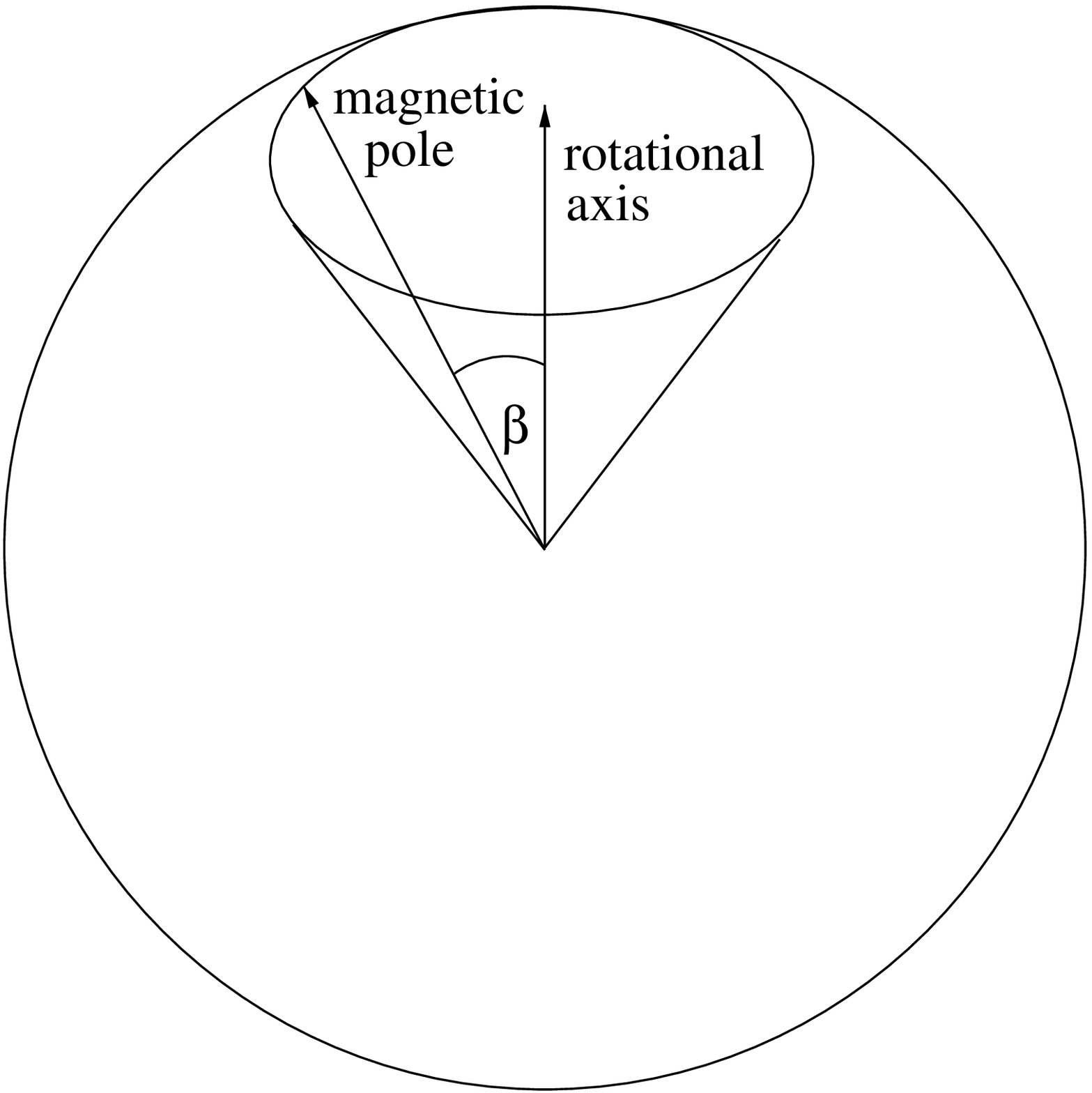,height=2.6in,width=2.6in}\hspace{1cm}
                  \psfig{figure=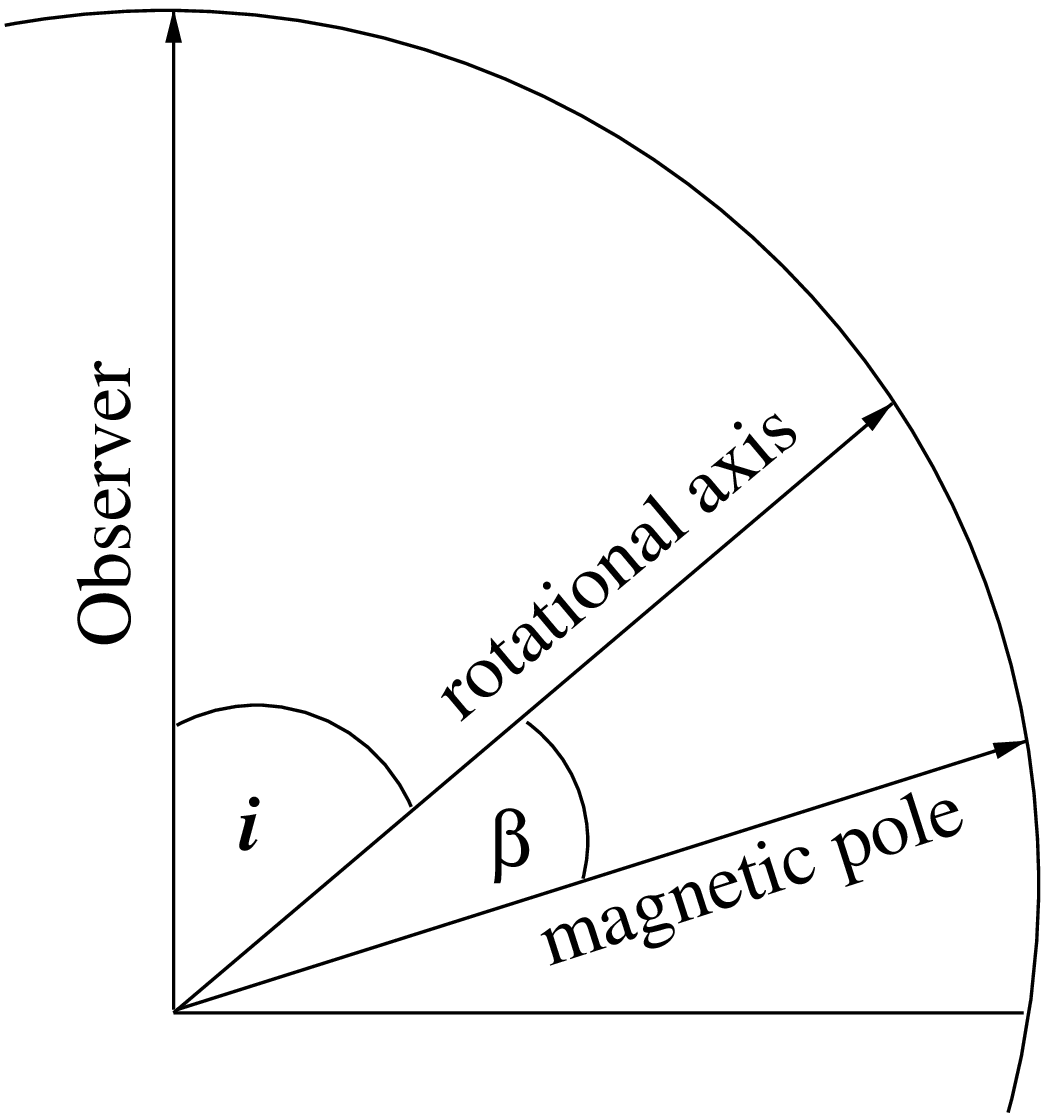,height=2.6in,width=2.6in}}
      \caption{Left figure: Projection of the conelike movement of the magnetic pole during the
               rotational period onto the celestial plane for $\beta \neq 0$. Right figure: 
	       Relative orientation of the line-of-sight to the dipole axis and the rotational axis.
               Adopted from Burleigh, Jordan and Schweizer \cite{bjo99}.}
    \end{figure}
 
    \newpage
    
  \begin{figure}[h]
    \centerline{
                \psfig{figure=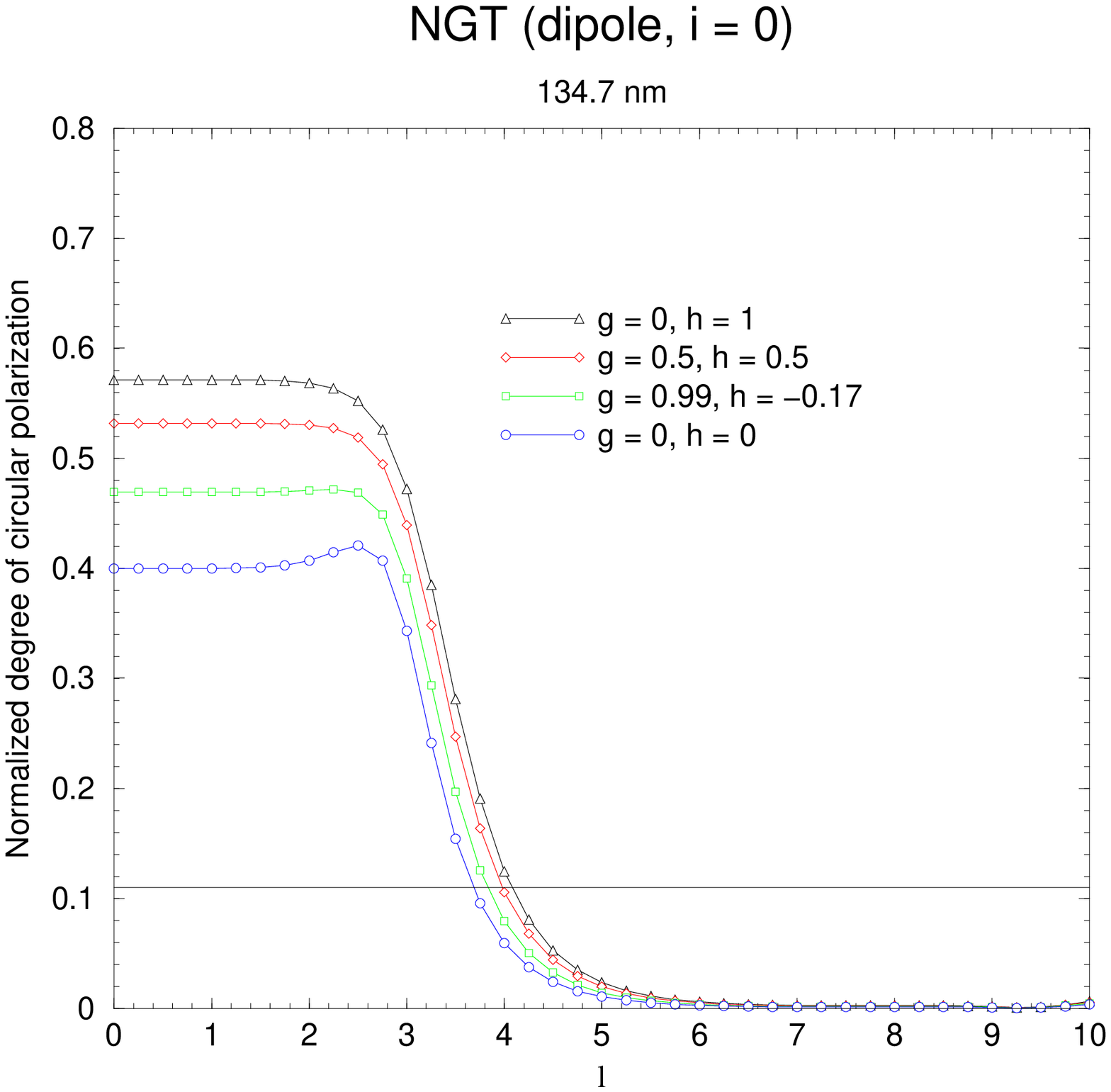,height=3.1in,width=3.1in,angle=270}
		\hspace*{0.3cm}
                \psfig{figure=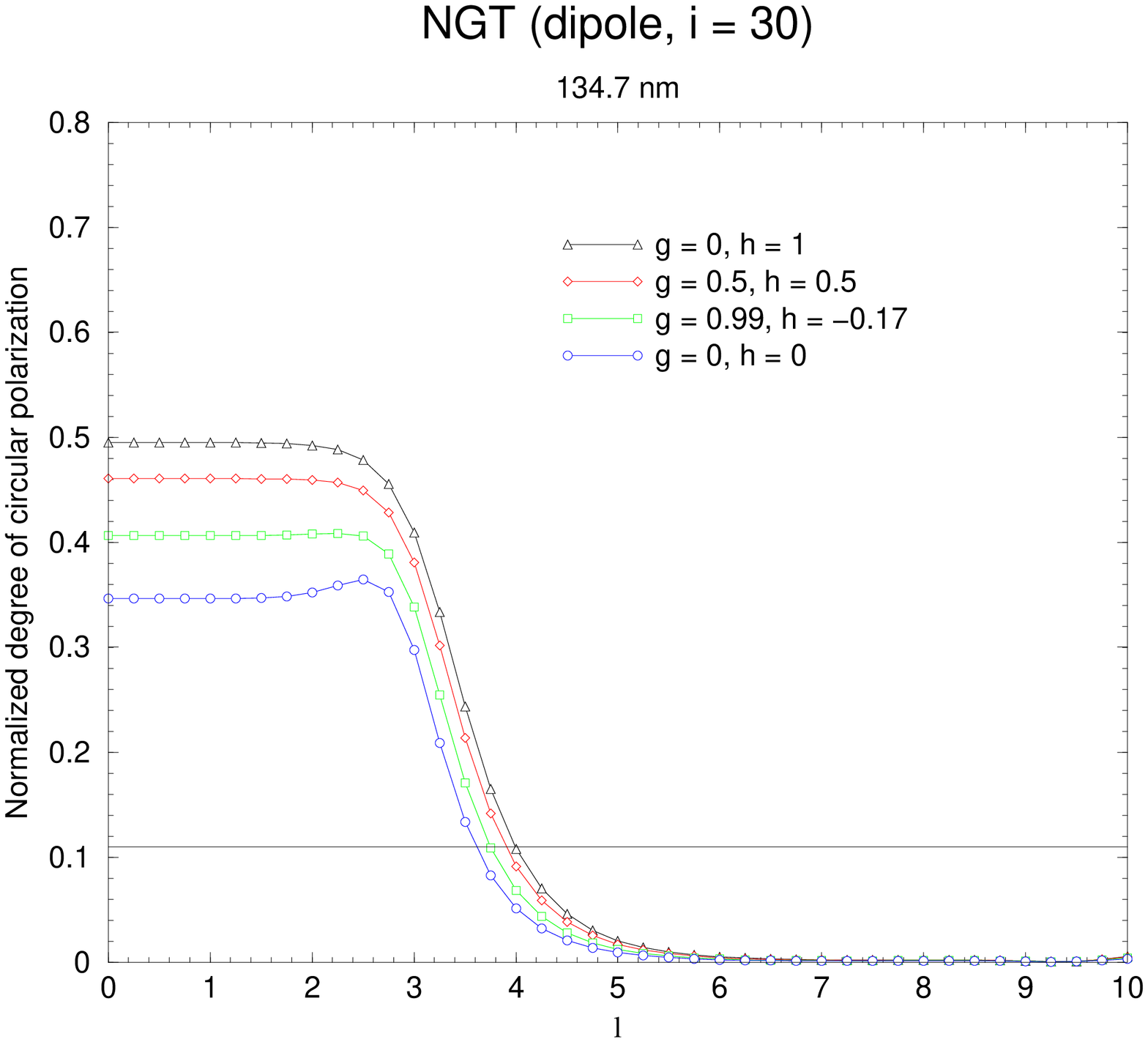,height=3.1in,width=3.1in,angle=270}}
    \caption{Observed degree of circularly polarized light for increasing
             values of $\ell_{\star}$ from NGT, normalized to full intensity $I = 1$.
	     Left figure: Dipole inclination $i=0^{\circ}$. Right figure: Dipole 
	     inclination $i=30^{\circ}$. The different curves are for different limb darkening
	     parameters. The horizontal line represents the observed
	     value of 11\%.}	
    \label{grw-ngt}	     	
    \vspace*{1cm}
    \centerline{            
		\psfig{figure=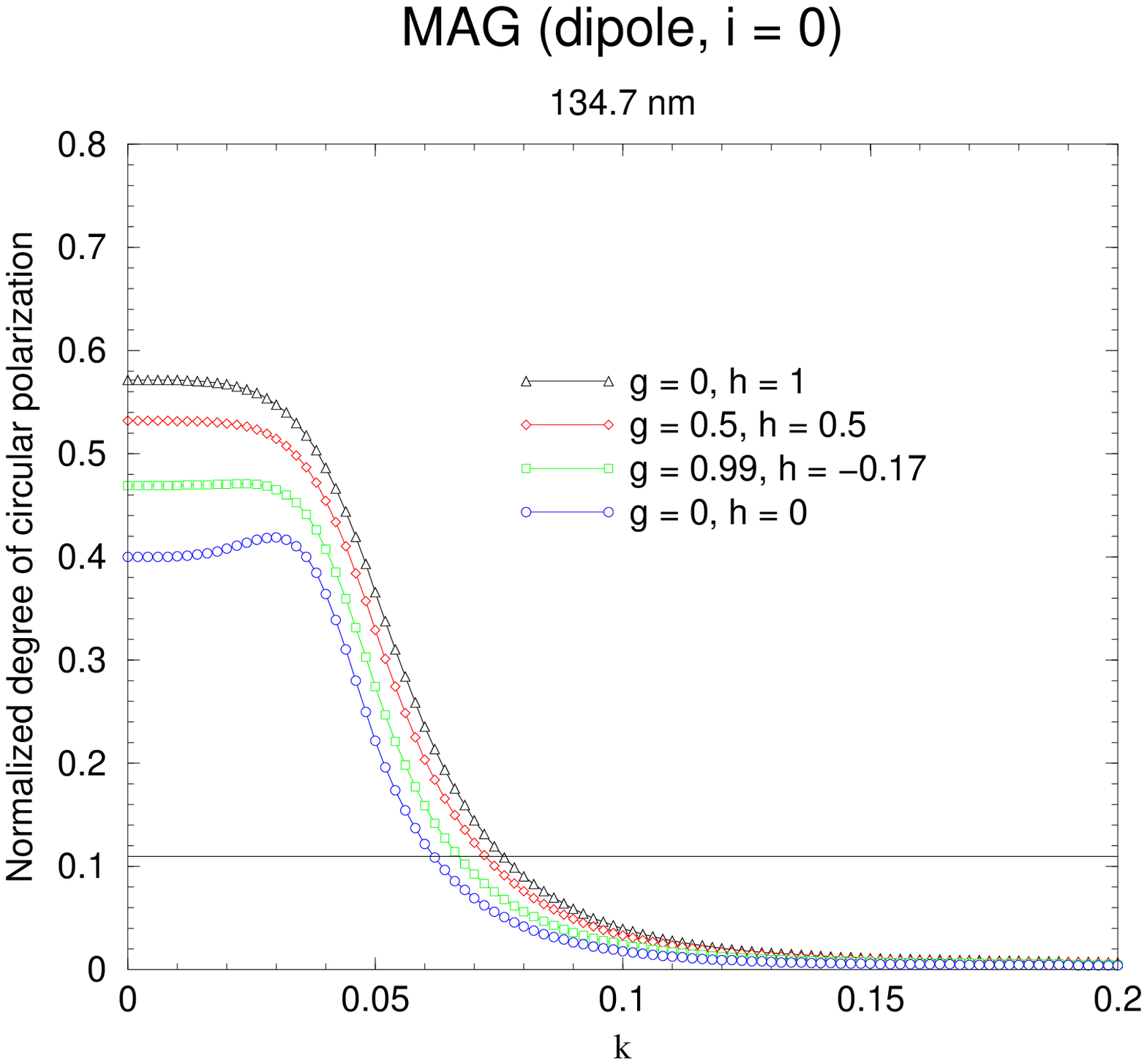,height=3.1in,width=3.1in}
		\hspace*{0.3cm}
                \psfig{figure=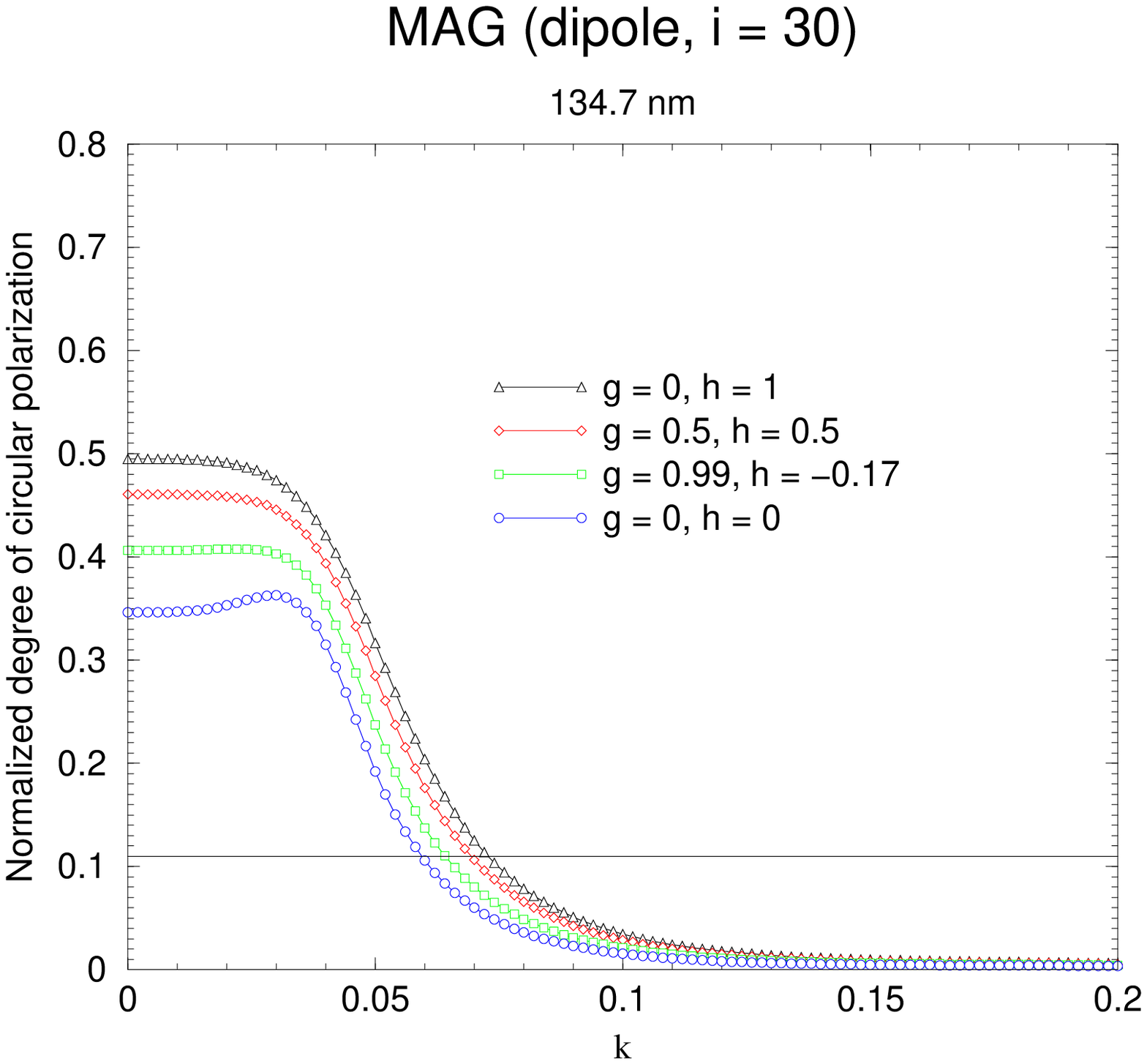,height=3.1in,width=3.1in}}
    \caption{Observed degree of circularly polarized light for increasing
             values of $k_{\star}$ from metric-affine gravity, normalized to 
	     full intensity $I = 1$. Left figure: Dipole inclination $i=0^{\circ}$. 
	     Right figure: Dipole inclination $i=30^{\circ}$. The different curves 
	     are for different limb darkening parameters. The horizontal line 
	     represents the observed value of 11\%.}
    \label{grw-mag}     		 
  \end{figure}
  \clearpage
  
\section{RE J0317-853}
  \subsection{Introduction}
    RE J0317-853 is certainly a highly unusual object which sets several records within the class 
    of isolated magnetic white dwarfs. Discovered in 1995 as an extreme-ultraviolet source during
    the {\sf ROSAT} Wide Field Camera (WFC) all-sky survey by Barstow et al. \cite{b95} (hereafter B95)
    the analysis of B95 revealed an unusual hot white dwarf with an effective temperature of
    $\approx 50000$ K and an exceptionally intense dipolar magnetic field of $\approx 340$ MG.
    B95 also reported nearly identical photometric and polarimetric variations of 725.4 s. Since 
    the most plausible explanation is that these modulations are due to the rotation of the star, this
    would lead to the fastest rotation period ever measured so far among the isolated magnetic white 
    dwarfs which have typical periods of several hours. Finally, the close proximity of the ordinary, 
    DA white dwarf LB09802, separated by only 16'' allowed B95 identify RE J0317-853 as a member of a
    double degenerate pair, which then enabled them to estimate the mass of RE J0317-853 at 
    $1.35\,M_{\odot}$, close to the Chandrasekhar limit, which makes it the so far most massive known 
    isolated white dwarf with a corresponding radius of only $0.0035\, R_{\odot}$. They found that 
    their data are best matched by a dipole model with a negative offset along the dipole axis of
    20\% viewed at an angle of $60^{\circ}$. Polarization measurements were not reported by B95.
  
    Ferrario et al. (hereafter F97) modelled phase-averaged spectropolarimetric data of RE J0317-853 
    in 1997 and obtained a dipole model with a polar field strength of $B_d=450$ MG and a negative 
    offset of 35\% of the stellar radius along the dipole axis. Their best fits were derived for 
    mean viewing angles to the dipole axis of $30^{\circ}-60^{\circ}$. Furthermore, they reported 
    variations in the wavelength-averaged circular polarization data at a period of $725 \pm 10 $ 
    s which confirmed the results from B95. More important for our work is that F97 also found 
    circular polarization in the continuum up to a maximum of 8\% in the wavelength range from 5600 
    {\AA} - 5800 {\AA}. To be conservative, we adopted for our purpose the maximum value of 5800 {\AA}
    and as for the B95 data the maximum limb darkening coefficients $g=0,\,h=1$. 
  
    The so far most detailed investigation regarding the magnetic field geometry of RE J0317-853 
    was published by Burleigh, Jordan and Schweizer in 1999 \cite{bjo99}. Using phase-resolved,
    far-UV HST Faint Object Spectrograph spectra they obtained the best fit for a dipole with
    $B_d=363$ MG and additional offsets perpendicular to the $z$-axis: $x_{\off}=0.057, \,
    y_{\off}=-0.04$ and $z_{\off}=-0.22$. In this model the rotation axis is viewed at an angle of 
    $i=50^{\circ}$ to the observer while the relative angle between rotation and dipole axis was 
    found to be $\beta=42^{\circ}$, leading to visible surface field strengths between 140 and 730 MG.
    In addition they reported the limb darkening coefficients $g=0.3$ and $h=0$.
    In 1999 Jordan and Burleigh (hereafter JB99 \cite{jb99}) reported a peak level of circular 
    polarization at $22$\% integrated over the wavelength range between 3400 {\AA} and 7000 {\AA} 
    measured in spectropolarimetric data taken with the AAT (Anglo-Australian Telescope). This
    is up to now the highest level of circular polarization ever recorded for a magnetic white dwarf.
    Although it seems very likely from the data in JB99 that the real wavelength range in which the high
    level of circular polarization was measured is between 5600 {\AA} and 5800 {\AA} this
    conjecture was so far not confirmed by the authors.          
  \subsection{Birefringence analysis}
    \begin{figure}[t]
      \centerline{\hspace*{-0.5cm}\psfig{figure=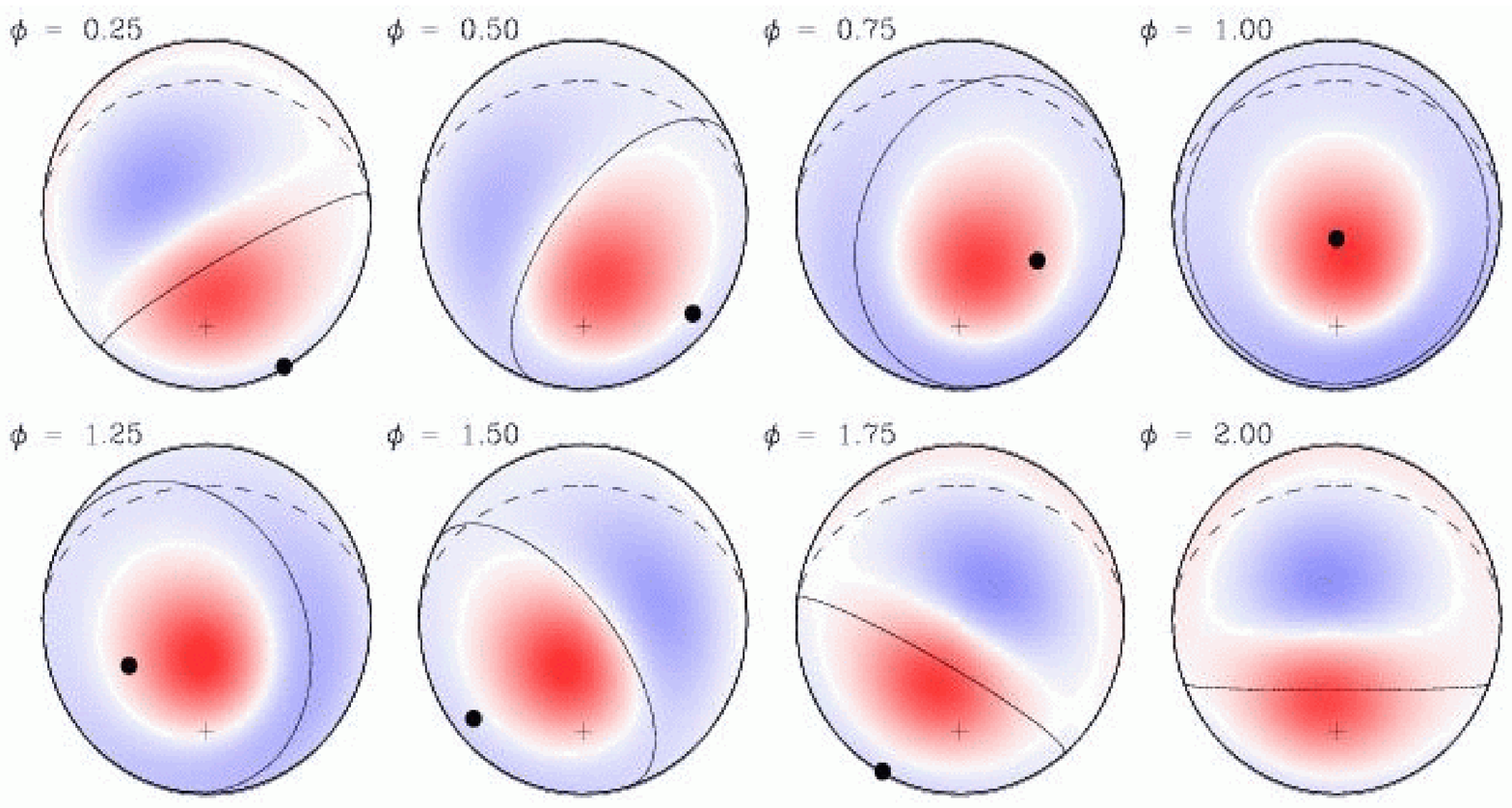,width=7in,height=3.3in}}
      \centerline{\psfig{figure=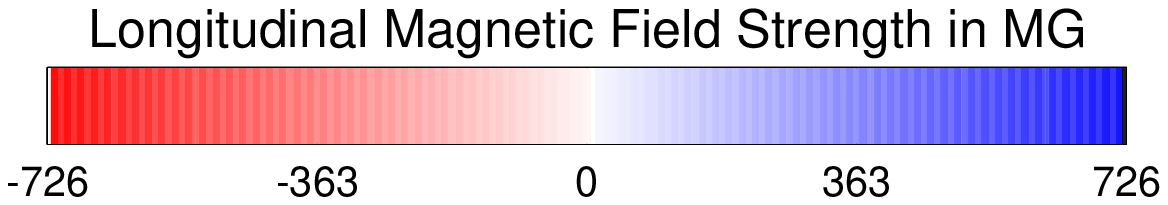,width=4in,height=0.5in}}
      \caption{Successive rotational phases of RE J0317-853 in steps of $0.25 \pi$, 
               beginning with $\phi=0.25 \pi$ (top left) to $\phi=2 \pi$ (bottom right). The dashed
	       line markes the stellar equator whereas the solid line shows the projection of the
	       magnetic equator on the stellar surface. The cross marks the position of the rotation
	       axis and the dot the position of the magnetic pole.}
      \label{phases}
    \end{figure} 

    Due to its small radius and high degree of circular polarization, RE J0317-853 is 
    a very suitably object concerning limits on gravitational birefringence. Unfortunately 
    the wavelength range in which the polarization was measured by F97 as well as by JB99 
    lies in the optical regime and additionaly is up to now not very well isolated in the 
    case of the 22\% measurement of JB99 \cite{jb99}. So, in order to get here conservative 
    and reliable upper limits on gravitational birefringence we use 7000{\AA} which marks 
    the upper end of the measured spectral range. For comparison we also present our estimates 
    in the case of 5800 {\AA} as the upper wavelength end. 
    
\begin{figure}[t]
      \centerline{\psfig{figure=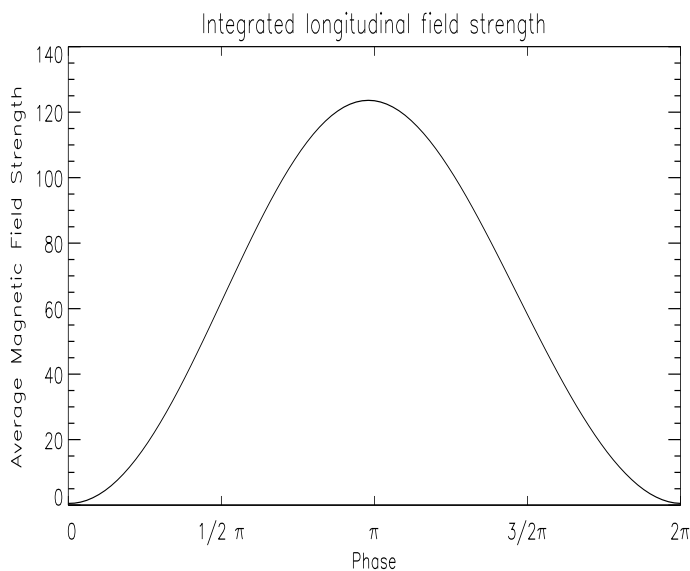,width=4in,height=3.in,angle=90}}
      \caption{Integrated longitudinal $({\bf B}_z)$ magnetic field strength of the visible
             hemisphere for successive rotational phases of RE J0317-853.}
      \label{polcurve}
    \end{figure}

    For our analysis of the JB99 data we also need information about the magnetic field 
    geometry at that rotation phase, when the highest degree of polarization is emitted 
    towards the observer. Since the highest degree is obtained for the highest average 
    longitudinal magnetic field strength over the visible hemisphere we have plotted in 
    Fig.(\ref{phases}) successive rotational phases of RE J0317-853 in steps of $0.25 \pi$
    based on the reported offset values $x_{\off}=0.057, \, y_{\off}=-0.04, \,z_{\off}=-0.22$,
    limb darkening coefficients $g=0.3,\,h=0$ from (\ref{limb}) and the angles $i=50^{\circ}$ and 
    $\beta=42^{\circ}$ by Burleigh, Jordan and Schweizer \cite{bjo99}. The longitudinal 
    field strength on the surface is colorcoded whereby the blue color denotes positive and 
    the red color negative polarity. It can be seen, that in phase $\phi = 0.25 \pi$ the 
    different polarities nearly cancel each other so that the disk averaged field strength is 
    approximately zero. During the progression of the rotation the average field strength 
    increases up to $\phi = \pi$ and then subsequently decreases. This is also quantitatively 
    shown in Fig.(\ref{polcurve}).      

   So, by using the oblique rotator model, described in Sec. 3.2 we obtain an upper limit of
   $k^2_{\star} \leq (0.038\, \mbox{km})^2$ for a wavelength of 7000{\AA} and 
   $k^2_{\star} \leq (0.036\, \mbox{km})^2$ in the case of 5800 {\AA}. For both wavelengths
   we have used the limb darkening coefficients $g=0.3$ and $h=0$.
   The F97 data yielded a result of $k^2_{\star} \leq (0.05\,\mbox{km})^2$ for 
   $i=30^{\circ}$ and $k^2_{\star} \leq (0.051\, \mbox{km})^2$ for the case of $i=60^{\circ}$ 
   and $g=0,\,h=1$ with a polarization level of 8\% at 5800 {\AA}. 
   
   The table below summarises the various upper limits for $k$ obtained for the data from B95, F97 and
   BJS99. 
  \begin{center}
  \renewcommand{\arraystretch}{1.4}
  \begin{tabular}{|c||c|c|c|}\hline
                                                & $i$ & $\beta$ &  $k$ (km) \\ \cline{2-4}
						&     &         &      \\  						 
       \raisebox{1.5ex}[-1.5ex]{F97 (8\%)}               
    &  \raisebox{1.5ex}[-1.5ex]{$30^{\circ}$} 
    &  \raisebox{1.5ex}[-1.5ex]{$(0^{\circ})$} 
    &  \raisebox{1.5ex}[-1.5ex]{$0.05$} \\ 
    \raisebox{1.5ex}[-1.5ex]{$x_{\off}=0$, $y_{\off}=0$ ,$z_{\off}=-0.35$}  
    &  \raisebox{1.5ex}[-1.5ex]{$60^{\circ}$} 
    &  \raisebox{1.5ex}[-1.5ex]{$(0^{\circ})$} 
    &  \raisebox{1.5ex}[-1.5ex]{$0.051$} \\ \hline
                                                &     &         &      \\
    \raisebox{1.5ex}[-1.5ex]{BJS99 (22\%)}      &  $50^{\circ}$ & $42^{\circ}$ &
    \raisebox{1.5ex}[-1.5ex]{$0.038$} \\
    \raisebox{1.5ex}[-1.5ex]{$x_{\off}=0.057$, $y_{\off}=-0.04$ ,$z_{\off}=-0.22$}  
                                                &     &         &      
    \raisebox{1.5ex}[-1.5ex]{$0.036$}\\  \hline	 
  \end{tabular}
  
  \vspace{0.5cm}
  \small{Tab.3.1: Upper limits on $k$ for the F97 and BJS99 data.}
  \end{center}

\section{PG 2329+267}
\begin{figure}[t]
    \centerline{\psfig{figure=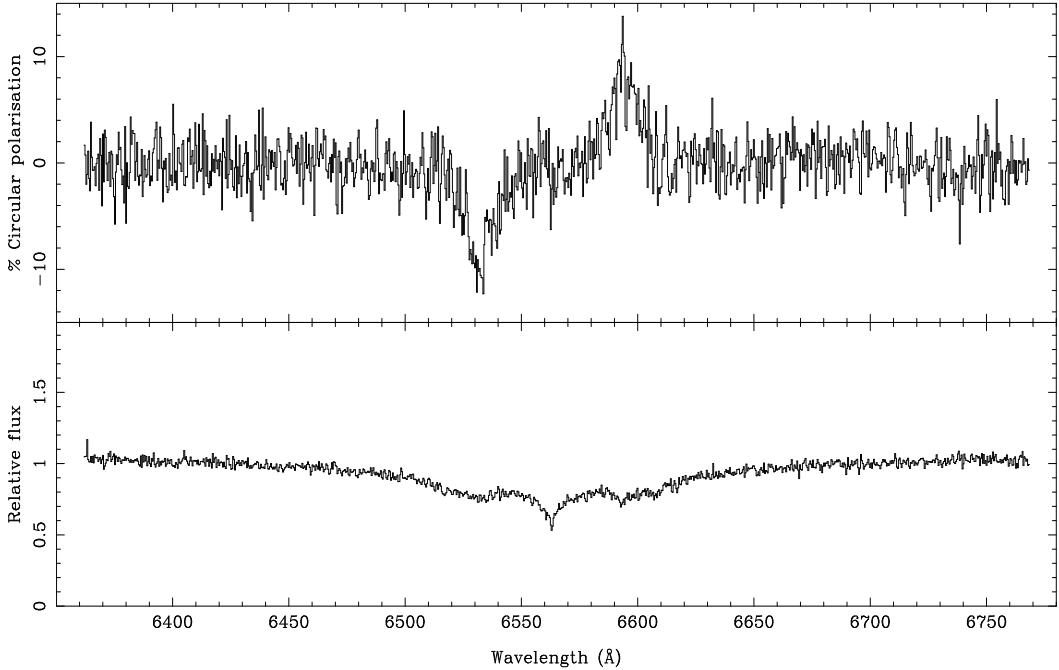,width=5.5in,height=3.5in}}
    \caption{Upper panel: Percentage of circularly polarized light, present in the $H\alpha$
    spectrum of PG 2329+267. Lower panel: The normalized spectrum. From: Moran, Marsh and 
    Dhillon (1998) \cite{mor98}.}
    \label{pg2329}
  \end{figure}

  The magnetic nature of the white dwarf PG 2329+267 was discovered in 1995 by Moran, Marsh
  and Dillon by observing the characteristic Zeeman splitting of Balmer lines \cite{mor98}. The 
  subsequent analysis revealed a centered dipole inclined at $i=60^{\circ}\pm 5^{\circ}$ to the 
  observer with a polar field strength of appoximately 2.3 MG. Rotation of the star was not 
  reported, so that we take $\beta=0$. The mass was determined at $\approx 0.9\, M_{\odot}$ which 
  means that PG 2329+267 is more massive than typical isolated white dwarfs. The radius was not 
  measured so far, so that we calculated it, using the mass-radius relation given by Weinberg 
  \cite{wb}. Asuming a helium star we got $R = 0.0156\,R_{\odot}$. This radius would decrease if we 
  asume heavier elements than helium which, in turn would lead to a stronger gravitational 
  birefringence, so that this is certainly a conservative limit on $R$. The results of the 
  circular spectropolarimetric measurements in the H$\alpha$ line are shown in Fig.(\ref{pg2329}).
  One can see very clear the S-shaped profile in the upper panel, indicating a magnetic field,
  with a peak level in the $\sigma$ components of approximately 10\% at 6530 {\AA}.   
  As in the case of Grw +70$^{\circ}$8247 we have calculated the observable degree of circular polarization
  for increasing values of $k$ and for different limb darkening parameters. For $i=55^{\circ}$ we
  found  $k^2_{\star} \leq (0.25\, \mbox{km})^2$ if limb darkening is neglegted and
  $k^2_{\star} \leq (0.31\, \mbox{km})^2$ for maximum limb darkening. In the case of
  $i=65^{\circ}$ we obtain $k^2_{\star} \leq (0.23\, \mbox{km})^2$ 
  and $k^2_{\star} \leq (0.28\, \mbox{km})^2$, respectively. 
    
\section{40 Eridani B}
  
  \begin{figure}[t]
    \centerline{\hspace*{-0.5cm}\psfig{figure=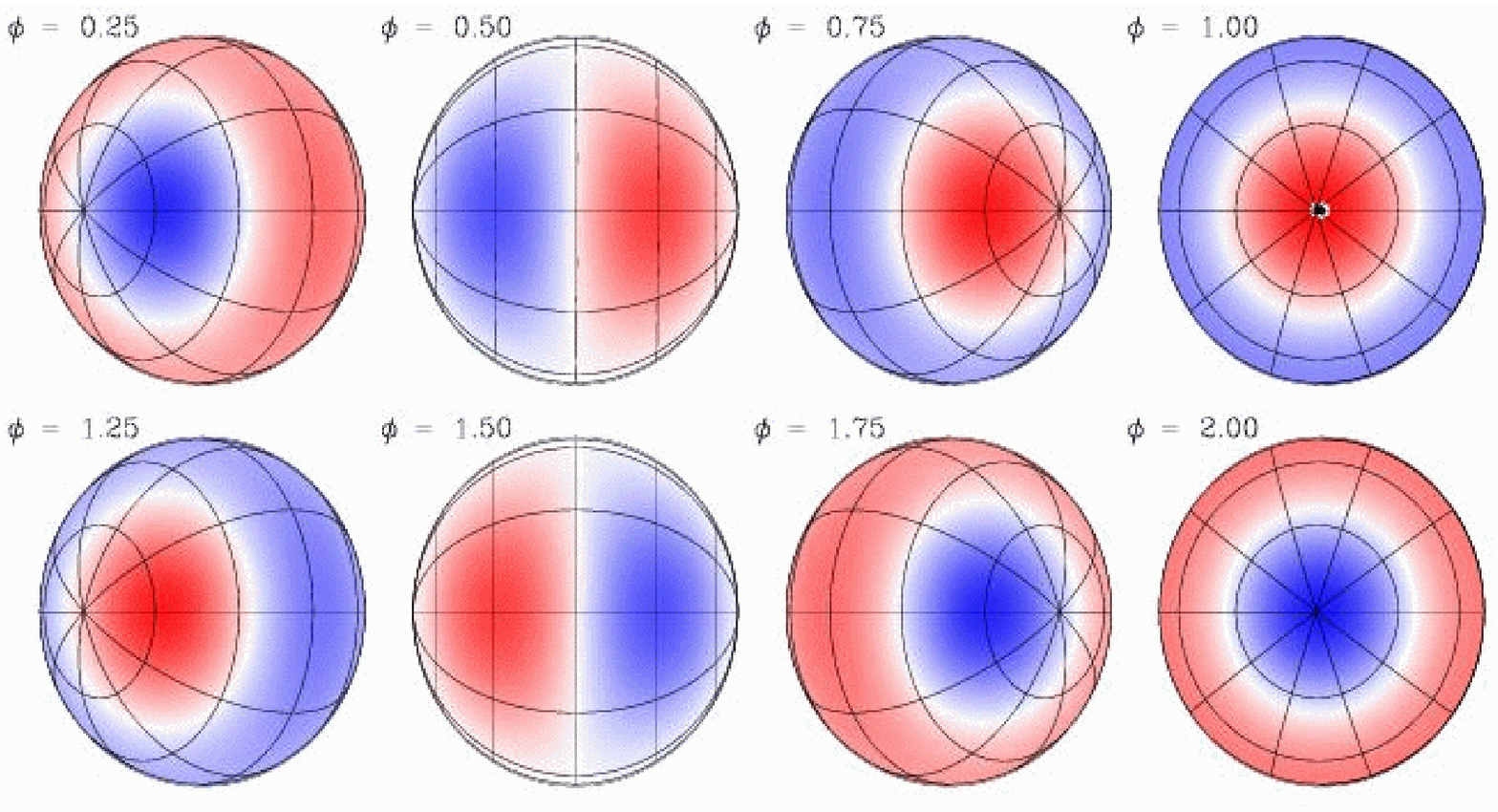,width=7in,height=3.3in}}
    \centerline{\psfig{figure=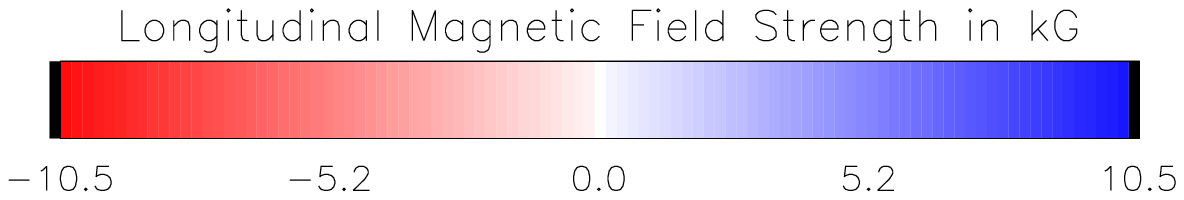,width=4in,height=0.5in}}
    \caption{Successive rotational phases of 40 Eridani B in steps of $0.25 \pi$, 
             beginning with $\phi=0.25 \pi$ (top left) to $\phi=2 \pi$ (bottom right). The 
	     poles of the grid match those of the magnetic dipole.}
    \label{eri-phases}	     
  \end{figure} 

\begin{figure}[t]
    \centerline{\psfig{figure=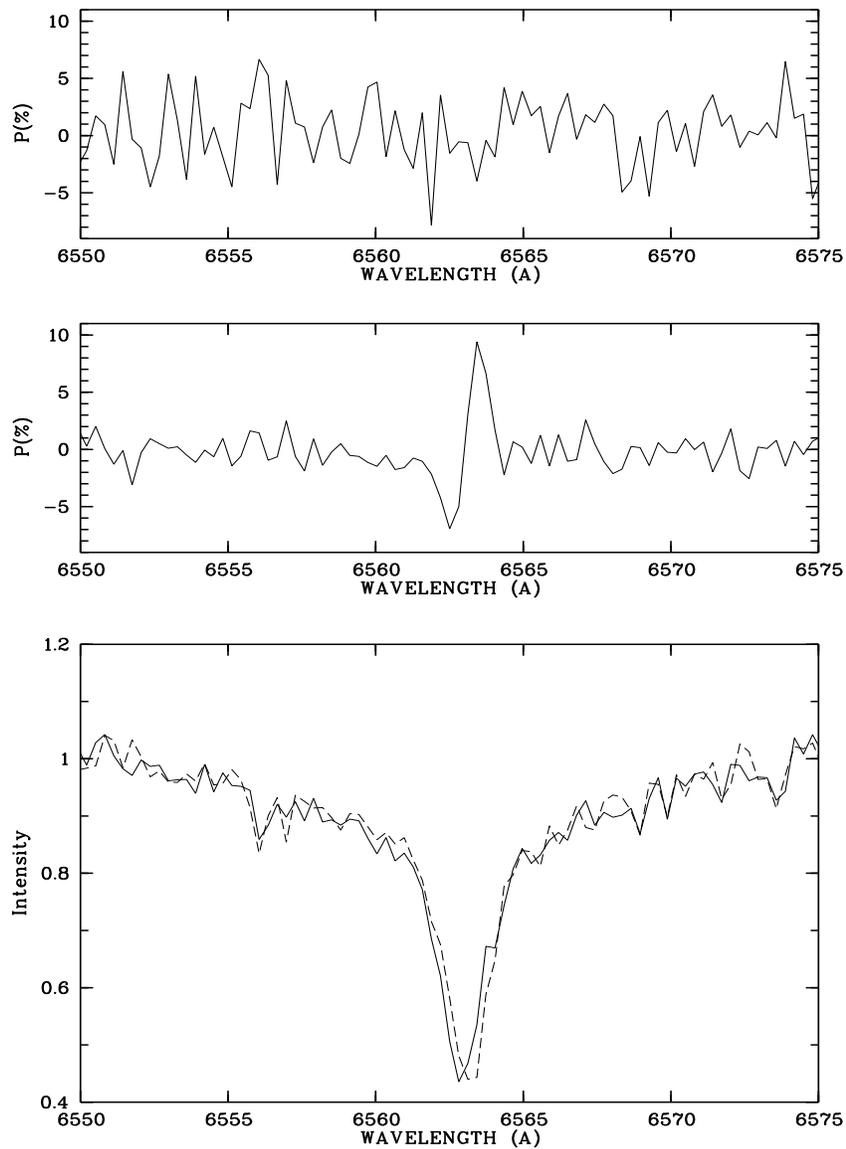,width=5.in,height=6.5in}}
    \caption{Top panel: Polarization obtained from individual spectra of the zero polarization
             ''cross-over point'' at $0.5 \pi$ or $1.5 \pi$. Bottom: Two spectra of different 
	     polarization in the region of the $H \alpha$ line core. The magnetic shift is clearly 
	     visible. Middle: The result of the substraction of these spectra, i.e. Stokes $V$.
	     From Fabrika et al. (2000) \cite{fbr00}.}
    \label{fab-circ}	     
  \end{figure}

  40 Eridani B is one of the most famous white dwarfs and represents together with Sirius B the first 
  known examples of this class of degenerate stars. The mass and radius of 40 Eri B were determined by 
  Koester and Weidemann in 1991 \cite{kw91} and more recently by Shipman et al. \cite{shp97} with the 
  result of $M=0.501\pm 0.011 M_{\odot}$ and $R=0.0136 \pm 0.00024 R_{\odot}$. Its magnetic field was 
  measured in 1999 by Fabrika et al. \cite{fbr00} using time-resolved Zeeman spectroscopy. From their 
  data they concluded the presence of a centered dipole with an average field strength of several kG. 
  The rotational axis was reported to be inclined at $i\approx 90^{\circ}$ to the observer with a 
  rotation period of about 4 hours. The magnetic axis inclination to the rotational axis is about 
  the same, $\beta \approx 90^{\circ}$. Fig.(\ref{eri-phases}) shows successive rotational phases of 
  40 Eri B using the given parameters. The circular polarization measurements reported by Fabrika et 
  al. \cite{fbr00} in the H$\alpha$ line core (see Fig.\ref{fab-circ} on the next page) yielded a level 
  of 10\% at $\approx 6564$ {\AA}. As for REJ0317-853 we have taken the magnetic field geometry of the 
  rotation phase $\phi=\pi$ with the highest average surface field strength for our analysis. With the 
  given parameters we obtained an upper limit $k^2_{\star} \leq (0.35\, \mbox{km})^2$ neglecting limb 
  darkening and $k^2_{\star} \leq (0.44\, \mbox{km})^2$ with maximum limb darkening.   
   
 \newpage

  \clearpage

\section{Comparison of the results}
  After we have obtained several upper limits on $k^2$ for various objects (including the sun) it is 
  interesting to ask, whether we can find a correlation between them. For this purpose we have listed 
  in Tab. 3.2 the masses and radii in solar units of the investigated objects, together with the
  corresponding highest and lowest upper limits on $k^2$ obtained from our analysis and their mean
  values in the fourth column. The last column shows the corresponding ratio $\Phi$ between the 
  Schwarzschildradius $r_S=2GM/c^2$ of the star and its physical radius $R$, i.e. $\Phi=2GM/(Rc^2)$.
  
  \vspace{0.5cm}
  \begin{center}
  \renewcommand{\arraystretch}{1.4}
  \begin{tabular}{|l|c|c|c|c|c|}\hline
    & $M/M_{\odot}$ & $R/R_{\odot}$ &  $k^2$ (km)$^2$ & $\overline{k^2}$ (km)$^2$ & $\Phi=2GM/Rc^2 \, (10^{-4})$ \\ \cline{2-6}
    &               &               &             &                 &  \\  						 
       \raisebox{1.5ex}[-1.5ex]{Sun}               
    &  \raisebox{1.5ex}[-1.5ex]{1} 
    &  \raisebox{1.5ex}[-1.5ex]{1} 
    &  \raisebox{1.5ex}[-1.5ex]{0.69 - 1.95}
    &  \raisebox{1.5ex}[-1.5ex]{1.32}
    &  \raisebox{1.5ex}[-1.5ex]{0.042} \\ 
       \raisebox{1.5ex}[-1.5ex]{40 Eri B}  
    &  \raisebox{1.5ex}[-1.5ex]{0.5} 
    &  \raisebox{1.5ex}[-1.5ex]{0.013} 
    &  \raisebox{1.5ex}[-1.5ex]{0.35 - 0.44} 
    &  \raisebox{1.5ex}[-1.5ex]{0.395}
    &  \raisebox{1.5ex}[-1.5ex]{1.614}\\ 
       \raisebox{1.5ex}[-1.5ex]{PG 2329}  
    &  \raisebox{1.5ex}[-1.5ex]{0.9} 
    &  \raisebox{1.5ex}[-1.5ex]{0.0156} 
    &  \raisebox{1.5ex}[-1.5ex]{0.23 - 0.28} 
    &  \raisebox{1.5ex}[-1.5ex]{0.255}
    &  \raisebox{1.5ex}[-1.5ex]{2.42}\\ 
       \raisebox{1.5ex}[-1.5ex]{GRW}  
    &  \raisebox{1.5ex}[-1.5ex]{1} 
    &  \raisebox{1.5ex}[-1.5ex]{0.0076} 
    &  \raisebox{1.5ex}[-1.5ex]{0.058 - 0.074}
    &  \raisebox{1.5ex}[-1.5ex]{0.066} 
    &  \raisebox{1.5ex}[-1.5ex]{5.52}\\ 
       \raisebox{1.5ex}[-1.5ex]{RE J0317}  
    &  \raisebox{1.5ex}[-1.5ex]{1.35} 
    &  \raisebox{1.5ex}[-1.5ex]{0.0035} 
    &  \raisebox{1.5ex}[-1.5ex]{0.036 - 0.051} 
    &  \raisebox{1.5ex}[-1.5ex]{0.0435}
    &  \raisebox{1.5ex}[-1.5ex]{16.2}\\ \hline 
  \end{tabular}
  \end{center}
  
  \vspace{0.5cm}
  {\small \noindent Tab.3.2: Relevant data of considered objects. First and second column: Mass and Radius 
         in solar units. Third column: Range of upper limits on $k^2$ obtained from our analysis. Mean values
	 $\overline{k^2}$ in the fourth column. The last column gives the ratio $\Phi$ between Schwarzschild and
	 physical radius $\times 10^{-4}$.}
  
\begin{figure}[h]
     \centerline{\psfig{figure=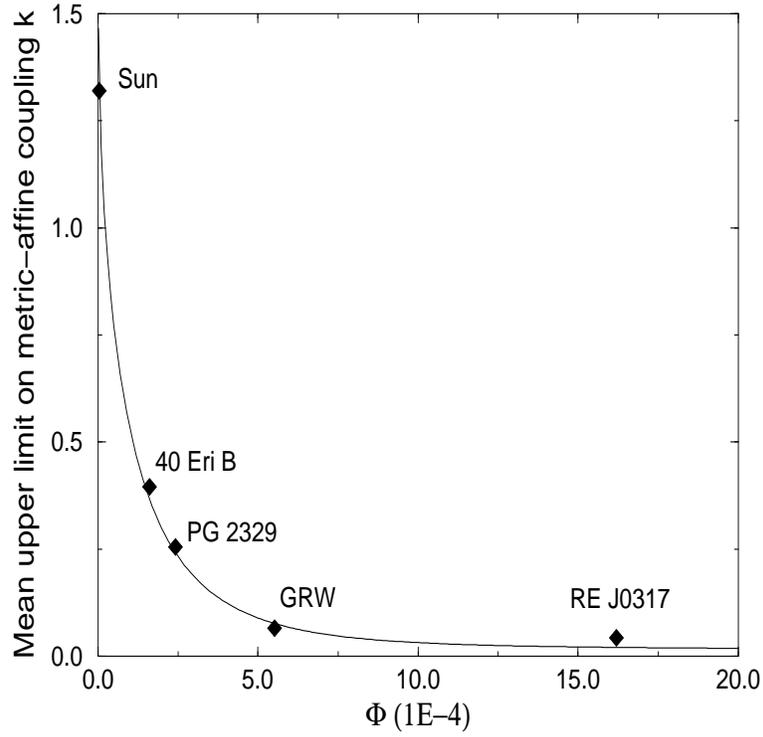,width=4.5in,height=4.5in}}
     \caption{Mean value of $k^2$ plotted against the corresponding ratio $\Phi$ in units of $10^{-4}$. The
              distribution can be fitted by a smooth curve, given in the text. The different ranges
	      of the $k^2$ values are of about the same size as the symbols and therefore omitted.}
     \label{pout}	     
   \end{figure}

  \vspace{0.5cm}
  \noindent The first thing which attracts attention is that the limits on $k^2$ decrease for increasing $\Phi$
  of the stellar objects. This correlation is graphically shown in Fig. \ref{pout} on the next page
  where we plot from Tab. 3.2 the mean value of $k^2$ for each object against $\Phi$, marked by solid diamonds. 
  The best fit to this distribution can be found with gaussian linear regression method. This yields
  
  \begin{equation}\label{k-phi}
    k^2(\Phi) = 1.363982\cdot (\exp(-\Phi^{0.7})+0.04\,\Phi^{-0.18}) \quad .
  \end{equation}  
  
  The agreement between the values $k^2(\Phi)$ of the fittet curve and the measured $\overline{k^2}$ are quite 
  good as one can see in Tab. 3.3. Therefore, (\ref{k-phi}) can be interpreted in that way, that it gives
  approximately those upper limits on $k^2$ which one could expect for celestial bodies of mass $M$ and radius 
  $R$ by using the polarization modelling technique used in this chapter. For example an object of about the mass
  and the size of the earth $(M_E=5.97\cdot 10^{24}\mbox{kg},\,R=6.378\cdot 10^6 \mbox{m})$ would give
  $k^2_E \lsim (1.771501 \,\mbox{km})^2$, while a supermassive black hole similar to the one suspected in the 
  center of our galaxy with $M_{BH}=2.5\cdot 10^6 M_{\odot}$ and a Schwarzschildradius of $r_S=2GM/c^2=2.969\cdot 
  10^9 \mbox{m}$ would give $k^2_{BH}\lsim (0.010396 \,\mbox{km})^2$. Finally we note that this would also be the
  upper limit for any other object which has reached its Schwarzschildradius, since $k^2_{\star}(\Phi)$ depends 
  only on the ratio $r_S/R_{\star}$.
  
  \clearpage
   
  \begin{center}
  \renewcommand{\arraystretch}{1.4}
  \begin{tabular}{|l|c|c|c|}\hline
    & $k^2(\Phi)$ & $\overline{k^2}$ & Deviation from $\overline{k^2}$ in \%   \\ \cline{2-4}
    &             &                  &               \\
       \raisebox{1.5ex}[-1.5ex]{Sun}               
    &  \raisebox{1.5ex}[-1.5ex]{1.320012} 
    &  \raisebox{1.5ex}[-1.5ex]{1.320}
    &  \raisebox{1.5ex}[-1.5ex]{$9\cdot 10^{-4}$} \\ 
       \raisebox{1.5ex}[-1.5ex]{40 Eri B}               
    &  \raisebox{1.5ex}[-1.5ex]{0.387055} 
    &  \raisebox{1.5ex}[-1.5ex]{0.395}
    &  \raisebox{1.5ex}[-1.5ex]{2.01} \\ 
       \raisebox{1.5ex}[-1.5ex]{PG 2329}               
    &  \raisebox{1.5ex}[-1.5ex]{0.259636} 
    &  \raisebox{1.5ex}[-1.5ex]{0.255}
    &  \raisebox{1.5ex}[-1.5ex]{1.81} \\ 
       \raisebox{1.5ex}[-1.5ex]{GRW}               
    &  \raisebox{1.5ex}[-1.5ex]{0.090102} 
    &  \raisebox{1.5ex}[-1.5ex]{0.066}
    &  \raisebox{1.5ex}[-1.5ex]{36.5} \\ 
       \raisebox{1.5ex}[-1.5ex]{RE J0317}               
    &  \raisebox{1.5ex}[-1.5ex]{0.034262} 
    &  \raisebox{1.5ex}[-1.5ex]{0.0435}
    &  \raisebox{1.5ex}[-1.5ex]{21.23} \\ \hline

  \end{tabular}
  \end{center}
  
  \vspace{0.5cm}
  {\small \noindent Tab.3.3: Comparison of $k^2(\Phi)$ values from the fitted curve with the measured
           mean values $k^2$. The last column gives the relative deviation in percent of the fitted from
	   the measured values.}
     
  \clearpage    
\section{Discussion and Conclusions}
  We have used spectropolarimetric observations of the white dwarfs Grw +70$^{\circ}$8247,
  RE J0317-853, PG 2329+267 and 40 Eridani B to impose new strong constraints on gravitational 
  birefringence. To improve earlier investigations of Grw +70$^{\circ}$8247 we employed a 
  dipole model for the magnetic field structure which leads to, on average, an 18\% sharper 
  limit.
 
  However, for all white dwarfs we have considerd our constraints on $k_{\star}$ and $l_{\star}$ 
  strongly depend on the sophisticated form of the weight function $w(\mu)$. At the moment, 
  $w(\mu)$ only gives very conservative results from a simple relation between the magnetic 
  field strength at a certain position on the visible stellar disc and the emitted degree of 
  cicular polarization. So, the next step to enhance $w(\mu)$ could consist of calculating 
  numerically the theoretical degree of circular polarization for each point on the stellar 
  disc by using simulations of radiative transfer in stellar atmospheres. A further, clear 
  improvement on the constraint on $k_{\star}$ should be possible within this approach.

  The possibiliy, given by (\ref{k-phi}) to calculate approximate values for upper limits on $k^2$
  depending on the ratio between the Schwarzschild and physical radius is certainly a useful result.
  Those values could be used as a guideline for the birefringence analysis of other objects. However,
  the constant numerical values in (\ref{k-phi}) possibly have to changed slightly when measured
  $k^2$ values of other objects has to be fitted into this scheme.

\chapter{Cataclysmic Variables}
  The method which we have used up to now for setting strong upper limits on gravity-induced 
  birefringence is very limited. As mentioned before, this is mainly due to the lack of a reliable 
  atmospheric model for magnetic white dwarf stars which could provide realistic estimates for 
  the emitted degree of polarized radiation as a function of field strengths and opacities.
  
  In this chapter we circumvent this restriction by focussing our attention on magnetic 
  white dwarfs which are members in close interacting binary systems. These are systems where 
  two stars move around the common center-of-mass in a close orbit so that matter can fall along a 
  ballistic stream from one star to the other. An important subclass of these cataclysmic variables are
  the AM-Her type binary systems where the white dwarf posses a strong magnetic field so that matter
  is prevented from forming an accretion disc and, instead, impacts directly on the stellar surface near 
  the magnetic poles. During this process highly polarized cyclotron radiation is emitted whose rotational 
  modulation provides a unique opportunity not only for setting upper limits on gravity-induced 
  birefringence but perhaps also for the very first direct observation of this effect.
  
  We therefore begin this chapter with a brief introduction into the basic physics of close interacting
  binary systems. The characteristics of the polarized radiation are determined by the various parameters
  like the temperature structure and height of the shock front, formed when the accreting matter hits the 
  stellar surface, as well as by the electron number density of the heated plasma and the angular extend 
  of the emission region. A realistic estimate of the emitted degree of polarization therefore depends 
  heavily on the quality of the shock front model. With this underlying physical structure, the theory of 
  radiative transfer of polarized radiation in the case of cyclotron emission is reviewed. Although the 
  theory is well known for decades now, we add a new feature by modifying the resulting Stokes vector 
  when we allow for gravity-induced birefringence. 
     
  Using this theoretical framework we present phase dependent light and polarization curves for the AM-Her 
  system VV Puppis which are compared with polarimetric data taken by Cropper \& Warner in 1986 \cite{cw86}. 
  Although the observed asymmetries are in agreement with our numerical curves, the flat topped character
  seen in observations cannot be achieved numerically. However, by allowing for sufficiently strong 
  birefringence we obtain a good conformity with the observations. 
 \newpage
 
  \section{Basic model of interacting binaries}
    Approximately $50 - 60$\% of all stars are assumed to be members of binary systems. By involving
    nearly the whole spectrum of stars from red giants to black holes as possible binary components, 
    these systems often provide examples of astrophysical processes which are among the most exciting 
    currently known. From a more practical point of view the most precise mass determinations in
    astronomy are available from binary systems where the period of revolution is precisely measurable 
    due to mutual occultations.
    
    The physics of binary systems depends of course on the masses of the involved stars and, more 
    importantly, on their spatial separation. If the distance between both components is big enough 
    so that they can interact at every evolutionary stage only gravitationally without any related 
    mass transfer, then each member of the binary will pass through its individual development without being 
    disturbed by its companion. An example for this is Sirius with the white dwarf Sirius B as the 
    second star. However, the situation becomes more interesting if the stars move so close to each
    other that tidal effects are no longer negligible. In this chapter we are interested in such systems
    which consist of a white dwarf as the primary component and an, evolutionary younger and more 
    massive, secondary component which can be represented by a main sequence or a red giant star.
    If both stars are not too close, their initial spherical shapes becomes teardrop-shaped with the tip
    pointing towards the companion. Basically a mass transfer stream between the components can now 
    be initialized in two ways, i.e. either by an evolutionary expansion of the secondary component 
    or by a continuous decrease of the distance between the stars. In both cases, the secondary component
    begins to fill its Roche lobe (the volume surrounding an object within matter is gravitationally 
    bound to it) and matter is transfered to the primary component via the inner Lagrange point L1.    
    The mechanism of how exactly the distance between the stars could be further decreased is not fully 
    understood at the moment. One possibility is that they already originated in close proximity or that 
    due to friction of the surrounding material from both components the rotational energy of the system 
    was partially converted into internal energy.  
    
    Once the mass transfer has started, the angular momentum of the system prevents
    the stream from falling directly onto the white dwarf surface. Instead the material forms, in most 
    cases, an accretion disc where the gas slowly spirals inwards while, again by friction, the 
    gravitational energy is converted into thermal energy. Finally, the matter falls onto the white dwarf 
    from the inner edge of the disc. However, this is not a continuous process since the material is
    mostly accumulated in the accretion disc until the disc becomes unstable and, then, is suddenly 
    deflated accompanied by a huge brightening of the disc due to conversion of gravitational to thermal 
    energy. Such an event is also known as a dwarf nova. Such close binary systems which interact by
    means of a mass transfer stream are called {\em cataclysmic variables}. 
    
    For our purpose, the most interesting among them are those where the white dwarf posseses a 
    ultrastrong magnetic field of the of the order $10 \lsim B \lsim 60$ megagauss (MG), which prevents
    the formation of a normal accretion disc and leads, instead, to a well-defined accretion flow
    from the secondary component to near the magnetic pole(s) of the white dwarf. These objects are
    often called AM-Her systems, after the prototype AM-Herculi discovered in 1923. 
     
    \newpage  
  
    \begin{figure}[t]
      \centerline{\psfig{figure=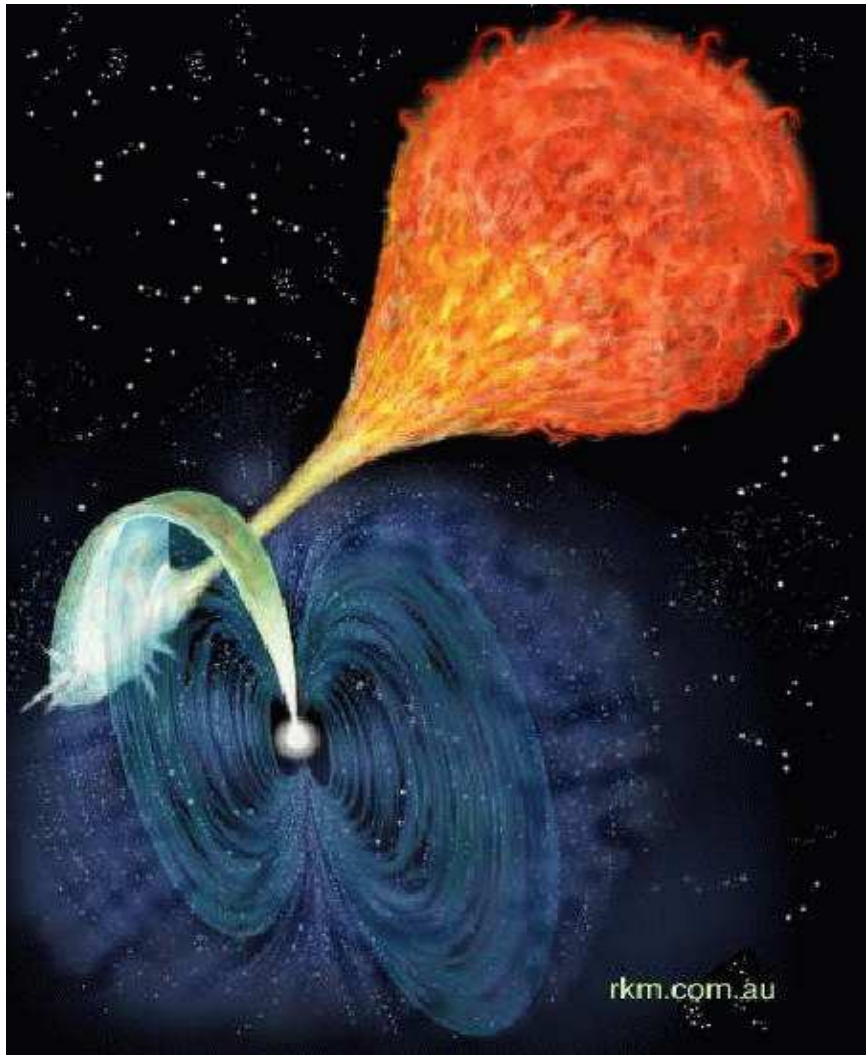,width=4.5in,height=5.5in}}
      \caption{Artist impression of an Am-Her type system.}
    \end{figure}   
    
    \vspace*{1cm} 
    \noindent The name ''Polar'' for AM-Hers and related systems was introduced by Krzeminski \& 
    Serkowski in 1977 \cite{ks77} because of the strong and variable circular $(\sim 10\% - 30\% \,
    \cite{cw86,tap77})$ and linear polarization which is typical for these objects. Certainly 
    this attribute makes the polars very appealing for utilizing them for possible measurements 
    of gravity-induced birefringence. Therefore, the next sections will give some insight into 
    the basic physical processes responsible for the emission of polarized radiation. This, in 
    turn provides the basis for a critical analysis of the observed polarization properties with respect to 
    gravity-induced birefringence.     
   
    \clearpage
    
    \begin{figure}[t]
      \centerline{\psfig{figure=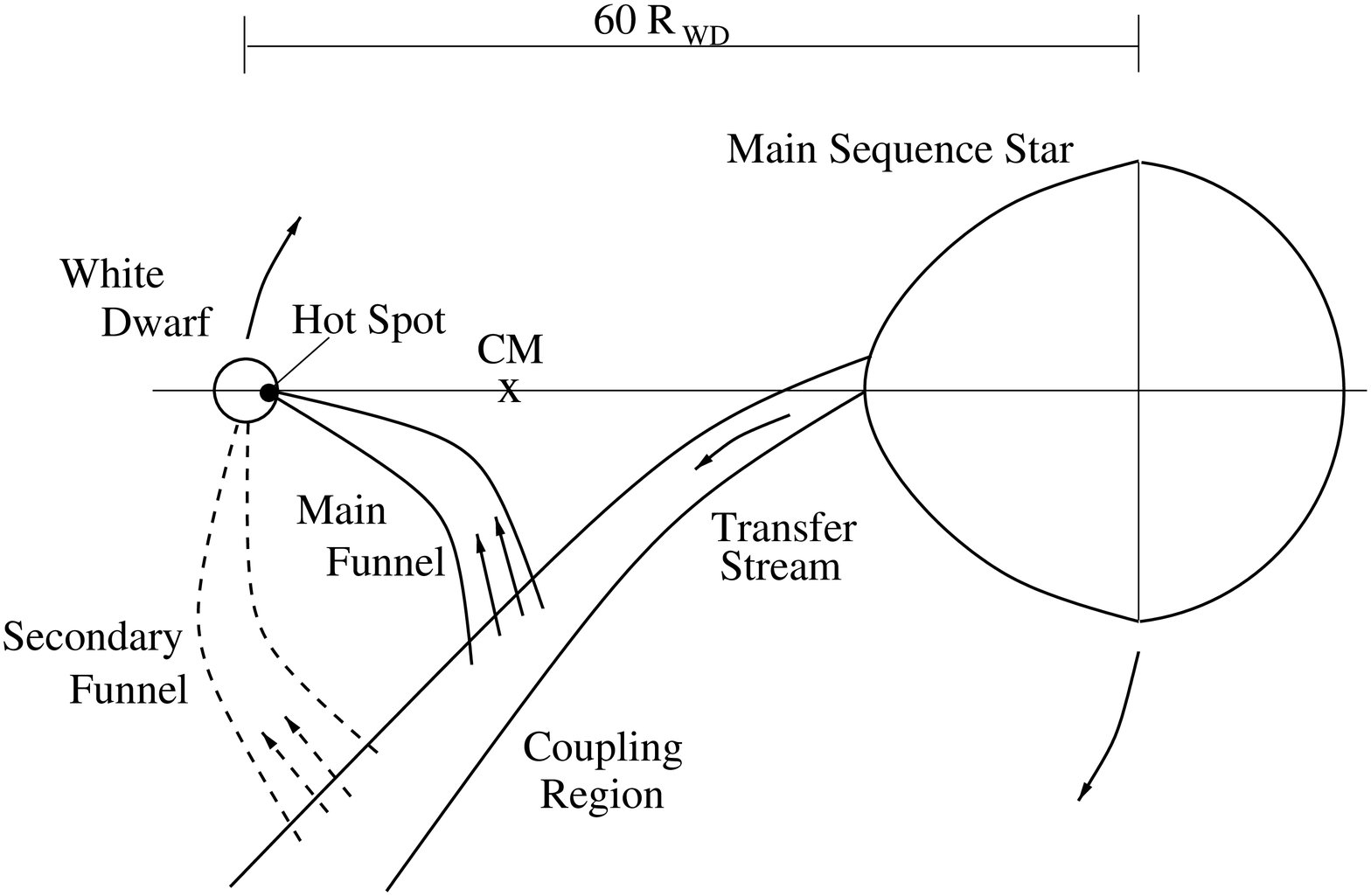,width=5.5in,height=3.5in}}
      \caption{Schematic representation of the accretion pattern in AM Hers (adopted from
               Wickramasinghe \& Ferrario \cite{wf00}).}
      \label{amher-scheme}       
    \end{figure}
    
  \section{Mass transfer and shock models}   
    The basic model which has been developed for AM-Her type systems is briefly sketched in 
    Fig.\ref{amher-scheme}. It is mainly based on X-ray observations \cite{beu98}, modelling 
    of the continuum and emission-line radiation from magnetically confined accretion funnels 
    \cite{fw99}, and polarized emission from shocks \cite{fwi90}. 
    
    For an ordinary, nonmagnetic primary component the matter streams from the secondary star and
    accumulates in a disc before impacting on the white dwarf surface. On the other hand the detailed 
    accretion mechanisms are much more complicated and even controversial (e.g. \cite{liw97}) if 
    the primary component posseses a strong magnetic field. However, it is generally accepted that 
    after the material has left the companion star near the inner Lagrangian point, a stream is 
    formed which falls freely towards the white dwarf on a ballistic trajectory. If the magnetic field
    of the accreting object is strong enough $(\gsim 1$MG) the increasing magnetic pressure will
    begin to dominate the material flow at some distance from the white dwarf in a region called the
    coupling region. Here, the radial velocity component is reduced and the flow aquires first 
    toroidal and then poloidal velocity components as the stream couples onto the field lines.
    From that point the material falls again freely towards the accreting object in magnetically confined 
    funnels that connect the coupling region with the white dwarf surface. Finally a static shock
    above one or both magnetic poles slows and heats the infalling material before it settles on the 
    surface with a temperature of $kT \approx 10$keV.
    \subsection{The shock region}
      The shock region can be regarded as the main source of emission in the optical and the X-ray wave
      band. Calculations of the circular polarization light curve presented in this chapter are therefore 
      based on a sophisticated model of the emission region, first invented by Wickramasinghe \& Megitt
      in 1985 \cite{wme85}.
      
      The first attempts to model the shock fronts of AM-Her type systems, given by Meggitt \&
      Wickramasinghe in 1982 \cite{mew82}, were based on the assumptions of constant temperature
      accretion columns. Although successful in explaining some of the gross properties
      like the continuum energy distribution, this model, based on cyclotron opacity, 
      failed to predict the correct degree of polarization (the predicted values were a factor of 
      about 2-3 higher than observed). The model which is discussed here can be regarded as a further 
      development of this early model since it includes a temperature structure (shock front) as 
      well as free-free opacity as a pure absorption process.  
     
      Assuming that the ionized material falls freely towards the magnetic poles, it moves a radial
      distance $r$ with the velocity       
      \begin{equation}
        V(r) = \left(\frac{2GM}{r}\right)^{1/2} = 5.2 \times 10^8\left(\frac{M}{M_{\odot}}\right)^{1/2}
	\left(\frac{10^9 \mbox{\footnotesize{cm}}}{R}\right)\left(\frac{R}{r}\right)^{1/2}
	\mbox{\footnotesize{cm}}\,s^{-1} \quad ,
      \end{equation}
      where $M$ and $R$ are the mass and the radius in (cm) of the white dwarf respectively. Before
      the material settles down on the surface, the density is increased in a strong shock by a factor
      4 and the velocity decreases by the same factor across the shock front. In the case that this 
      shock is formed near the white dwarf surface, the shock temperature is given by 
      \begin{equation}
        T_S = \frac{3\mu m_H GM}{8 kR} = 6\times 10^8 \mu\left(\frac{M}{M_{\odot}}\right)
	\left(\frac{10^9 \mbox{\footnotesize{cm}}}{R}\right)\mbox{K} \quad ,
      \end{equation} 
      where $\mu$ denotes the mean molecular weight. The postshock electron density is      
      \begin{equation}
        N_e = 3.8 \times 10^{17}\left(\frac{F_s}{10^2\mbox{\footnotesize{g cm}}^{-2}s^{-1}}\right)
	\left(\frac{M}{M_{\odot}}\right)^{-1/2}\left(\frac{R}{R_{\odot}}\right)^{1/2}
	\mbox{\footnotesize{cm}}^{-3} \quad ,
      \end{equation}  
      where $F_s$ is the specific accretion rate. In this context it is important to note that the scope
      of this model only allows accretion within a small fraction of the stellar surface namely on
      a point like structure. This restriction will later be removed by a more general approach 
      that also allows for more extended shock regions.
      
      The bulk of the translational energy is carried by the ions which achieve a Maxwellian velocity
      distribution by ion-ion collisions at a distance $h_{\mbox{\tiny{ion}}}\sim V_{\mbox{\tiny{ion}}}
      t_{\mbox{\tiny{ion}}}$, where $t_{\mbox{\tiny{ion}}}$ denotes the ion-ion collision time scale
      \cite{mew84}, while the electrons are mainly heated by Coulomb collisions with the ions. The
      electron temperature usually differs from the ion temperature and both vary with height in the
      shock \cite{ima87,wuc90}. As the gas then settles on the stellar surface it is cooled by
      bremsstrahlung and cyclotron radiation. In the case that cyclotron radiation dominates, the
      electrons cool faster than they can be heated by collisions with ions, with the consequence that 
      ions and electrons have different temperatures, so that the gas in the ''two-fluid'' regime. 
%
      However, it is important to note that the accretion flow in AM-Her systems is not continuous.
      Occasionally these systems drop to low states of reduced brightness which could be explained
      by interruptions or reductions in the material stream. The reason for this accretion modulation
      is not fully understood at the moment, but it may be related to solar-type magnetic activity
      in connection with starspots of the secondary component \cite{hgm00}.
    \subsection{Extended emission regions}    
      \begin{figure}[t]
        \centerline{\psfig{figure=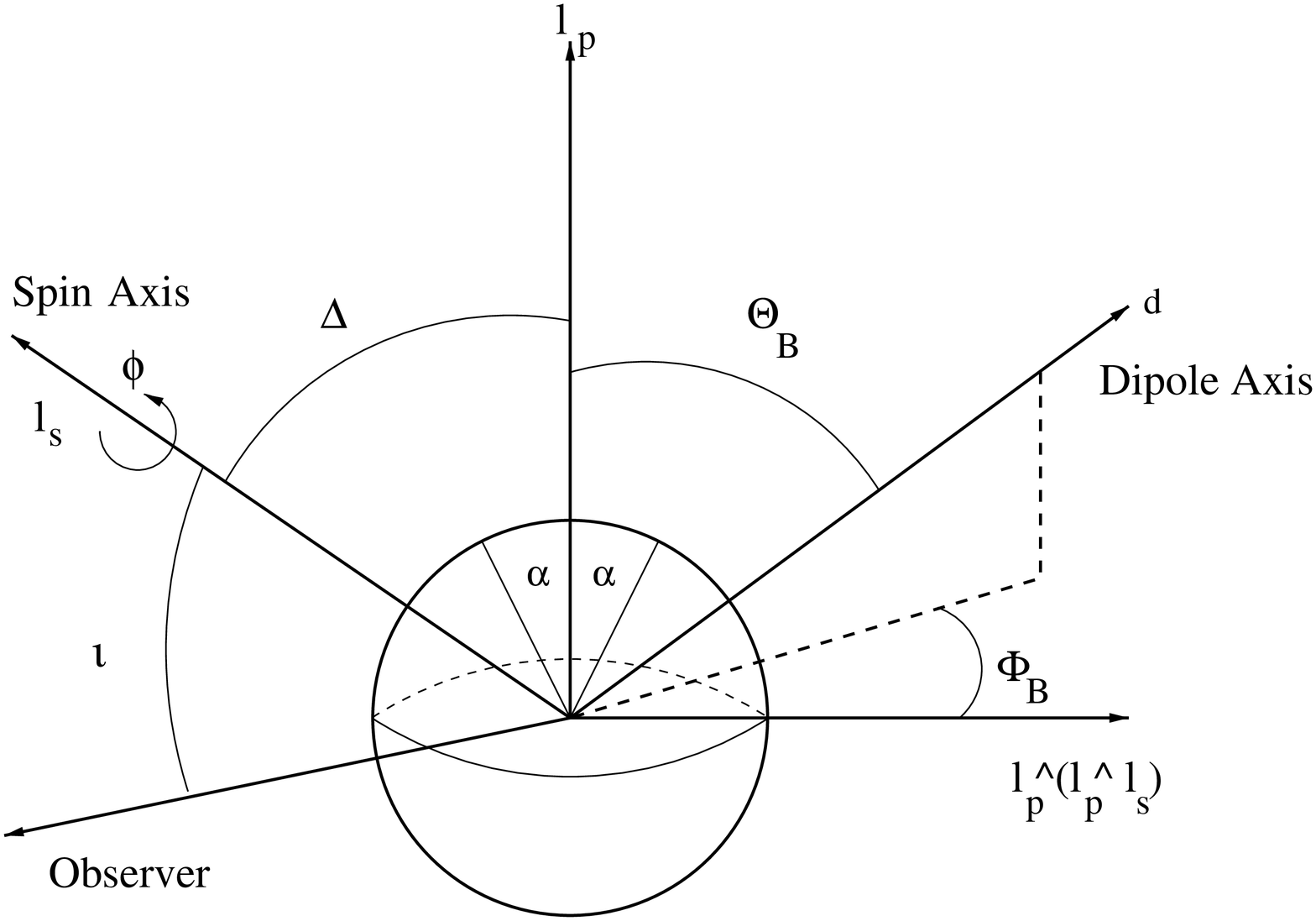,width=4.8in,height=3.5in}}
        \caption{Orientation of the dipole axis $d$ and the spin axis $l_s$ relative to the symmetry axis
	         $l_p$ of the emission region and the line-of-sight.}
        \label{axes}
      \end{figure}
      The idea that the emission regions are pointlike sources located at the diametrically opposite magnetic 
      poles is insufficient and often contradicted by observations. For example the asymmetries seen in 
      polarization light curves of most systems or the wavelength dependence of intensity and polarization in 
      AM-Her itself are hardly reproduced by these models. Therefore, Wickramasinghe \& Ferrario \cite{wf87,wf88} 
      introduced emission regions which extend across and above the stellar surface. Consequently, this 
      new concept allowed for field spread, density and temperature structure, and for displacement from 
      the magnetic poles. 
      
      The field geometry is that of a centered dipole. Using a cartesian coordinate system, the spin axis 
      $l_s$ is inclined with respect to the line-of-sight by an angle $i$ in the $y-z$-plane, while the 
      symmetry axis of the emission region $l_p$ is identified with the positive $z$-axis making an angle 
      $\Delta$ with respect to the spin axis. The emission region itself extends over a spherical cap which 
      subtends an angle $2\alpha$ at the center of the star. In this coordinate system, the dipole axis has a
      polar angle $\Theta_B$ with $l_p$ and an azimuthal angle $\Phi_B$ with respect to a line perpendicular
      to $l_p$ in the same plane as $l_p$ and $l_s$. We can therefore identify this line with the (positive)
      $y$-axis of the cartesian system in the direction $l_p\times(l_p \times l_s)$. The relative orientations
      of the various axes are shown in Fig.\ref{axes}. 
      
      For our purpose, the best agreement with observations was achieved by assuming uniform electron 
      temperature $T_e$ and fixed electron number density $N_e$. The emission region has a constant radial 
      thickness $H$ so that the characteristics of the emission region can be summarized by the optical depth 
      parameter 
      \begin{equation}
        \Lambda_H = 2.01\times 10^8(H/10^8\mbox{\footnotesize{cm}})(N_e/10^{16}\mbox{\footnotesize{cm}}^{-3})
	(3\times 10^7 G/B)
      \end{equation}
      in the radial direction \cite{wf88}. Of course $\Lambda_H$ varies over the emission region due to changes
      in $B$. The magnetic field strength $B$ in the above expression is evaluated at the magnetic pole.
  \section{Cyclotron radiation and birefringence}
      Since the cyclotron emission is highly polarized, the AM-Her systems provide an exellent opportunity
      to test for gravity-induced birefringence. For this purpose we have used a FORTRAN program,
      written by D.T. Wickramasinghe which calculates the Stokes vector of cyclotron emission in the 
      'point source' approximation of Wickramasinghe and Meggitt \cite{wme85} using a fixed electron 
      temperature $T_e$ and a fixed optical depth parameter $\Lambda$. This section will present the 
      underlying mathematical formalism with the additional new feature of a gravity-induced alteration
      of the Stokes vector. Extended emission regions, introduced in the previous section, are easily
      obtained from this concept as a homogeneous composite of pointlike sources.
  
      Consider an electron in helical motion in a uniform magnetic field directed along
      the z-axis in a cartesian coordinate system. The energy emitted per unit 
      solid angle $\Omega$ making an angle $\theta$ with the magnetic field and per frequency 
      interval $d\omega$ is given by
      \begin{equation}\label{cyc1}
        \epsilon_0d\omega\,d\Omega=\frac{e^2\omega^2}{2\pi c}\sum_{n=1}^{\infty}F_n\delta_n(y)
      \end{equation}
      where $\epsilon_0$ is also the coefficient of spontaneous emission, with
      \begin{equation}
        F_n=(\cot\theta-\beta_{\|}\mbox{cosec}\,\theta)^2 J^2_n(n\xi)+\beta_{\perp}^2 J^{'2}_n(n\xi)
      \end{equation}
      where $J_n$ is the Bessel function of order $n$ and $J_n^{'}(n\xi)\equiv dJ_n(n\xi)/d(n\xi)$
      \cite{bekefi}.
      Further, $\beta_{\|}$ and $\beta_{\perp}$ are the parallel and perpendicular components of
      the dimensionless velocity $\beta=v/c$, so that we define $\xi=\beta_{\perp}\sin\theta/(1-
      \beta_{\|}\cos\theta)$. The argument of the $\delta$ function is
      \begin{equation}
        y = n\omega_c/\gamma-\omega(1-\beta_{\|}\cos\theta)
      \end{equation}
      where $\omega_c=eB/mc$ is the electron cyclotron frequency and $\gamma=(1-\beta^2)^{-1/2}$ the
      usual Lorentz factor. From this one can easily see, that the radiation spectrum consists of
      spectral lines occuring at frequencies
      \begin{equation}
        \omega = \frac{n\omega_c}{\gamma(1-\beta_{\|}\cos\theta)} \quad ,
      \end{equation}
      which also gives a relation between $\gamma$ and harmonic number $n$.
      
      In order to describe a uniform plasma with temperature $T$ and electron number density $N$,
      one has to introduce a relativistic Maxwellian distribution in (\ref{cyc1}) and integrate over
      $\beta_{\|}$  
      \begin{equation}\label{cfi}
        \epsilon_I\,d\omega\,d\Omega=\frac{e^2\omega^2}{4\pi c}\frac{N\mu}{K_2(\mu)}\sum_{n=1}^{\infty}
	\int_{-1}^1F_I\exp(-\mu\gamma)(\gamma^4/n\omega_c)d\beta_{\|}\,d\omega\,d\Omega
      \end{equation}
      where the dimensionless electron temperature $\mu=mc^2/kT$ serves as the argument of the modified 
      Bessel function of the second kind, $K_2(\mu)$.
      
      To obtain the emission coefficients $\epsilon_Q$ and $\epsilon_V$ for Stokes $Q$ and Stokes $V$,
      one defines a coordinate system where the z-axis serves as the line-of-sight and the uniform
      magnetic field lies in the x-z plane, making an angle $\theta$ with the z-axis. Then, the complex
      Jones vector for the electric field of the radiation of a single electron is proportional to 
      (\cite{bekefi})
      \begin{equation}
        \left(
	  \begin{array}{c}
	     (\cot\theta-\beta_{\|}\mbox{cosec}\,\theta)J_n(n\xi)\\
	     -i\beta_{\perp} J^{'}_n(n\xi)\\
	     0
	  \end{array}
	\right)\quad .
      \end{equation}
       The emission coefficients $\epsilon_Q$ and $\epsilon_V$ for the Stokes parameters $Q$ and $V$ are 
       obtained by replacing $F_I$ in (\ref{cfi}) with
      \begin{equation}
        F_Q = (\cot\theta-\beta_{\|}\mbox{cosec}\,\theta)^2 J^2_n(n\xi)-\beta_{\perp}^2 J^{'2}_n(n\xi)
      \end{equation}
      and
      \begin{equation}
        F_V = -2(\cot\theta-\beta_{\|}\mbox{cosec}\,\theta)J_n(n\xi)\beta_{\perp} J^{'}_n(n\xi)
      \end{equation}      
      respectively, while $\epsilon_U=0$. Since $0\leq \xi \leq 1$ the Bessel function $J_n$ and its
      derivative can be replaced with the Wild-Hill approximations \cite{whill} which are more easy to
      implement in a computer program
      \begin{eqnarray}
        J_n(n\xi) &\approx& \frac{\xi^n\exp(nw)}{\sqrt{2\pi n}(1+w)^n}[w^3+0.5033/n]^{-1/6} \\
	J^{'}_n(n\xi)&\approx& \frac{\xi^{n-1}\exp(nw)}{\sqrt{2\pi n}(1+w)^n}[w^3+1.193/n]^{1/6}
	[1-1/5n^{2/3}]
      \end{eqnarray}
      with $w=\sqrt{1-\xi^2}$. The cyclotron absorption coefficients are given by Kirchhoff's law,
      since the electrons are in thermal equilibrium, i.e.
      \begin{equation}
        \epsilon_I = \kappa_{\mbox{\tiny{cyc}}} B_{\omega}, \quad \epsilon_Q = q_{\mbox{\tiny{cyc}}}
	 B_{\omega}, \quad \epsilon_V = v_{\mbox{\tiny{cyc}}} B_{\omega},
      \end{equation}
      with the Rayleigh-Jeans law
      \begin{equation}
        B_{\omega} = \omega^2 k T/4\pi^3 c^2 \quad .
      \end{equation}
      In contrast to the first model, given by Meggitt \& Wickramasinghe in 1982 \cite{mew82} the new
      scheme also includes free-free opacity as a pure absorption process. The corresponding opacities
      are
      \begin{eqnarray}
        \kappa_{\mbox{\tiny{ff}}} &=& \frac{\omega_p^2(2\omega^4+2\omega^2\omega_c^2-
	                              3\omega^2\omega_c^2\sin^2\theta+\omega_c^4\sin^2\theta)}
	                              {2c\omega^2(\omega^2-\omega_c^2)^2}\,\nu_c \\
        q_{\mbox{\tiny{ff}}}      &=& \frac{\omega_p^2\omega_c^2\sin^2\theta(\omega_c^2-3\omega^2)}
	                              {2c\omega^2(\omega^2-\omega_c^2)^2}\,\nu_c \\
        u_{\mbox{\tiny{ff}}}      &=& 0 \\
	v_{\mbox{\tiny{ff}}}      &=& \frac{2\omega_p^2\omega\omega_c\cos\theta}
	                              {c\omega^2(\omega^2-\omega_c^2)^2}\,\nu_c 
      \end{eqnarray}
      with the plasma frequency $\omega_p=(4\pi N_e e^2/m)^{1/2}$ and the approximately collision 
      frequency $\nu_c=3.63 N_e T^{-3/2}\ln(2.95\times 10^{11}T/\omega)$. It is assumed that the 
      opacity can be treated as a sum of cyclotron and free-free components, hence
      \begin{eqnarray}
        \kappa = \kappa_{\mbox{\tiny{cyc}}}+\kappa_{\mbox{\tiny{ff}}}, \quad
	     q = q_{\mbox{\tiny{cyc}}} + q_{\mbox{\tiny{ff}}}, \quad
	     v = v_{\mbox{\tiny{cyc}}} + v_{\mbox{\tiny{ff}}} \quad .
      \end{eqnarray}
      The transfer equation now becomes
      \begin{equation}\label{tfr}
        \frac{d}{ds}\left(\begin{array}{c}I\\Q\\U\\V\end{array}\right)=
	\left(\begin{array}{c}\epsilon_I\\\epsilon_Q\\0\\\epsilon_V\end{array}\right)+
	\left(\begin{array}{cccc}-\kappa&-q&0&-v\\-q&-\kappa&-f&0\\0&f&-\kappa&-h\\
	-v&0&h&-\kappa\end{array}\right)
	\left(\begin{array}{c}I\\Q\\U\\V\end{array}\right)
      \end{equation}
      with the Faraday mixing coefficients
      \begin{eqnarray}
        f &=& (\omega_p^2/c\omega_c)\cos\theta/(\omega^2/\omega^2_c-1) \\
        h &=& (\omega_p^2/c\omega_c)\sin^2\theta/2(\omega^3/\omega^3_c-\omega/\omega_c) \quad .
      \end{eqnarray}
      For uniform conditions these coefficients are constant along a ray. In order to solve the 
      transfer equation (\ref{tfr}) set
      \begin{eqnarray}
        m &=& 0.5(f^2+h^2), \, n = 0.5(q^2+v^2), \, p = qf-vh, \, r=qh+vf, \\
	R &=& [(m+n)^2-p^2]^{1/2}, \, a_1 = (m+n+R)/p, \, a_3 = 1/a_1, \\
	\lambda &=& (n-m+R)^{1/2}, \, \mu = (m-n+R)^{1/2}, \\
	b_1 &=& (f-qa_1)/\lambda, \, b_3 = (f-qa_3)/\mu \\
	c_1 &=& -(h+va_1)/\lambda, \, c_3 = -(h+va_3)/\mu \quad .
      \end{eqnarray}
      Setting all intensities equal to zero at $s=0$, the Stokes parameters of the radiation after
      passing through a distance $s$ are (see \cite{mew82})
      \begin{eqnarray}
        I/B_{\omega} &=& 1-(p/2R)(a_1\cosh \lambda s - a_3\cos\mu s)\exp(-\kappa s) , \label{icyc}\\
	Q/B_{\omega} &=& -(p/2R)(b_1\sinh\lambda s - b_3\sin\mu s )\exp(-\kappa s) , \\
	U/B_{\omega} &=& (p/2R)(\cosh \lambda s - \cos\mu s)\exp(-\kappa s) , \label{ucyc}\\
	V/B_{\omega} &=& -(p/2R)(c_1\sinh\lambda s - c_3\sin\mu s)\exp(-\kappa s) \label{vcyc}\quad .
      \end{eqnarray}
      The question of how gravitational birefringence could be implemented in this scheme depends
      heavily on the shock height above the stellar surface where the polarized radiation is emitted. 
      Since the phase shift $\Delta\Phi$ in (\ref{mag-form}) is proportional to $1/R$ where $R$ denotes 
      the radial distance from the stellar center, an emission region which is extended of approximately
      one stellar radius above the surface would require an integration process along the shock height
      where a gravitationally modified transfer equation has to be solved for each plane parallel slab.
      An upper limit on the shock height is given in the 'single-fluid' approximation when bremsstrahlung
      is the dominating cooling mechanism. In this case the height $h_{\mbox{\tiny{br}}}$ is simply 
      proportional to the product between the free-fall velocity $V_{\mbox{\tiny{ff}}}$ (the index 'ff', 
      of course, must not be confused with the index for free-free absorption) and the specific cooling 
      time $t_{\mbox{\tiny{ff}}}$ for electrons due to bremsstrahlung. The explicit expression for 
      $h_{\mbox{\tiny{br}}}$, given by Wickramasinghe \& Ferrario \cite{wf00} is
      \begin{equation}
        h_{\mbox{\tiny{br}}}=9.6\times 10^7\left(\frac{10^{16}\mbox{\footnotesize{cm}}^{-3}}{N_e}\right)
	\left(\frac{M}{M_{\odot}}\right)\left(\frac{10^9\mbox{\footnotesize{cm}}}{R}\right)\,
	\mbox{\footnotesize{cm}} \quad .
      \end{equation} 
      Assuming a white dwarf mass of $M_{\odot}$ with a corresponding radius of $0.014\, R_{\odot}$
      this yields a height of $\sim 10^8$cm with $N_e=10^{15}$. Since we focus on a 'two-fluid' model 
      when the electrons are cooled by cyclotron radiation, this means that we have for the cooling time
      $t_{\mbox{\tiny{cyc}}} \ll t_{\mbox{\tiny{ff}}}$ and, therefore, $h_{\mbox{\tiny{cyc}}} \ll
      h_{\mbox{\tiny{br}}}$. For this reason it is justified to view the polarization as emitted from 
      near the surface because $h_{\mbox{\tiny{cyc}}} \ll 10^2$km is certainly a sufficient upper limit.
      We can therefore assume that birefringence acts directly on (\ref{ucyc}) and (\ref{vcyc})
      according to
      \begin{equation}
        \left(\begin{array}{c}U_{\mbox{\tiny{grav}}}\\V_{\mbox{\tiny{grav}}}\end{array}\right)=
	\left(\begin{array}{cc}\cos\Delta\Phi & 0 \\ 0 & \cos\Delta\Phi\end{array}\right)
	\left(\begin{array}{c}U_{\mbox{\tiny{cyc}}}\\V_{\mbox{\tiny{cyc}}}\end{array}\right) \quad .
      \end{equation}      
      Our objective is to include birefringence given by these equations in the calculations of synthetic
      polarization light curves from AM-Her type systems. In this way we can set sharp upper limits
      on this effect with a new method, independent from techniques presented in chapter two.    
  \section{Polarimetry of VV Puppis}  
    VV Puppis serves as a key system for our purpose of either detecting weak gravitational birefringence or
    setting strong upper limits on it by means of gravitationally modified polarization curves. We therefore 
    present here some of the most important results that have been achieved for this system within the 
    last decades and which provide the basis for the subsequent analysis.
     
    VV Puppis was discovered in 1931 by van Gent \cite{gent} as a faint $(V = 14.5-18)$ variable with
    a period of 100 min. Sinusoidal variations in radial velocity of the H and He II line, correlated
    with the 100 min photometric variation led Herbig in 1960 to the conclusion that the observations 
    could best be explained by means of a binary system where the emission lines originate on the brighter 
    (primary) component \cite{herb}. Herbig also concluded that changes in the shape of the light curve
    could have its origin in variations of the emission area on the visible stellar surface which can be
    realized by self eclipses of the relevant area due to rotation. This picture was refined in 1977 by 
    Tapia's discovery of strong linear and circular polarization with a maximum of $\sim 16$\%\cite{tap77a} 
    in each polarization state of the optical light curve, confirming the suggestion by Bond and Wagner 
    (IAUC 3049) that this object is similar to the ''original'' AM-Her system. Two years later, 
    Visvanathan and Wickramasinghe identified a series of absorption features in the spectrum with 
    the 6th, 7th and 8th harmonics of cyclotron absorption in a nearly uniform magnetic field of 
    $\sim 3\cdot 10^7$G. This discovery renewed the general interest in astrophysical cyclotron emission 
    models, since it was the first detection of resolvable cyclotron harmonics in the optical spectrum
    of a stellar object. However, these results had an uncertainty of one harmonic in the identification 
    since the observed features did not have well defined cores. As a consequence the corresponding 
    uncertainty in $B$ was $\sim 20$\%. Concerning the visibility of cyclotron harmonics, numerical 
    computations, based on the calculations presented in section 4.3 revealed that very special conditions 
    are required to produce consistency with the observed spectrum \cite{wf87}. 
    \begin{figure}[t]
      \centerline{\psfig{figure=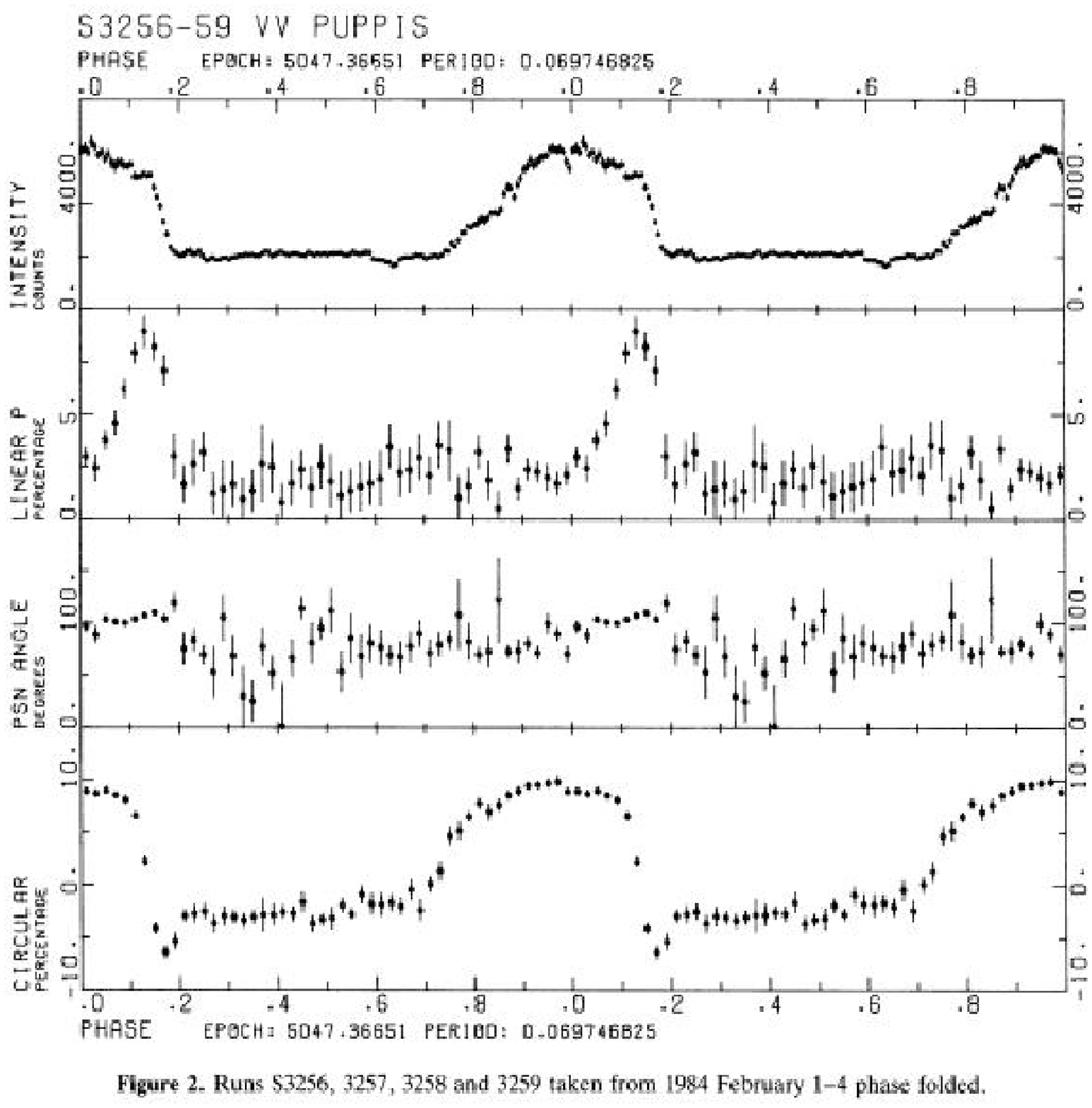,width=5.5in,height=5.in}}
      \caption{Intensity and polarization light curves of VV Puppis. From: Copper \& Warner (1986) \cite{cw86}.}
      \label{pol-curv}
    \end{figure}            
    For the shock front model related to VV Puppis one can therefore conclude a viewing angle 
    $i \sim 70^{\circ} - 90^{\circ}$ with respect to the magnetic field, low optical depths 
    $\Lambda \sim 10^5$ and high temperatures of $T \sim 10$ keV in a magnetic field of $\sim 3\cdot 10^7$G. 
    The fact that VV Puppis is the only stellar object to show resolvable cyclotron harmonics in its 
    optical spectrum is difficult to explain, although it might be possible that magnetic field broadening 
    often renders the features undetectable \cite{wf87}.                          
    \begin{figure}[t]
      \centerline{\psfig{figure=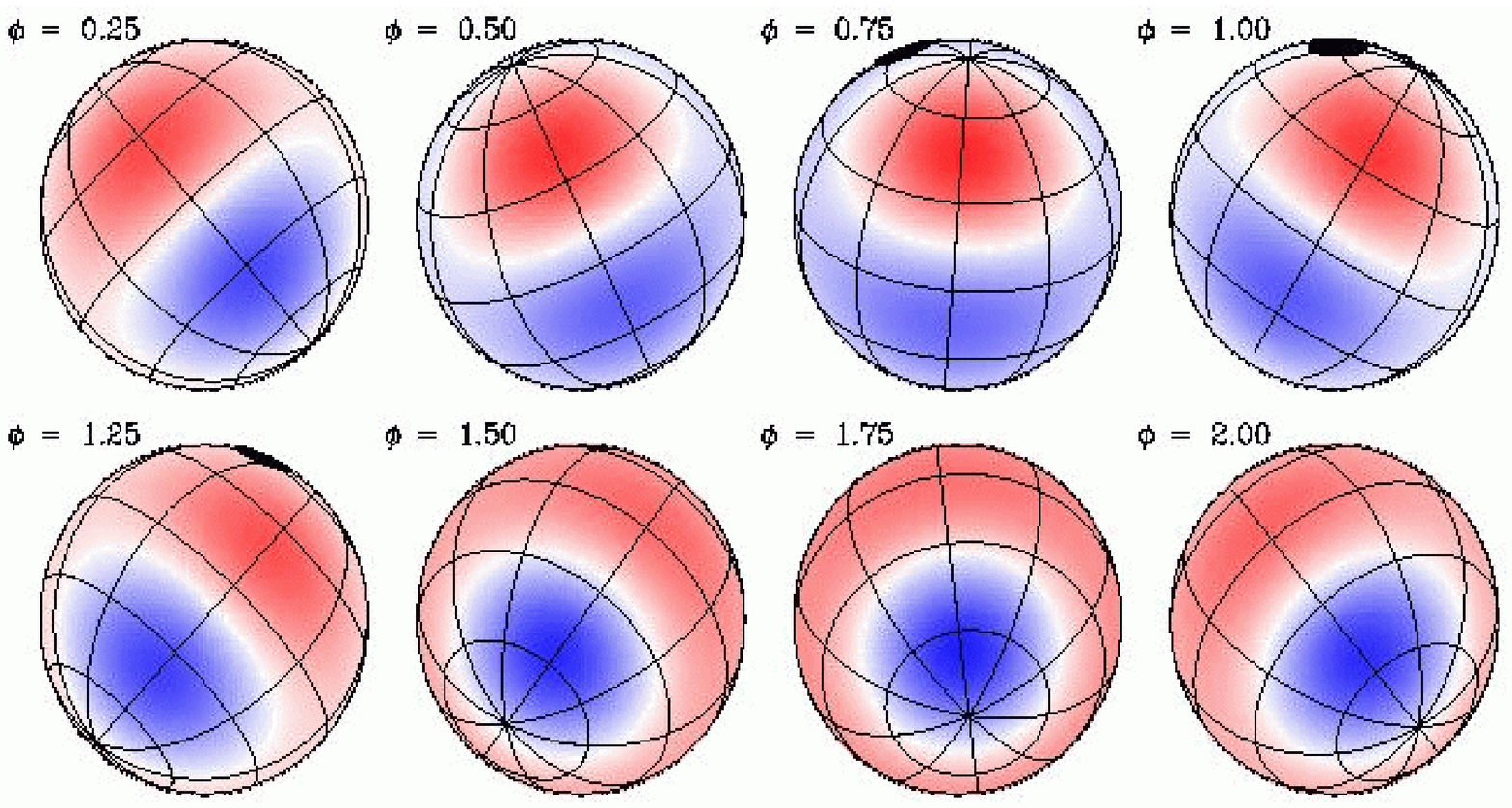,width=7in,height=3.3in}}
      \centerline{\psfig{figure=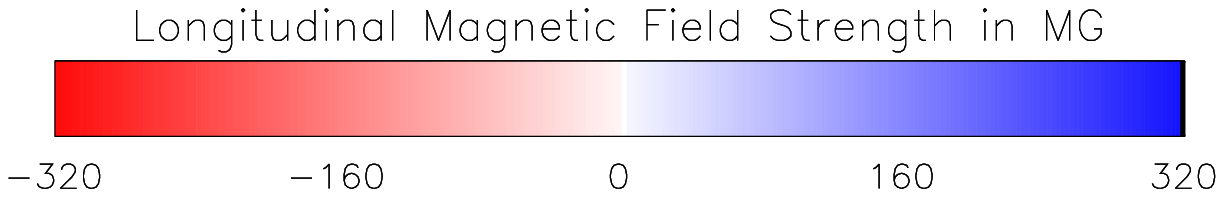,width=4in,height=0.5in,angle=90}}
      \caption{Visualization of the longitudinal magnetic field components of VV-Puppis during successive 
               rotational phases. The black spot which is visible from $\phi\approx 0.75$ to $\phi\approx 1.25$
	       markes the position of an emission region with size $\alpha = 10^{\circ}$ and displacements
	       $\Phi_B=90^{\circ}$, $\Theta_B=30^{\circ}$ from the dipolar axis.}
      \label{vphases}
    \end{figure}    
   
    Fig.\ref{pol-curv} shows phase dependend, wavelength averaged polarization curves of VV Puppis from 
    Cropper \& Warner (1986) \cite{cw86}. As one can see, the light curves are strongly modulated at the 
    orbital period and the accreting pole is visible only for $\sim 40$\% of the period. A common feature is 
    the slow rise after eclipse egress and a more rapid decline before eclipse ingress. Although linear
    polarization is present during the entire bright phase, with a second, lower peak at phase 0.7, the
    strong linear polarization peak of $\sim 9$\% is expected to be caused by a magnetic field
    configuration that can be seen only briefly at a viewing angle of $\sim 90^{\circ}$ with respect to 
    the line-of-sight. In this case linear polarization due to cyclotron radiation is strongest and would 
    therefore explain the observed feature. The circular polarization reaches a maximum value of $\sim 10$\% 
    during the bright phase after a more gradual rise. The steeper descent of the circular polarization is 
    followed by a sign reversal and a short dip, lasting 0.05 of a cycle. One of the most important results 
    is that VV Puppis shows steady negative circular polarization throughout the entire eclipse phase, which 
    could be explained by accretion onto a second region of the white dwarf \cite{ls79}, that is visible at 
    all rotation phases \cite{wfb89}. However, the negative polarization dip at eclipse ingress is probably 
    related to the shape of the primary accretion region. This conjection is supported by synthetic light 
    curves of only one accretion region as we will later show.
    
    On the basis of the light curves in Fig.\ref{pol-curv}, Cropper \& Warner reported for the system 
    geometry an inclination of $i = 77^{\circ}\pm 7^{\circ}$ and a magnetic colatitude of 
    $\Delta = 155^{\circ}\pm 6^{\circ}$. In 1987 Wickramasinghe \& Ferrario used $i=75^{\circ}$ and 
    $\delta=150^{\circ}$ for the calculation of phase dependent light curves of a displaced accreting pole.
    Fig.\ref{vphases} shows a visualization of the longitudinal magnetic field component during a rotation
    cycle of VV Puppis. The emission region with size $\alpha = 10^{\circ}$ has an azimuthal and polar 
    angle of $\Phi_B=90^{\circ}$ and $\Theta_B=30^{\circ}$, respectively, marked by the black spot. 
    One can see that in this configuration the emission region merely grazes the limb during the 
    bright phase, which is extraordinarily useful for our purpose, since gravitational birefringence 
    is most pronounced for sources located at the stellar limb. Further, the asymmetry in the 
    linear pulse and the sign reversal in circular polarization, prior to eclipse ingress can now 
    be explained by means of the system geometry. Calculating the angles $\theta$ between the field 
    directions in the spot and the line-of-sight one finds, that the angles are mostly greater than 
    $90^{\circ}$ at eclipse ingress and mostly less than $90^{\circ}$ at eclipse egress. The observed 
    reversal in the sign of circular polarization occurs just shortly before eclipse ingress when 
    $\theta > 90^{\circ}$ while the stronger linear pulse as well as the maximum intensity occurs at 
    $\theta \approx 90^{\circ}$ as expected from the beaming properties of cyclotron radiation. 
    
   \subsection{Gravitationally modified lightcurves}
     The programm which we have developed for calculations of gravitationally modified polarization curves
     is based on a FORTRAN routine written by D.T. Wickramasinghe which calculates cyclotron emission in 
     the ''point source'' approximation of Wickramasinghe \& Meggitt \cite{wme85} for fixed optical depth 
     parameter $\Lambda$ and fixed electron temperature. Since the point source approximation is insufficient
     to account for most of the observed polarization curve characteristics as explained in section 4.2.2,
     our code describes an extended emission region as a composition of $\sim 3\times 10^2$ point sources,
     depending on the size of the emission region. This emission region is placed on a rotating sphere
     according to the coordinate system described in sect. 4.2.2, so that the polarization values (circular
     and linear) taken at each rotational phase yields the final polarization curve. 
     
\begin{figure}[h]
       \centerline{\psfig{figure=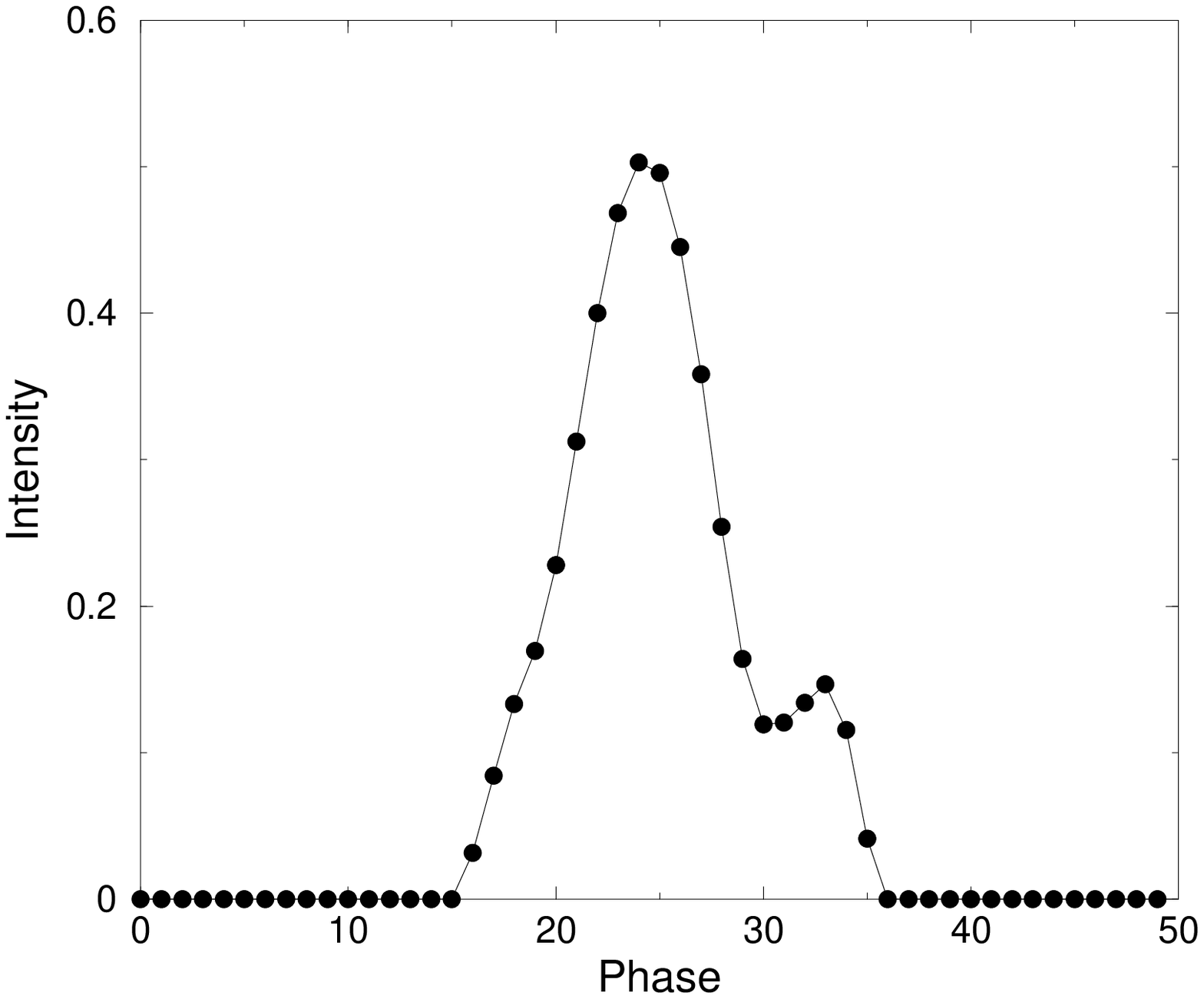,width=3.08in,height=2.64in}
                   \psfig{figure=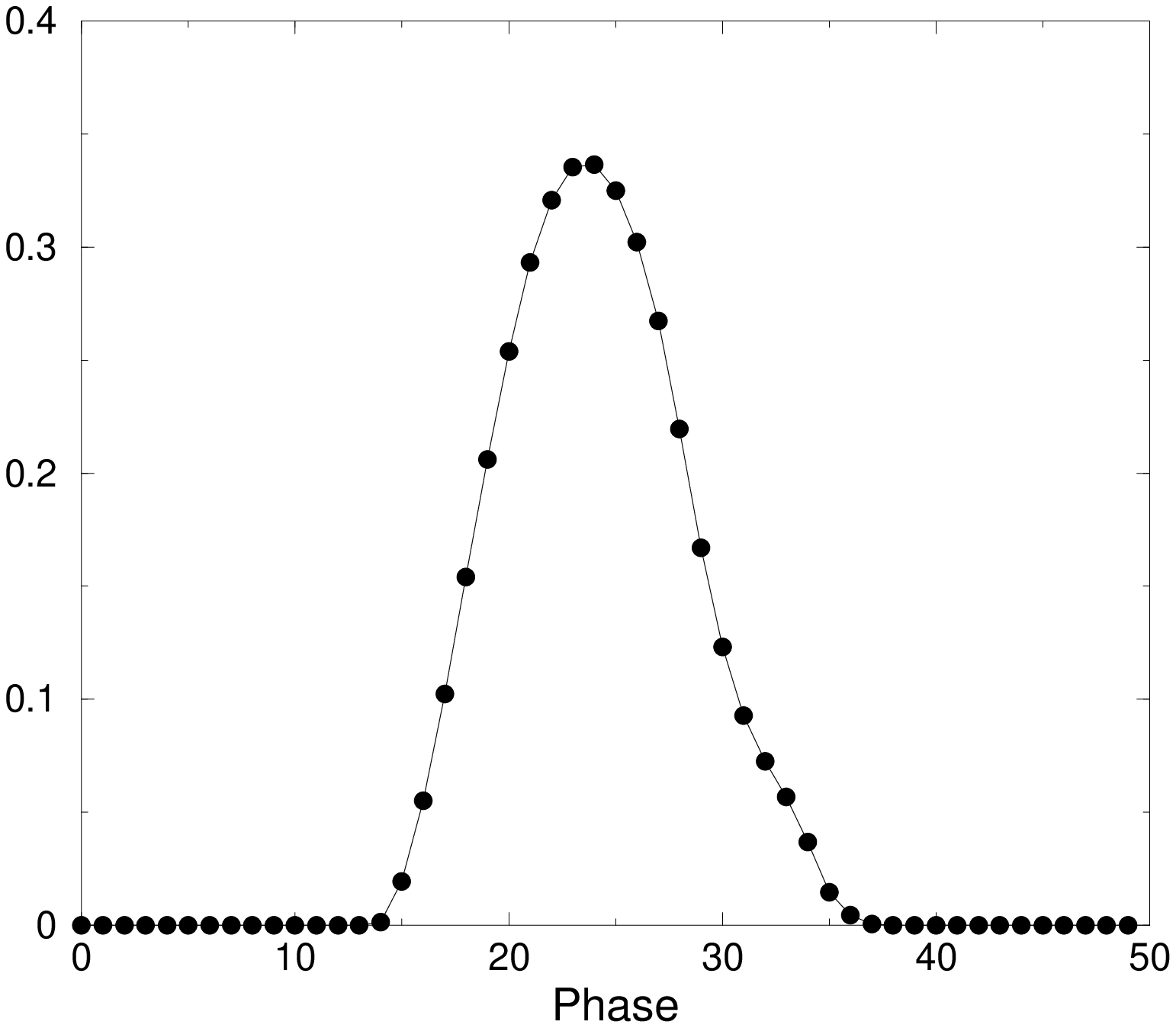,width=3.08in,height=2.64in}}
       \centerline{\psfig{figure=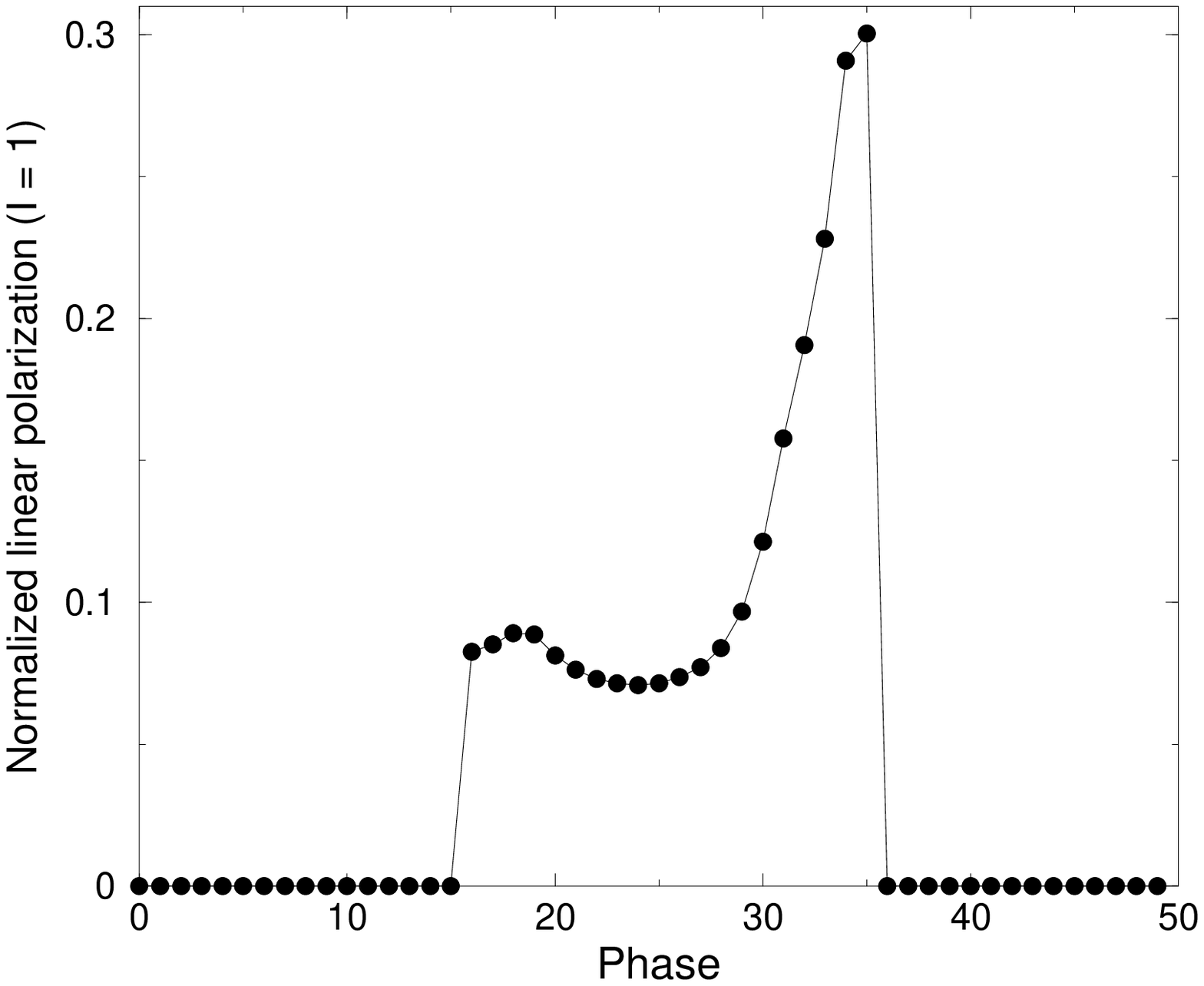,width=3.08in,height=2.64in}
                   \psfig{figure=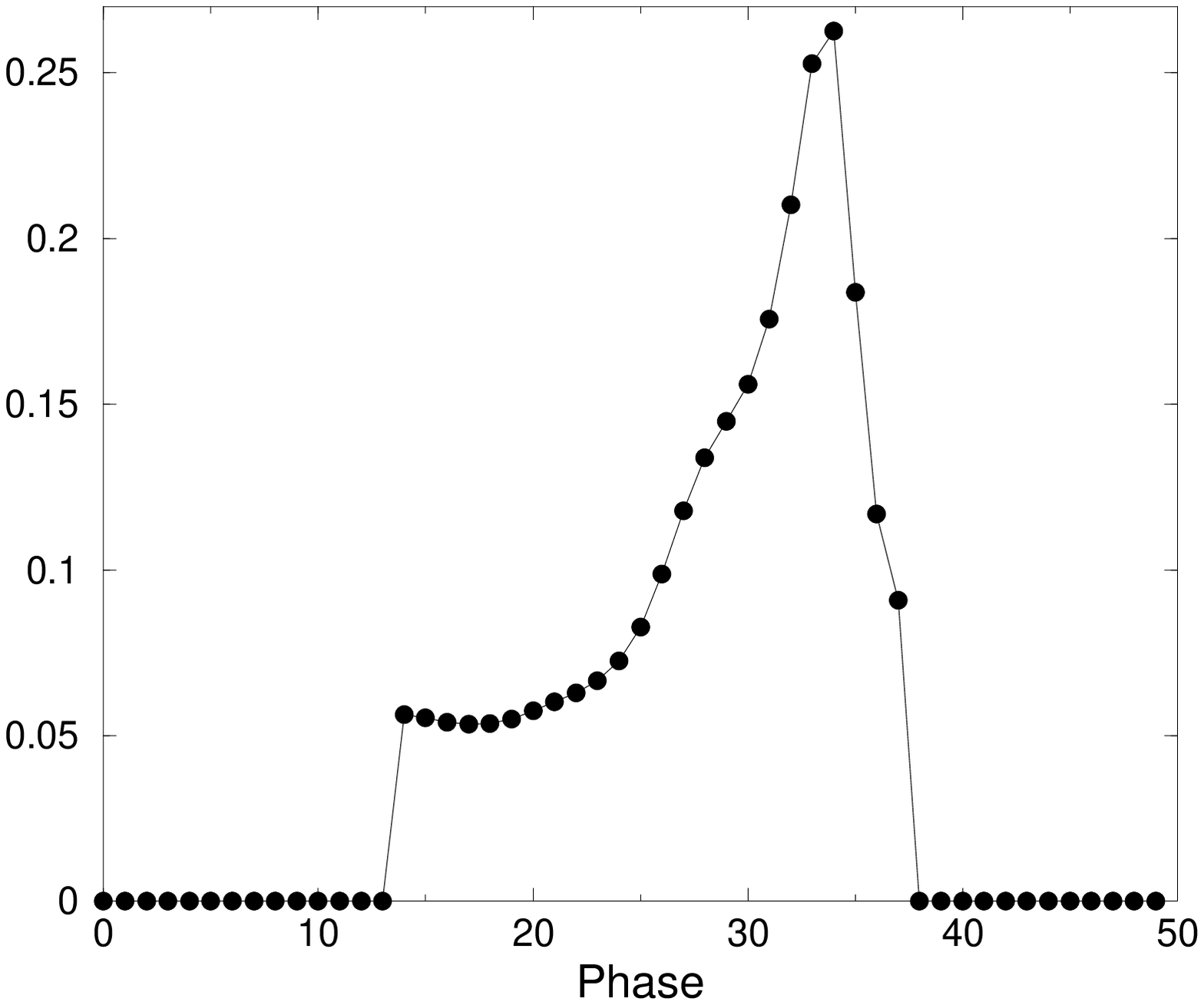,width=3.08in,height=2.64in}}
       \centerline{\psfig{figure=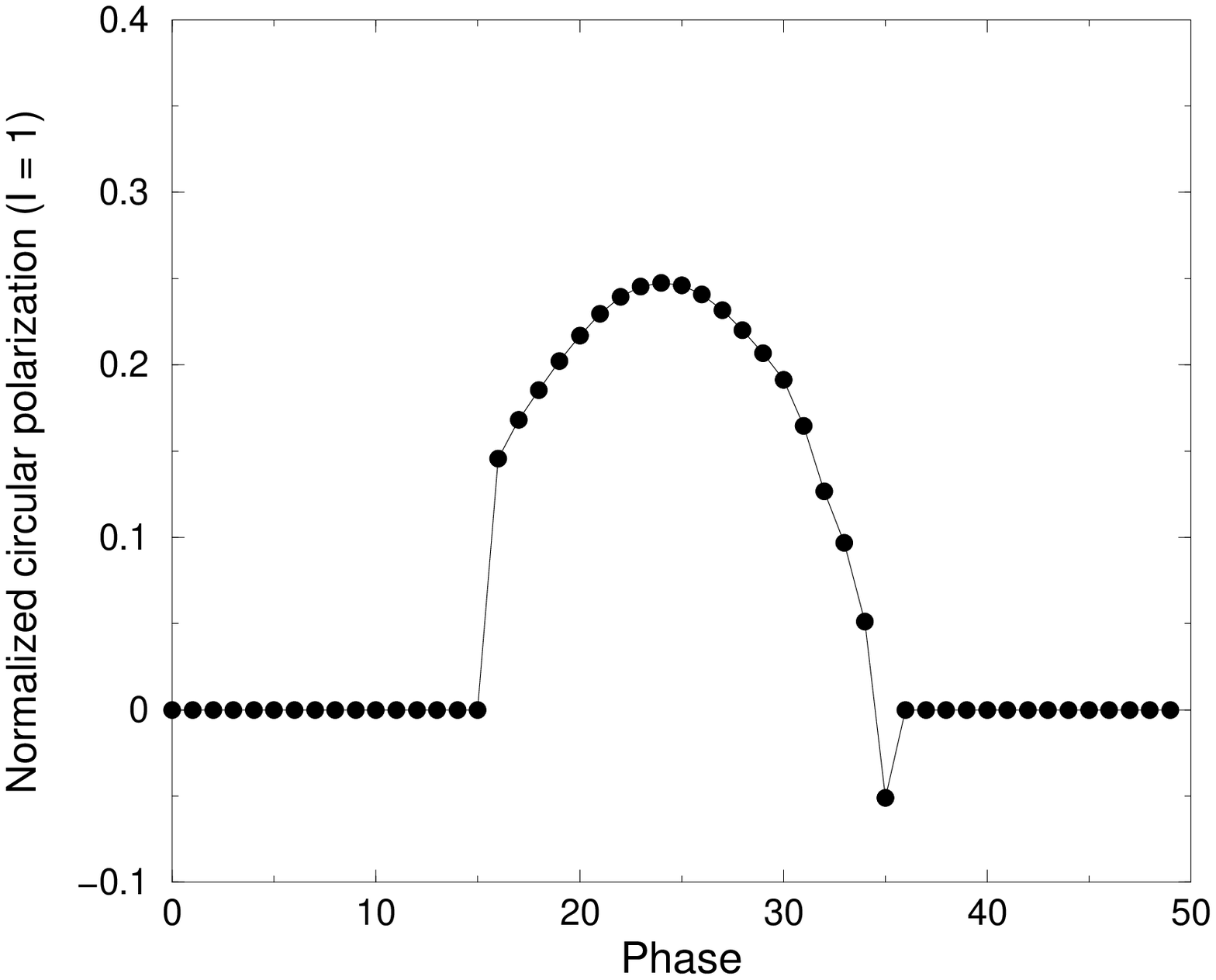,width=3.08in,height=2.64in}
                   \psfig{figure=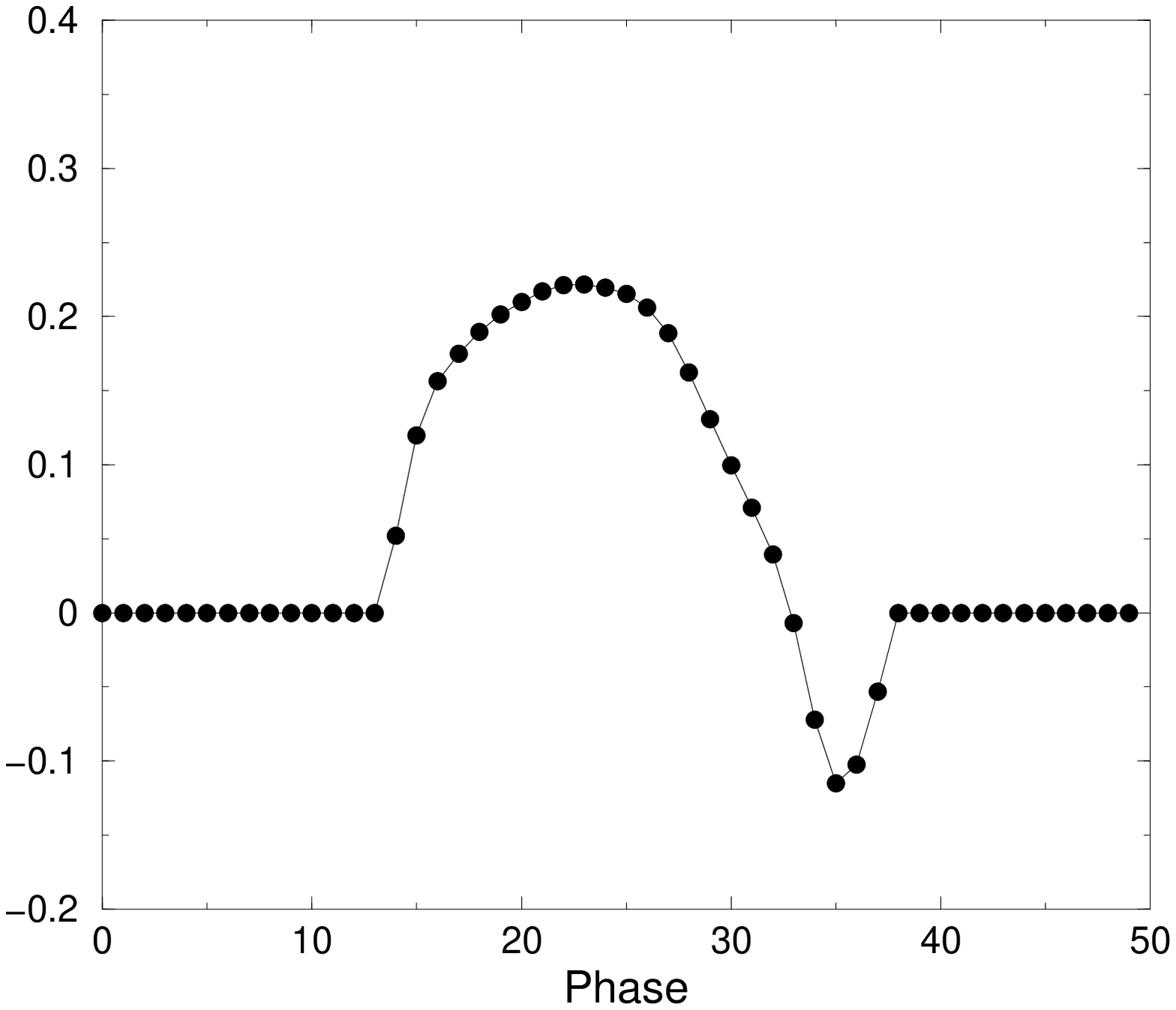,width=3.08in,height=2.64in}}
       \caption{Phase dependent circular polarization curve of VV Puppis for an emission region centered at
                latitude $\Delta=150^{\circ}$. The orbital inclination is $i=75^{\circ}$ and 
		$\omega/\omega_c = 6.2$. The curves in the left column correspond to $\alpha=5^{\circ}$ and
		$\theta_B=10^{\circ},\,\Phi_B=90^{\circ}$, while the right column corresponds to 
		$\alpha=10^{\circ}$ and $\theta_B=30^{\circ},\,\Phi_B=90^{\circ}$.}
       \label{vvp-circ1}
     \end{figure}

     We first present light curves without gravitational modification in order to have a comparison to the 
     observed curves of Cropper \& Warner \cite{cw86}.     
     Fig.\ref{vvp-circ1} shows the phase dependent polarization and light curves for two different models. 
     While both models have the same inclination $i=75^{\circ}$ and magnetic colatitude $\Delta=150^{\circ}$ 
     for the position of the spot, the left column shows the curves for polar cap size $\alpha=5^{\circ}$ 
     and a dipole axis orientation with $\theta_B=10^{\circ}$ and $\Phi_B=90^{\circ}$. The curves in the 
     right column have $\alpha=10^{\circ}$, $\theta_B=30^{\circ}$ and $\Phi_B=90^{\circ}$. We take note that
     these curves are in good agreement with the curves presented by Wickramasinghe \& Ferrario in 1988 
     \cite{wf88} (hereafter WF88) for extended emission regions using similar parameters. At this point it 
     is important to note, that these curves (including those in WF88 \cite{dayal}) require an additional constant 
     background of unpolarised light in order to reproduce observations at least approximately. This constant, 
     unpolarized continuum probably comes from a component of the binary system which is always visible, like 
     the accretion stream, white dwarf photosphere or something similar - up to now the genuine origin is 
     unknown as well as how to correctly estimate it. Currently the only restriction is placed by the observed 
     polarization levels and, so, the continuum has to be introduced by hand to get the observed levels right.
      
     Since our code only allows for emission from one pole, the circular polarization curve does not show 
     a constant negative polarization throughout the hole period which is seen in the observations.       
     Basic characteristics like the asymmetry in the linear pulse and the negative dip in circular 
     polarization prior to eclipse ingress stand comparison to what is observed by Cropper \& Warner. 
     Nevertheless we note that the observations shows a more gradual rise in intensity and polarization 
     whereas the rapid declines are in agreement with our models.    
       
\begin{figure}[t]
       \centerline{\psfig{figure=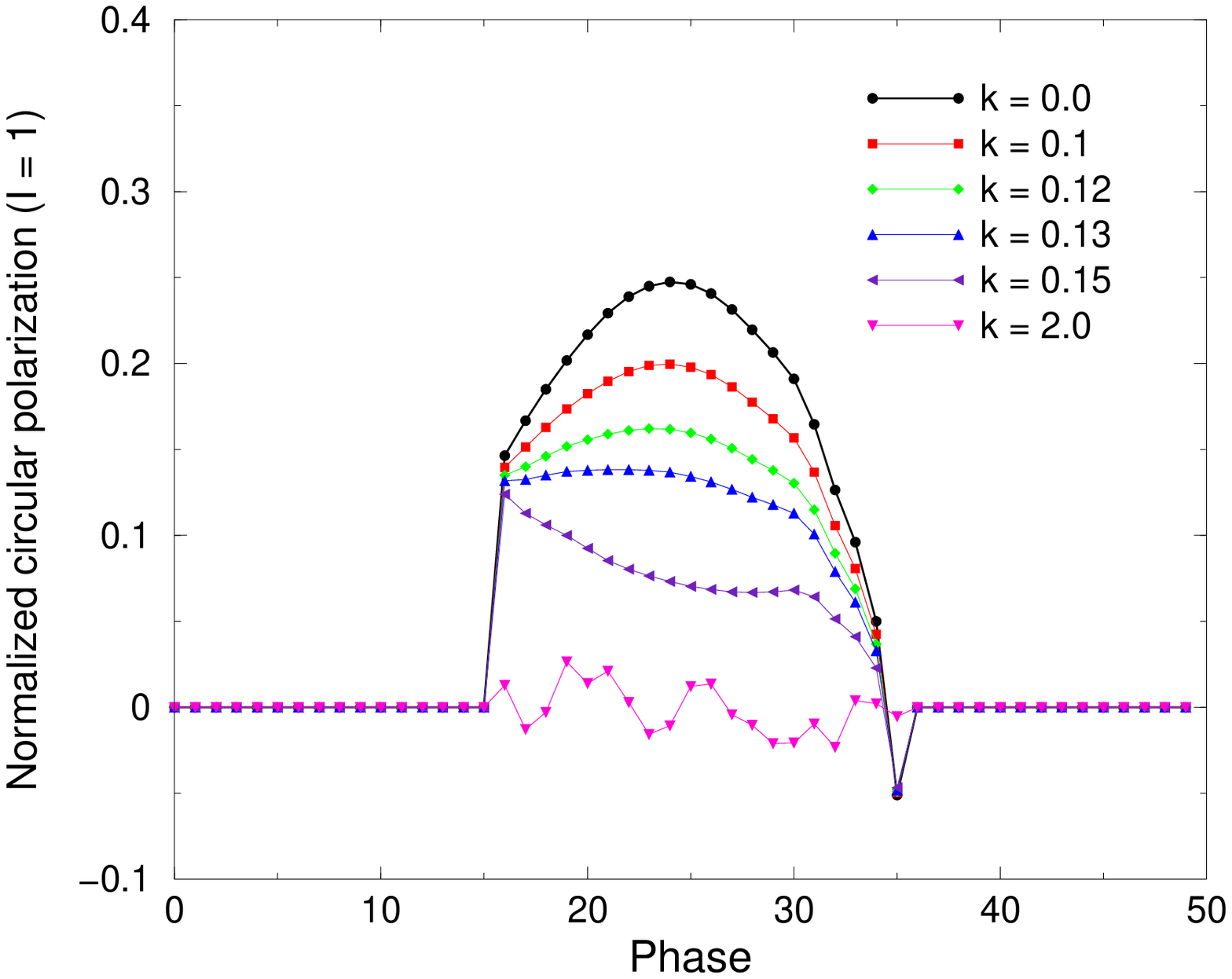,width=4.in,height=3.5in}}
       \centerline{\psfig{figure=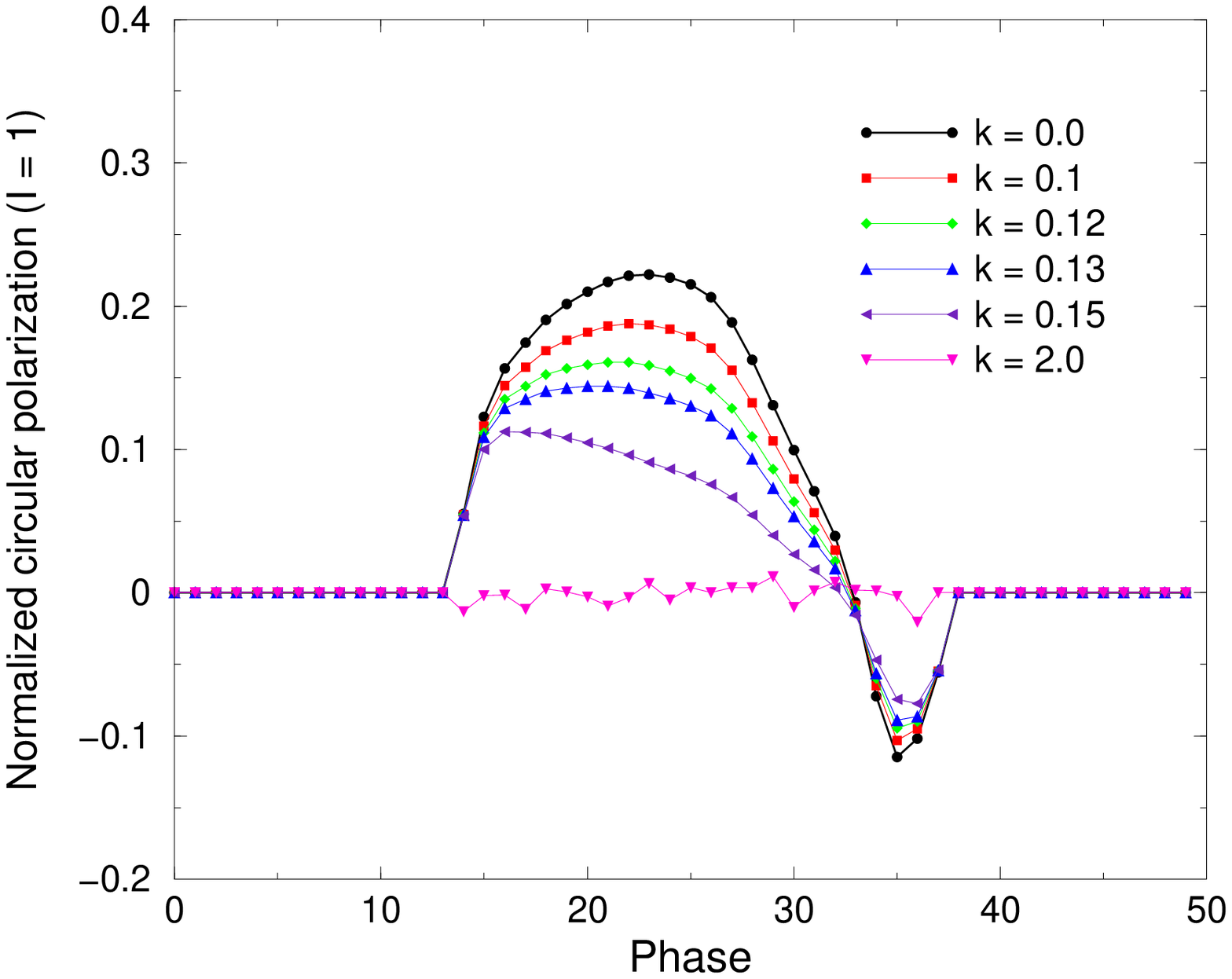,width=4.in,height=3.5in}}
       \caption{Gravitationally modified, phase dependent cicular polarization curves. The physical parameters 
       are the same as in Fig.\ref{vvp-circ1}. Upper curves: $\alpha=5^{\circ}$, $\theta_B=10^{\circ},\,
       \Phi_B=90^{\circ}$. Lower curves: $\alpha=10^{\circ}$, $\theta_B=30^{\circ},\,\Phi_B=90^{\circ}$.}
       \label{gcurve-circ}
     \end{figure}

     These ''pure'' curves (without any birefringence influence) could possibly better fit\-ted to observations 
     by a different geometrical shape of the emission region. While the present curves are based on a cylindrical 
     symmetric source, a more elongated, arc-like structure could perhaps account for the observed gradual rise. 
     Observational hints for arc-like emission regions has been presented by Beuermann et al. in 1987 \cite{beu87} 
     and also by Cropper \cite{crp87}. We will come to this point again after the discussion of the gravitationally 
     modified polarization curves. In accordance with WF88 we emphasize that the characteristics of the 
     polarization curves depend sensitively on $\Phi_B$ for a given $\alpha$ and $\theta_B$. For instance, for 
     $\Phi_B=0^{\circ}$ we obtain no polarization reversals while for $\Phi_B=180^{\circ}$ we observed two 
     symmetrical reversals at eclipse ingress and eclipse egress.      
     \clearpage       
     \clearpage  
     One of the main problems that these curves and also the curves presented by WF88 have is the 
     sinusoidal form of the circular polarization and intensity, in contrast to what is observed.
     For this reason it is certainly interesting to see how gravity-induced birefringence modifies
     the circular polarization curve. Fig.\ref{gcurve-circ} on the next page shows the phase dependent 
     circular polarization using the same system parameters as in Fig.\ref{vvp-circ1} and 
     $M_{\star}=0.4\,M_{\odot}$, $R_{\star}=0.014\,R_{\odot}$ but this time additionally with 
     increasing values of the metric-affine coupling constant $k$. As expected, for $k^2=0$ the curves 
     are identical with the unmodified polarization regime while for a sufficiently big $k^2$, the light
     is almost completely depolarized. In between these two ''boundary'' states, one can adjust $k^2$ to a 
     value $(0.13\,\mbox{km})^2 \leq k^2 \leq (0.14\,\mbox{km})^2$ so that the circular polarization curve 
     is nearly flatten as seen in observations. At the same time, the negative dip which arises
     prior to eclipse ingress is only marginally influenced by birefringence, although it but also decreases 
     somewhat with increasing $k^2$. 
     
     In order to see if this range for $k^2$ is in agreement with upper limits, derived with the methods 
     of the chapter 3, we have made two consistency checks. By using the oblique dipolar rotator 
     technique we got an upper limit of $k^2_{VV\,Pup} \leq (0.35\, \mbox{km})$ for VV Puppis without limb darkening 
     and $k^2_{VV\,Pup} \leq (0.42\, \mbox{km})$ with maximum limb darkening and an observed polarization level of 10\%. 
     These limits are similar to those of 40 Eridani B from the previous chapter, i.e. $k^2_{40\, Eri\,B} \leq (0.35\, 
     \mbox{km})$ and $k^2_{40\, Eri\,B} \leq (0.44\, \mbox{km})$, respectively with $M_{40\, Eri\,B}=0.5 M_{\odot}$ and 
     $R_{40\, Eri\,B}=0.0136\,R_{\odot}$ comparable to $M_{VV\,Pup}=0.4M_{\odot}$ and $R_{VV\,Pup}=0.014\,R_{\odot}$.
     The other white dwarfs of chapter 3 have higher masses and/or smaller radii so that their upper values 
     on $k^2$ are correspondingly lower. The same is true for the $k^2$ value of the sun.
     However, one has to note that this dipolar rotator technique assumes that the emitted degree of 
     polarization is direct proportional to the longitudinal magnetic field strength, whereas we have seen 
     that this relation becomes much more complicate in the case of cyclotron radiation. It is therefore 
     more reliable to use the ''plain'' model proposed by Solanki, Haugan and Mann in 1999 \cite{shm99} which 
     assumes that the polarized radiation is emitted 
     homogeneously over the visible stellar surface. Since this is a rather conservative restriction 
     the resulting upper limit is congruously $k^2 \leq (0.4\, \mbox{km})^2$. Nevertheless 
     it is important that the range of $k^2$-values for VV Puppis is consistent with the limits from 
     the dipolar rotator model as well as with the plain model.
  
\begin{figure}[t]
       \centerline{\psfig{figure=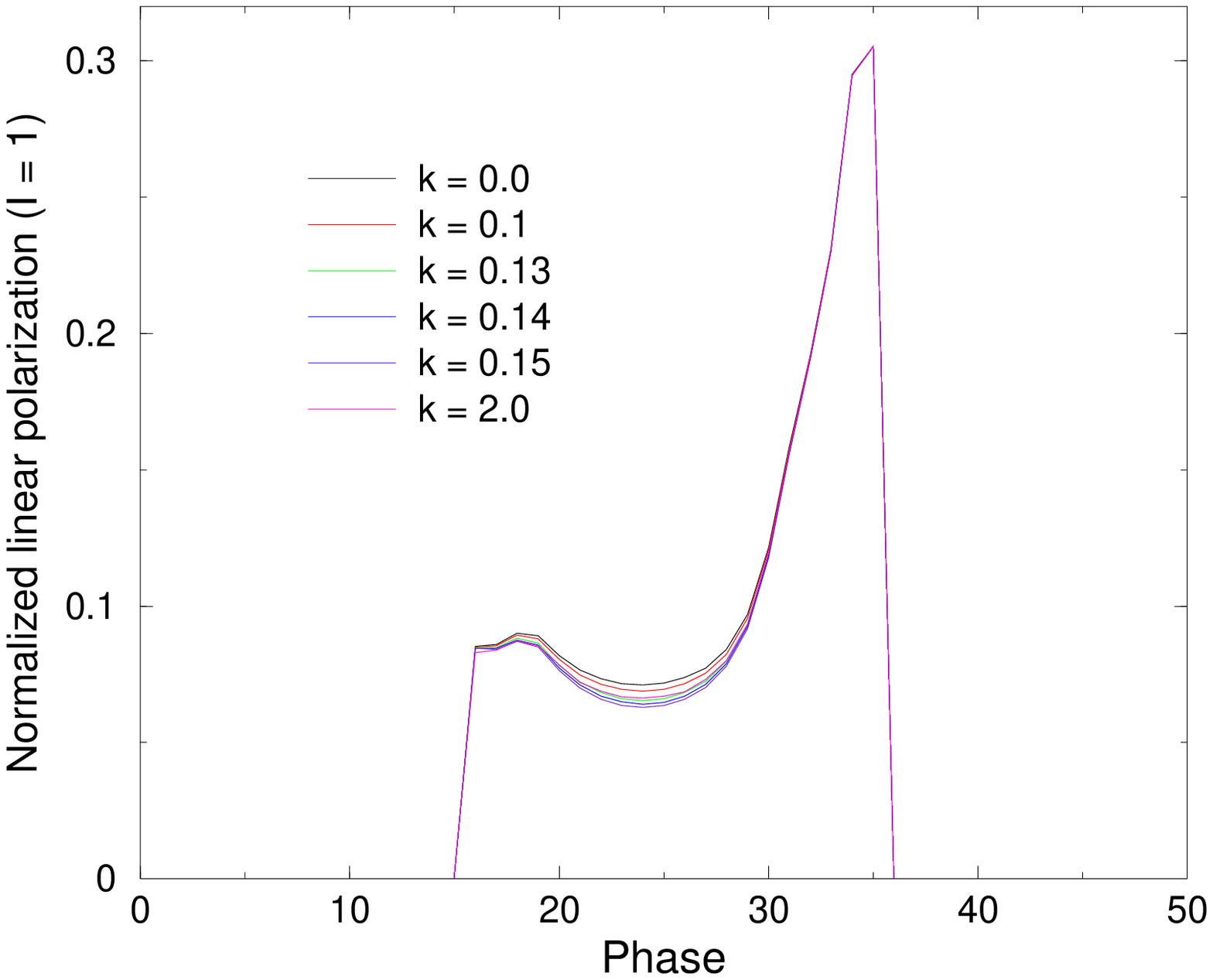,width=4.in,height=3.5in}}
       \centerline{\psfig{figure=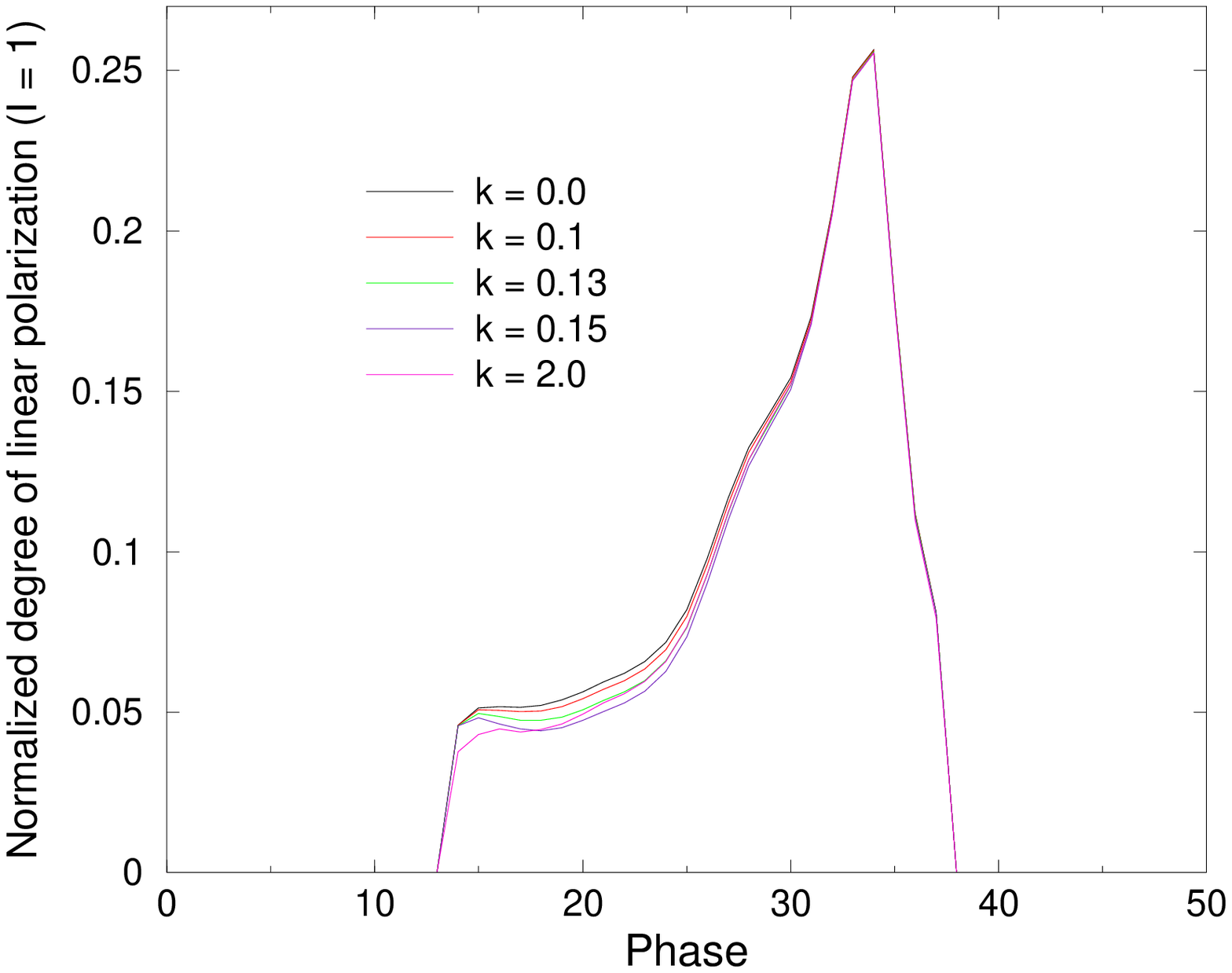,width=4.in,height=3.5in}}
       \caption{Gravitationally modified, phase dependent linear polarization curves. The physical parameters 
       are the same as in Fig.\ref{vvp-circ1}. For comparison the same $k$-values as for circular polarization
       have been plotted. Upper curves: $\alpha=5^{\circ}$, $\theta_B=10^{\circ},\,\Phi_B=90^{\circ}$. 
       Lower curves: $\alpha=10^{\circ}$, $\theta_B=30^{\circ},\,\Phi_B=90^{\circ}$.}
       \label{gcurve-lin}
     \end{figure}

     Since the amount of Stokes $U$ in the linear polarization signal is always $\lsim 5$\% the linear 
     polarization curve is hardly affected by gravity-induced birefringence which only acts on Stokes $V$ and
     Stokes $U$ as one can see in Fig.\ref{gcurve-lin}. Nevertheless, the first smaller peak becomes sharper 
     and more pronounced for $\alpha=5^{\circ}$ while for $\alpha=10^{\circ}$ the peak is washed out.
     \clearpage

     \clearpage
     
     \clearpage
 \section{Comparison of the results}
\begin{figure}[h]
     \centerline{\psfig{figure=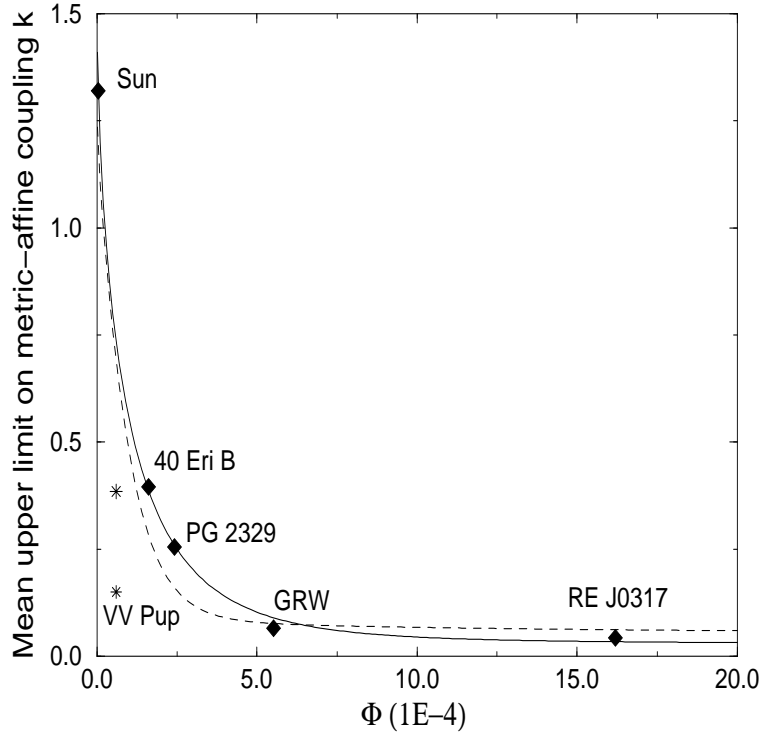,width=4.5in,height=4.5in}}
     \caption{$k^2(\Phi)$ curve of chapter 3 together with the VV Puppis limits. The upper star $(0.6|0.385)$ 
     marks the limit obtained by the polarization modelling technique. The lower star $(0.6|0.15)$ marks the
     limit from lightcurve fitting. A better fit to the VV Puppis value is provided by the dashed curve.}
     \label{poutII}	     
   \end{figure}

   Finally, we want to check how the limits on $k^2$ for VV-Puppis, obtained by lightcurve fitting and
   the polarization modelling technique from chapter 3 fit into the empirical scheme, given by (\ref{k-phi}).
   For this purpose, Fig. \ref{poutII} shows again the curve from Fig. \ref{pout} additionally with the 
   $k^2_{VV\, Pup}$ values, obtained in this chapter.
   The limit which is obtained by the polarization modelling technique deviates slightly from the curve. 
   Nevertheless the fitting can be improved with a minimally different function 
   \begin{equation}\label{k-phiII}
     k^2(\Phi) = 1,026262\cdot (\exp(-\Phi^{1.1})+0.1\,\Phi^{-0.18}) \quad ,
   \end{equation}  
   which yields the dashed line in the above figure. The bigger deviation of the lower point $(0.6|0.15)$ from
   the curve is not surprising, since its intention is not to serve as an upper but as a lower limit on $k^2$.
   Certainly more datapoints measured at different celestial bodies are needed to see which curve can best
   approximate the upper limits on $k^2$.   
     
 \clearpage
 \section{Conclusions}
    The main difference between the results of this chapter and those of chapter 3 is obvious. While for
    ''ordinary'' magnetic white dwarfs we merely set upper limits on $k^2$ we now additionally have to
    assume a lower bound on $k^2$ in order to link the presented theory of cyclotron radiation in AM-Her type 
    systems with observations. However one has to be very careful when making an interpretation of these 
    results since the underlying model of interacting binary systems is afflicted by many uncertainties:
    \begin{enumerate} 
      \item The fact that a constant unpolarized continuum background has to be introduced by hand to 
            reproduce observations is unsatisfactory. A certain aspect of the physics of binary systems is
            therefore probably not understood and although it is rather unlikely that this aspect could 
	    lead to a modification of the present lightcurves similar to the gravitationally modified 
	    curves, it cannot be completely ruled out.
      \item More importantly, Wickramasinghe \& Ferrario \cite{wf90} have shown that an arc-shaped emission 
            region is also able to replace the sinusoidal shape seen in the circular polarization curve by a 
	    flat top.
    \end{enumerate}
    At a first glance, the second point seems to discard any claim for a lower bound on $k^2$. Indeed, 
    we are here confronted with two alternative models which try to explain the same observed polarization 
    properties with different approaches. Currently there is no reliable and convincing possibility to judge 
    in favour of one or the other approach. Although the agreement of the conventional approach with observations 
    seems to be very appealing, the new question which now arises is, if the assumption of an arc-shaped emission 
    region for VV-Puppis is justified any longer, taking into account the new results from gravitational 
    birefringence. One of the main arguments supporting the arc-shaped emission regions is the flat top character 
    of the calculated circular polarization curve as well as the correctly predicted degree of polarization - 
    both also provided by gravity-induced depolarization. Concerning future projects it is therefore highly 
    desirable to look for specific predictions of the conventional model which do not involve polarized radiation 
    and are, as a result, not affected by birefringence. 
    
    However, as an important conclusion of this chapter we can say that, as long as gravitational birefringence
    cannot be excluded, the interpretation of polarized radiation from compact astrophysical objects is always
    afflicted with many uncertainties since the real nongravitational source properties may be hidden behind a
    birefringent curtain.

\chapter{Future Projects}   
  Since astronomical tests of the Einstein equivalence principle with respect to gravitational
  birefringence are a relatively new area of research, we are left with several promising but so far
  unexplored ideas which have the potential not only for setting sharp upper limits on this effect 
  but also for possible direct detections. We present in this chapter a selection of two elaborated 
  but not yet finished examples where gravity-induced birefringence could either serve as a possible 
  important contribution for a satisfying interpretation of current data or could show up as a new 
  effect in atomic spectra under certain conditions.
  
  The first example is concerned with the problem of circular polarization from active galactic nuclei 
  (AGN). Under the reasonable assumption that synchrotron radiation is the main source for polarized 
  radio emission, the degree of circular polarization measured in those AGN is on average an order of
  magnitude higher than expected from synchrotron theory. The suggested classical solution in terms 
  of circular repolarization is not free from contradictions which leads to the conclusion that a 
  gravity-induced crosstalk between linearly and circularly polarized light could provide an important 
  contribution for solving this puzzle. The second example is concerned with gravitational red shift
  measurements in nonmetric theories of gravity. While in metric theories of gravity the ticking rates
  of (atomic) clocks are independent on their internal structure, i.e. the red shift is universal, this
  universality is violated for clocks which are affected by nonmetric gravitational fields. Defining
  three atomic clocks by different transitions of hydrogen one gets three different red shift predictions
  which could be tested in astronomical spectra of compact objects or under well defined laboratory conditions.
  
  In the last part of this chapter we briefly discuss two further ideas for promising future projects.
  We suggest to compare the polarization properties of multiple images generated by gravitational lenses to
  look for traces of gravity-induced alterations due to the propagation of light through different gravitational
  potentials. The last issue picks up the current controversy about an intrinsic anisotropy of space time
  where the underlying data could be explained in terms of cosmological birefringence.

   \newpage                                      
   \section{Circular Polarization of AGN}
     Active galactic nuclei (AGN) are characterized by a huge variety of high energy phenomena
     and nonthermal radiation processes which cannot be ascribed to normal stellar activity.
     Instead it is commonly accepted, that the main central engine is driven by a super-massive 
     black hole, whose strong gravitational field in combination with magnetic fields and 
     relativistic jets feeds the processes which supply the kind of radiation which is central 
     to our current investigation. Of special interest are in this regard the compact, 
     extragalactic radio sources which belong to the class of active galaxies. Typical 
     representatives are Quasars and BL Lac objects, named after the prototype BL-Lacertae with 
     strong, variable polarization and weak emission lines (Blazar). A further important example 
     is provided by our own galaxy with the central radio source in Sagittarius A*. The optical 
     brightness of all these objects is often clearly exceeded by emission in the radio regime with a 
     nonthermal spectrum and non-zero degrees of polarization.         
      
     The most plausible, dominant mechanism for such radio emission in AGN is usually considered  
     to be synchrotron radiation from an ensemble of relativistic electrons. Concerning the 
     often differently used notations it is important to mention, that usually the name cyclotron 
     radiation is reserved for emission from low-energy electrons whereas synchrotron radiation 
     traditionally describes emission from highly relativistic electrons because it was first 
     observed in 1948 in electron synchrotrons. While it is usual to use the notion 'cyclotron 
     radiation' with respect to magnetic white dwarfs without regarding different energies, the 
     traditional distinction between cyclotron and synchrotron radiation is used in most of the 
     relevant literature concerning AGN and therefore also used here.
     
     Our special interest in this context applies to the origin of circular polarization observed
     in only a few AGNs. Analysis of the collation of circular polarization data presented by 
     Weiler and De Pater in 1983 \cite{wdp83} revealed that out of 50 radio galaxies and 43 quasars, 
     circular polarization was detected for only 6 radio galaxies and 15 quasars.     
     Although at a first glance it seems very likely that a synchrotron process has to be
     considered as the source, the small degrees of circular polarization observed in many AGN
     has a frequency dependence and time variability that is not consistent with the simple 
     predictions based on the intrinsic polarization of synchrotron emission (see \cite{sasa} also
     for a huge list of references). Likewise the observed degrees of circular polarization are
     often indeed small but nevertheless too high for a simple synchrotron source. It was therefore 
     suggested that the circular polarization is due to a propagation effect where the 
     elliptic eigenmodes of a relativistic plasma effects cyclic conversion of linear polarization 
     into circular polarization \cite{kme98}. However, this approach is again not free of 
     difficulties so we suggest that a conversion mechanism due to gravitational birefringence 
     might solve these problems and contribute a significant fraction to the observed circular 
     polarization.
     
     In the following we give a brief overview on some of the relevant data, mainly focussed on
     observations of the galactic center in Sagittarius A* (Sgr A*) and discuss the conventional,
     propagation induced polarization mechanisms. We outline the idea how gravity-induced birefringence
     could account for the observed polarization.   
     
     \subsection{Polarization properties of synchrotron emission}
       Following the current understanding, the radio emission of AGN is mainly produced by highly 
       relativistic electrons which move on helical paths along magnetic field lines. A single electron 
       moving with the synchro-cyclotron frequency $\omega_c$ emitts its radiation in a narrow cone 
       of half-angle $\gamma^{-1}$ along the instantaneous trajectory, where $\gamma$ denotes the 
       usual Lorentz factor. For a distant observer this gives a continuous spectrum with a  
       peak at the frequency $\omega_m \propto \gamma^3 \omega_c$. Considering an isotropic and 
       homogeneous ensemble of relativistic 
       electrons within a magnetic field having a power-law energy distribution such that the 
       particle density between $E$ and $E+dE$ can be written as $N(E)\,dE\propto E^{-\Gamma}dE$,
       one finds that the total intensity observed from the source is $I(\nu)\propto a(\Gamma)
       (B\sin\theta)^{(\Gamma+1)/2}\nu^{-(\Gamma-1)/2}$, where $a(\Gamma)$ is a slowly varying 
       function of $\Gamma$ and $\theta$ denotes the pitch angle with respect to the magnetic field. 
       The spectral energy distribution, i.e. the dependence of intensity $I_{\nu}$ or flux $F_{\nu}$
       on the frequency can often be approximated by a power law $I_{\nu}\propto \nu^{-\alpha}$ at least
       over limited frequency ranges. The exponent $\alpha$, called the spectral index is related to 
       $\Gamma$ by $\Gamma=2\alpha+1$.
       
       In the optically thin case, the polarization plane is perpendicular to the projected magnetic 
       field with a linear polarization degree of $m_L=(\Gamma+1)/(\Gamma+7/3)$. In the optically thick 
       regime, the total intensity spectrum reaches a $\nu^{5/2}$ dependency, while the degree of
       linear polarization now becomes $m_L=3/(6\Gamma+13)$, with the electric vector being maximum
       parallel to the projected field. The degree of linear polarization can reach theoretically up 
       to $\sim 70$\% but this high level is only rarely observed. Currently, there are two accepted
       mechanisms which can cause depolarization: Unresolved inhomogenities within the source and
       the presence of Faraday rotation internal to the source. The last point will be discussed in 
       more detail later.
       
       Besides linear polarization, the theory of synchrotron radiation predicts that there should
       also be a small amount of circular polarization \cite{legg}. Considering again a single 
       electron gyrating in a magnetic field ${\bf B_0}$ which is directed along ${\bf k}$, the circular
       polarization is right-handed when the direction of motion passes close to the line-of-sight (LoS)
       on the opposite side of ${\bf B_0}$, while it is left-handed if it passes close to the LoS
       on the same side as ${\bf B_0}$. Therefore, in an ensemble of electrons with an isotropic 
       distribution of pitch angles one finds approximately as many electrons contributing to right-
       and left-handed polarization within the emission cone so that, to first order, the circular
       polarization cancels out, leaving the emission linearly polarized. However, Legg \& Westfold \cite{legg}
       proposed that the total emission from an ensemble of gyrating electrons could have a net circular
       polarization if the number of electrons contributing to the observed radiation with right- and 
       left-handed polarization is not equal.      
       For an isotropic distribution of electrons, Legg \& Westfold showed that this condition is satisfied.
       The standard equation for the degree of circular polarization for an optically thin, homogeneous
       synchrotron source as given by Melrose in 1971 \cite{mel71} reads
       \begin{equation}\label{mc}
         m_c = \frac{\cot\theta}{3}\left(\frac{\nu}{3\nu_B\sin\theta}\right)^
	 {-1/2}f(\alpha) \approx \cot\theta/\gamma\quad ,
       \end{equation} 
       with $\theta$: angle between the field and the line of sight; $\nu$: emission frequency in Hz and
       $\nu_B$ is the electron gyrofrequency. $f(\alpha)$ is a weak function of the spectral index $\alpha$;
       for optically thick emission in the limit of strong Faraday rotation, $f(\alpha)$ varies monotonically
       between 0.6 and 2.0 for $\alpha$ between 0 and 2 \cite{mel71}.
       Using typical source parameters like $B\approx 1$mG, $\sin\theta\approx 1$ and $\nu=1$GHz one can
       expect a level of $m_c\approx 0.1$\%. Usually, $m_c$ is smaller than 0.1\% in AGN with only few
       cases for which CP approaches 0.5\% \cite{rns00}. In an optically thick source the degree of CP remains
       similar, only modified by a term of order unity dependent on the distribution of relativistic electrons
       \cite{mel71}. Equation (\ref{mc}) implies that CP is mainly produced by particles with low
       Lorentz factor $\gamma$, or by emission nearly parallel to the magnetic field (small $\theta$).
       Furthermore, it is important to note that CP produced by a synchrotron process is expected to show
       a $\nu^{-1/2}$ dependence as can be seen from (\ref{mc}).    
     \subsection{Observations}
       This section will present a brief summary of the most important results concerning circular polarization
       measurements in AGN over the last few years. We do not lay claim to completeness and we apologies
       for omissions.
       \subsubsection{Sagittarius A*}
       Studies of stellar proper motion in the vicinity of the nonthermal radio source Sagittarius 
       A* \cite{ghez} revealed a highly compact object with a mass of $\sim 2.5\times 10^6 M_{\odot}$ 
       on a scale less than 0.01 pc. Hence, the most conservative interpretation for this is given 
       today by a super-massive black hole with a synchrotron emission region, fed through accretion. 
       
       The first detection of CP radiation from Sgr A* was made by Bower, Falke, \& 
       Backer in 1998 \cite{bow99}. Using the Very Large Array (VLA) they measured an average fractional 
       polarization of $m_c=-0.36\%\pm 0.05\%$ at 4.8 GHz and of $m_c=-0.26\%\pm 0.06\%$ at 8.4 GHz, 
       respectively. The average spectral index $(m_c\propto \nu^{\alpha})$ was $\alpha = -0.6\pm  0.3$. 
       Since the special off-axis design makes the VLA a poor instrument for CP 
       measurements, it is important that this result was later confirmed by Sault and Macquart 
       \cite{sma99} who used two archival and one new ATCA (Australia Telescope Compact Array) 
       observation.
       
       Multifrequency observations of CP and LP in Sagittarius A* were reported 
       recently in 2002 by Bower et al. \cite{bow02}. They obtained 13 epochs of VLA observations at 1.4, 
       4.8, 8.4 and 15 GHz in summer 1999 and 11 epochs from ATCA at 4.8 and 8.5 GHz in the same year. 
       Their measured mean fractional polarizations are given in the following table, together with data 
       taken from the VLA  archives.
      
       \begin{center}
         \begin{tabular}{|l|cccc|}\hline
	   & & & & \\
           Data from & 1.4 GHz & 4.8 GHz & 8.4 GHz & 15 GHz  \\
	   & (\%) & (\%) & (\%) & (\%) \\ \hline
	   VLA 1999 & -0.21 $\pm$ 0.10 & -0.31 $\pm$ 0.13 & -0.34 $\pm$ 0.18 & -0.62 $\pm$ 0.26 \\
	   ATCA 1999 & ... & -0.37 $\pm$ 0.08 & -0.27 $\pm$ 0.10 & ... \\
	   VLA archive & ... & -0.31 $\pm$ 0.13 & -0.36 $\pm$ 0.10 & ...\\ \hline           
         \end{tabular}
	 
	 \vspace{0.3cm}
	 \small {Tab. 5.1: Mean fractional circular polarization from Bower et al. 2002 \cite{bow02}.}
       \end{center}
       For the low frequencies they repoted a spectral index $\alpha\approx -0.5$ while for higher
       frequencies the spectrum was best matched by $\alpha\approx 1.5$. Changes in CP were accompanied
       by small changes in the total intensity, suggesting that the processes driving these phenomena
       are fundamentally linked. Concerning LP, Bower et al. reported no detection with an upper limit 
       of 0.2\% at 8.4 GHz, less than 0.2\% at 22 GHz and less than $\sim 1\%$ at 112 GHz. Aitken et 
       al. \cite{ait} have claimed 10\% of LP at $\nu > 150$GHz, on the basis of low-resolution James 
       Clerk Maxwell Telescope (JCMT) observations, but this detection is so far not confirmed.
     \subsubsection{Quasars and BL Lacs}
       A recent ATCA-survey by Rayner et al. \cite{rns00} for CP in radio-loud Quasars, BL Lacs and 
       Radio Galaxies has revealed fractional CP at 5 GHz between 0.05\% and 0.5\% in 11 out of 13 
       sources at a spatial resolution of 2 arcsec. Also, VLBI measurements from Homan \& Wardle in 1999 
       \cite{howa} yielded localized CP of 0.3\% up to 1\% in the jet-cores of 3C273, PKS 0528+134 and 
       3C279 (1\%), while a few cases may be as high as the local linear polarization. Brunthaler et al.
       \cite{brun} measured CP in the compact radio jet of the nearby spiral galaxy M81*. They reported 
       a value of $m_c=0.54\%\pm 0.06\%\pm 0.07\%$ at 8.4 GHz and $m_c=0.27\% \pm 0.06\% \pm 0.07\%$ at
       4.8 GHz with an error separation into statistical and systematic terms. Similar to Sgr A* they 
       detected no LP im M81* at a level of 0.1\%.
   
    \subsection{Problem}
       Although LP observed in AGN is generally accepted as being produced by a synchrotron process, the
       correct interpretation of the high levels of CP in AGN is still unclear. The theoretically predicted 
       degree of LP due to synchrotron emission is about 70\% - a level which is rarely observed so that
       several depolarization mechanisms like Faraday rotation or strongly tangled magnetic fields have 
       been discussed in order to explain that the observed LP is often at least an order of magnitude lower. 
       By taking these various mechanisms into account, the low LP is not really a surprise. But, what is a 
       surprise are the high levels of CP given these stringent limits on LP. Applying the same depolarization 
       mechanisms also to CP, the observed levels should also be an order of magnitude lower than the 
       predicted 0.1\% from equation (\ref{mc}). Additionally, one must take into account that the 
       $m_c\propto\nu^{-1/2}$ dependence as expected from synchrotron emission is generally not observed 
       \cite{sasa}. For this reasons it is commonly recognized, that simple synchrotron models cannot account 
       for the full polarization characteristics without depolarization or repolarization, discussed later 
       here, in the source or the accretion region. We will therefore briefly recall the circular repolarization 
       mechanism, first invented by Pacholczyk in 1970 \cite{pch70} where CP is explained as being due to a 
       propagation effect. We point out some difficulties of this approach and suggest a perhaps more suitable 
       propagation effect: Gravitational birefringence.       
    
      \subsubsection{Circular repolarization}
        For a given polarization mode with Stokes parameters $I$, $Q$, $U$ and $V$ it is possible to define
        a second polarization mode with the same $I$ but opposite $Q$, $U$ and $V$ as an equivalent
        solution of the Maxwell equations. While these so-called orthogonal modes can travel 
        independently with the same propagation velocity through empty space and homogeneous 
        isotropic media, in astrophysical magnetized plasmas, however different polarization 
        modes have different propagation velocities. In such a plasma for a given propagation 
        direction with respect to the magnetic field, one can always find two such eigenmodes
        which travel through this medium with different propagation velocities, but without 
        changing their polarization vector. Such a medium with two refractive indices is called
        birefringent. In general, birefringence will be elliptical, i.e. the eigenmodes have linear
        and circular contributions.
        The transfer equation for a Stokes vector, subjected to magnetooptical effects due to 
        different refractive indices in an anisotropic medium without absorption is of the form
        (Melrose \& McPhedran 1991 \cite{mel91}, p.188)
        \begin{equation}\label{trans}
          \frac{d}{ds}\left(\begin{array}{c} I \\ Q \\ U \\ V \end{array}\right) = 
          \left(\begin{array}{cccc} 0 & 0 & 0 & 0 \\ 0 & 0 & -\rho_V & \rho_U \\
                                    0 & \rho_V & 0 & -\rho_Q \\ 0 & -\rho_U & \rho_Q 
		           	    & 0 \end{array}\right)
				    \left(\begin{array}{c} I \\ Q \\ U \\ V \end{array}\right)       
        \end{equation}
        with
        \begin{equation}
          \rho_Q = -\Delta k \frac{T^2-1}{T^2+1}\cos(2\psi),\quad
 	  \rho_U = -\Delta k \frac{T^2-1}{T^2+1}\sin(2\psi),\quad
	  \rho_V = -\Delta k \frac{2T}{T^2+1}\quad .
        \end{equation}
        Here, $s$ denotes the distance along the ray path, $\Delta k$ the difference in wavenumber 
        between the eigenmodes and $\psi$ the polarization angle. $T$ is the axial ratio of the 
        polarization ellipse of one of the eigenmodes. Regarding the physical relevance, we basically
        distinguish between circularly and linearly polarized eigenmodes.
       
        In the first case, circular polarization implies a polarization ellipse with an axial ratio
        equal to unity, i.e. $|T_{\pm}| = 1$, '+' and '-' denoting the orthogonal modes. Hence,
        $\rho_Q=\rho_U=0$, so that (\ref{trans}) reduces to
        \begin{equation}
          \frac{d}{ds}\left(\begin{array}{c}  Q \\ U  \end{array}\right) = \rho_V
	              \left(\begin{array}{cc} 0 & -1 \\ 1 & 0 \end{array}\right)
		      \left(\begin{array}{c}  Q \\ U  \end{array}\right) \quad .
        \end{equation}
        Here, the transfer equation leaves $I$ and $V$ unaffected, while $U$ and $Q$ changes only 
        such that the polarization angle $\psi = 1/2 \arctan U/Q$ rotates according
        \begin{equation}
          \frac{d\psi}{ds} = \frac{1}{2}\rho_V \quad ,
        \end{equation}
        which is commonly known as Faraday rotation. Otherwise, if the natural modes are linearly
	polarized this corresponds to $T=0$ or $T=\infty$ so we get from (\ref{trans})
	\begin{equation}
	  \frac{d}{ds}Q = \rho_U V\, , \quad \frac{d}{ds}U = -\rho_Q V \quad .
        \end{equation}
	This means that if the natural modes are linear or also elliptical then radiation that is initially
	linearly polarized develops a circularly polarized component as the polarization changes in a periodic
	manner along the ray path \cite{pch73}.	        
	Following Kennett \& Melrose \cite{kme98} we can therefore write
	\begin{eqnarray}
	  V(\nu)     &=& U_0(\nu)\sin(\lambda^3\mbox{RRM}), \\[0.5cm]
	  \mbox{RRM} &=& 3\times 10^4\left(\frac{L}{1\,\mbox{pc}}\right)\left\langle{\cal E}_L\left(\frac{n_r}
	  {1\,\mbox{cm}^{-3}}\right)\left(\frac{B}{1\,\mbox{G}}\right)^2\sin^2\theta\right\rangle\mbox{rad m}^{-3}
	\end{eqnarray}
	The relativistic rotation measure RRM describes quantitatively the strength of circular repolarization,
	namely the phaseshift $\Delta\Phi$ between the two natural modes, depending on the special details of 
	the distribution of relativistic particles and the local magnetic field ${\bf B}$. $n_r$ denotes the 
	particle density, $L$ the path length within the source region and ${\cal E}_L$ the minimum Lorentz 
	factor of the plasma. 	 	
    \subsection{Gravitational birefringence and repolarization} 
      Although the conditions required for the observed circular polarization of some synchrotron sources  
      due to circular repolarization are indeed not impossible, they are rather very restrictive. We present two 
      situations in which this mechanism might operate, point to some difficulties and discuss the possible
      alternative mechanism of gravitational birefringence which could circumvent several problems.
      
      In the first situation where circular repolarization might operate, the relativistic particles dominate 
      within the plasma, so that the natural modes are linearly polarized. Requiring either $\mbox{RRM}\lambda\sim 1$
      in a small part of the source or $\mbox{RRM}\lambda\sim 10^{-3}$ over a large part leads to
      significant circular polarization, comparable to the degree emitted from synchrotron radiation.
      However, the frequency dependence is more like $m_c \propto \omega^{-3}$ than $m_c \propto \omega^{-1/2}$
      predicted for a uniform, optically thin synchrotron source \cite{kme98}. The second situation
      assumes that the plasma is a mixture of cold and relativistic particles causing the natural modes to have
      a small circular component. The relative phase shift $\Delta\Phi$ between the two natural modes is then
      determined by the cold plasma, and provided that $\Delta\Phi \ll 1$, the resulting circular polarization
      is of the order of the eccentricity of the modes. This situation implies $m_c \propto \omega^{-1}$. 
      
      Even though these scenarios seem to be quite realistic, the existing data do not appear consistent with
      $m_c \propto \omega^{-a}$ with either $a=0.5$ or $a=1$ \cite{kme98}. Another problem concerns the ratio
      of linear to circular polarization. For a homogeneous optically thin synchrotron source and a highly
      relativistic plasma one expects $m_l/m_c\gsim 100$ \cite{jod77} while Ryle and Brodie \cite{rb81} reported 
      sources with even $m_l/m_c\gsim 1$. This problem is also known as the ''circular polarization excess''.
       
      To summarize up we can say, that despite the different possibilities which have been discussed so far, the
      mechanism for the production of circular polarization in AGNs is still not known with absolute certainty.  
      The question therefore arises if gravity-induced birefringence could serve as an important source for
      circular polarization which could solve some of the problems, mentioned above. Such a gravitational conversion
      of linearly to circularly polarized light might be possible for rays emitted from the same pointlike source
      and is independent of the special conditions within the plasma, e.g. the amount of cold electrons and the 
      optical thickness of the source. However, one obstacle is that the birefringence models we have used up to 
      now are appropriate for stars with spherical symmetry where one can, in principle, assign to each pointsource
      on the surface a heliocentric angle $\theta=\arccos\mu$. Consequently, the quality of our limits on 
      birefringence strongly depends on the amount of available information about the source. It is clear that
      the situation is now quite different in the case of an active galactic nuclei. Of course a first primitive
      ansatz could consist in putting the nuclei inside an imaginary sphere and locate the relativistic jet as the 
      source of linearly polarized light at the limb of this sphere. Using the metric-affine phase shift formula
      (\ref{mag-form}) with the parameters of Sagittarius A*, e.g. $M_{SgrA}=2.5\times 10^6\, M_{\odot}$, 
      $\lambda=6\times 10^{-2}m$ and $R_{SgrA} = 0.01$pc we get a phase shift of $\Delta\Phi=0.056\times 10^{-11}
      \,k^2$. It is currently not possible to decide, whether this value of $\Delta\Phi$ is sufficient to 
      convert a reasonable amount of linearly into circularly polarized light because of the complete lack of a 
      reliable estimate for $k^2$ of a supermassive black hole. Furthermore it is currently unclear how the
      observed wavelength is correlated to the emission radius. A further important aspect which could test for a 
      possible influence of birefringence is the frequency dependence of the polarized radiation, described by the
      spectral index. As already mentioned above, the synchrotron model as well as the circular repolarization
      model fail to predict the correct frequency dependence. The hope is, that a combined model which incorporates
      repolarization as well as birefringence might be able to provide here a better fit to the observations. 
       
 \section{Gravitational redshift measurements}
   Since electromagnetism was the only fundamental interaction besides gravity that was known 
   in 1915 and for which this generalization was relevant, Einstein himself suggested the first 
   test of this new principle by showing that an electromagnetic wave propagating between points 
   with different gravitational potentials as a consequence must suffer a red shift.
   Therefore, measurements of the gravitational redshift belong to the first classical tests 
   of the EEP. 
   
   In every metric theory of gravity, the frequencies of two identical atomic clocks at rest at different locations
   ${\bf x_1}$ and ${\bf x_2}$ in a static gravitational potential differ according to
   \begin{equation}
     z = \frac{\nu_1-\nu_2}{\nu_2}=\sqrt{\frac{g_{00}({\bf x_1})}{g_{00}({\bf x_2})}}-1 \quad ,
   \end{equation}     
   so that for weak gravitational fields, e.g. $g_{00}=1+2\Phi \, (c\equiv 1)$, where $\Phi$ is the Newtonian 
   gravitational potential to first order, this yields
   \begin{equation}\label{redsft} 
     z = \Phi({\bf x_1}) - \Phi({\bf x_2}) \quad .
   \end{equation}
   Since the derivation of (\ref{redsft}) is only based on the validity of the weak equivalence principle and the 
   conservation of energy, this prediction is independent of the special form of the field equations of a certain 
   theory and therefore valid for every metric theory of gravity. Congruously, this result was put in the form of 
   a conjecture by Will \cite{will74}:   
   
   \vspace{0.5cm}
   \noindent\fbox{\begin{minipage}{\textwidth}   
   \begin{center} \vspace{0.5cm}
     {\bf Universal gravitational redshift conjecture (UGR):}
   
     \vspace{0.5cm}
     Any complete, self-consistent, and relativistic theory of gravity that embodies the universality of 
     gravitational redshift is neccesarily a metric theory.      
   \vspace{0.5cm} 
   \end{center}
   \end{minipage}}
   \vspace{0.5cm}
   
   \noindent A simple plausibility argument supporting this statement can be given by means of the metric postulates 
   (see Sec.1.2.3, p.7). Since in a local free-falling reference frame the physical laws are those of special
   relativity, the frequencies of free-falling atoms are functions only of the universal atomic constants
   and therefore independent of the external gravitational field. Consequently, a comparison of the frequencies
   of the free-falling atoms involves only a comparison of their space-time trajectories, but since these
   trajectories are universal, the measured red shift is universal. Nonmetric theories of gravity violate, per
   definition, one or more of the metric postulates and therefore do not predict a universal redshift.
   On the contrary, Will \cite{will74} has shown that in nonmetric theories of gravity the ticking rates of 
   atomic clocks in a gravitational field and, so, the gravitational redshift are affected in a manner that 
   depends on their internal structure. By applying the gravitationally modified Dirac equation in the 
   $TH\epsilon\mu$ formalism to the hydrogen atom, he found that the predicted redshifts of clocks, defined 
   by transitions between principal levels, between fine-structure levels and between two hyperfine levels are
   given by
   \begin{eqnarray}
     z_P &=& T^{-1/2}(T\epsilon^2/H) - 1 \label{th-red_P}\\
     z_F &=& T^{-1/2}(T\epsilon^2/H)^2 - 1 \\
     z_H &=& T^{-1/2}(T\epsilon^2/H) (\epsilon/\mu) - 1 \label{th-red_H}\quad ,
   \end{eqnarray}
   with one clock being far away from the gravitational potential $(\Phi({\bf x_1})=0,\,{\bf x_1}\to\infty)$.
   In order to explore the physical consequences of these predictions, we focus attention in a first step on
   weak gravitational fields $(\Phi \lsim 10^{-5})$ so that the results can be used for tests within the solar 
   system $(\Phi \sim 10^{-6})$ and also in good approximation for white dwarfs. In the weak field limit, the
   functions $T$, $H$, $\epsilon$ and $\mu$ are expanded in power series in $\Phi$ according to \cite{ll73}
   \begin{eqnarray}
     T &=& 1 - 2\alpha\Phi + 2\beta\Phi^2 + \cdots \\
     H &=& 1 + 2\gamma\Phi + \frac{3}{2}\delta\Phi^2 + \cdots \\ 
     \epsilon &=& 1 + \epsilon_1\Phi +\epsilon_2\Phi^2 + \cdots \\
     \mu &=& 1 + \mu_1\Phi + \mu_2\Phi^2 + \cdots   \quad .
   \end{eqnarray}
   By requiring that $\alpha=1$ we achieve that $T = 1-2\Phi$ to first order, so that the theory is in agreement 
   with Newtonian gravitation at lowest order. Now, the redshift predictions (\ref{th-red_P})-(\ref{th-red_H}) 
   take the form
   \begin{eqnarray}
     z_P &=& (1-2\Gamma_0)\Phi+\left(\frac{3}{2}-\beta-\Gamma_P\right)\Phi^2 \\
     z_F &=& (1-4\Gamma_0)\Phi+\left(\frac{3}{2}-\beta-\Gamma_F\right)\Phi^2 \\
     z_H &=& (1-4\Gamma_0+\Upsilon_1)\Phi+\left(\frac{3}{2}-\beta-\Gamma_H\right)\Phi^2
   \end{eqnarray}
   where
   \begin{eqnarray}
     \Gamma_0 &=& 1+\gamma-\epsilon_1 \\
     \Gamma_1 &=& \frac{3}{2}\delta-4\gamma^2-2\epsilon_2-2\beta+2\epsilon_1^2+\gamma\epsilon_1+\mu_1-5\gamma+
                  \epsilon_1-1 \\
     \Upsilon_1 &=& 2(\gamma+1)-(\epsilon+\mu) \\
     \Upsilon_2 &=& \frac{3}{2}\delta-2\beta+4(\gamma+1)-\epsilon_1\mu_1-(\epsilon_2+\mu_2)\\
     \Gamma_P &=& \Gamma_1 + 3(\gamma+1)\Gamma_0-3\Gamma_0^2+\Gamma_0+\Upsilon_1 \\
     \Gamma_F &=& 2\Gamma_P-4\Gamma_0^2 \\
     \Gamma_H &=& 2\Gamma_P-4\Gamma_0^2 -\Upsilon_2+(2\gamma+1)\Upsilon_1+\Upsilon_1(4\Gamma_0-\Upsilon_1)  
   \end{eqnarray}
   
   \vspace{1cm}
   \noindent The exact expression for the $\Gamma$'s and $\Upsilon$'s are different for different nonmetric 
   theories. In
   metric theories of gravity, all $\Gamma$'s and $\Upsilon$'s vanish identically. In a further project we
   would therefore like to derive the values of these parameters for metric-affine gravity and see how the
   validity of this theories could be constrained by experiments. Since the possible measurable differences 
   in the predictions between universal gravitational redshift and those of nonmetric theories becomes more
   pronounced the stronger the involved gravitational fields are, it is obvious to use redshift measurements
   of solar spectral lines or of white dwarf lines for such tests. On the Sun Manganese lines show rich
   hyperfine structure which could be used for setting strong limits on nonmetric theories of gravity.
   Unfortunately in the case of white dwarfs these measurements are seriously complicated by extreme pressure 
   broadening which currently renders the signals of finestructure and hyperfine structures hardly detectable.
   
   A possible alternative could be provided by measurements of various redshifts under controlled laboratory 
   conditions. Measuring the energy levels of hydrogen in gravitational potentials different from those on the
   earth's surface, for example onboard the International Space Station (ISS), give the chance of revealing 
   differences in the levels of hydrogen to those of standard textbooks. Since the relevant parameters which
   could influence the measurements can be determined with high accuracy, the probability of detecting possible
   deviations is not insignificant.      
  
  \clearpage   
   
 \section{Further ideas}     
   The consequences of the idea that gravitational birefringence might have an important impact on the 
   interpretation of astronomical polarimetric data are so far a rather unexplored topic. We have shown
   that gravity-induced depolarization has the potential to bridge the gap between former AM-Her models and
   observations and that it also might help to solve the puzzle of circular polarization from active
   galactic nuclei. However, since we are involved in a relatively new area of research, many other systems
   or issues in astrophysics may also be better understandable in terms of gravity-induced birefringence or could,
   at least, provide strong upper limits on this effect. We therefore provide here two brief examples of such systems
   which it might be worth to turn one's attention to and which we have not yet explored as deeply as the foregoing
   topics in this chapter.         
       
   \vspace{0.5cm}
   \noindent {\bf Gravitational lensing:} In a gravitational lense system, light emitted by a distant astronomical
   object is deflected by the gravitational field of a massive object (star, galaxy, cluster of galaxies) which 
   lies along the line of sight, allowing for multiple source images. Depending on the relative orientations 
   of source, lens and observer multiple source images correspond to different trajectories of the light around the
   lense. Consequently, light rays of different images could have propagated through different gravitational 
   potentials, so that the polarizations of these images becomes different if gravitational birefingence is not
   negligible, taking into account the time delay. A systematic survey of the polarization properties in 
   gravitational lens systems therefore not only has the potential for strong upper limits on this effect but also 
   for a direct detection of birefringence in principle.

   \vspace{0.5cm}
   \noindent {\bf Polarization rotation over cosmological distances:} We have already shown in this chapter that
   gravity-induced birefringence might help to understand the polarization properties of active galactic nuclei.
   Since many of these objects are at high redshifts, the polarization properties of light could therefore also 
   be altered by propagating over cosmological distances in a nonmetric background gravitational field. 
   
   Recently, several authors claimed to have found a systematic rotation of the plane of polarization of light 
   emitted by distant radio galaxies\cite{clk80,hav75,nr97}. By measuring the angle $\theta(\lambda)$ between
   a fixed reference direction in space and the plane of polarization of a radio wave of wavelength $\lambda$,
   they found that their data could be fitted by
   \begin{equation}   
     \theta(\lambda)=\alpha\lambda^2 + \chi \quad .
   \end{equation}
   The linear dependence of the angle $\theta$ on $\lambda^2$ is a characteristic feature of Faraday rotation. The
   fitting parameter $\alpha$ depends upon the magnetic field and the electron density along the line of sight.
   Therefore, the angle $\chi$ describes the orientation of the polarization plane after Faraday rotation is
   taken out, i.e. the orientation of the plane before the rotation. Now, Nodland and Ralston \cite{nr97} claimed
   that the observed $\chi$ angles could be reproduced by assuming that the polarization plane of a wave emitted 
   by a galaxy is initially oriented at a fixed angle relative to the galaxy's major axis, and then undergoes a 
   rotation, specified by an angle $\beta$, that depends on the direction of the galaxy in the sky. Defining
   an angle $\gamma$ between the line of sight to the galaxy and a fixed direction is space they found
   \begin{equation}
     \beta=\frac{1}{2}\Lambda^{-1}r\cos\gamma \quad ,
   \end{equation}
   where $r$ is the distance to the galaxy and $\Lambda$ denotes a rotation measure. This dependency of $\beta$
   on a fixed direction in space, provided by $\gamma$, is an indication of anisotropy. The importance of this effect 
   lies in the fact that it can probe the fundamental structure of space time at very early times and over length 
   scales commensurate with the size of the universe. Because of the clear importance that the confirmation of an 
   intrinsic anisotropy of space would have, much effort have been put into explaining these results. At least 
   three mechanisms have been proposed which can lead to this cosmological birefringence
   \begin{itemize}
     \item Modification of conventional electrodynamics by introducing pseudoscalar axion fields \cite{axion}. 
     \item Modified dispersive Maxwell equations due to semi-classical space time with polymer-like structure
           at microscales in loop quantum gravity \cite{gam98}.
     \item Rotation of the polarization plane due to a nonmetric gravitational background field.
   \end{itemize}
   The polarization rotation therefore serves as a test for alterations of various basic physical theories
   which should be thoroughly studied to analyse the distinguishing features and measurable differences between 
   them with the hope of finding true modifications of fundamental physics. A high priority in this list of possible 
   alterations should be given to the coupling of torsion to electromagnetism which appears natural in the 
   framework of metric-affine gauge theories of gravity. The main advantage of this approach is that its predictions
   are already testable and even perhaps verifiable in other, more accessible branches of astrophysics, as 
   was presented in this thesis.
   
   It is important to note that the importance of further tests of these different modifications is unaffected by the
   current controversy over the correctness of the interpretation of the data \cite{carol,eb97,lfw97}.
   Even if Nodland and Ralston \cite{nr97} could be proven wrong, these studies can be useful to uncover the source 
   of any polarization rotation effect in the future.

\chapter{Conclusions}

  The main purpose of this thesis was the development of new astronomical tests of the 
  Einstein equivalence principle (EEP) in terms of gravity-induced birefringence. So, 
  the very first question we had to answer in the beginning was how justified it is to 
  proceed with testing a principle which belongs to the physical predictions with the 
  most accurate empirical underpinning. However, this principle was invented at a time 
  when physicists discussed about the existence of atoms and, so, it is obvious that
  the Einstein and also the Weak equivalence principle are blind to the microscopic 
  structure of matter. Regarding all kinds of charges and spins in modern quantum field 
  theory it would be more than remarkable if this statement holds up for all times.
  Therefore the desired unification of quantum mechanics with general relativity provides
  the strongest clue for violations of the Einstein equivalence principle as was
  shown in chapter one. 
  
  An appropiate framework for developing and analysing experimental tests of the EEP is
  given by lagrangian based nonmetric theories of gravity. Within this class we have 
  focused our attention on two prototypical representatives, the nonsymmetric gravitation
  theory (NGT) and the metric-affine gauge theory of gravity (MAG), both build upon a 
  non-Riemannian geometry of space time. While it was known for several years that NGT 
  predicts violations of EEP in terms of gravitational birefringence where the strength
  of birefringence is detemined by a material dependent coupling constant $\ell^2$ we 
  have shown that this is also the case for MAG. The latter couples torsion to the 
  electromagnetic field in the $\chi g$-formalism, giving rise to a new coupling constant
  $k^2$. In this context it is important to note that the question of how the electromagnetic
  field exactly couples to gravity, in particular to torsion, is an age-old, still 
  unresolved problem. Certainly, there may be other couplings than those we have proposed, 
  which might seem no less natural so that a lot of work is left for the future.  
  However, our objective of constraining possible EEP violations in these theories is 
  therefore equivalent with setting strong limits on $\ell^2$ and $k^2$.         
  
  The search for traces of gravity-induced birefringence in polarized light from an astronomical 
  object is promising only if we have a sufficient knowledge about its source properties so
  that we can compare the observed polarization signal with its theoretically predicted 'original'
  properties without any birefringence influence. For this reason we have used in our first project
  polarization measurements in solar spectral lines where the underlying theory about the generation
  of polarization profiles within the solar photosphere is well developed. Using the Stokes asymmetry 
  technique and the new profile difference technique, our new limits on $\ell^2_{\odot}$ are 3 - 7 orders 
  of magnitude smaller than previous results. Since birefringence is most pronounced for short 
  wavelengths these limits could be improved with the same methods by using UV polarization 
  measurements where we can expect a maximum gain relative to the current analysis of a factor 4.
  Currently, however, it is not possible with our methods to decide whether the observed asymmetries 
  of solar Stokes profiles are partially due to gravitational birefringence since possible influences 
  must be less than or equal to the asymmetries induced by nongravitational mechanisms. 
  
  At this point a very important issue must be mentioned: Conventional models which try to explain 
  the creation of Stokes profiles in the solar atmosphere are {\em not} axiomatic models from first 
  principles. Instead the main guideline is the agreement with observations to which the model has to be 
  fitted. Consequently, the predictions of these models which are based on interpretations of
  polarized signals must not be valid if gravitational birefringence has a non negligible influence.
      
  The limits on $k^2$ were the first so far made and therefore could not be compared with other, older 
  values. Although the idea of a possible utilization of future space missions is very appealing because 
  of its high potential for further improvements of our current limits, the technological realization 
  of such an experiment is currently out of reach.
  
  In our second project we have used polarization measurements from isolated magnetic white dwarfs
  to constrain birefringence. By using a dipole model for the magnetic field geometry we were able to 
  sharpen previous results, given by Solanki, Haugan and Mann for GRW $+70^{\circ}8247$ by $\sim 18$\%.
  In addition to GRW we found three other white dwarfs which complied with our restriction of available
  wavelength resolved polarimetric data and also of known mass, radius and magnetic field geometry.
  The resulting limits indicate, that $k^2$ for a particular star only depends on the ratio between
  its Schwarzschild radius and the physical, stellar radius. Whether $k^2$ also depends on the chemical
  composition of a celestial body like $\ell^2$ cannot be decided conclusively. The source of the torsion
  field is not specified in most papers concerning metric-affine gravity and, therefore, they do not make
  firm predictions of differences between the torsion fields generated by, for example, different kinds of
  stars. It is certainly possible that such differences exist. For this reason it would be interesting
  to compare limits on $k^2$ obtained from a large sample of stars with different chemical abundances.
  
  The polarization characteristics of AM-Her systems provide a fundamentally different analysis for our
  purpose. By calculating the gravitationally modified polarization curves for VV Puppis we could achieve 
  a good agreement with observations only for a nonzero $k^2$. However, since other models which assume
  different source properties are able to achieve the same good conformity with observations, gravitational
  birefringence can only serve currently as one possibility among other more conventional approaches.
  
  This thesis represents the first systematic, extensive search for gravitational birefringence in
  astrophysical spectropolarimetric data. Therefore several promising projects are left for further
  investigations. In this regard, one the most interesting open question is concerned with the origin
  of circular polarization from active galactic nuclei. Gravity-induced birefringence seems to be the
  ideal candidate since its influence is not afflicted by entangled magnetic fields or different
  compositions with respect to the amount of relativistic or cool electrons in hot plasmas, which are 
  the main sources of problems for the conventional repolarization approach. Nevertheless, red shift 
  measurements, gravitational lenses and the question of intrinsic cosmic anisotropy have also the 
  potential for making major progress in this field of research. 
  
  The quest for possible violations of the Einstein equivalence principle still remains an open problem.
  Although we have several theoretically convincing arguments as well as experimentally gained hints, the
  results are so far not conclusive. However, the results of this thesis have shown that this possibly 
  could change, not too far away in the future.

\begin{appendix}

\chapter{Gravity-induced birefringence in the $\chi g$-formalism}
      In this chapter the propagation of polarized light is investigated under the assumption
      of a relatively weak gravitational background field, that vary on length and time scales which are 
      large compared to the light's wavelength. Since the electromagnetic lagrangian density of special 
      relativity has the form (\ref{lchi-g}) with $\chi^{\alpha\beta\gamma\delta} 
      = \frac{1}{2}(\eta^{\alpha\gamma}\eta^{\beta\delta} - \eta^{\alpha\delta}\eta^{\beta\gamma})$
      it is possible to find a quasi-Lorentzian coordinate system in the weak-field limit where
      $\chi^{\alpha\beta\gamma\delta}$ has the form
      \begin{equation}\label{chigrav}
       \chi^{\alpha\beta\gamma\delta} = \frac{1}{2}(\eta^{\alpha\gamma}\eta^{\beta\delta} - 
       \eta^{\alpha\delta}\eta^{\beta\gamma}) + \delta\chi^{\alpha\beta\gamma\delta}
      \end{equation}
      with $\delta\chi^{\alpha\beta\gamma\delta} \ll 1$. 
      
      The amplitude and phase representation of a plane, electromagnetic wave is given by
      \begin{equation}\label{emwaves}
        {\bf E} = {\bf A}_E e^{i\Phi}, \quad {\bf B} = {\bf A}_B e^{i\Phi} \quad .
      \end{equation}
      The main feature of this representation is that the temporal and spatial derivatives of the 
      vector amplitudes ${\bf A}_E$ and ${\bf A}_B$ are small compared to derivatives of the rapidly
      varying phasefunction $\Phi$. Since the gravitational background field varies only slowly
      in space and time too, its derivatives are also small compared to derivatives of $\Phi$.
      By considering the propagation of a high-frequency electromagnetic wave this means that
      all derivatives other than those of the phasefunction can be ignored. Therefore, locally the 
      gravitational field is treated as a homogeneous medium, which is known as the {\em high-frequency 
      approximation}.
      So, the electromagnetic field equations which follow from (\ref{lchi-g}) are 
      \begin{equation}
        \chi^{\alpha\beta\gamma\delta} F_{\gamma\delta,\beta} = 0 \quad.
      \end{equation}
      Defining electric and magnetic fields in the usual way as $F_{0i}\equiv E_i$ and $F_{jk}\equiv 
      \epsilon_{jkl}B_l$ together with the decomposition (\ref{chigrav}) one gets the modified 
      Maxwell equations describing electromagnetic fields in a background gravitational field.
      Written in a somehow more transparent way this reads as
      \begin{equation}\label{max1}
        \nabla \cdot {\bf E} + \mbox{terms proportional to $\delta \chi$ and {\bf E} or {\bf B}}
      \end{equation}
      and
      \begin{equation}\label{max2}
        \nabla \times {\bf B} - \frac{\partial {\bf E}}{\partial t}
	+ \mbox{termes proportional to $\delta \chi$ and {\bf E} or {\bf B}} \quad.
      \end{equation}
      The homogeneous Maxwell equations are unaltered 
      \begin{equation}\label{max3}
        \nabla \times {\bf E} + \frac{\partial {\bf B}}{\partial t} = 0
      \end{equation}
      and
      \begin{equation}\label{max4}
        \nabla \times {\bf B} = 0 \quad .
      \end{equation}
      
      The objective target of the following analysis is the derivation of the eikonal equation
      which describes the propagation of a locally plane electromagnetic wave and, therefore, 
      provides information about the local coordinate velocities of wave propagation.
      For this purpose (\ref{emwaves}) is inserted into the field equations (\ref{max1}) - (\ref{max4}) 
      neglegting all derivatives other than those of the phase function $\Phi$. Denoting
      $k_{\mu}$ as the gradient of this function
      \begin{equation}
        k_{\mu} \equiv \partial \Phi \equiv (\partial \Phi/\partial t, \nabla \Phi) \equiv 
	(-\omega, {\bf k})
      \end{equation}
      the homogeneous Maxwell equations yield 
      \begin{equation}\label{hommax}
        {\bf A}_B = \frac{{\bf k}\times {\bf A}_E}{\omega}
      \end{equation}
      and
      \begin{equation}
       {\bf k} \cdot {\bf A}_B = 0 
      \end{equation}
      respectively. While the latter of these implies that the magnetic field vector of a locally
      plane wave is transverse to the direction in which the wave propagates, it follows from (\ref{max1})
      that
      \begin{equation}\label{egrav}
       {\bf k} \cdot {\bf A}_E = \mbox{Terme proportional zu $\delta \chi$ und ${\bf A}_E$ oder 
       ${\bf A}_B$}
      \end{equation}
      i.e. electric field vector is transverse to the direction of propagation if and only if no
      gravitational field is present!
      Together with equation (\ref{hommax}) this implies for $A_E$ expressed via the two independent
      components of $A_B$
      \begin{equation}
       {\bf A}_E = -\frac{\omega}{k^2} {\bf k} \times {\bf A}_B + \mbox{terms proportional to 
                                                                        $\delta \chi$ and ${\bf A}_B$}
      \end{equation}  
      so that this equation together with(\ref{max2}) and (\ref{hommax}) implies the eikonal equation
      \begin{equation}\label{eikonal}
         \left(1 - \frac{\omega^2}{k^2}\right){\bf A}_B = \mbox{terms proportional to $\delta \chi$ and 
	                                                        ${\bf A}_B$} \quad .
      \end{equation}
      Since the magnetic amplitude ${\bf A}_B$ can be regarded as the superposition of two independent
      polarized components with the coordinate phase velocity $\omega / k$, finding the two
      independent polarization states is equivalent to solve a two-dimensional eigenvalue problem.
      For this purpose one makes a decomposition of $\delta \chi^{\alpha\beta\gamma\delta}$ in a set of 
      SO(3) tensor objects
      \begin{eqnarray}
       \xi^{ij}    &=& -\delta \chi^{0i0j} \\
       \gamma^{ij} &=& \frac{1}{2} \epsilon^{jlm}\delta\chi^{0ilm} \\
       \zeta^{ij}  &=& \frac{1}{4} \epsilon^{ilm}\epsilon^{jpq}\delta\chi^{lmpq}
      \end{eqnarray}
      where $\epsilon^{ijk}$ is the Levi-Civit\`a antisymmetric symbol. This decomposition of $\delta 
      \chi^{\alpha\beta\gamma\delta}$ is now rotated from the original, quasi-Lorentzian coordinate system 
      $(t,x,y,z)$ into a new system where the background gravitational field is represented to a set of 
      $(t,x',y',z')$ coordinates and the light propagates in the $z'$-direction. 
      In this representation ${\bf A}_B$ has only $x'$ and $y'$ components so that equation (\ref{eikonal}) 
      reduces to a system of two equations
      \begin{eqnarray}
       \left(1 - \frac{\omega^2}{k^2}\right)A^{1'}_B &=&  {\cal A}A^{1'}_B - {\cal B} A^{2'}_B \label{eikonal2} \\                                                                                 
       \left(1 - \frac{\omega^2}{k^2}\right)A^{2'}_B &=& -{\cal B}A^{1'}_B + {\cal C} A^{2'}_B 
       \quad .\label{eikonal3}
      \end{eqnarray}
      The matrix which defines the structure of the right-hand sides of (\ref{eikonal2}) and (\ref{eikonal3})
      is real valued and symmetric when $\delta \chi^{\alpha\beta\gamma\delta}$ is real. The coefficients 
      ${\cal A}$, ${\cal B}$ and ${\cal C}$ depend on the location in space-time and on the direction in 
      which the wave propagates. This can be expressed in terms of the tensor components $\xi^{i'j'}$, 
      $\zeta^{i'j'}$ and $\gamma^{i'j'}$ in the $(t,x',y',z')$ coordinate system. In particular Haugan \& 
      Kauffmann \cite{hkf95} has shown
      \begin{eqnarray}
        {\cal A} &=& \xi^{2'2'}-2\gamma^{2'1'}-\zeta^{1'1'} \\
	{\cal B} &=& \xi^{1'2'}+(\gamma^{2'2'}-\gamma^{1'1'})+\zeta^{1'2'} \\
	{\cal C} &=& \xi^{1'1'}+2\gamma^{1'2'}-\zeta^{2'2'} \quad .
      \end{eqnarray}       
      The eigenvalues of (\ref{eikonal2}) and (\ref{eikonal3}) are given by
      \begin{equation}\label{a-c}
       \lambda_{\pm} = \frac{{\cal A} + {\cal C}}{2} \pm \frac{1}{2}\sqrt{({\cal A} - {\cal C})^2 + 
                                                                           4{\cal B}^2} \quad .
      \end{equation}
      From this it follows that the corresponding eigenvectors define the polarization states which
      propagate with well-defined phase velocities
      \begin{equation}
        c_{\pm} = 1 - \frac{1}{2}\lambda_{\pm} + O(\delta \chi^2)
      \end{equation}
      Denoting the fractional difference between $c_+$ and $c_-$ by $\delta c/c$ one gets a local, 
      dimensionless observable
      \begin{equation}\label{deltac}
       \frac{\delta c}{c} = \frac{1}{2}\sqrt{({\cal A} - {\cal C})^2 + 4{\cal B}^2} \quad .
      \end{equation}
      Therefore, during the propagation of wave with circular frequency $\omega$ this effects yields an
      accumulated phase shift
      \begin{equation}\label{phaseint}
        \Delta \Phi = \omega \int \frac{\delta c}{c} \; dt + O(\delta \chi^2)
      \end{equation}
      e.g. the difference in the local velocities leads to a change in the relative phase between the two
      independent components which in turn implies an alteration of the initial polarization state 
      Since this analysis revealed that the speed of an electromagnetic wave depends on its orientation 
      within the background gravitational field this implies that gravity-induced birefringence is a
      direct consequence of a violation of EEP.

      This scheme is now applied to the special case of metric-affine gravity. As explained in the 
      introduction the nonmetricity independent, tensorial part of the torsion as spherically symmetric 
      solution of the metric-affine field equations was given by Tresguerres in 1995 \cite{tre95a,tre95b}
      \begin{equation}\label{torsion}
        T^{\alpha}=k_0\left[\frac{1}{r}(\theta^0-\theta^1)+\left(\frac{m}{r^2}-\frac{\Lambda r}{3\chi}
        -\frac{\kappa b_4 N_0^2}{\chi r^3}\right)(\theta^0+\theta^1)\right]\wedge \theta^{\alpha}
      \end{equation}
      with the dilatation charge $N_0$, the torsion mass $m'$ and $k_0=1$ for $\alpha=0,1$ and 
      $k_0=-1/2$ for $\alpha=2,3$. Setting 
      \begin{equation}\label{def-A}
        A \equiv \left(\frac{m}{r^2}-\frac{\Lambda r}{3\chi}-\frac{\kappa b_4 N_0^2}{\chi r^3}\right)
      \end{equation}
      this gives
      \begin{eqnarray}
        T^0 &=& \left[\frac{1}{r}\,\theta^0\wedge\theta^1-A\,\theta^0\wedge\theta^1\right] =
                \left(\frac{1}{r}-A\right)\,\theta^0\wedge\theta^1 \label{t0a}\\
        T^1 &=& \left[\frac{1}{r}\,\theta^0\wedge\theta^1+A\,\theta^0\wedge\theta^1\right] =
    	        \left(\frac{1}{r}+A\right)\,\theta^0\wedge\theta^1 \\
        T^2 &=& -\frac{1}{2}\left[\frac{1}{r}\left(\theta^0\wedge\theta^2-\theta^1\wedge\theta^2\right)
                +A\left(\theta^0\wedge\theta^2+\theta^1\wedge\theta^2\right)\right] \\
	    &=& -\frac{1}{2}\left(\frac{1}{r}+A\right)\theta^0\wedge\theta^2+
	         \frac{1}{2}\left(\frac{1}{r}-A\right)\theta^1\wedge\theta^2 \\	
        T^3 &=& -\frac{1}{2}\left[\frac{1}{r}\left(\theta^0\wedge\theta^3-\theta^1\wedge\theta^3\right)
                +A\left(\theta^0\wedge\theta^3+\theta^1\wedge\theta^3\right)\right] \\
	    &=& -\frac{1}{2}\left(\frac{1}{r}+A\right)\theta^0\wedge\theta^3+
	         \frac{1}{2}\left(\frac{1}{r}-A\right)\theta^1\wedge\theta^3 \label{t3a} 
      \end{eqnarray}
      which can be expressed in terms of the most general structure of the torsion in the case of spherical 
      symmetry \cite{tre95a} 
      \begin{eqnarray}
        T^0 &=& \alpha(r)\,\theta^0\wedge\theta^1 \\
        T^1 &=& \beta(r)\,\theta^0\wedge\theta^1  \\
        T^2 &=& \gamma_{(1)}\,\theta^0\wedge\theta^2+\gamma_{(3)}\,\theta^1\wedge\theta^2 \label{t2}\\
        T^3 &=& \gamma_{(1)}\,\theta^0\wedge\theta^3+\gamma_{(3)}\,\theta^1\wedge\theta^3 \label{t3}       	       	    
      \end{eqnarray}
      with $\gamma_{(2)}=\gamma_{(4)}=0$. According to (\ref{mag-lem}) and equation (3.11) in \cite{tre95a}
      together with the usual scalar-valued two-form $F$, representing the electromagnetic field one can write
      \begin{eqnarray*}
        T_0\wedge F &=& \alpha \, \theta^0\wedge\theta^1\wedge F = \alpha\,
                        F_{23}\,\theta^0\wedge\theta^1\wedge\theta^2\wedge\theta^3 \\
        -T_1\wedge F &=& \beta \, \theta^0\wedge\theta^1\wedge F = \beta\, 
                         F_{23}\,\theta^0\wedge\theta^1\wedge\theta^2\wedge\theta^3 \\
        -T_2\wedge F &=& \gamma_{(1)}\,\theta^0\wedge\theta^2\wedge F
                         +\gamma_{(2)}\,\theta^0\wedge\theta^3\wedge F +\\
		     & &  \gamma_{(3)}\,\theta^1\wedge\theta^2\wedge F
		         +\gamma_{(4)}\,\theta^1\wedge\theta^3\wedge F \\
		     &=&  \gamma_{(1)}\,F_{13}\,\theta^0\wedge\theta^2\wedge\theta^1\wedge\theta^3 
		         +\gamma_{(2)}\,F_{12}\,\theta^0\wedge\theta^3\wedge\theta^1\wedge\theta^2 +\\
		     & &  \gamma_{(3)}\,F_{03}\,\theta^1\wedge\theta^2\wedge\theta^0\wedge\theta^3 
		         +\gamma_{(4)}\,F_{02}\,\theta^1\wedge\theta^3\wedge\theta^0\wedge\theta^2 \\
		     &=& [\gamma_{(2)}\,F_{12}-\gamma_{(1)}\,F_{13}+\gamma_{(3)}\,F_{03}-\gamma_{(4)}\,F_{02}]\,
		         \theta^0\wedge\theta^1\wedge\theta^2\wedge\theta^3 \\
		     &=& [\gamma_{(2)}\,B_3+\gamma_{(1)}\,B_2+\gamma_{(3)}\,B_3-\gamma_{(4)}\,E_2]\,
		         \theta^0\wedge\theta^1\wedge\theta^2\wedge\theta^3\\
        -T_3\wedge F &=& -\gamma_{(2)}\,\theta^0\wedge\theta^2\wedge F 
                         +\gamma_{(1)}\,\theta^0\wedge\theta^3\wedge F \\
                     & & -\gamma_{(4)}\,\theta^1\wedge\theta^2\wedge F 
		         +\gamma_{(3)}\,\theta^1\wedge\theta^3\wedge F \\
		     &=& -\gamma_{(2)}\,F_{13}\,\theta^0\wedge\theta^2\wedge\theta^1\wedge\theta^3 
		         +\gamma_{(1)}\,F_{12}\,\theta^0\wedge\theta^3\wedge\theta^1\wedge\theta^2 \\
		     & & -\gamma_{(4)}\,F_{03}\,\theta^1\wedge\theta^2\wedge\theta^0\wedge\theta^3 
		         +\gamma_{(3)}\,F_{02}\,\theta^1\wedge\theta^3\wedge\theta^0\wedge\theta^2 \\
		     &=& [\gamma_{(2)}\,F_{13}+\gamma_{(1)}\,F_{12}-\gamma_{(4)}\,F_{03}-\gamma_{(3)}\,F_{02}]\,
		         \theta^0\wedge\theta^1\wedge\theta^2\wedge\theta^3 \\
		     &=& [-\gamma_{(2)}\,B_2+\gamma_{(1)}\,B_3-\gamma_{(4)}\,E_3-\gamma_{(3)}\,E_2]\,
		         \theta^0\wedge\theta^1\wedge\theta^2\wedge\theta^3    \quad .
      \end{eqnarray*}
      Since the lagrangian density we refer to reads 
      \begin{equation}\label{lem}
        \delta{\cal L}_{\mbox{\small EM}} =  k^2 \,{}^*(T_{\alpha}\wedge F)^*(T^{\alpha}\wedge F)
      \end{equation}  
      this leads to
      \begin{eqnarray*}
         ^*(T_{\alpha}\wedge F)^*(T^{\alpha}\wedge F) &=& [^*T^0\wedge F]^2 - [^*T^i\wedge F]^2 \\
         &=& (\alpha\, B_1)^2 - (\beta\, B_1)^2 - (\gamma_{(2)}\,B_3+\gamma_{(1)}\,B_2+\gamma_{(3)}\,E_3
              -\gamma_{(4)}\,E_2)^2 - \\
         & & (\gamma_{(1)}\,B_3-\gamma_{(2)}\,B_2-\gamma_{(4)}\,E_3-\gamma_{(3)}\,E_2)^2  \\	  
         &=& (\alpha^2 - \beta^2)B^2_1 - (\gamma_{(2)}^2\,B_3^2+2\gamma_{(1)}\gamma_{(2)}\,B_2B_3+2\gamma_{(2)}
              \gamma_{(3)}\,B_3E_3- \\
         & & 2\gamma_{(2)}\gamma_{(4)}\,B_3E_2+\gamma_{(1)}^2\,B_2^2+2\gamma_{(1)}\gamma_{(3)}\,B_2E_3-
             2\gamma_{(1)}\gamma_{(4)}\,B_2E_2+\\
         & & \gamma_{(3)}^2\,E_3^2-2\gamma_{(3)}\gamma_{(4)}\,E_2E_3+\gamma_{(4)}^2\,E_2^2)- 
             (\gamma_{(1)}^2\,B_3^2-2\gamma_{(1)}\gamma_{(2)}\,B_2B_3-\\
         & & 2\gamma_{(1)}\gamma_{(4)}\,B_3E_3-2\gamma_{(1)}\gamma_{(3)}\,B_3E_2+
             \gamma_{(2)}^2\,B^2_2+2\gamma_{(2)}\gamma_{(4)}\,B_2E_3+ \\
         & & 2\gamma_{(2)}\gamma_{(3)}\,B_2E_2+\gamma_{(4)}^2\,E_3^2+2\gamma_{(3)}\gamma_{(4)}\,E_2E_3+ 
             \gamma_{(3)}^2\,E_2^2) \\
         &=& (\alpha^2 - \beta^2)B^2_1 - (\gamma_{(1)}^2+\gamma_{(2)}^2)B_3^2 + 2(\gamma_{(1)}\gamma_{(4)}-\\
         & & \gamma_{(2)}\gamma_{(3)})B_3E_3 + 2(\gamma_{(1)}\gamma_{(3)}+\gamma_{(2)}\gamma_{(4)})B_3E_2 - \\
         & & (\gamma_{(1)}^2+\gamma_{(2)}^2)B_2^2 - 2(\gamma_{(1)}\gamma_{(3)}+\gamma_{(2)}\gamma_{(4)})B_2E_3+\\
         & & 2(\gamma_{(1)}\gamma_{(4)}-\gamma_{(2)}\gamma_{(3)})B_2E_2-(\gamma_{(3)}^2+\gamma_{(4)}^2)E_3^2-
         (\gamma_{(3)}^2+\gamma_{(4)}^2)E_2^2 
      \end{eqnarray*}
      Rearranging the terms yields
      \begin{eqnarray*}
        ^*[T_{\alpha}\wedge F]^*[T^{\alpha}\wedge F] 
        &=& (\alpha^2 - \beta^2)B_1^2-(\gamma_{(1)}^2+\gamma_{(2)}^2)[B_2^2+B_3^2]-(\gamma_{(3)}^2+\gamma_{(4)}^2)\\
        & & [E_2^2+E_3^2] + 2(\gamma_{(1)}\gamma_{(4)}-\gamma_{(2)}\gamma_{(3)})[B_2E_2+B_3E_3]+\\
        & & 2(\gamma_{(1)}\gamma_{(3)}+\gamma_{(2)}\gamma_{(4)})[B_3E_2-B_2E_3]\\
      \end{eqnarray*}
      which can be written in terms of matrix elements
      \begin{eqnarray*}
        ^*(T_{\alpha}\wedge F)^*(T^{\alpha}\wedge F) &=& \zeta_{11}\,B_1^2 + \zeta_{22}\,B_2^2 + \zeta_{33}\,B_3^2 - 
	\xi_{22}\,E_2^2 - \xi_{33}\,E_3^2 + 2\gamma_{22}\,E_2B_2 + \\
        & & 2 \gamma_{33}\,E_3B_3 + 2\gamma_{32}\,E_3B_2 + 2\gamma_{23}E_2B_3  \quad .   
      \end{eqnarray*} 
      The symmetries of $\delta \chi^{\alpha\beta\gamma\delta}$ implies that $\xi^{ij}$ and $\zeta^{ij}$ are symmetric.
      
      \vspace{3cm}
      \noindent The explicit matrices are 
      \begin{equation}
        \xi^{ij}=\left(\begin{array}{ccc}
              0 & 0 & 0 \\
	      0 & (\gamma_{(3)}^2+\gamma_{(4)}^2) & 0 \\
	      0 & 0 & (\gamma_{(3)}^2+\gamma_{(4)}^2)  
        \end{array}\right)
      \end{equation}
      \begin{equation}
        \zeta^{ij}=\left(\begin{array}{ccc}
              (\alpha^2 - \beta^2) & 0 & 0 \\
	      0 & -(\gamma_{(1)}^2+\gamma_{(2)}^2) & 0 \\
	      0 & 0 & -(\gamma_{(1)}^2+\gamma_{(2)}^2)  
         \end{array}\right)
      \end{equation}
      and
      \begin{equation}
          \gamma^{ij}=\left(\begin{array}{ccc}
              0 & 0 & 0 \\
	      0 & (\gamma_{(1)}\gamma_{(4)}-\gamma_{(2)}\gamma_{(3)}) &
	      (\gamma_{(1)}\gamma_{(3)}+\gamma_{(2)}\gamma_{(4)}) \\
	      0 & -(\gamma_{(1)}\gamma_{(3)}+\gamma_{(2)}\gamma_{(4)}) & (\gamma_{(1)}\gamma_{(4)}-\gamma_{(2)}\gamma_{(3)})  
           \end{array}\right)
      \end{equation}
      The expressions ${\cal A}-{\cal C}$ and ${\cal B}$ from (\ref{a-c}) can now be expressed in terms of
      the spherical components of the tensors $\xi^{ij},\,\zeta^{ij}$ and $\gamma^{ij}$. In terms of components
      in the $(t,x',y',z')$ coordinate system Haugan \& Kauffmann \cite{hkf95} have shown that
      \begin{equation}\label{exp.a-c}
        {\cal A}-{\cal C} = \frac{2}{\sqrt{6}}((\xi^{(2)}_{2'}+\xi^{(2)}_{-2'})+2i(\gamma^{(2)}_{2'}-\gamma^{(2)}_{-2'})
	+(\zeta^{(2)}_{2'}+\zeta^{(2)}_{-2'}))
      \end{equation}
      and
      \begin{equation}\label{exp.b}
        {\cal B} = -\frac{1}{\sqrt{6}}(i(\xi^{(2)}_{2'}-\xi^{(2)}_{-2'})+2(\gamma^{(2)}_{2'}+\gamma^{(2)}_{-2'})+i
	(\zeta^{(2)}_{2'}-\zeta^{(2)}_{-2'})) \quad .
      \end{equation}
      Only $l=2$ components appear. Expressions for ${\cal A}-{\cal C}$ and ${\cal B}$ in terms of components in the
      original $(t,x,y,z)$ coordinate system follow from the transformation law for spherical tensor components 
      \cite{edm}, e.g.
      \begin{equation}\label{transd}
        \xi^{(l)}_{m'}={\cal D}^{(l)}_{m'm}(\phi,\theta,\psi)\xi^{(l)}_m
      \end{equation}
      where $\phi,\,\theta$ and $\psi$ are the Euler angles specifying the rotation from $(t,x,y,z)$ to $(t,x',y',z')$
      and the rotation matrix ${\cal D}$ given in terms of spherical harmonics. It is now useful to introduce a local
      quasi-Lorentzian $(t,x,y,z)$ coordinate system at each point of the light ray's path oriented so that the $x$ axis
      lies in the ray's plane because the spherical tensors introduced above are simple in these local coordinate 
      systems. Specifically, $\xi^{(2)}_m,\,\zeta^{(2)}_m$ and $\gamma^{(2)}_m$ are nonzero only for $m=0$.
      
      At each of these points along the ray path, the local $(t,x,y,z)$ coordinate system is rotated about the $y$ axis 
      through an angle $\theta$ to obtain a local $(t,x',y',z')$ system so that now the ray runs in the $z'$ direction.  
      The local value of $\delta c/c$ in (\ref{deltac}) is expressed in terms of ${\cal A}-{\cal C}$ and ${\cal B}$ 
      which are, in turn, according to (\ref{exp.a-c}) and (\ref{exp.b}) expressed in terms of $\xi^{(2)}_{\pm 2},\,
      \zeta^{(2)}_{\pm 2}$ and $\gamma^{(2)}_{\pm 2}$. Since the Euler angles of the rotation from $(t,x,y,z)$ to 
      $(t,x',y',z')$ are $\theta$ and $\phi=\psi=0$, the transformation law (\ref{transd}) implies
      \begin{equation}
        \xi^{(2)}_{\pm 2}=\sin^2\theta\xi_0^{(2)} \quad ,
      \end{equation}
      with the same relationship between $\zeta^{(2)}_{\pm 2}$ and $\zeta^{(2)}_0$ and between $\gamma^{(2)}_{\pm 2}$
      and $\gamma^{(2)}_0$. Again, according to Haugan \& Kauffmann the form of (\ref{exp.a-c}) and (\ref{exp.b}) and the
      transformation law (\ref{transd}) implies that ${\cal B}$ is proportional to $\gamma^{(2)}_0$ while ${\cal A}-
      {\cal C}$ is proportional to $(\xi^{(2)}_0 + \zeta^{(2)}_0)$. In order to get explicit values for $\delta c/c$ one
      can express $\xi_0^{(0)},\,\xi_0^{(2)},\,\zeta_0^{(0)},\,\zeta_0^{(2)}$ and $\gamma^{(0)}_0,\,\gamma^{(2)}_0$
      by the matrix elements $\xi^{ij},\,\zeta^{ij}$ and $\gamma^{ij}$
      \begin{eqnarray*}
        \xi_0^{(0)} &=& \xi^{11}+\xi^{22}+\xi^{33} = 2(\gamma_{(3)}^2+\gamma_{(4)}^2) \\
        \xi_0^{(2)} &=& \xi^{11}-\frac{1}{2}(\xi^{22}+\xi^{33}) = -(\gamma_{(3)}^2+\gamma_{(4)}^2)\\
        \zeta_0^{(0)} &=& \zeta^{11}+\zeta^{22}+\zeta^{33} = (\alpha^2 - \beta^2) - 2 (\gamma_{(1)}^2+\gamma_{(2)}^2) \\
        \zeta_0^{(2)} &=& \zeta^{11}-\frac{1}{2}(\zeta^{22}+\zeta^{33}) = (\alpha^2 - \beta^2) 
                         + (\gamma_{(1)}^2+\gamma_{(2)}^2) \\
        \gamma^{(0)}_0 &=& \gamma^{11}+\gamma^{22}+\gamma^{33}=2(\gamma_{(1)}\gamma_{(3)}-
                           \gamma_{(2)}\gamma_{(4)})\\
        \gamma^{(2)}_0 &=& \gamma^{11}-\frac{1}{2}(\gamma^{22}+\gamma^{33})=-(\gamma_{(1)}\gamma_{(4)}
                           -\gamma_{(2)}\gamma_{(3)}) 	\quad .	           	     
      \end{eqnarray*}
      Since, according to (\ref{t2},\ref{t3}) we have $\gamma_{(2)}=\gamma_{(4)}=0$ and, therefore, ${\cal B}=0$.
      So, in this case we have
      \begin{equation}
        {\cal A}-{\cal C}=\frac{4}{\sqrt{6}}(\xi_0^{(2)}+\zeta_0^{(2)})\sin^2\theta \quad.
      \end{equation}
      Using the expressions (\ref{t0a})-(\ref{t3a}) one gets
      \begin{eqnarray}
        \xi_0^{(2)}   &=& -\frac{1}{4}\left(\frac{1}{r^2}-\frac{2A}{r}+A^2\right) \\
	\zeta_0^{(2)} &=& -\frac{4A}{r}+\frac{1}{4}\left(\frac{1}{r^2}+\frac{2A}{r}+A^2\right)
      \end{eqnarray} 
      which yields
      \begin{equation}
        \xi_0^{(2)} + \zeta_0^{(2)} = -\frac{4A}{r}+\frac{A}{r} = -\frac{3A}{r} = -\frac{3m}{r^3} \quad .
      \end{equation}
      In this equations we have dropped the last two terms in (\ref{def-A}) since the dilatation charge $N_0$
      vanishes if the nonmetricity field does and the small observed value of the cosmological constant $\Lambda$
      means that effects of its term can be neglected on galactic and smaller scales.  
      Together with (\ref{torsion}) and $k^2$ from (\ref{lem}) this yields
      \begin{equation}
        {\cal A}-{\cal C}=-\frac{12\,k^2\,m}{\sqrt{6}\,r^3}\sin^2\theta \quad .
      \end{equation}
      Therefore the fractional difference between the velocities of the two polarization states is given by
      \begin{equation}
        \frac{\delta c}{c} = -\sqrt{6}\frac{k^2m}{r^3}\sin^2\theta \quad .
      \end{equation}
      For the total phase shift $\Delta\Phi$ which accumulates between the source and the observer one now has 
      to calculate
      \begin{equation}
        \omega\int\frac{\delta c}{c}\,dt = -\sqrt{6}\omega k^2m\int\frac{\sin^2\theta}{r^3}\, dt \quad .
      \end{equation}
      Using the definitions explained in Fig.B1 with the ray parametrization ${\bf x}(t)={\bf b}+{\bf k_0} t$ and
      $\sin^2\theta(t)=b^2/R^2(t)\,\,(R(t)\equiv r)$, we can write
      \begin{equation}
        \Delta\Phi = -\sqrt{6}\omega k^2\,m R_0^2(1-\mu^2)\int\limits_{t_0=R_0\mu}^{\infty}\frac{dt}{(R_0^2(1-\mu^2))^{5/2}}
      \end{equation}
      where we have used $R(t)=\sqrt{b^2+t^2}$ and $b=R(t)\sqrt(1-\mu^2)$ with $\mu=\cos\theta(t)$. $R_0$ denotes the radius
      of the star.
      
      The integral could be solved with the substitution
  
      \begin{equation}
        \overline{R}=at^2+bt^2+c, \quad a=1,\,b=0,\,c=R_0^2(1-\mu^2) \quad .
      \end{equation}  
      So that we have
      \begin{eqnarray*}     
        I_5 &=& \left.\int\limits_{t_0=R_0\mu}^{\infty}\frac{dt}{\sqrt{\overline{R}^5}}=
	\frac{4t}{3(4R_0^2(1-\mu^2))(R_0^2(1-\mu^2)+t^2)^{3/2}}\right|_{R_0\mu}^{\infty}+
	\frac{8}{12 R_0^2(1-\mu^2)}I_3 \\
	I_3 &=& \int\limits_{t_0=R_0\mu}^{\infty}\frac{dt}{\sqrt{\overline{R}^3}}= \frac{4t}
	{4R^2_0(1-\mu^2)(R_0^2(1-\mu^2)+t^2)^{1/2}}\Bigg|_{t_0=R_0\mu}^{\infty}=\frac{1}{R_0^2(1+\mu)}\quad .
      \end{eqnarray*}
      Evaluating the above term for $I_5$ yields
      \begin{equation}
        I_5 = -\frac{\mu}{3R_0^4(1-\mu^2)}+\frac{2}{3R_0^4(1-\mu^2)(1+\mu)} \quad ,
      \end{equation}
      so that the accumulated phase shift becomes
      \begin{eqnarray*} 
        \Delta\Phi &=& -\sqrt{6}\,\omega k^2\,m\omega R_0^2(1-\mu^2)I_5 \\[0.5cm]
	           &=& \sqrt{\frac{2}{3}}\frac{k^2 m \omega}{R_0^2}\left(\mu - \frac{2}{1+\mu}\right) \quad .
      \end{eqnarray*} 
      Finally we get
      \begin{equation}         
        \Delta\Phi = \sqrt{\frac{2}{3}}\,\frac{2 \pi\, k^2\, m}{\lambda\,R_0^2}\left(\frac{(\mu+2)(\mu-1)}{\mu+1}\right)
      \end{equation}
      Whether $k^2$ also depends on the chemical composition of a celestial body like $\ell^2$ cannot be 
      decided conclusively. The source of the torsion field is not specified in most papers concerning 
      metric-affine gravity and, therefore, they do not make firm predictions of differences between the 
      torsion fields generated by, for example, different kinds of stars. It is certainly possible that 
      such differences exist.
\chapter{Tests using the {\sl Solar Probe} spacecraft}
  
  \begin{figure}[hhh]
    \centerline{\psfig{figure=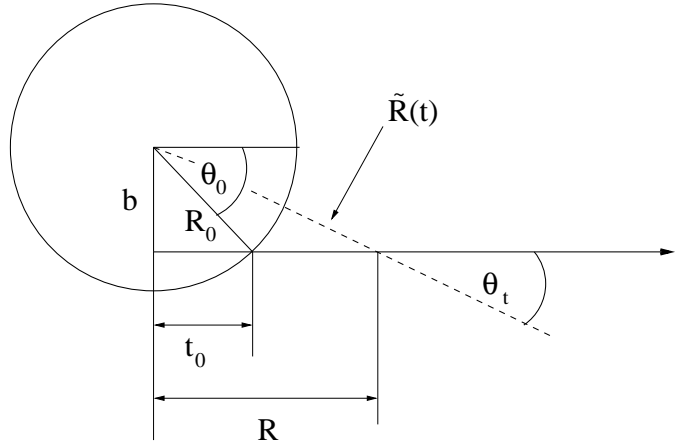,width=3.5in,height=2.3in}}
    \caption{Schematic representation of used definitions.}
  \end{figure}
  \noindent In this chapter we derive the formula which gives the accumulated phase shift between
  a point, located at a distance $R$ from the sun and an observer in an infinite distance. 
  The principles of this calculation are mainly based on the calculation that Gabriel et al. 
  \cite{gea91} used to derive the phase shift formula (\ref{ngt-form}).
   
  We consider a ray which originate on the Sun's surface and use the ray parametrization
  ${\bf x}(t)={\bf b}+{\bf k_0} t$. As usual the unit vector ${\bf k_0}$, for convenience 
  not explicitly shown in the figure specifies the direction of the (unperturbed) ray. By
  demanding that ${\bf k_0}\cdot {\bf b} =0$ we define ${\bf b}$ as the impact vector which connects
  the center of the sun with the closest point on the ray. Those points on the ray which are 
  inside the Sun for the case ${\bf b}<R_0$ are irrelevant in the calculation. The integration 
  \begin{equation}
    \Delta\Phi(\mu)=\frac{1}{2}\omega\int\limits_{t_0=R_0\mu}^{\infty}\Omega\sin^2\theta(t)\,dt
  \end{equation}  
  begins at the Sun's surface, $t_0=(R_0^2 - b^2)^{1/2}=R_0\cos\theta_0\equiv R_0\mu$ with 
  $b=\tilde{R}(t)\sin\theta(t)$. The integration extends to an infinite observer in $t_1=\infty$. 
  For convenience we use for the speed of light $c\equiv 1$ in our calculations. Using the
  NGT relation $\Omega=\frac{\ell^4_{\odot}}{\tilde{R}^4(t)}$ this yields
  \begin{equation}	    
    \Delta\Phi(\mu)=\frac{1}{2}\omega\int\limits_{t_0=R_0\mu}^{\infty}\Omega\sin^2\theta(t)\,dt
    =\frac{\pi\ell^4_{\odot}}{\lambda}\int\limits_{t_0=R_0\mu}^{\infty}\frac{R_0^2(1-\mu^2)}
    {(R_0^2(1-\mu^2)+t^2)^3}\,dt \quad .
  \end{equation}
  This integral could be solved with the substitution
  
  \begin{equation}
    \overline{R}=at^2+bt^2+c, \quad a=1,\,b=0,\,c=R_0^2(1-\mu^2) \quad .
  \end{equation}  
  So that we have
  \begin{equation}\label{asub}
    \int\frac{dt}{\overline{R}^3}=\frac{2at+b}{\Delta}\left(\frac{1}{2\overline{R}^2}+
    \frac{3a}{\Delta\cdot\overline{R}}\right)+\frac{6a^2}{\Delta^2}\left(\frac{2}{\sqrt{\Delta}}
    \arctan\frac{2at+b}{\sqrt{\Delta}}\right)
  \end{equation}
  with
  \begin{equation}
    \Delta = 4ac-b^2
  \end{equation}
  This solution is valid as long as $\Delta > 0$, i.e. $4R_0^2(1-\mu^2) > 0$ which is always
  fulfilled. Performing the integration in (\ref{asub}) yields
  \begin{eqnarray}
    \int\limits_{t_0=\mu R_0}^{\infty}\frac{dt}{\overline{R}^3} 
    &=& \frac{2t}{4R_0^2(1-\mu^2)}\left(\frac{1}{2(R_0^2(1-\mu^2)+t^2)^2}\;+ \right.\\\nonumber
    & & \left.\frac{3}{4R_0^2(1-\mu^2)(R_0^2(1-\mu^2)+t^2)}\right)+\frac{6}{16R_0^4(1-\mu^2)^2}\\\nonumber
    & &	\left(\frac{1}{R_0(1-\mu^2)^{1/2}}\left.\arctan 
        \frac{t}{R_0(1-\mu^2)^{1/2}}\right)\right|_{\mu R_0}^{\infty} \\\nonumber
    &=& \frac{3\pi}{16R_0^5(1-\mu^2)^{5/2}}-\frac{R_0\mu}{2R_0^2(1-\mu^2)}\left(
        \frac{1}{2R_0^4}+\frac{3}{4R_0^4(1-\mu^2)}\right) \\\nonumber
    & & -\frac{6}{16R_0^5(1-\mu^2)^{5/2}}\arctan\frac{\mu}{(1-\mu^2)^{1/2}}\\\nonumber
    &=& \frac{3\pi}{16R_0^5(1-\mu^2)^{5/2}}-\frac{\mu}{4R_0^5(1-\mu^2)} \\\nonumber
    & & -\frac{3\mu}{8R_0^5(1-\mu^2)^2}-\frac{3}{8R_0^5(1-\mu^2)^{5/2}}\arcsin\mu	 
  \end{eqnarray}
  so that we get the usual function for the phase shift between the solar surface
  and an observer in infinite distance
  \begin{eqnarray}\label{r0inf}
        \Delta\Phi(\mu)_{\mu R_0 \to\infty} 
    &=& \frac{\pi\ell^4_{\odot}}{\lambda}R_0^2(1-\mu^2)\int\limits_{t_0=\mu R_0}^{\infty}
        \frac{dt}{\overline{R}^3} \\\nonumber
    &=& \frac{\pi\ell^4_{\odot}}{\lambda R_0^3}\left(\frac{3\pi}{16(1-\mu^2)^{3/2}}
        -\frac{\mu}{4}-\frac{3\mu}{8(1-\mu^2)}\right.\\\nonumber
    & &	\left.-\frac{3}{8(1-\mu^2)^{3/2}}\arcsin\mu\right)	 
  \end{eqnarray} 
  The phase shift which is accumulated between the solar surface and $R$ is given by
  \begin{eqnarray}\label{rinf}
        \Delta\Phi(\mu)_{\mu R_0 \to R}
    &=& \frac{\pi\ell^4_{\odot}}{\lambda}R_0^2(1-\mu^2)\int\limits_{t_0=\mu R_0}^R
        \frac{dt}{\overline{R}^3} \\\nonumber
    &=& \frac{\pi\ell^4_{\odot}}{\lambda}\left(\frac{R}{2}\left(\frac{1}
        {2(R_0^2(1-\mu^2)+R^2)^2}\right.\right.\\\nonumber
    & & \left.\left.+\frac{3}{4R_0^2(1-\mu^2)(R_0^2(1-\mu^2)+R^2)}\right)\right.\\\nonumber
    & & \left.+\frac{3}{8(1-\mu^2)^2R_0^3}\arctan\frac{R}{R_0(1-\mu^2)^{1/2}}-
        \frac{\mu}{4R_0^3}\right.\\\nonumber
    & &	\left.-\frac{3\mu}{8(1-\mu^2)R_0^3}-\frac{3}{8R_0^3(1-\mu^2)^{3/2}}\arcsin\mu\right)
  \end{eqnarray}
  One can see that, if the signal travels in the opposite direction from $R$ towards $t_0=\mu R_0$
  we have to switch the integration limits which gives a negative sign for the phase shift.
  
  At last, we get the phase shift between $R$ and the infinite distant observer by 
  substracting (\ref{rinf}) from (\ref{r0inf}).
  \begin{eqnarray}
        \Delta\Phi(\mu)_{\mu R \to\infty}
    &=& \frac{\pi\ell^4_{\odot}}{\lambda}R_0^2(1-\mu^2)\left\{\int\limits_{t_0=\mu R_0}^{\infty}
        \frac{dt}{\overline{R}^3}\quad - \quad\int\limits_{t_0=\mu R_0}^R
	\frac{dt}{\overline{R}^3}\right\}\\\nonumber  
    &=& \frac{\pi\ell^4_{\odot}}{\lambda}\left(\frac{3\pi}{16(1-\mu^2)^{3/2}R_0^3}-
        \frac{R}{2}\left(\frac{1}{2(R_0^2(1-\mu^2)+R^2)^2}+ \right.\right.\\\nonumber
    & & \left.\left.\frac{3}{4R_0^2(1-\mu^2)(R_0^2(1-\mu^2)+R^2)}\right)-
        \frac{3}{8R_0^3(1-\mu^2)^{3/2}}\right.\\\nonumber
    & &	\left.\arctan\frac{R}{R_0(1-\mu^2)^{1/2}}\right)	
  \end{eqnarray}
  Since for our purposes the best spacecraft position is given when it passes the solar limb,
  we can focus on the case $\mu=0$ which yields
  \begin{eqnarray}
    \big.\Delta \Phi\;\big|_{ \atop {\mu=0 \atop R \rightarrow \infty}} &=& 
    \frac{\pi\ell^4_{\odot}}{\lambda}\left(\frac{3\pi}{16 R_0^3}-\frac{R}{2}\left(
    \frac{1}{2(R_0^2+R^2)^2}+\frac{3}{4R_0^2(R_0^2+R^2)}\right)\right.\\\nonumber
    & &\left.-\frac{3}{8R_0^3}\arctan\frac{R}{R_0}\right)
  \end{eqnarray}
\end{appendix}

\chapter*{Thanks}
\addcontentsline{toc}{chapter}{Thanks}
  Von Dezember 1999 bis November 2002 hatte ich die Gelegenheit am Max-Planck-Institut
  f\"ur Aeronomie die hier vorliegende Dissertation zu verfassen. Mein besonderer Dank gilt
  hierf\"ur meinem Betreuer Herrn Prof. Dr. Sami K. Solanki dessen stets freundliche, unkomplizierte 
  und geduldige Art ein angenehmes und kreatives Klima geschaffen haben wie es besser kaum sein 
  kann. Seine Intuition und Erfahrung haben dieser Arbeit in zahlreichen Diskussionen wichtige
  Impulse geliefert und sie so entscheidend mitgepr\"agt. Einzigartig sind sein erfolgreiches
  Bem\"uhen um das Zustandekommen internationaler Kontakte und die Unterst\"utzung bei der
  Teilnahme an nationalen und internationalen Konferenzen - Danke!  
  
  Danken m\"ochte ich ebenso Herrn Prof. Dr. Bengt Petersson f\"ur die Betreuung dieser Dissertation 
  seitens der Universit\"at Bielefeld, f\"ur sein stetes Interesse an den Ergebnissen dieser Arbeit
  sowie seiner freundlichen Unterst\"utzung bei der Abwicklung der Pr\"ufungsformalit\"aten.

  I am grateful to Prof. Dr. Mark P. Haugan from the Purdue University, Indiana, USA for his constant
  important advice and assistance. Due to his long experience in theoretical gravitational physics and his
  pioneering feats on the subject of the Einstein equivalence principle he has provided the basis
  for this Ph.D. project which would have been impossible without him.
  
  I am also grateful to Prof. Dr. Dayal T. Wickramasinghe from the Australian National University,
  Canberra, Australia for his invaluable advice with respect to the physics of cataclysmic variables. 
  His sympathetically support and profound knowledge on the subject of magnetic white dwarfs and
  related systems has not only provided the basis for many interesting results in this thesis but
  also increased the motivation for continuing the research in this field.
  
  Stellvertretend f\"ur all diejenigen welche an der Aufnahme sowie der ersten Verarbeitung der Solaren 
  Polarimetriedaten beteiligt waren, m\"ochte ich Frau Dr. Katja Stucki und Herrn Dr. Achim Gandorfer von
  der ETH Z\"urich f\"ur ihre wertvolle Arbeit danken.
  
  Ein grosser Dank geht an meinen ehemaligen B\"urokollegen Dr. Stefan Ploner f\"ur seine freundliche 
  Unterst\"utzung bei der Einarbeitung in die Anwendung der IDL Software. Die zahlreichen interessanten
  Diskussionen im B\"uro und in den Restaurants der Umgebung werden mir in guter Erinnerung bleiben.

  Danken m\"ochte ich ebenso den Herren Prof. Dr. Manfred Sch\"ussler, Dr. habil. Dieter Schmidt sowie den
  Drs. Volkmar Holzwarth, Matthias Rempel, Peter Vollm\"oller und Alexander V\"ogler sowie der
  Arbeitsgruppe 15:00 Uhr (Kaffeerunde) f\"ur Diskussionen \"uber physikalische und nicht-physikalische
  Themen die eine angenehme Abwechslung in den Arbeitstag brachten. 

  Stellvertretend f\"ur alle Mitarbeiter des Instituts welche bei Organisatorischen Angelegenheiten
  stets behilfreich waren, moechte ich mich bei Frau Barbara Wieser f\"ur ihre wertvolle, sympathische
  Hilfe bedanken.
  
  Ein letztes, besonderes Dankesch\"on geht an meine Freundin Katrin David welche in den letzten Wochen
  und Monaten vor der Pr\"ufung nicht besonders viel von mir hatte, dies aber stets mit grossem
  Verst\"andnis mitgetragen hat.

\thispagestyle{empty}

\chapter*{Curriculum Vitae}
\addcontentsline{toc}{chapter}{Curriculum Vitae}

\setlength{\topskip}{2cm}
\noindent{\large {\bf Name:}}\\[0.1cm]
\begin{tabular}{ll}
Oliver Preu\ss &  
\end{tabular}

\vspace{0.5cm}
\noindent{\large {\bf Date of birth:}}\\[0.1cm]
\begin{tabular}{ll}
24th. June 1968  
\end{tabular}

\vspace{0.5cm}
\noindent{\large {\bf Citizenship:}}\\[0.1cm]
\begin{tabular}{ll}
  German &
\end{tabular}

\vspace{0.5cm}            
\noindent{\large {\bf Appointments:}}\\[0.1cm]
\begin{tabular}{l@{ : }l}
  12.99 - now & Ph.D. - Student of Prof. Dr. Sami K. Solanki at the \\
              & Max-Planck-Institut f\"ur Aeronomie, Katlenburg-Lindau  \\
  1997 - 1999 & University of Bielefeld. Tutor for student courses    \\
\end{tabular} 

\vspace{0.5cm}
\noindent{\large {\bf University:}}\\[0.3cm] 
\begin{tabular}{ll}  
  1997 - 1999 :&Diploma thesis at the Department of Theoretical Physics,\\
               &University of Bielefeld. Title: ''Untersuchung der Phasenstruktur\\
	       &in modifizierter 3-dimensionaler Simplizialer Quantengravitation \\
	       &mit Hilfe der Strong-Coupling-Entwicklung'' \\
  1992 - 1999 :&Study of Physics, University of Bielefeld \\ 
\end{tabular}

\vspace{0.5cm}
\noindent{\large {\bf Alternative Service:}}\\[0.3cm]
\begin{tabular}{ll}
  1991 - 1992 :&German Red Cross (DRK) Bielefeld \\
               &Hospital Rosenh\"ohe
\end{tabular}

\vspace{0.5cm}
\noindent{\large {\bf Education:}}\\[0.3cm]
\begin{tabular}{l@{ : }l}
  1988 - 1991 & Westfalenkolleg Bielefeld, general matriculation standard \\
  1985 - 1988 & Vocational training as turner, FAG Kugelfischer \\
  1979 - 1985 & Secondary school, Realschule Halle/Westf. \\
  1974 - 1979 & Primary school Langenheide, Werther/Westf.
\end{tabular}

\end{document}